\documentclass[acmtog,timestamp,nonacm]{acmart}

\AtBeginDocument{%
  \providecommand\BibTeX{{%
    \normalfont B\kern-0.5em{\scshape i\kern-0.25em b}\kern-0.8em\TeX}}}

\setcopyright{acmcopyright}
\copyrightyear{2020}
\acmYear{2020}
\acmDOI{10.1145/1122445.1122456}
\acmConference[Preprint Under Review]{Preprint Under Review}{05, 2020}{}
\acmBooktitle{Preprint Under Review}
\acmPrice{15.00}
\acmISBN{978-1-4503-XXXX-X/18/06}

\usepackage{hyperref}
\hypersetup{colorlinks=true}
\usepackage{graphicx}
\usepackage{subcaption}
\usepackage{multirow}
\usepackage{amsmath}
\usepackage[export]{adjustbox}
\usepackage{enumitem}
\hypersetup{draft}
\usepackage{xcolor}

\newcommand{\ourmodel}{UniCon}

\begin{document}

% \title{Towards Fully Interactive Physics-based Character Animation}
\title{UniCon: Universal Neural Controller For Physics-based Character Motion}

\author{Tingwu Wang}
\affiliation{\institution{NVIDIA, University of Toronto, Vector Institute}\city{Toronto}\country{Canada}}
\email{tingwuw@nvidia.com}
\author{Yunrong Guo}
\affiliation{\institution{NVIDIA}\city{Toronto}\country{Canada}}
\author{Maria Shugrina}
\affiliation{\institution{NVIDIA, University of Toronto, Vector Institute}\city{Toronto}\country{Canada}}
\author{Sanja Fidler}
\affiliation{\institution{NVIDIA, University of Toronto, Vector Institute}\city{Toronto}\country{Canada}}




%keywords

\begin{teaserfigure}
\centering
% \vspace{-0.4cm}
\includegraphics[width=\textwidth]{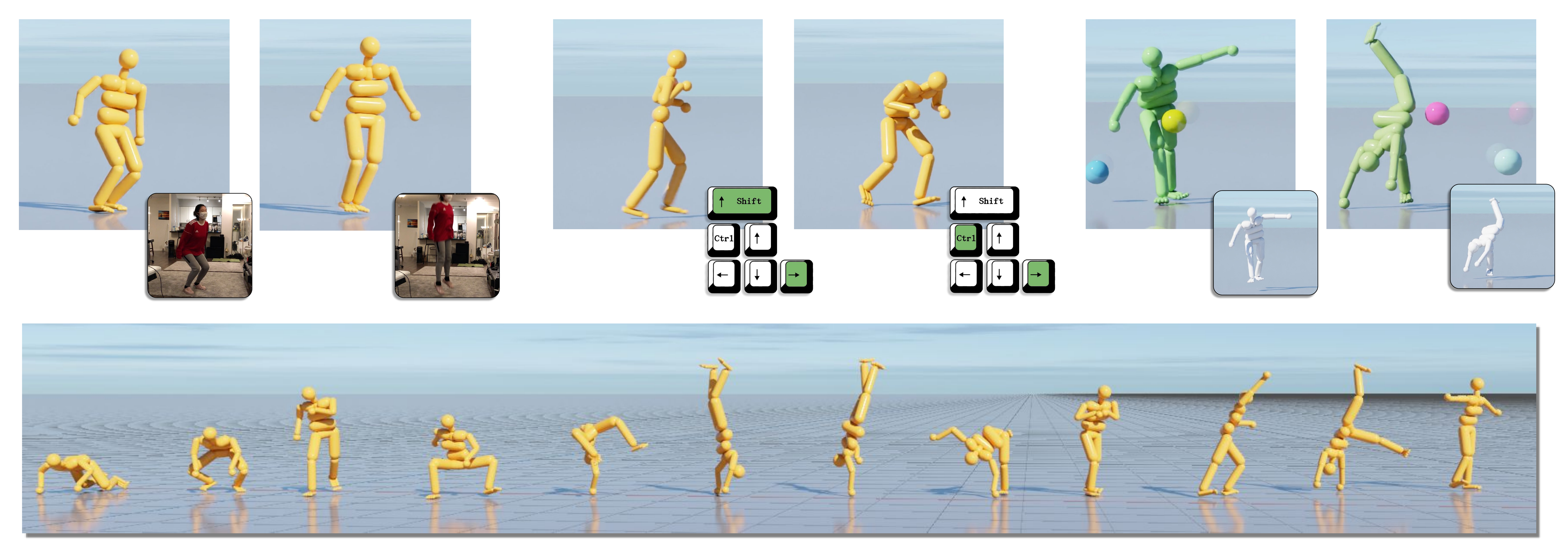}
% \vspace{-0.7cm}
    \caption{UniCon provides physically valid control for thousands of diverse motions with a single trained model and can accept a variety of interactive inputs. The top figures show interactive control examples using video stream, keyboard input and reference motions with environment perturbations. The bottom figure shows UniCon's ability to generate smooth transitions and adapt on-the-fly to interactive motion stitching.
    Project page is {\color{magenta}\href{https://nv-tlabs.github.io/unicon/}{https://nv-tlabs.github.io/unicon/}}.}\label{fig:teaser}
% \vspace{-0.05cm}
\end{teaserfigure}

\begin{abstract}

The field of physics-based animation is gaining importance due to the increasing demand for realism in video games and films, and has recently seen wide adoption of data-driven techniques, such as deep reinforcement learning (RL), which learn control from (human) demonstrations. While RL has shown impressive results at reproducing  individual motions and interactive locomotion, existing methods are limited in 
their ability to generalize to new motions and their ability to compose a complex motion sequence interactively. In this paper, we propose a physics-based universal neural controller (\ourmodel) that learns to master thousands of motions with different styles by learning on large-scale motion datasets. {\ourmodel} is a two-level framework that consists of a high-level motion scheduler and an RL-powered low-level motion executor, which is our key innovation. By systematically analyzing existing multi-motion RL frameworks, we introduce a novel objective function and training techniques which make a significant leap in performance. Once trained, our motion executor can be combined with different high-level schedulers without the need for retraining, enabling a variety of real-time interactive applications. We show that {\ourmodel} can support keyboard-driven control,
compose motion sequences drawn from a large pool of locomotion and acrobatics skills and teleport a person captured on video to a physics-based virtual avatar.
Numerical and qualitative results demonstrate a significant improvement in efficiency, robustness and generalizability of {\ourmodel} over prior state-of-the-art,
showcasing %extreme 
transferability to unseen motions, unseen humanoid models and unseen perturbation.

\end{abstract}
\maketitle

\section{Introduction}

Physics-based animation can provide more realistic motions and richer interactions with the environment compared to traditional keyframe-based animation methods.
However, film and video game industries have not yet adopted physics-based animation since the available controllers are still very limited in the number of supported motions, animation quality, interactiveness and efficiency.
% . \SF{Can you talk about what these limitations are?}
 
The field of physics-based animation has recently seen a plethora of data-driven techniques using deep reinforcement learning (RL). Data-driven techniques promise scalability to a wide variety of motions by learning directly from human demonstrations. 
Most of the RL physics-based animation algorithms fall under the umbrella of imitation learning~\cite{liu2010sampling,liu2015improving,merel2017learning,peng2018deepmimic,chentanez2018physics,bergamin2019drecon},
where the reward signals are given based on the distance or similarity between the generated and the target motions.
A policy network, which maps the current state of the character to the torques applied to the joints, is then optimized to minimize this distance.
Policies trained with imitation learning have shown success in driving a virtual character to naturally follow the reference target motions in a physically plausible way.

However, existing physics-based RL controllers are extremely inefficient to train, requiring hours or days of training to reproduce even a single motion.
% \masha{Note I removed motion planning, b/c some motion planners are quite on the fly and this may not come off as accurate.}
Furthermore, most of the existing controllers have shown little  diversity in motions,
and usually lack  robustness to perturbations in the environment.
To increase the range of  supported motions and allow for better user interaction,
new methods are proposed by, for example,
combining the controller with a powerful reference motion generator~\cite{liu2016guided,liu2018basketball,bergamin2019drecon},
or utilizing a hierarchical control system~\cite{merel2018neural,merel2018hierarchical,peng2019mcp,park2019learning}.
However, these methods usually specialize towards certain applications and support limited number of motions.
These controllers are typically trained and tested on the same motion dataset
 and have not demonstrated generalization to unseen motions. 

In this paper, we propose a physics-based \emph{universal} neural controller (\ourmodel)
which greatly improves training efficiency, robustness, motion capacity and generalizability,
allowing for a wide range of real-time interactive applications.
\ourmodel{} consists of two components:
1) a low-level motion executor that generates physics-based control signal
which drives the character to follow a target reference motion, and 2) a high-level motion scheduler
which converts various high-level inputs (for example, keyboard commands)
into a target reference motion.  
A powerful and robust motion executor is the key innovation of our work. 
We introduce several components that allow us to train our motion executor on large-scale motion datasets of diverse motion styles using reinforcement learning. 
In particular, we utilize a constrained multi-objective reward optimization,
a motion balancer and a policy variance controller,
which enable efficient and robust policy learning.
%\ourmodel{} benefits from a powerful low-level executor,
%which contains skill knowledge of thousands of motions.
Once the low-level motion executor is trained,
\ourmodel{} can utilize different motion schedulers for real-time interactive applications.  \ourmodel{} can be used to perform keyboard-driven control,
compose user-specified motion sequences,
and supports teleporting a person captured on video to a physics-based virtual avatar.

Our experiments demonstrate key improvements of \ourmodel{} over previous work:
\begin{itemize}[leftmargin=*]
    \item \textbf{Generalization:} \ourmodel{} can be used to imitate motions which are unseen during training. Natural transition skills between motions are learnt automatically without the need of recording specific training samples in the dataset. % \SF{not sure i understand this}.
    \item \textbf{Robustness:} \ourmodel{} produces robust control even when the source motion is of poor quality.
        \ourmodel{}  demonstrates \textit{zero-shot} robustness to environment obstacles that are not seen during training, such as projectiles.
        Our model can also adapt to characters with widely varying mass, or slower or faster motion than the motions seen during training.
	\item \textbf{Interactive Applications:} Generalization and robustness allow many modes of interactive control, ranging from keyboard commands, noisy pose tracking from video capture, and user-specified sequences of locomotion or acrobatic motions, without having to retrain or fine-tune separate models for each application.
	\item \textbf{Learning Efficiency:} Our learning algorithm has better sample efficiency and asymptotic performance compared to existing baselines.
\end{itemize}

\section{Related Work}\label{sec:related}
%In this section, we review existing related research in character animation.
The problem of generating plausible human motion on the fly based on user input or a target goal is a long standing problem. We first discuss methods of motion generation which do not adhere to the laws of physics in Section~\ref{ssec:rel_kinematic},
then cover physics-aware methods in both non-interactive and interactive settings in Section~\ref{section:related_noninter_physics} and~\ref{section:related_inter_physics}.
We do not discuss all methods in character animation
and refer readers to~\cite{geijtenbeek2012interactive} for a comprehensive overview.

\subsection{Keyframe Based Animation}\label{ssec:rel_kinematic}
% \masha{Introduce this section by putting it in context with your approach. E.g. something along the lines of: .} 
Keyframe-based animation systems or kinematic systems  can be used to interactively animate  virtual characters by exploiting pre-recorded motions or human authored data.
%have been employed to interactively generate human motion in the presence of recorded or human authored data
%by directly controlling the pose of the character through a forward or inverse kinematics model.
While often enabling better interactive control compared to their physics-based counterparts,
these systems result in motions that are not physically accurate and brittle to perturbations in the environment.
% The keyframes could either obtained in a non-parametric way,
% where the algorithm selects and modifies keyframes from a motion dataset,
% or generates the future keyframes in a parametric way by training on the dataset. \masha{You seem to only refer to a specific subset of keyframe-based animation methods. In a typical keyframe-based animation there's neither a dataset nor a trained model. Maybe you mean kinematic systems? At least that's how they are categorized in Deep Mimic.}
% Keyframe based methods are well studied and applied in real life.
% They usually provides much better interactiveness compared with their physics-based counterparts,
% but at same time they are non-responsive to the environment and produce less realistic motions.
% \masha{I would condense, since this is not as relevant. Briefly mention non-interactive approaches, then include interactive approaches and state clearly what mode of user interaction is available.}

Motion graph~\citep{kovar2002motion,arikan2002interactive,lee2002interactive},
as a pioneer in interactive animation,
connects motions from the dataset based on the character's states and user specifications.
However, the discretized connection in motion graph can result in non-smooth transitions,
and the generated motion is not very responsive to changes in direction or perturbations. %disturbances\SF{is there a better word than disturbances?}.
Follow up works improve motion graph by 
introducing parameterization, planning or semantic analysis~\citep{kovar2004automated,safonova2007construction,min2012motion,agrawal2016task}. 
%the authors show that the parameterization of motion allows more precise and flexible motion query.
In motion field~\citep{motionfield2010},
the authors use reinforcement learning to choose the most appropriate next motion.
Motion matching~\citep{buttner2015motion,clavet2016motion} on the other hand,
utilizes unstructured motion data to search for the most appropriate future frames given the character's current states and user control input.
Motion matching is considered by many as the state-of-the-art keyframe based animation technique due to its flexibility and ability to support a large variety of motions ~\citep{buttner2019, GeoffHarrower2018}.

In \citep{grochow2004style,levine2012continuous}, more focus is given to motion synthesis by learning a latent variable model.
With advances in data-driven modeling via deep neural networks, %data-driven training,
powerful generative motion models, such as phase-functioned neural network (PFNN)~\cite{holden2017phase} and auto-conditioned recurrent neural network~\cite{zhou2018auto} were proposed.
PFNN trains a neural network on a labeled dataset, which predicts future character states based on the trajectory control signal and the character's current states. The idea can be extended to quadrupeds with different walking phases such as dogs~\cite{zhang2018mode}. Furthermore, 
data-driven auto-regressive method is capable of animating character-scene interactions with impressive quality~\citep{starke2019neural}.
In~\cite{lee2018interactive},
the authors use a recurrent neural network to enable complex combinatorial basketball motion generation.
In~\cite{ling2020character}, an auto-regressive conditional variational auto-encoder is used to generate complex soccer motions.
In~\cite{kwon2020fast}, model-predictive control is applied to generate locomotion step plans.
Motion retargeting across different skeletons can be obtained via data-driven training with skeleton-aware operators, as shown in~\cite{aberman2020skeleton}.
%As we will later discuss in the paper,

While not physically plausible, 
keyframe-based algorithms are capable of producing a wide variety of complex motions,
and can all be theoretically used as a high-level scheduler for our algorithm, which we showcase in our experiments. 
% \textbf{Phase-Functioned Neural Networks for Character Control
% (PFNN)~\cite{holden2017phase}}:
% Non-physics based method that is data-driven, which predict next motion given current motion and trajectory. Our method can use output from this algorithm.
\subsection{Non-interactive Physics-based Methods}\label{section:related_noninter_physics}
% Physics-based animation is generally considered much harder to control than keyframe based system,
% and therefore we first start with the vast majority of physics-based research,
% which does not focus on interactive control, but are inspiring and related to our paper.
% Physics-based animation shares a lot of similarity with reinforcement learning,
% optimal control and robotics.
% Therefore physics-based animation are affected by and affecting closely these research areas.
% \masha{All of the above is too general; make specific. E.g., we first consider physics-based methods that aim to imitate a pre-recorded motion in a physically valid way without affording any interactive control.} 
We first consider physics-based methods that aim to imitate a pre-recorded motion in a physically valid way without consideration of interactive control.
One common approach is to formulate animation as an optimal control problem, or a reinforcement learning problem,
which aims to minimize the distance between the reproduced and the recorded motions.

Some of the early attempts include~\cite{liu2010sampling}, where the authors reconstruct and animate a diverse set of captured motions with randomized sampling,
which is mathematically similar to the random shooting algorithms in model-based reinforcement learning~\cite{richards2005robust,rao2009survey,wang2019benchmarking}.
Other methods include~\cite{tassa2012synthesis}, which uses model-predictive control to obtain natural walking motions by specifying a reward function for walking.
These methods all assume knowledge of the forward dynamics
and can therefore be categorized as model-based reinforcement learning algorithms.
The performance can be further improved by designing better sampling schemes
as shown in~\cite{liu2015improving,hamalainen2015online}.
However, these model-based RL methods are usually time consuming to train.
They are either not real-time during test time~\cite{tassa2012synthesis},
or require several hours of training time per motion through an iterative planning process~\cite{liu2015improving}.
% \SF{you mean training?}.
%\masha{(at run time? can you cite methods that suffer from this here?)} and
These methods are also more prone to fall into local minimas during complex motion planning and are
less resistant to external perturbations, which makes them difficult to adapt to new environments.
% making it hard to be interactive and . % \masha{What exactly do you mean by turbulence, ability to recover, adapt to environment, or?}

In contrast, model-free reinforcement learning 
allows more capacity for turbulence and test-time efficiency.
In generative adversarial imitation learning (GAIL)~\cite{ho2016generative,merel2017learning},
a neural controller is optimized with generative adversarial training~\cite{goodfellow2014generative},
where a discriminator is used to distinguish the target motion from the generated ones. 
The neural controller is trained iteratively to compete with the discriminator by generating motions that are closer to the target ones.
% a generative adversarial network~\cite{goodfellow2014generative} \masha{(modify to something like: is trained to output X based on input Y at every step by competing with a discriminator that distinguishes target motion from generated motion.)} is used to discriminate between the motion to animate and the motion generated by the agent.
% The similarity score is then used as the reward function to drive the learning of model-free neural network controller such as the one in proximal policy optimization~\cite{ppo}.
In~\cite{NIPS2017_7116}, GAIL is further extended with a variational auto-encoder, 
which encodes global motion information, such as style.
Results demonstrate that the controller can reproduce walking motions with several different styles.
% is capable of reproducing several walking motion with different style with one neural network. \masha{Clarify input and output.}
In DeepMimic~\cite{peng2018deepmimic}, 
the authors show that deep model-free reinforcement learning can be used to train animation controllers for a wide range of skills.
Their observation function includes the current states of the humanoid agents
and a phase variable to encode time information of the frame.
A policy network generates control signal to minimize the distance between the resulting next frame
and the corresponding target frame.
% 
% \SF{you dont really discuss in what way you are better/different from these approaches?}
% \TW{I think they are non-interactive methods. our method is interactive method so we can probably just brush them through?} 
%
% animate \masha{(be specific; what does it mean to "animate" motion clips? They say they learn controllers for a wide array of skills.)} a broad range of captured motion clips efficiently and robustly.
In~\cite{peng2018sfv},
the authors further show that DeepMimic can be used to reproduce motions captured by pose estimators.
Similar to the original DeepMimic method, their method requires training with RL on the motion frame data and is not real-time.
In~\cite{merel2020catch}, the idea is further extended such that the agent can use first-person perception as input observation.
Model-predictive control is used to model realistic eye and head movements in~\cite{eom2019model}.
% \masha{You need to specify exactly what deep mimic is doing for the next sentence to make sense; state what is the input to the policy, etc.}
%In the next section,
%we also show how DeepMimic can be extended for interactive control.
% It can also be used for interactive control as we show in our experiments\SF{rewrote, is this ok?}.
% which utilizes an one-hot encoding of several motions.
% it is limited in the number of motions supported. \masha{(DeepMimic belongs in interactive methods; any controller where the user can set target and generate motion allows some form of interaction.)}

The work most related to ours is~\cite{chentanez2018physics}, which feeds a set of future frames from a dataset into the network to produce control signal that tracks these future frames. % \SF{track future motion? not clear}.
% and transforms the problem into a goal conditioned reinforcement learning.
Being one of the earlier attempts to obtain a multi-motion animation controller,
this algorithm is not used for interactive control
due to its performance limitations on motion imitation and lack of robustness. % \SF{what does it mean degenerated performance?}.
% does not have good sample-efficiency.
% and the performance and robustness needed for interactive control.
% While this algorithm can be trained on and be used to reproduce multiple motions,
% the training schemes are rather straight-forward,
% the motion quality, number of motions supported are rather limited,
% and cannot be used in interactive session \masha{(why not, if multiple motions are supported?)}.
In our paper,
we study and identify several key issues that prevent successful training of an animation system on large-scale motion datasets, and propose a technique to significantly improve performance. 
%
% \SF{you mean the previous approach? not clear}.
% \TW{it seems that i didn't mention the previous approach?}
%
% We argue that this was because of the severe overfitting from training and testing on the same motion data and uncareful choice in network design and training schemes.
% We later show that overfitting is a commonly overlooked yet existing problem in recent physics-based animation systems driven by reinforcement learning.

Recently, several works on constructing complex task-driven long motion sequences using a hierarchical RL framework have been proposed~\cite{heess2016learning,merel2018neural,merel2018hierarchical,peng2019mcp}. These methods usually focus on solving one or two specific tasks,
and the number of supporting motions required is usually quite limited.
% \SF{some comments/discussion?}
% However the number of motions remains again limited in a hierachical setting,
% and they are usually not used for interactive control.
% \masha{One sentence is enough for these methods, since you just brush them off here.} 
% \kelly{I think MCP showed heading control, and carrying and dribble tasks with goal targets, which could technically be all done by keyboard, not sure if it should be considered as interactive.}

\subsection{Interactive Physics-based Methods}\label{section:related_inter_physics}
% \masha{First things first, summarize various ways interactive input is supported. It is \textbf{very important} to make this clear. Also, add some methods like DeepMimic here, and make it clear that mode of interactivity is limited.}
% \kelly{I agree, would be good to list some examples, like keyboard/gamepad control, video, live actor...}
Interactive control is typically done via a keyboard or a gamepad input.
% Interactive physics-based control requires higher level of skills.
Earlier attempts such as~\cite{wu2010terrain,peng2017deeploco} aimed to solve bipedal locomotion for humanoids
using a high-level footstep planner that takes control and terrain information as input.
In recent work~\cite{bergamin2019drecon} (DReCon),
complex interactive locomotion skills of the full humanoid are enabled,
where motion matching is used as a high-level planner that sends future kinematics states to a proportional-derivative (PD) controller.
A neural network is used to further generate corrective control
on top of PD control targets generated by the motion matching system.
This method produces impressive and realistic locomotion results 
and can be regarded as the state-of-the-art interactive physics-based method for locomotion.
Similarly, the idea can also be extended to quadruped locomotion problems as shown in~\cite{luo2020carl}.
In~\cite{liu2016guided}, the authors demonstrate interactive control of several complex acrobatic skills 
by designing a high-level control graph that schedules the reasonable motion fragments.
In~\cite{liu2017learning}, deep Q-learning was combined into the fragment scheduler to further improve the scheduling performance.
The authors also extend the proposed method to handle complex basketball controls in~\cite{liu2018basketball}.
However, as mentioned in their papers and later shown in our experiments,
% the above mentioned methods are limited to the training motion dataset\SF{what do you mean?},
the above methods cannot generalize to unseen motions due to severe overfitting.
The number of motions supported by these methods is also limited.
Our method, on the other hand,
can be applied to unseen motions and is capable of generalizing to a larger number of motions of drastically different styles.

Notable recent, concurrent work~\cite{won2020scalable} also considers learning large number of motions.
In~\cite{won2020scalable}, the authors first divide the motions into separate clusters,
then use different networks for different clusters.
{\ourmodel{}} instead focuses on transferability, zero-shot robustness,
and exploits capacity by systematically rethinking the training framework and proposing new techniques.

Besides using the keyboard control to generate future states,
interactiveness can also be established by directly encoding control information into the observation function, 
such as using a user-specified one-hot clip-selection vector~\cite{peng2018deepmimic}.
In~\cite{park2019learning},
a recurrent neural network is used to predict task-driven future frames.
A motion or intention vector can be interactively controlled by the user.
The amount of motions supported in this encoding scheme is usually quite limited.
In our experiments, we show that
the performance drops catastrophically with increasing number of motions
and the learnt controller exhibits virtually no ability to transfer to new motions.

Our proposed {\ourmodel{}} can also be used for physics-based interactive video control,
which has not been widely studied in existing research.
In real-time interactive video control,
we directly teleport a person captured by camera into a virtual physics-based avatar.

% \cite{,lee2019scalable}
% \textbf{Drecon} : It uses a motion-matching system to generate the target torque state. And a policy network that looks at torque states and correct the target torque state.
% It is different from our methods in that 1) our algorithm does not reply on a interactive motion matching system to generate target pose for us, making our algorithm more flexible and generalized. 2) We use target observations instead of target pose, which enable use to solve more complex tasks other than walking behavior.
% \masha{Include Deep Mimic here as well, since they claim they have some interactive capabilities. In fact, if you list it with interactive methods you are both being generious, and you can explain all their limitations here.} \masha{Also include any non-learning competitors, e.g. \cite{liu2016guided} do much of what we plan to demonstrate; explain how your method relates.}
% \masha{Should also mention work specific to locomotion, which tends to be interactive (user can direct where the character goes), but is obviously limited to walking, e.g. \cite{peng2017deeploco}.}

% \masha{In the related work you should also list methods that have been developed for specific skills (i.e. goal-based control, which can be considered interactive). For example, list \cite{peng2018sfv,liu2018basketball,lee2018interactive}. Somewhere should mention that your approach belongs to a family of imitation learning. }
% \cite{park2019learning,lee2019scalable}.

\section{Overview}
\begin{figure}[!ht]
    \centering
    \begin{subfigure}{.48\textwidth}
        \centering
        \includegraphics[width=\linewidth]{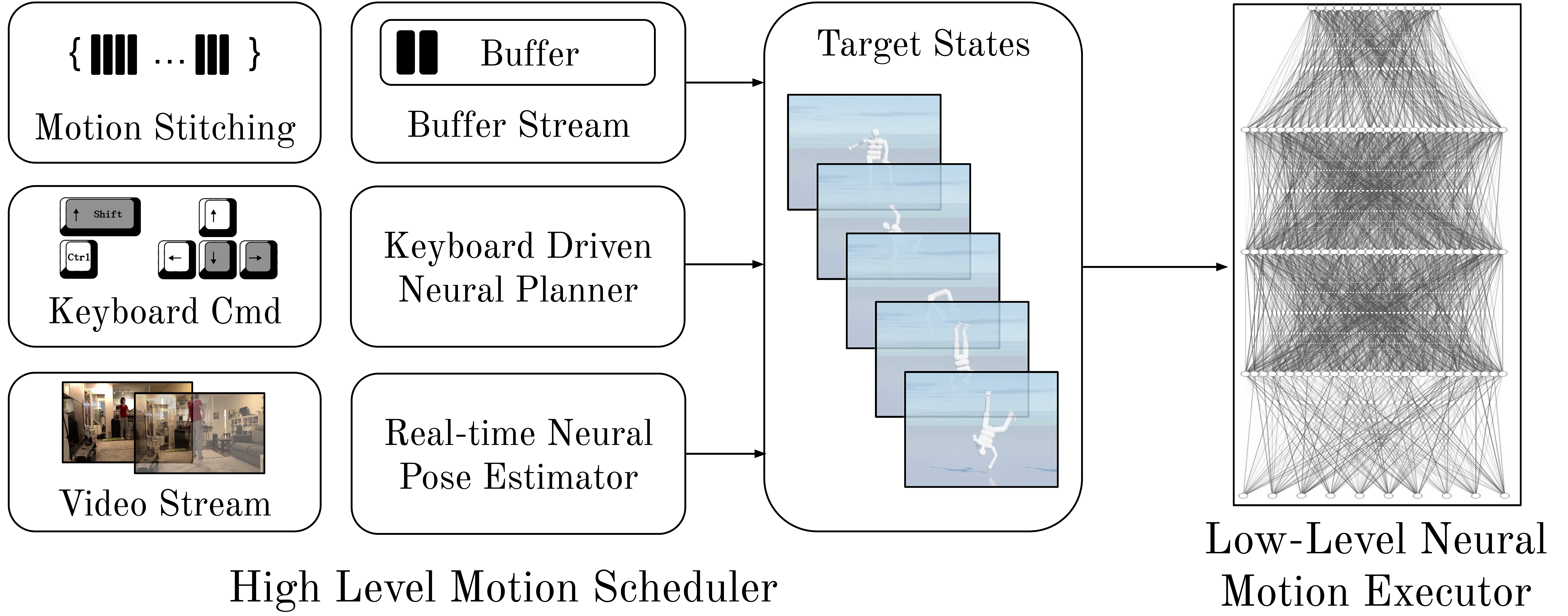}
    \end{subfigure}
    \caption{{Overview of \ourmodel}. Our model consists of ({\bf right}) an RL-powered low-level motion executor that is able to physically animate a given (non-physically plausible) sequence of target motion frames. The low-level motion executor can work in conjunction with a plethora of ({\bf left})  high-level motion schedulers which produce target motion frames. }
    \label{fig:overview_image}
\end{figure}

The \ourmodel{} framework consists of two levels: a \textbf{high level motion scheduler} which takes as input interactive high-level control such as keyboard command or video 
and generates target motion frames, and a \textbf{low-level motion executor} which produces physics-based animation based on the target motion frames.  The  low-level motion executor presents the key innovation of our work, and can be used in conjunction with a variety of different high-level motion schedulers. 
Figure~\ref{fig:overview_image} shows the overview of our model. 

We first introduce the low-level executor in section~\ref{section:lowlevel}, and  
then separately discuss important design decisions that enable its robust training in a multi-task setting in section~\ref{section:training}. In section~\ref{section:highlevel}, we describe how different high level motion schedulers can interact with our low-level executor, resulting in various interactive applications. Section~\ref{section:environment} provides implementation details, including information about the humanoid model and physics engine used in our work. In section~\ref{section:experiments}, we showcase our method both qualitatively and quantitatively, and provide benchmarks against existing work.

\section{Low Level Motion Executor}\label{section:lowlevel}
The low-level motion executor assumes input in the form of target character states. These states come, for example, from mocap data, or can be produced by a high level motion scheduler. The low-level controller  is implemented as a policy neural network that
outputs a physics-based control signal, which drives the character to closely follow the target states.

We denote the state of the character as $\mathcal{X}$. To describe the state of a character, we consider the following information:
the root position $p^{r} \in \mathcal{R}^{3}$,
the root rotation quaternion $q^{r} \in \mathcal{R}^{4}$,
the joint position $p^{j} \in \mathcal{R}^{3J}$,
and the joint rotation quaternion $q^{j} \in \mathcal{R}^{4J}$ where $J$ is the number of joints.
We note that this information is redundant in that, for example,
the joint position $p^{j}$ can be inferred from $p^{r}, p^{j}, q^{r}$.
We also consider the first order information from the  four states,
i.e.,
the root translation velocity $\dot{p}^{r}$,
the root angular velocity $\dot{q}^{r}$,
the joint translation velocity $\dot{p}^{j}$,
and the joint angular velocity $\dot{q}^{j}$.
The state thus takes the following form: 
\begin{equation}
    \mathcal{X} =
    \left[p^{r}, q^{r}, p^{j}, q^{j}, \dot{p}^{r}, \dot{q}^{r}, \dot{p}^{j}, \dot{q}^{j}\right].
\end{equation}
We denote the time steps in the physics engine as $t$.
At time step $t$, the state of the character is denoted as $\tilde{\mathcal{X}}_t$.
The use of $\mathcal{X}$ with or without tilde symbol $\,\tilde{}\,$ is to differentiate between the actual state and target state.
The target states, which are input to the low-level controller, are denoted as $\left[\mathcal{X}_{t+1}, ..., \mathcal{X}_{t+\tau}\right]$, where $\tau$ is the length of the target frames. 
In section~\ref{section:highlevel}, we describe how to generate the target states from training datasets, or from interactive high-level schedulers that take control signal $c_t$. We here focus on our low-level motion executor. 

\subsection{Observation Function}

We first introduce our observation function $s_t$,
which encodes information about both the current state $\tilde{\mathcal{X}_t}$ and the target future states $\left[\mathcal{X}_{t+1}, ..., \mathcal{X}_{t+\tau}\right]$.
In our RL formulation, the agent's controller takes input the observation function,
and generate the corresponding control signal as output.
We propose to use an agent-centric state encoding operator $T_{p^r, q^r}$,
which transforms the quaternion, translation and the corresponding velocity
with respect to the root $p^r,q^r$.
The agent-centric local state observation function $o(\mathcal{X})$ can be written as:
\begin{equation}\label{equation:local_state}
    o(\mathcal{X}) = \left[
        p^{r}(z), T_{p^r, q^r}\left(q^{r}, p^{j}, q^{j}, \dot{p}^{r},
                               \dot{q}^{r}, \dot{p}^{j}, \dot{q}^{j}\right)\right],
\end{equation}
where $p^{r}(z)$ here indicates that we only use the $z$ value out of all the $xyz$ coordinate values of $p^{r}$.
We note that in equation~\ref{equation:local_state},
we transform the target state $\mathcal{X}$ information with respect to the root of each target state $\mathcal{X}$,
rather than with respect to the root of the current actual state $\tilde{\mathcal{X}}$.
Empirically, the two choices do not make noticeable performance difference.
% We would want the agent discard information such as the absolute $x,y$ position in the plane,
% \kelly{This is slightly confusing, I think we still have x,y position info, but just relative to the root? But we remove the rotation around the up-axis IIRC}
% and process all the remaining information from a agent centric way.
% Since our agent needs to simultaneously master a giant amount of skills,
% it is very important not to use a global observation that will greatly decrease controller's generalization. 
By processing the character's actual state in an agent-centric way,
we increase the generalization of the observation function.
The second part of the observation function is a relative information function $y_{\tilde{\mathcal{X}}}(\mathcal{X})$ that extracts the relative root coordinate offset between future target states and the current actual state, i.e.,
\begin{equation}
    y_{\tilde{\mathcal{X}}}(\mathcal{X}) =
    \left[T_{\tilde{p^r}, \tilde{q^r}}\left(p^{r}, q^{r} \right)\right].
\end{equation}
By combining both the local state information $o(\mathcal{X})$ and the relative information between states,
 the observation vector $s_t$ takes the following form:
\begin{equation}
    s_t = \left[
        o(\mathcal{\tilde{X}}),
        o(\mathcal{X}_{t+1}), 
        ...,
        o(\mathcal{{X}}_{t+\tau}),\,\,\,\,
        y_{\tilde{\mathcal{X}}}(\mathcal{X}_{t+1}),
        ...,
        y_{\tilde{\mathcal{X}}}(\mathcal{X}_{t+\tau})
    \right].
\end{equation}
We avoid the use of absolute information in world space,
which helps the agent to learn generalized features for control.

\subsection{Controller}
The choice of a controller is an important one. In existing RL-based animation algorithms, 
the proportional-derivative (PD) controller has been commonly used 
instead of the alternative torque-based controller that decides how much force to apply on each joint. We here argue in favour of torque-based controllers for our setting.
%as later we show that torque-based based controller has better transferablity.
In particular, in the experimental section we show empirically that,
while PD controller usually has good sample efficiency and performance in training,
it also tends to severely overfit.
Although we do not present a mathematical explanation for this phenomena,
a general rule of thumb we consider here is to avoid providing the controller with additional observation features, pre-processing or post-processing.
In~\cite{greydanus2017visualizing},
the authors show that the Atari RL agents actually strongly focus on the scoreboard or the timer in their policies instead of paying attention to the actual game screen,
suggesting the risk of RL agents using unexpected features from observation function.
We also refer to~\cite{peng2017learning} for comparison study.

% we draw attention to two observations:
% {\bf 1)} PD controller cannot typically be applied directly to the target kinematics state without a trained corrective controller
% as shown  in existing research including~\cite{liu2016guided,bergamin2019drecon}; 
% {\bf 2)} Reinforcement learning is extremely prone to producing policies based on unexpected features.
% In~\cite{greydanus2017visualizing},
% the authors show that the Atari RL agents actually strongly focus on the scoreboard or the timer in their policies 
% instead of paying attention to the actual game screen.
% We suspect that this also happens to neural PD controllers.
% The PD controller overfits heavily on the target kinematic states,
% but the kinematic state itself is not a valid source to drive the PD controller.
% Previous work trains and tests on the same distribution.
% However, with different train and test distributions,
% the overfitting becomes evident.
% \kelly{I'm not sure if this is a good thing to mention in the paper, but my understanding is that PD controllers can generally give smoother motion, which could be a reason why our results can seem to have more noise?}
% TW maybe we can include this in the appendix, or just ignore this?

Therefore, we use a torque based controller for \ourmodel{}, which we denote as $\pi(a_t|s_t)$,
where $a_t$ is the concatenation of torques applied to each joint.
We use a fully-connected neural network (Multilayer Perceptron, MLP),
and denote the network weights as $\theta_{\pi}$.
Since our controller is tasked to master a much larger number of skills compared to~\cite{peng2018deepmimic,bergamin2019drecon},
we also use a much larger neural network. Specifically, in this paper we use  three hidden layers of 1024 units, and ablate this choice
 in section~\ref{section:ablation_lowlevel_executor}.
\subsection{Constrained Multiobjective Reward Optimization}\label{section:multireward_function}
Similar to ~\cite{peng2018deepmimic,bergamin2019drecon},
we define the reward as a sum of several terms that measure the difference of the target state and the actual state on various statistics, i.e.,
\begin{equation}
    r(s_t) = r_t(\tilde{\mathcal{X}},\mathcal{X}) =
    w_{p^r}r_{p^r} + 
    w_{q^r}r_{q^r} + 
    w_{p^j}r_{p^j} + 
    w_{q^j}r_{q^j} + 
    w_{\dot{q}^j}r_{\dot{q}^j},
\end{equation}
where the weight coefficients $(w_{p^r}, w_{q^r}, w_{p^j}, w_{q^j}, w_{\dot{q}^j})$
are respectively $(0.2, 0.2, 0.1, 0.4, 0.1)$,
which were empirical numbers first used in DeepMimic~\cite{peng2018deepmimic}.
We use the exact same reward function $r_{q^j}$ for joint quaternion deviation,
$r_{\dot{q}^j}$ for joint angular velocity deviation,
and root position deviation $r_{p^r}$ (centor-of-mass), as the ones in DeepMimic~\cite{peng2018deepmimic}.
However instead of penalizing only the mismatch of positions for end-effectors such as hands and feet,
we penalize all joints,
as accurate representations of many motions require attention to all joints rather than hands and feet only.
We also separately penalize the root rotation similar to how we penalize joint rotations as: 
$r_{q^r} = \mbox{exp}(-2||\tilde{q^r} - q^r||^2)$.

It is common practice in RL-based animation algorithms to optimize the following objective:
$J(\theta_{\pi})=\mathbb{E}_{s_t\sim\rho_{\theta_{\pi}}(s)}\left(r(s_t)\right)$.
However, we show that directly optimizing the sum of the reward is problematic.
This objective is essentially a multi-objective reward function,
and mathematically every separate reward term is competing against the others in training.
The competing effect is not apparent when training on very few motions, yet it is almost guaranteed to degenerate our learnt controller when the reward is dominated by certain reward terms~\cite{vamplew2011empirical,liu2014multiobjective}.
A common resulting suboptimal behavior is the \textit{moon-walk} behavior,
where the agents get high reward on matching joint and root rotations,
but almost completely ignore the reward for matching root position.
Increasing the reward weight for matching the root position, however,
would cripple the joint matching score and lead to ``blurred" motions.

Therefore we propose the following constrained optimization objective:
\begin{equation}\label{equation:multireward_obj}
\begin{aligned}
    \max_{\theta_{\pi}} J(\theta_{\pi})=&\mathbb{E}_{s_t\sim\rho_{\theta_{\pi}}(s)}\left(r(s_t)\right),\\
    s.\,t.\,\,\, &r_i(s_t) > \alpha_i, \,\, \forall r_i \in
    \left[r_{p^r}, r_{q^r}, r_{p^j}, r_{q^j}, r_{\dot{q}^j}\right],
\end{aligned}
\end{equation}
where $\alpha_i$ is the tolerance coefficient ensuring that
no reward term is dominated by other reward terms.

For a practical algorithm, it is impossible to directly optimize equation~\ref{equation:multireward_obj} using existing RL algorithms off-the-shelf.
However, we can maintain a soft version of the constraint $r_i(s_t) > \alpha_i$ by enforcing early termination separately for each term, i.e., we terminate the episode if any reward term drops below the tolerance threshold.
Empirically, we find that $\alpha_i = 0.1\, \forall\, i$ works well.

\section{Training}\label{section:training}
We use proximal policy optimization (PPO)~\cite{ppo} to optimize \ourmodel.
PPO has been commonly used in recent physics-based research including~\cite{peng2018deepmimic,bergamin2019drecon}.
We refer readers to the original paper for the detailed PPO algorithm~\cite{ppo}. 
We here use the following surrogate objective denoted as $L_{PPO}$ from PPO\footnote{PPO is commonly referred to as a Policy Gradient (PG) method in current research.
While PPO shares a lot of similarities with the original PG~\cite{sutton2000policy},
it is considered as a trust region method by its authors and optimizes a slightly different loss function~\cite{ppo}.}:
\begin{equation}\label{equation:ppo}
    \max_{\theta_{\pi}} \mathbb{E}\left[ \frac{\pi(a_t|s_t)}{\pi_{old}(a_t|s_t)}A_t -
    \beta\cdot\mbox{KL}\left[\pi(\cdot|s_t) || \pi_{old}(\cdot|s_t)\right] \right],
\end{equation}
where $A_t$ is the estimated advantage function, $\pi_{old}$ represents the policy weights which are fixed during the update,
and $\beta$ is the weight for the Kullback-Leibler divergence penalty that discourages over-confident updates, as proposed in~\cite{ppo}.
The iterative update in equation~\ref{equation:ppo} is sample-based,
and we describe how we sample the motion and initial state in section~\ref{section:training_motionbalancer} and~\ref{section:reset_state}.
During training, we simultaneously use 4096 workers to generate training samples.
The number of samples per iteration per worker is 64.
We also point out that other potentially more powerful RL algorithms can be applied as well,
such as for example~\cite{haarnoja2018soft,fujimoto2018addressing}.

\subsection{Motion Balancer}\label{section:training_motionbalancer}
As later described in detail in section~\ref{section:data_drive},
the training motion dataset contains classes with an imbalanced number of samples.
Random sampling during training can lead to a policy that is dominated by one specific skill,
for example walking, which makes for around 35.4\% of the dataset.
It is difficult to control the balance of the dataset if,
for example, we wish to extend \ourmodel{} to train on a dataset generated from vastly available Youtube videos in the future.
We do not want to discard unbalanced data samples 
which may contain useful information that can improve generalization to a broader set of skills.
Furthermore, the class labels for motions are usually a mixture of rough and fine-grained labels,
posing additional challenges in maintaining the balance of training samples.

We thus propose to use a \emph{motion balancer}.
In particular, we first build a hierarchical tree structure for class labels,
similar to what was done in ImageNet~\cite{deng2009imagenet}.
For every motion, starting from class root node,
we label their high-level class (major style such as walking),
and then move downwards in the tree structure to the low-level classes (minor style such as forward or backward for walking).
We do not limit the depth of this class hierarchy,
such that for complicated motion  we are able to generate more fine-grained class labels.
For example, we can label a zombie-walking motion as \texttt{root-walking-forward-specialStyle-zombie}.

During training, we sample the motion by going down the hierarchical tree and uniformly sampling from all sub-node (child node) classes of the current node.
We denote each node as $v$ and its sub-nodes or child-nodes as $\mathcal{C}(v)$.
The sampling process can be described as
\begin{equation}
    P(v_i | v) = \frac{1}{|\mathcal{C}(v)|},\, \forall v_i\in\mathcal{C}(v).
\end{equation}
We note that the sampling probability for each motion can be calculated off-line once for the entire dataset.
We also point out that motion balancer can be extended with other adaptive sampling schemes in the future.
Later in section~\ref{section:data_drive}, we explain the collection of hierarchical dataset by using human-labor and existing pose estimators.
In the experimental section~\ref{section:experiments} we show that with the motion balancer 
our controller is able to obtain significantly better performance for a variety of tasks.

\subsection{Reactive State Initialization Scheme}\label{section:reset_state}
In DeepMimic, reference state initialization (RSI) was proposed to sample the state of a particular frame as the initial state.
In our work, we further propose \textit{reactive state initialization scheme} (RSIS),
where the agent is initialized with a state of the frame $k$ time-steps away from the actual target frame it is supposed to track.
RSIS also includes a much larger noise added to the velocity and translation of the initialized stateas mentioned in later experiment sections.
% In~\cite{chentanez2018physics},
% a similar strategy was used to train a separate recovering agent.
With RSIS,
the agent learns how to self-adjust and catch up with the target states.
In \ourmodel{}, we show that recovering skills can be learnt automatically 
without the need to train a separate recovery network as in~\cite{chentanez2018physics},
or adding recovering motions in the dataset.
When used with a motion stitching scheduler 
our agent naturally transitions between stitched motions that may have large discontinuities.
The robustness of our agent is also greatly increased,
demonstrating extreme recovering skills from perturbations as shown in the accompanied videos.
Empirically, we choose a frame offset with 5-10 time-steps,
which does not lead to early termination of episodes while encouraging the character to learn recovering skills.

\subsection{Policy Variance Controller}
Another important consideration in training is avoiding bad local minima.
Specifically, we consider the variance of the stochastic policy network $\pi(a_t|s_t)$.
In PPO, a trainable vector $\hat{\mathbf{\sigma}} \in \mathbb{R}^{|a_t|}$ is used to represent the diagonal Gaussian policy standard deviation.
Typically for single-motion or few-motion training, $\hat{\mathbf{\sigma}}$ can be automatically learnt by optimizing equation~\ref{equation:ppo}.
However, with a large-scale dataset 
the variance is much more fragile during training.
There is a design dilemma for the initial value.
% During training with a giant motion dataset,
Initializing with a large variance can make the training hard to converge and thus generates divergent behaviors.
Initializing with a small variance can lead to premature convergence and thus hurt the performance.
For \ourmodel, we are inspired by~\cite{ppo} and use a similar exponential annealing on the policy's variance.
We also notice that for a character different joints require variance of different scales.
For example, the control variance of the toes should be intuitively and empirically smaller than the ones of the legs.
We want to preserve the learnt differences between joints 
and therefore propose an adaptive variance update scheme as follows:
\begin{equation}
    \begin{aligned}
        \hat{\mathbf{\sigma}}' &\gets \hat{\mathbf{\sigma}} -
            \alpha_{lr}\nabla_{\hat{\mathbf{\sigma}}} {L_{PPO}}, \\
        \hat{\mathbf{\sigma}}' &\gets \left\{\begin{array}{rcl}
            \mathcal{Z}(\hat{\mathbf{\sigma}}'), & & {l < \mathcal{L}}\\
            \hat{\mathbf{\sigma}}', & & {\mbox{Else}}\\
        \end{array}\right.
    \end{aligned}
\end{equation}
where $L_{PPO}$ is the loss defined in equation~\ref{equation:ppo},
$\alpha_{lr}$ is the learning rate, and $l$ is the current PPO training iteration.
We define a control iteration range from $0$ to $\mathcal{L}$,
during which we linearly anneal the target average of $\log(\hat{\mathbf{\sigma}})$
from the hyperparameters $\mbox{logstd}_0$ and $\mbox{logstd}_F$.
$\mathcal{Z}$ is the operation where we linearly increase or decrease
the log of each component of $\hat{\sigma}$ by the same amount,
so that the average value matches the linear annealed target value.
By doing this we preserve the learnt variance structure,
which we show in section~\ref{section:ablation_lowlevel_executor}.

\section{High Level Motion Scheduler}\label{section:highlevel}
The high-level motion scheduler in our framework outputs target reference states from interactive control signal or from a motion dataset. We here discuss several different high level motion schedulers, and note that other existing schedulers can work in conjunction with our low-level executor. 
We denote the high-level motion scheduler as $\phi$. 

The high-level motion scheduler outputs states of $\tau$ future frames as
\begin{equation}\label{equation:motion_scheduler}
    \left[\mathcal{X}_{t+1}, ..., \mathcal{X}_{t+\tau}\right] =
    \phi_{\theta}\left(\{c_i\}_{i=t-\tau_c}^{t}, \{\tilde{\mathcal{X}}_i\}_{i=t-\tau_x}^{t}, t\right),
\end{equation}
where $c_i$, $\tilde{\mathcal{X}_i}$ represent the control signals and actual robot state,
with $\tau_c$ and $\tau_x$ representing respectively the history length to consider.
$\theta$ in equation~\ref{equation:motion_scheduler} represents the parameters or configurations of the scheduler.
The planning length $\tau$ needs to be carefully chosen: 
it cannot be too long since this can make the executor hard to train and is more prone to overfitting.
% and more dangerously overfits the executor to the specific scheduler, as shown in the experiments for algorithms such as~\cite{chentanez2018physics}.
For real-time applications such as animation from video,
it is also challenging to generate frames for the unforeseen future.
We therefore choose a relatively short output length $\tau$, with a value of $1$ or $2$.
% for example, weights of the neural network or configurations of motion graph systems.

% In \ourmodel{}, multiple high-level motion scheduler instances are unified as in equation~\ref{equation:motion_scheduler}.
By unifying input observation to our low-level motion executor in \ourmodel{},
we establish a universal and transferable framework for different sources of control input.
In experiments, we show that the low-level executor trained with a large-scale motion dataset
can be used directly with a scheduler for specific applications without the need of retraining.
% We show that by carefully design training for the executor,
% extremely recovering skills can be learnt automatically when the target $\mathcal{X}_{t+1}$ is very different from the actual starting state $\tilde{\mathcal{X}}_t$.

\subsection{Motion Dataset Training Scheduler}\label{section:data_drive}
In this subsection, we discuss how one can utilize a motion dataset for the low-level executor.
The motion dataset training scheduler is different from other schedulers which enable interactive applications,
but serves as the backbone of \ourmodel{},
as it trains a powerful low-level executor to support other schedulers.
% The first scheduler we introduce is the motion dataset scheduler,
% where we simply generate future target states from the dataset.
During training we randomly sample a motion $m$ and frame ID $j$ and set this frame's state as the initial state for our agent, i.e. $\tilde{\mathcal{X}}_{t=0} = m(j)$.
For time step $i$, our training scheduler outputs the state from motion $m$ as
\begin{equation}
    \phi_{\mbox{MocapData}}\left(t\right) =
    \left[m(t+j+1), m(t+j+2), ..., m(t+j+\tau)\right].
\end{equation}
The scheduler terminates and reschedules a new motion only when the motion has reached the end,
or the target and actual states have deviated significantly,
as discussed in section~\ref{section:multireward_function}.

\begin{table}[!t]
\resizebox{0.47\textwidth}{!}{\begin{tabular}{l|c|c|c|c}
\toprule
    Dataset & Split & Num of Motions & Num of Frames & Avg Length\\
\midrule
% 3352
    \multirow{2}{*}{All} & Train & 2758 & 752648 & 272.8\\
    & Test & 594 & 187130 & 315 \\
\midrule
    \multirow{2}{*}{RunJumpWalk} & Train & 1376 & 290210 & 210.9 \\
    & Test & 238 & 71262 & 299.4 \\
\midrule
    \multirow{2}{*}{Casual}  & Train & 590 & 176698 & 299.5 \\
    & Test & 108 & 42934 & 397.5 \\
\midrule
    \multirow{2}{*}{SportDance} & Train & 576 & 148182 & 257.2\\
    & Test & 66 & 35314 & 535.1 \\
\midrule
    \multirow{2}{*}{Animal} & Train & 244 & 109396 & 448.3 \\
    & Test & 62 & 25954 & 418.6 \\
\midrule
    \multirow{2}{*}{Walk} & Train & 180 & 23322 & 129.5 \\
    & Test & 48 & 4812 & 100.2 \\
\bottomrule
\end{tabular}}
\caption{In this table we show the statistics of each sub dataset we created.}\label{table:mocap_data}
%\vspace{-0.3cm}
\end{table}
\subsubsection{Data Preparation}
We utilize the widely available CMU Graphics Lab Motion Capture Database (CMU Mocap dataset)~\cite{hodgins2015cmu}.
While motion dataset is commonly used in a physics-based system,
the choice of train and test set splits has been generally overlooked.
In our work, we carefully study the overfitting and transferability of the algorithms.
We clean up the CMU Mocap dataset,
and divide the data samples into training and testing sets based on several categories as presented in Table~\ref{table:mocap_data}.
We create several smaller datasets by grouping via style and tasks.
For example, \texttt{Casual} dataset contains casual motions such as sweeping the floor,
cleaning dishes, and the majority of the motions are standing motions.
\texttt{Animal} dataset contains motions where the actors imitate animals such as cats and dogs.
The \texttt{All} dataset contains all remaining motions after filtering out infeasible motions or motions that are highly dependent on external objects such as stairs.

We split the dataset into a training set that has 80\% of the frames and a test set that has the remaining 20\%.
One issue with the CMU Mocap dataset is that the motions are extremely unbalanced.
% We label the motions by their major style, and minor style,
% such that we can for example classify a motion as \texttt{walking-forward},
% \texttt{walking-veer}, \texttt{jumping-up}.
We find that for all of the motions available, 35.4\% of them are walking motions, and 25.5\% of all motions are walking-forward motions.
On the other hand,  there are lots of classes with very few motion clips.
Unbalanced data is a very common~\cite{lemaitre2017imbalanced},
which may bias the neural network and detriment performance.
% It could be extremely dangerous to split the training and testing set randomly. \kelly{Maybe describe why it could be an issue with an imbalanced data set and random splitting?}
To build the training and test sets,
we split motion classes individually, starting with least frequent categories.
% Therefore, we first allocate the classes with fewer number of motions,
% so that they are allocated equally in the testing and training sets.
If there is a class with only one motion, we always place it in the test set.
We then allocate large classes such as walking-forward to fill the remaining training and testing sets.
By doing this, we make sure that the test set has a good variability of motions.
While the training set is unbalanced across classes,
our motion balancer discussed in section~\ref{section:training_motionbalancer}
helps in training our motion executor effectively. %  with an unbalanced training set.
For {\ourmodel{}} we manually label the motions.
While we used human labeled dataset in our experiments,
we point out crawled unlabeled videos can be used to generate future datasets.
Fine-grained hierarchical labels can be automatically generated with video action classifiers,
which also produce hierarchical labels as shown in~\cite{shao2020finegym,feichtenhofer2020x3d}.

\begin{figure}[!t]
    \centering
%     \begin{subfigure}{0.48\textwidth}
%         \centering
%         \includegraphics[width=\linewidth]{video/massive/train_dataset/cropped_image-014.jpg}
%     \end{subfigure}
% 
%     \begin{subfigure}{0.48\textwidth}
%         \centering
%         \includegraphics[width=\linewidth]{video/massive/train_dataset/cropped_image-015.jpg}
%     \end{subfigure}
% 
%     \begin{subfigure}{0.48\textwidth}
%         \centering
%         \includegraphics[width=\linewidth]{video/massive/train_dataset/cropped_image-016.jpg}
%     \end{subfigure}
%     \begin{subfigure}{0.48\textwidth}
%         \centering
%         \includegraphics[width=\linewidth]{video/massive/train_dataset/cropped_image-017.jpg}
%     \end{subfigure}
%     \begin{subfigure}{0.48\textwidth}
%         \centering
%         \includegraphics[width=\linewidth]{video/massive/train_dataset/cropped_image-018.jpg}
%     \end{subfigure}
%     \begin{subfigure}{0.48\textwidth}
%         \centering
%         \includegraphics[width=\linewidth]{video/massive/train_dataset/cropped_image-019.jpg}
%     \end{subfigure}
    \begin{subfigure}{0.48\textwidth}
        \centering
        \includegraphics[frame,width=\linewidth]{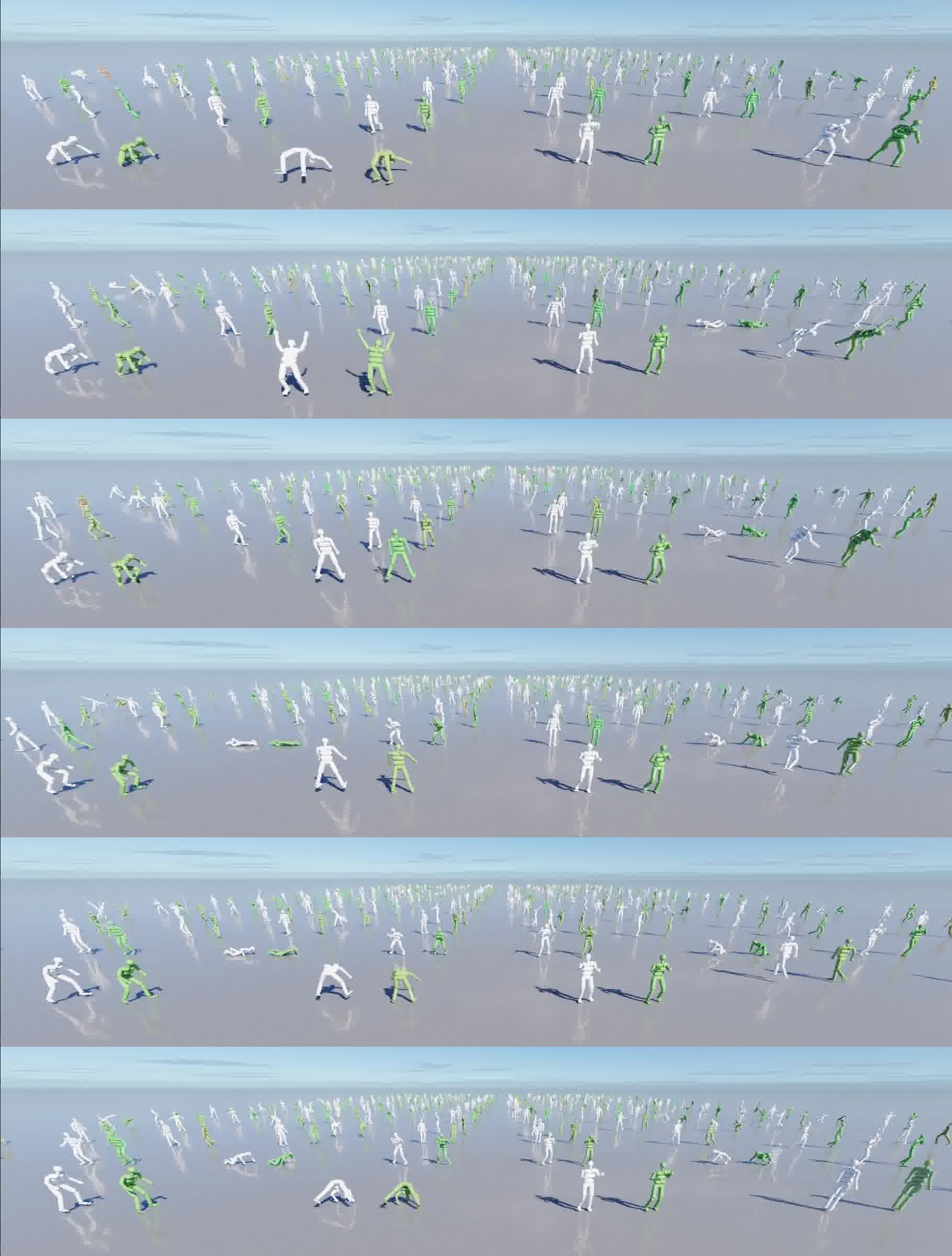}
    \end{subfigure}
    \caption{{The figures here visualize the training process, where the white characters are driven by ground-truth motions, and the green ones are trained characters. We use the motion dataset training scheduler to train our low-level motion executor.
    4096 characters are used to generate training samples.}}
    \label{fig:motion_datasetdata_driven}
\end{figure}

\subsection{Video Stream Scheduler}\label{section:video_driven}
\begin{figure}[!h]
    \centering
    \begin{subfigure}{0.45\textwidth}
        \centering
        \includegraphics[width=\linewidth]{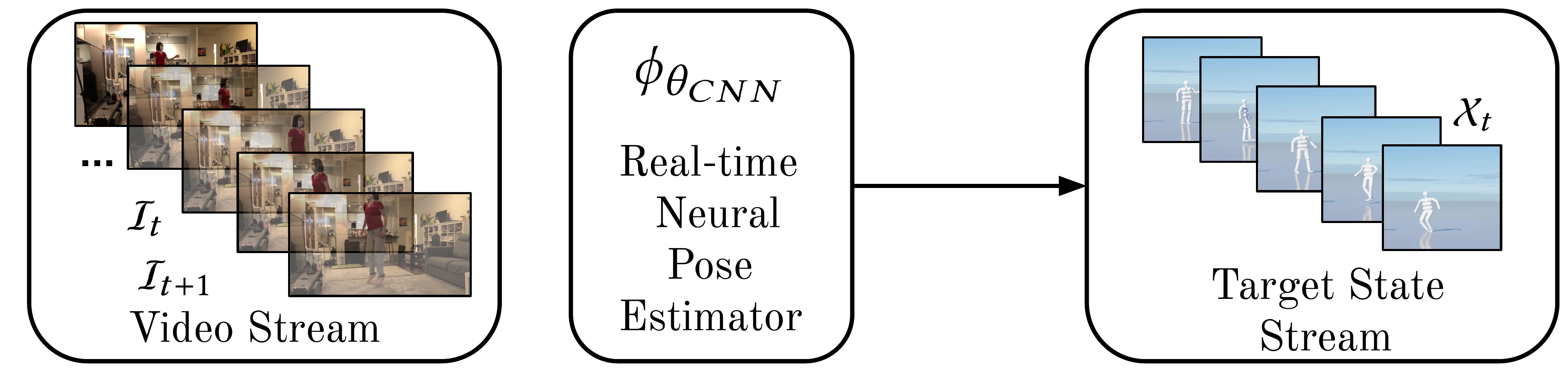}
    \end{subfigure}

    \caption{{Diagram of interactive control from video.}}
    \label{fig:diagram_video_estimation}
\end{figure}

Video can be used as an interactive control input for \ourmodel{}.
Specifically, we consider a human who is captured by a camera and wants to teleport her/his motion onto a virtual avatar.
We use a real-time pose estimator to estimate the 3D pose from video
as shown in figure~\ref{fig:diagram_video_estimation}.
We use~\cite{iqbal2020weakly} in our work but any other 3D pose estimation approach can be utilized, such as~\cite{kanazawa2018end,guler2018densepose}. 
Without the loss of generality,
we assume that the pose estimator is parameterized by a convolutional neural network with weights $\theta_{CNN}$,
and we represent the pose estimator as $\mathcal{F}_{\theta_{CNN}}$.
We denote the video frame at time step $t+1$ as $\mathcal{I}_{t+1}$.
The estimator generates the following prediction: 
\begin{equation}
    \mathcal{P}_{t+1} = \left[p^{r}_{t+1}, q^{r}_{t+1}, p^{j}_{t+1}, q^{j}_{t+1}\right] =
    \mathcal{F}_{\theta_{CNN}}(\mathcal{I}_{t+1})
\end{equation}
% \kelly{Actually, the pose estimator only generates the joint positions, so we can technically use any pose estimator that can provide 3D keypoint positions. We do an IK step via a swing-twist decomposition method to compute the joint rotations. The linear and angular velocities are then just computed with the delta changes divided by time between frames (but I think dt is fixed at 1/60 atm).}
% \TW it seems that we need to condense the scheduler section, will put this into the appendix
A user datagram protocol (UDP) system connects the animation engine and the pose-estimator,
sending $\mathcal{P}_{t+1}$ in a real-time fashion.
Assuming we maintain a pose buffer of length $\tau_p$,
the first order information such as the linear and angular velocities are then interpolated from previous poses,
i.e., $\left[\tilde{p}^{r}_{t+1}, \tilde{q}^{r}_{t+1}, \tilde{p}^{j}_{t+1}, \tilde{q}^{j}_{t+1}\right] = \mbox{Interp}(\{\mathcal{P}_{i=t+1-\tau_p}^{t+1}\})$.
% \tingwu{maybe Kelly can say more on the UDP system and velocity interpolation?} 
% \kelly{I'm not sure if UDP is a really important implementation detail? Technically, we should be able to integrate the pose estimator directly into our framework as well, UDP is just a simple way to pass the data since it was easier to set it up as a separate repo, and also allows us to run the pose estimator and our framework in parallel. We can also describe it as maybe locally passing data in real-time from the pose estimator into our framework or something like that?}
% TODO: put the UDP stuff into appendix
We can write the video stream scheduler as
\begin{equation}
    \mathcal{X}_{t+1} =
    \left[\phi_{\theta_{CNN}}\left(\mathcal{I}_{t+1}\right),
    \mbox{Interp}\left(\{\mathcal{P}_{i=t-\tau_i}^{t}\}\right)\right].
\end{equation}
%We note that the frame rate and estimation accuracy is not stable.
% which was caused by the unstable inference time in $\mathcal{F}_{\theta_{CNN}}$.

Note that the accuracy of pose estimation approaches is not perfect. 
Our experiments show that
our low-level executor can handle very noisy estimated poses with reasonable accuracy.
% video driven states efficiently,
% even though the noise and error in pose detection is rather big.

\subsection{Keyboard Driven Interactive Control Scheduler}\label{section:keyboard_driven}
\begin{figure}[!h]
    \centering
    \begin{subfigure}{0.45\textwidth}
        \centering
        \includegraphics[width=\linewidth]{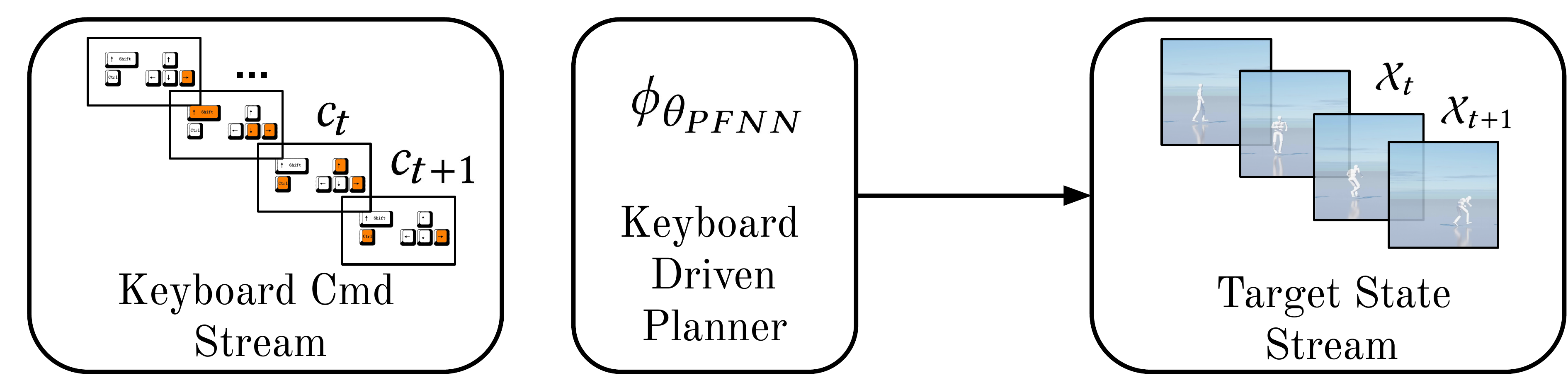}
    \end{subfigure}

    \caption{{Diagram of keyboard driven interactive control scheduler.}}
    \label{fig:diagram_pfnn}
\end{figure}

% To demostrate our interactive control from keyboard command,
% we utilize algorithms such as 
For Unicon, instead of training a hierachical interactive scheduler from scratch,
we use Phase-functioned neural networks (PFNN)~\cite{holden2017phase} to process the keyboard commands and generate future states, as shown in figure~\ref{fig:diagram_pfnn}.
We note that the future states generated by PFNN is not physics-based.
% In PFNN, a set of future frames are generated from the current keyboard command,
% as well as the command history.
In PFNN, one can control the walking direction of the agent,
and choose the walking style from walking, jogging, crouching, etc.
We refer readers to~\cite{holden2017phase} for details regarding PFNN. 
We write the target state generation of PFNN as
\begin{equation}
    \mathcal{X}_{t+1} =
    \phi_{\theta_{PFNN}}\left(\mathcal{X}_{t}, \{c_i\}_{t=t-\tau_c}^{t}\right).
\end{equation}
Note that PFNN is an auto-regressive method 
where the previously generated state $\mathcal{X}_t$ is also an input for the generation of $\mathcal{X}_{t+1}$.
Our experiments show that
we do not require feeding the actual state $\tilde{\mathcal{X}_{t}}$ back into PFNN.
Our low-level executor can automatically correct the accumulating mistakes in PFNN.

We also point out that besides PFNN,
our algorithm can also be extended to use other keyframe based animation systems such as~\cite{starke2019neural}.

\subsection{Motion Stitching Scheduler}\label{section:stitching_driven}
\begin{figure}[!h]
    \centering
    \begin{subfigure}{0.45\textwidth}
        \centering
        \includegraphics[width=\linewidth]{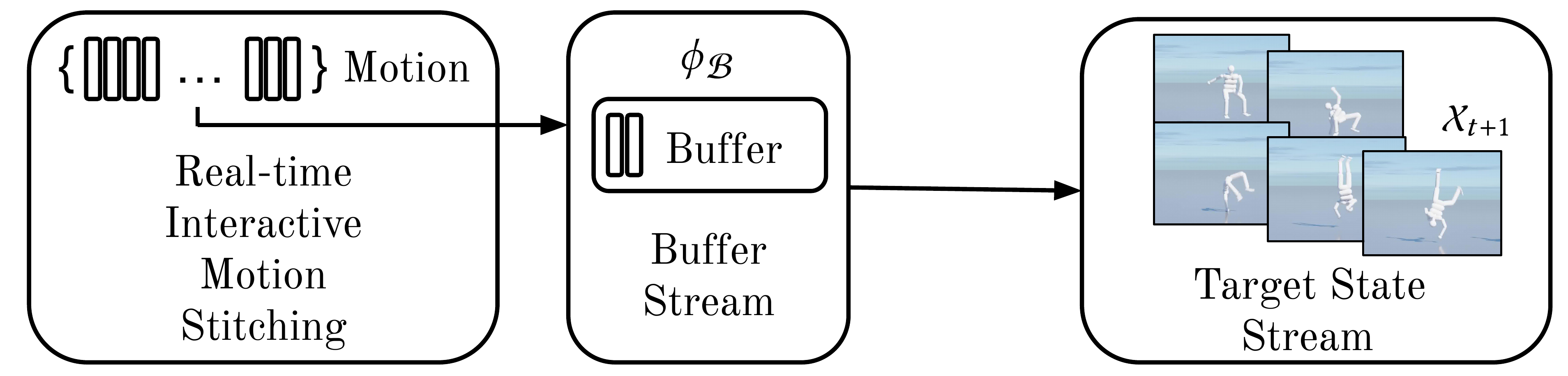}
    \end{subfigure}
    \caption{{Diagram of motion stitching scheduler.}}
    \label{fig:motion_stitching}
\end{figure}

% In this section, we propose motion stitching scheduling.
Users can also interactively specify the motion for the character 
using a scheduler we call motion stitching scheduler.
Motion stitching refers to directly stitching motions from database consecutively without worrying about proper transitions, as shown in figure~\ref{fig:motion_stitching}.
We maintain a stitched target motion buffer $\mathcal{B}$.
Before the current buffer finishes,
one can interactively add another motion $m$ with $|m|$ frames into the buffer, i. e.
\begin{equation}
    \mathcal{B}.\mbox{push}(\{m(k)\}), \, \mbox{where}\, k=0, ..., |m|-1.
\end{equation}
We use spherical linear interpolation to add several transition target frames.
At every time-step, motion stitching scheduler generates the next target state by popping a state like a FIFO buffer:
\begin{equation}
    \mathcal{X}_{t+1} = \phi_{\mathcal{B}} = \mathcal{B}.\mbox{pop}()
\end{equation}
We note that the stitching scheduler can be regarded as the simplest version of motion graph,
where motion is animated one-by-one.
However, we show that our controller can animate a wide variety of highly difficult acrobatic motions on-the-fly in a physically plausible way, some of which are unseen in training.
Our character automatically makes smooth transitions between motions 
even though we do not have transition skills recorded in the training set.
Empirical results suggest that by plugging in a more powerful interactive motion graph system,
our controller can generate an even more diverse set of physically plausible motion sequences.

\begin{figure*}[!th]
    \centering
    \begin{subfigure}{.16\textwidth}
        \centering
        \quad All
        \includegraphics[width=\linewidth]{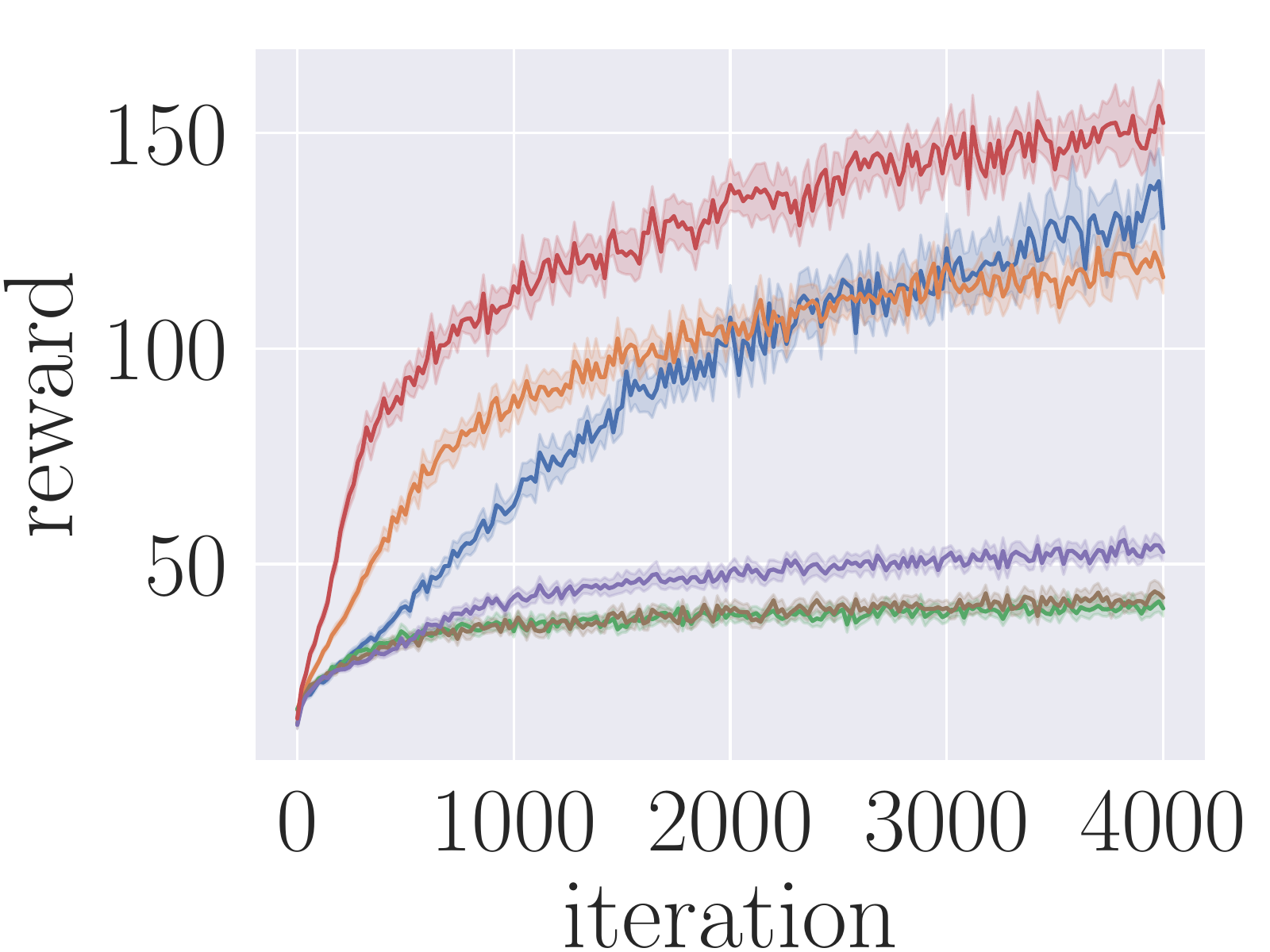}
    \end{subfigure}
    \begin{subfigure}{.16\textwidth}
        \centering
        \quad\, RunJumpWalk
        \includegraphics[width=\linewidth]{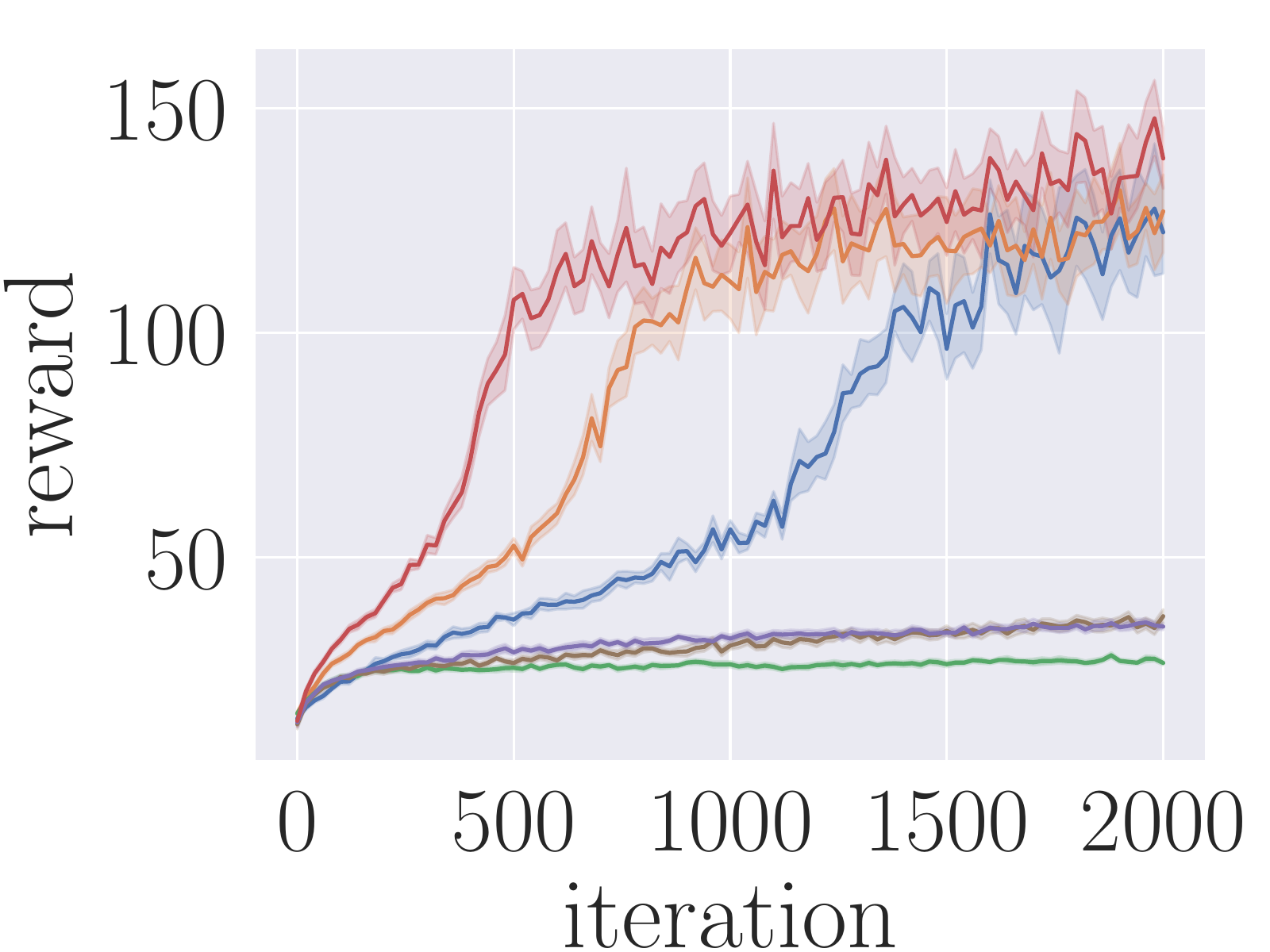}
    \end{subfigure}
    \begin{subfigure}{.16\textwidth}
        \centering
        \quad Casual
        \includegraphics[width=\linewidth]{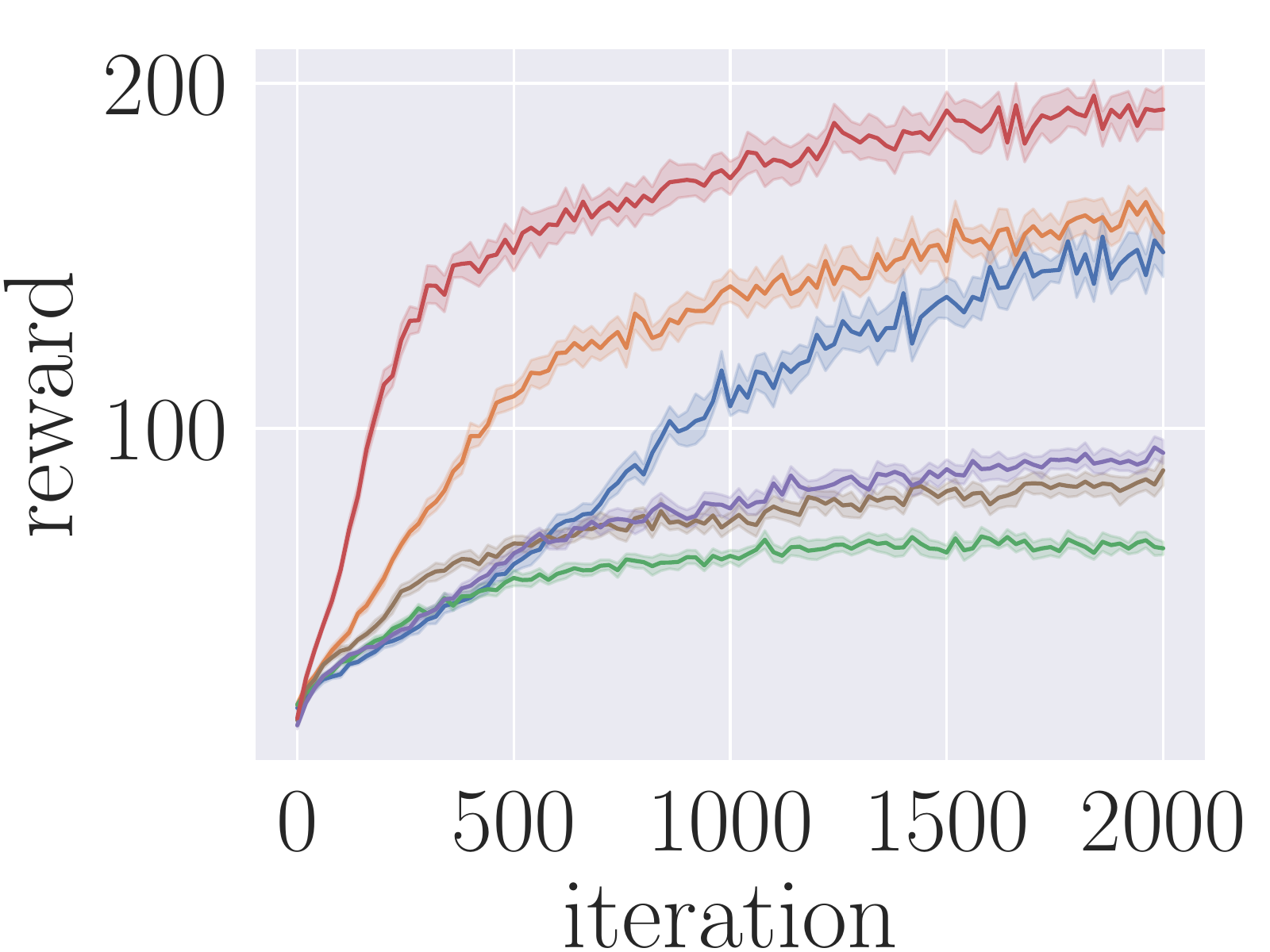}
    \end{subfigure}
    \begin{subfigure}{.16\textwidth}
        \centering
        \quad Animal
        \includegraphics[width=\linewidth]{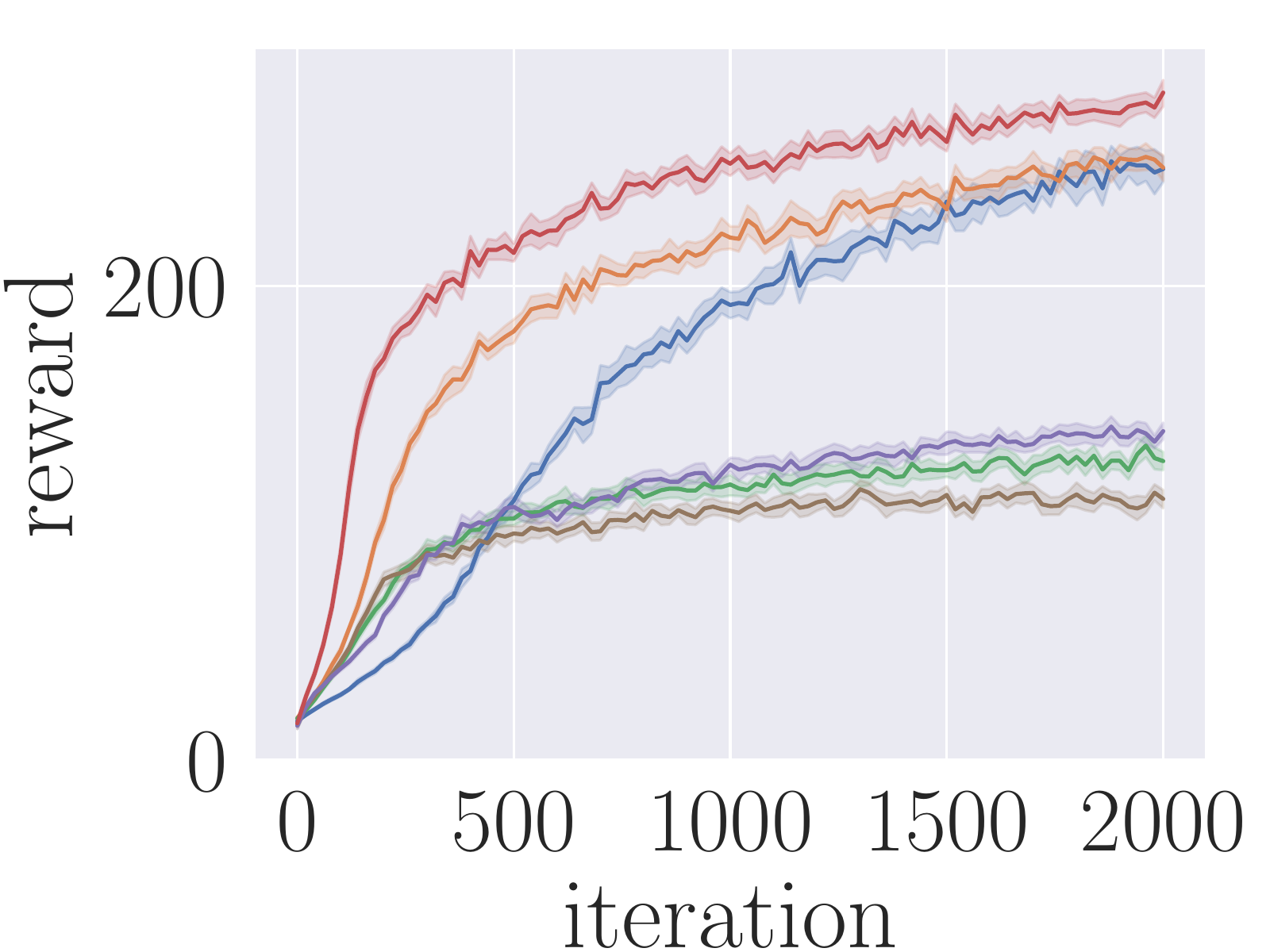}
    \end{subfigure}
    \begin{subfigure}{.16\textwidth}
        \centering
        \quad SportDance
        \includegraphics[width=\linewidth]{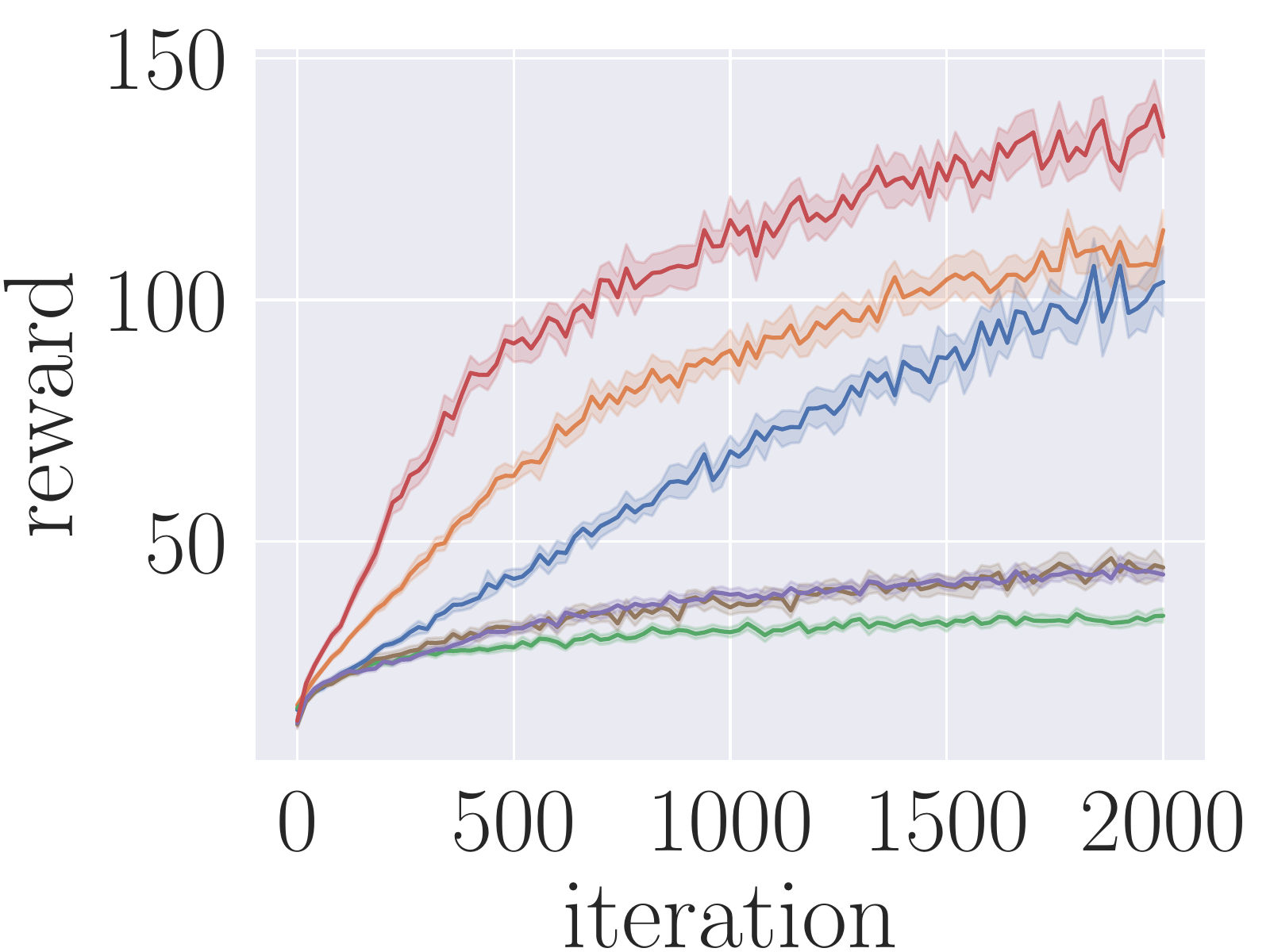}
    \end{subfigure}
    \begin{subfigure}{.16\textwidth}
        \centering
        \quad Walk
        \includegraphics[width=\linewidth]{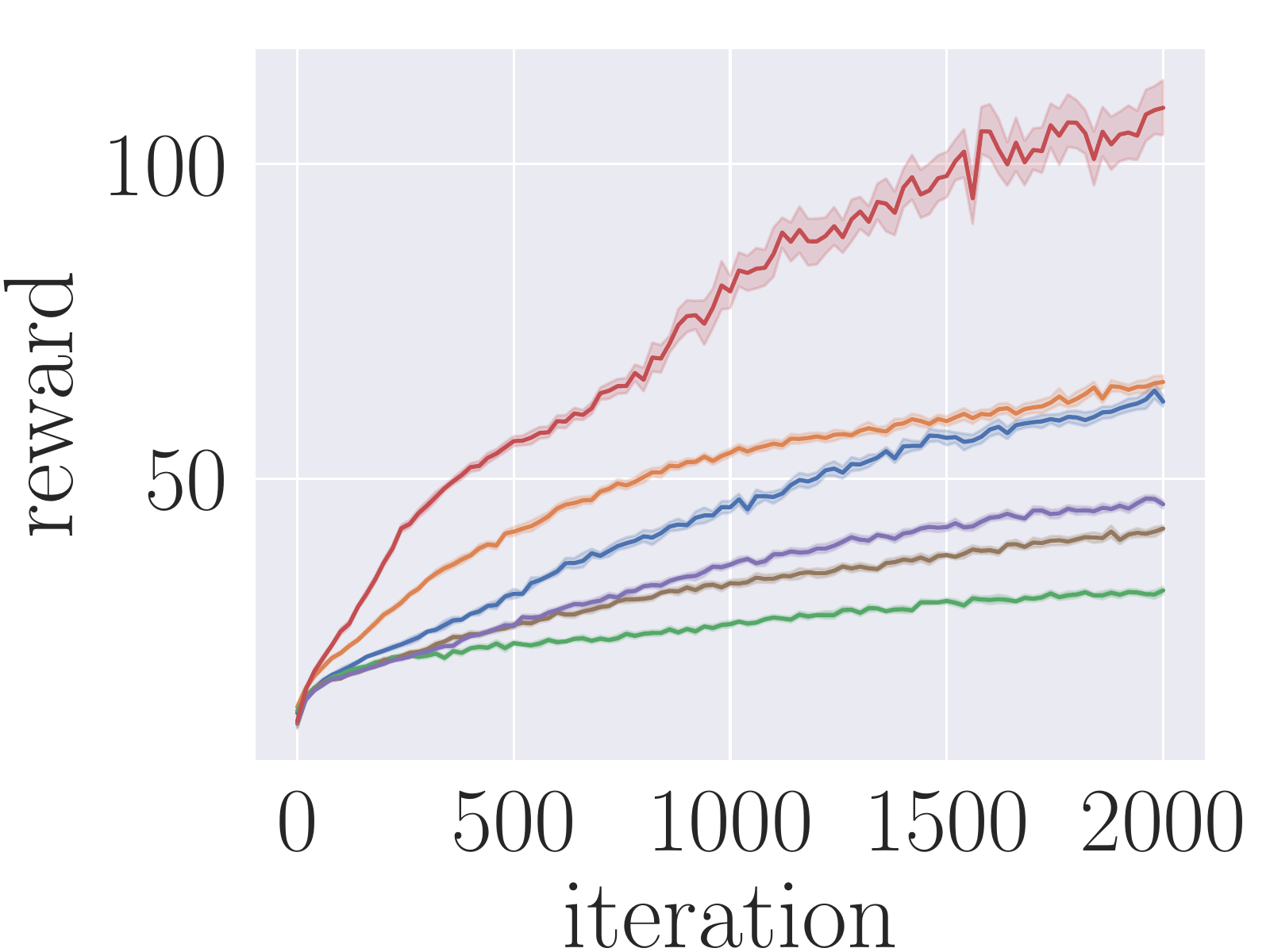}
    \end{subfigure}
    \begin{subfigure}{.95\textwidth}
        \centering
        \includegraphics[width=\linewidth]{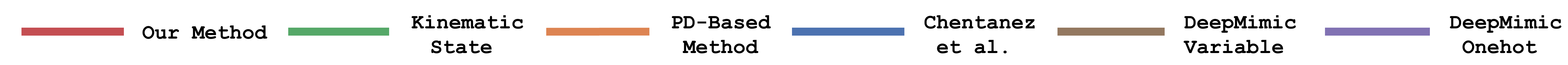}
    \end{subfigure}
    \caption{{The numerical training performance for our algorithm and baselines.
              We show the average sum of reward per episode.
              Our algorithm obtains better sample efficiency and performance.}}
    \label{fig:multitask_performance}
\end{figure*}
\section{Environment}\label{section:environment}
\subsection{Physics Engine}
Our experiments are performed with a reinforcement learning simulator similar to OpenAI Gym~\cite{gym}. The simulator is powered by the GPU-accelerated Flex physics engine as the core backend for physics simulation. We refer the reader to~\cite{liang2018} for more implementation details of the simulator and ~\cite{macklin2019} for the Flex physics engine solvers. 

In our experiments, we use the CUDA-based Newton Preconditioned Conjugate Residual Method (PCR) solver for rigid-bodies provided by the Flex physics engine, with a simulation timestep of $1/60s$ and a simulation substep value of $4$. We set the number of iterations taken by the solver per simulation step as $4$, and the number of inner loop iterations taken by the solver per simulation step as $15$. For scene simulation parameters, we set gravity to $9.8m/s^{2}$ downwards and use a value of $1.0$ for coefficient of static and dynamic friction.

\subsection{Humanoid Model}
We design our humanoid model with the basic topology of a rigid-body representation modeled after the human body. Our humanoid includes 20 rigid-bodies and 35 degrees-of-freedom. Each degree-of-freedom is assigned an effort factor within the range of 50 to 600, to simulate the difference in strength of joints in the human body. This effort factor is taken into consideration when torque control is applied. The height and mass of the humanoid model resembles a realistic proportion of the human body, at $1.8m$ and $70kg$ respectively. The mass of each rigid-body in the humanoid model is proportionally distributed based on a rough estimate of the human body mass distribution. We use a fully symmetric humanoid model with respect to the left and right rigid-bodies and joints.
We do not tune the parameters of the humanoid,
and the learned policy generalizes across different models, which we refer to section~\ref{section:results_robustness}.

\begin{figure}[!t]
    \centering
    \begin{subfigure}{.23\textwidth}
        \centering
        \quad Walk
        \includegraphics[width=\linewidth]{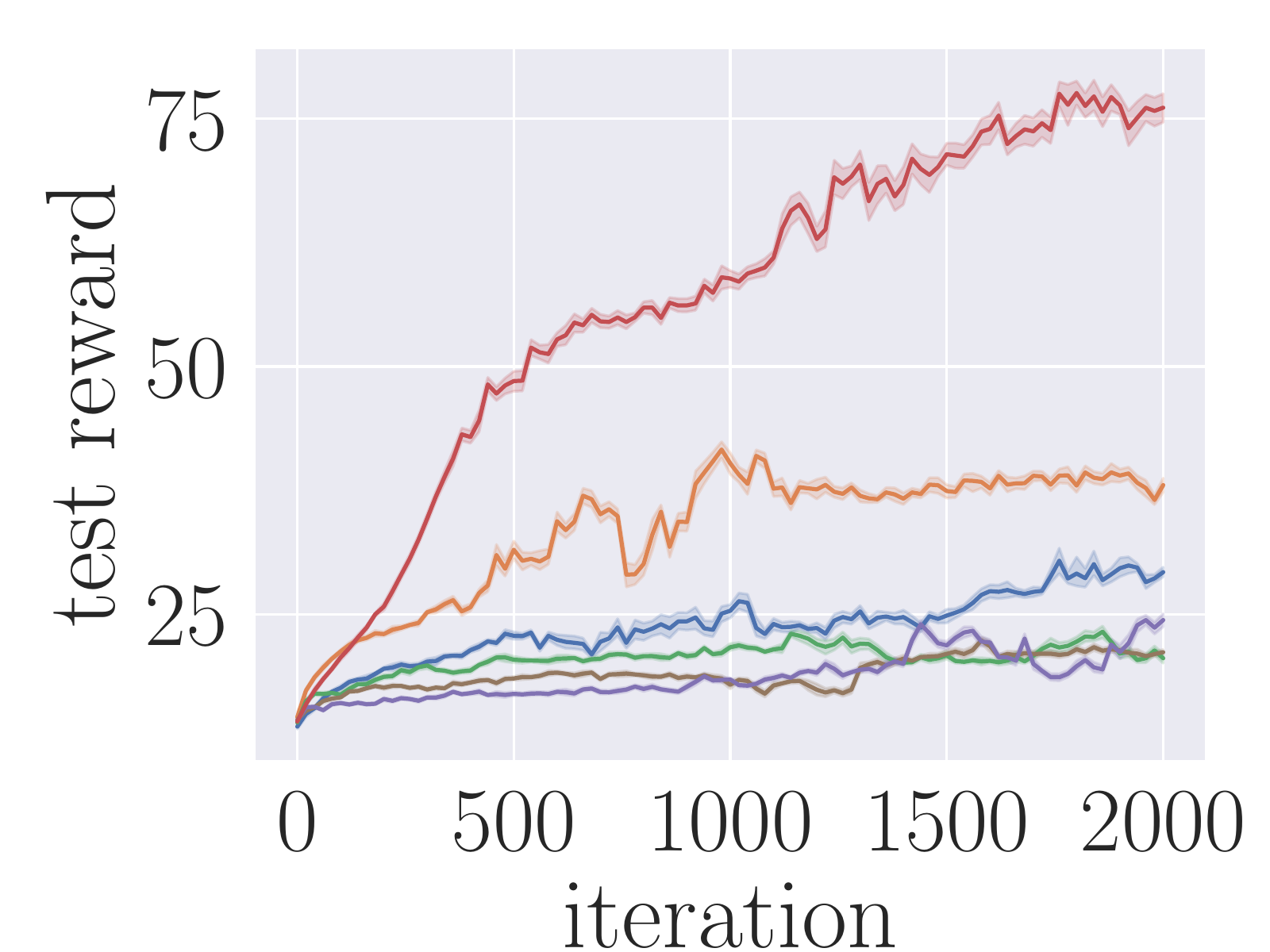}
    \end{subfigure}
    \begin{subfigure}{.23\textwidth}
        \centering
        \quad RunJumpWalk
        \includegraphics[width=\linewidth]{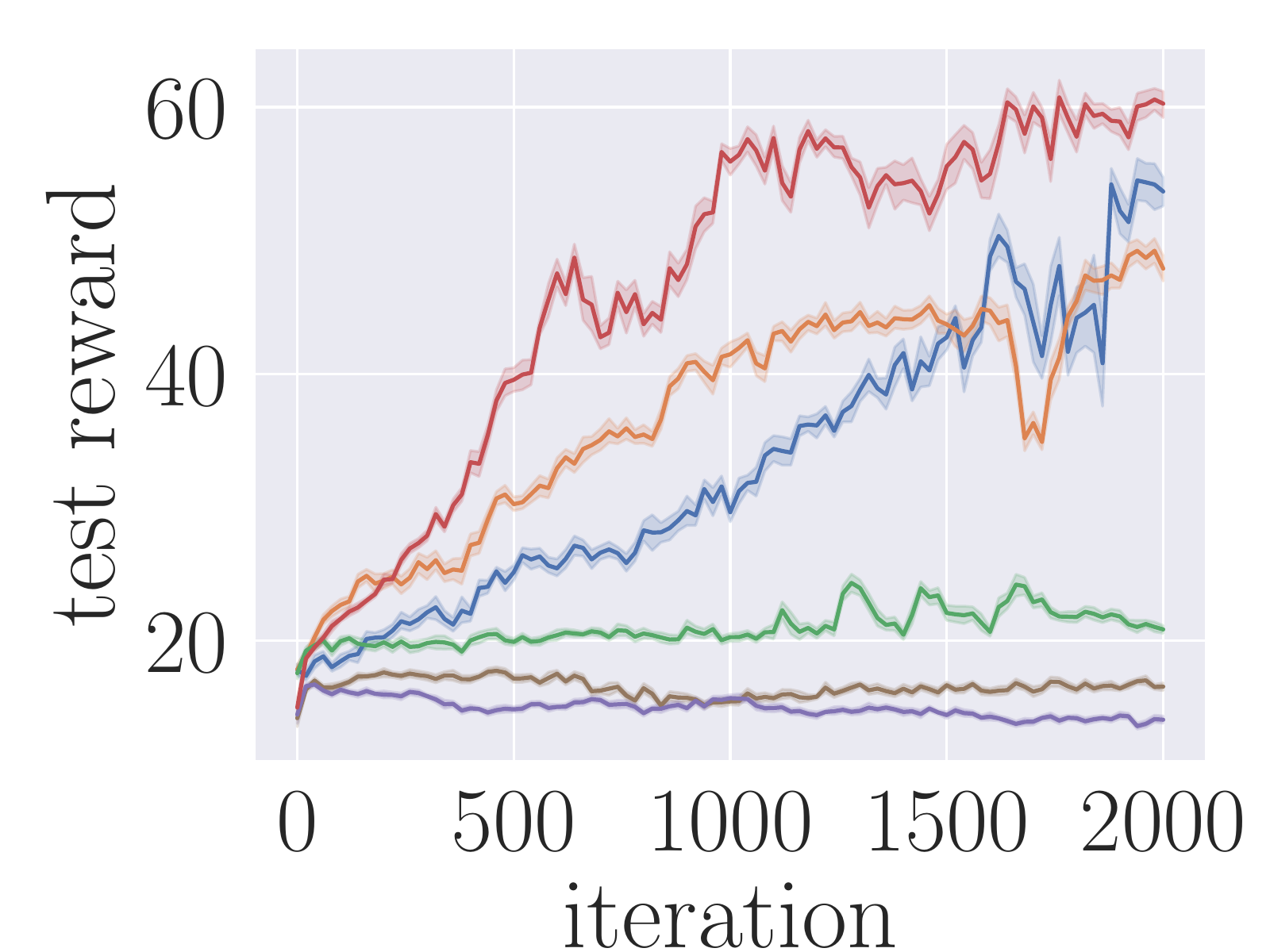}
    \end{subfigure}
    \begin{subfigure}{.40\textwidth}
        \centering
        \includegraphics[width=\linewidth]{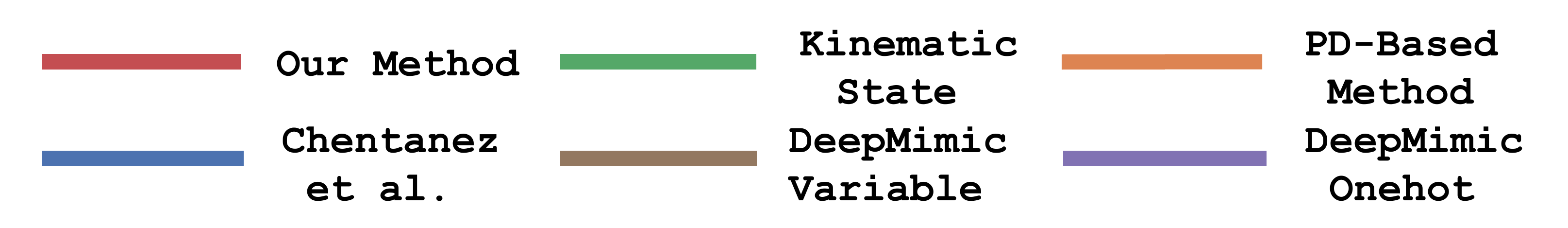}
    \end{subfigure}
    \caption{{Performance on the test set during training.}}
    \label{fig:test_performance}
\end{figure}

\section{Experiments}\label{section:experiments}

In this section we study the performance of the low-level motion executor (section~\ref{section:lowlevel}), the backbone of our algorithm, on our motion dataset. More specifically, we numerically compare our low-level executor with existing algorithms
by modifying them to be trainable on a large-scale motion dataset.
Results show that our low-level executor performs better in almost every metric (section~\ref{section:results_learning_performance}),
and we analyze the key factors behind its success in section~\ref{section:ablation_lowlevel_executor}. In addition, we demonstrate that, unlike prior work, our low-level executor exhibits zero-shot robustness to environment perturbations not seen during training (section~\ref{section:results_robustness}). These qualities of the low-level executor enable interactive applications outlined in section~\ref{section:results_interactive}.
The effectiveness of motion balancer and variance controller is shown with numerical ablation study.
We also show visual ablation study on constrained multi-objective reward optimization and RSIS in the attached demo video.
% \masha{Put this somewhere more appropriate, e.g Environment.} During the training,
% we use 4096 workers to simultaneously generate sample rollouts for the PPO algorithm.

\subsection{Low-level Controller Baselines}\label{section:results_baselines}

We first introduce the baselines we use for comparisons here.
We emphasize that constrained multi-objective reward optimization
and the policy variance controller are crucial to training,
without which, neither of our nor the baseline algorithms can be trained successfully.
Therefore we apply these techniques equally to all baselines,
and focus only on the controller design.
We use the same network structure (1024x3) for every algorithm.
We also tried the original structures specified in the papers,
but the performance is worse than or similar to the ones with 1024x3.

\textbf{PD-based Method}:
PD-based methods utilize a PD controller instead of a torque controller.
During training,
a neural policy network takes as input the current state and the target states from the dataset,
and outputs the corrective offsets to the PD-control targets.
This method is most similar to~\cite{bergamin2019drecon},
except for the fact that~\cite{bergamin2019drecon} uses an online motion matching system to generate target states, whereas we directly use the target states from the dataset.
We also use similar hyper-parameters from~\cite{bergamin2019drecon}.
We show another simple variant of removing the actual state from the observation function and performing open-loop control, which we name as \textbf{Kinematic-State} baseline.
%We use Kinematic-State baseline to test the performance of open-loop in complex high-dimensional animation problems.

\begin{figure*}[!t]
    \centering
    \begin{subfigure}{.26\textwidth}
        \centering
        \quad\quad (a) Training Performance
        \includegraphics[width=\linewidth]{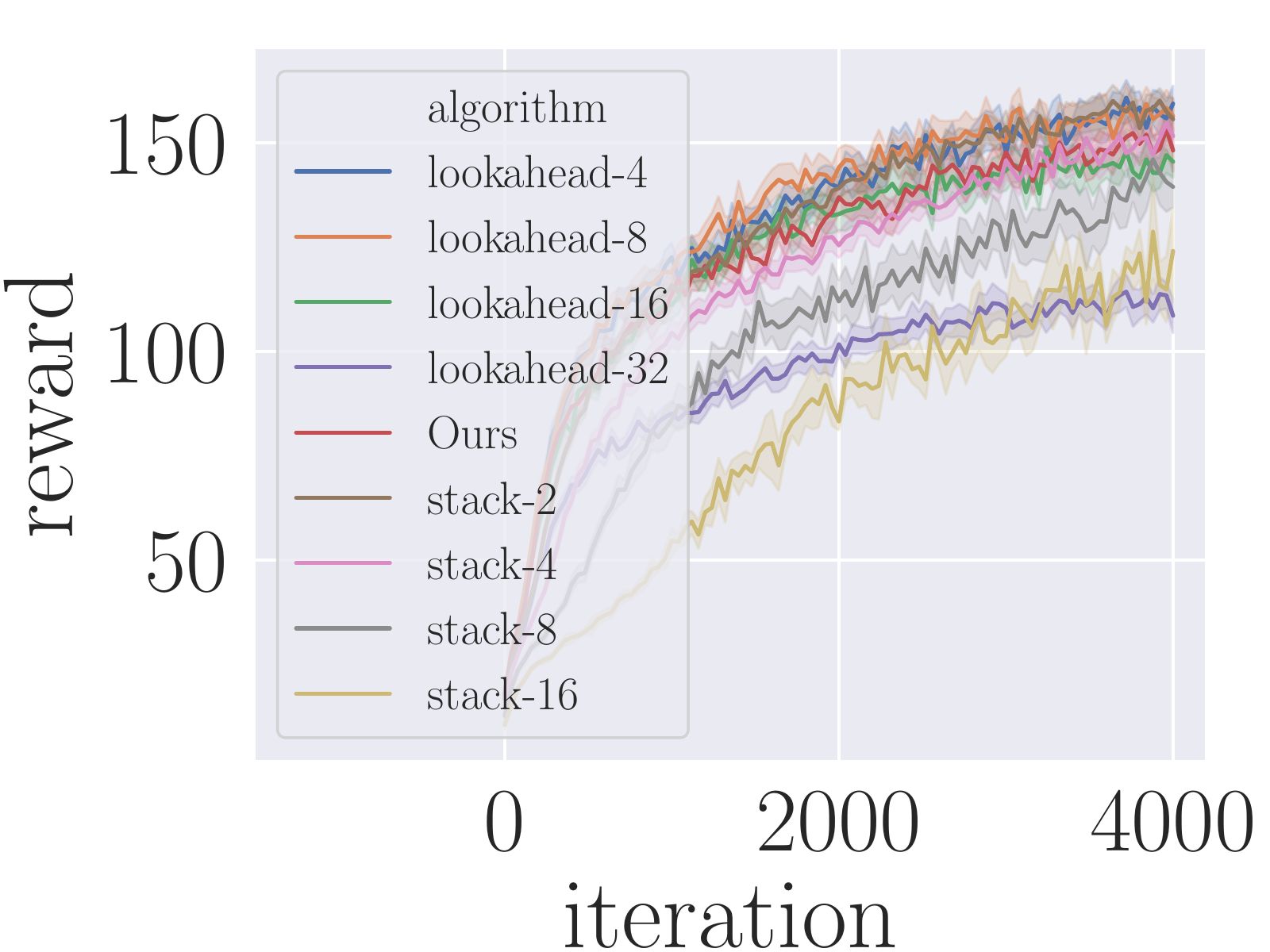}
    \end{subfigure}
    \hfil
    \begin{subfigure}{.26\textwidth}
        \centering
        \quad\quad\, (b) Testing Performance
        \includegraphics[width=\linewidth]{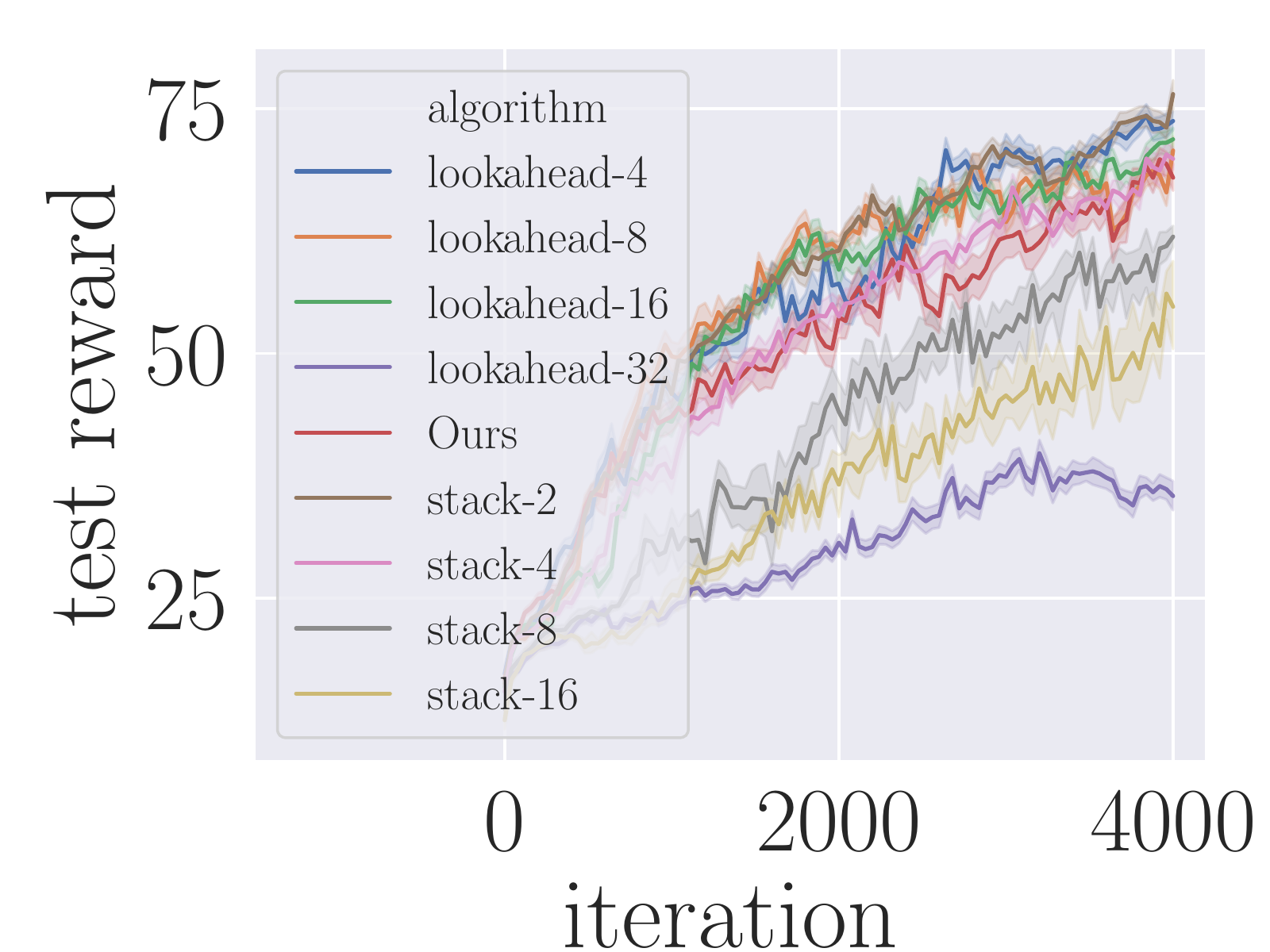}
    \end{subfigure}
    \hfil
    \begin{subfigure}{.26\textwidth}
        \centering
        \quad\quad (c) Network Structures
        \includegraphics[width=\linewidth]{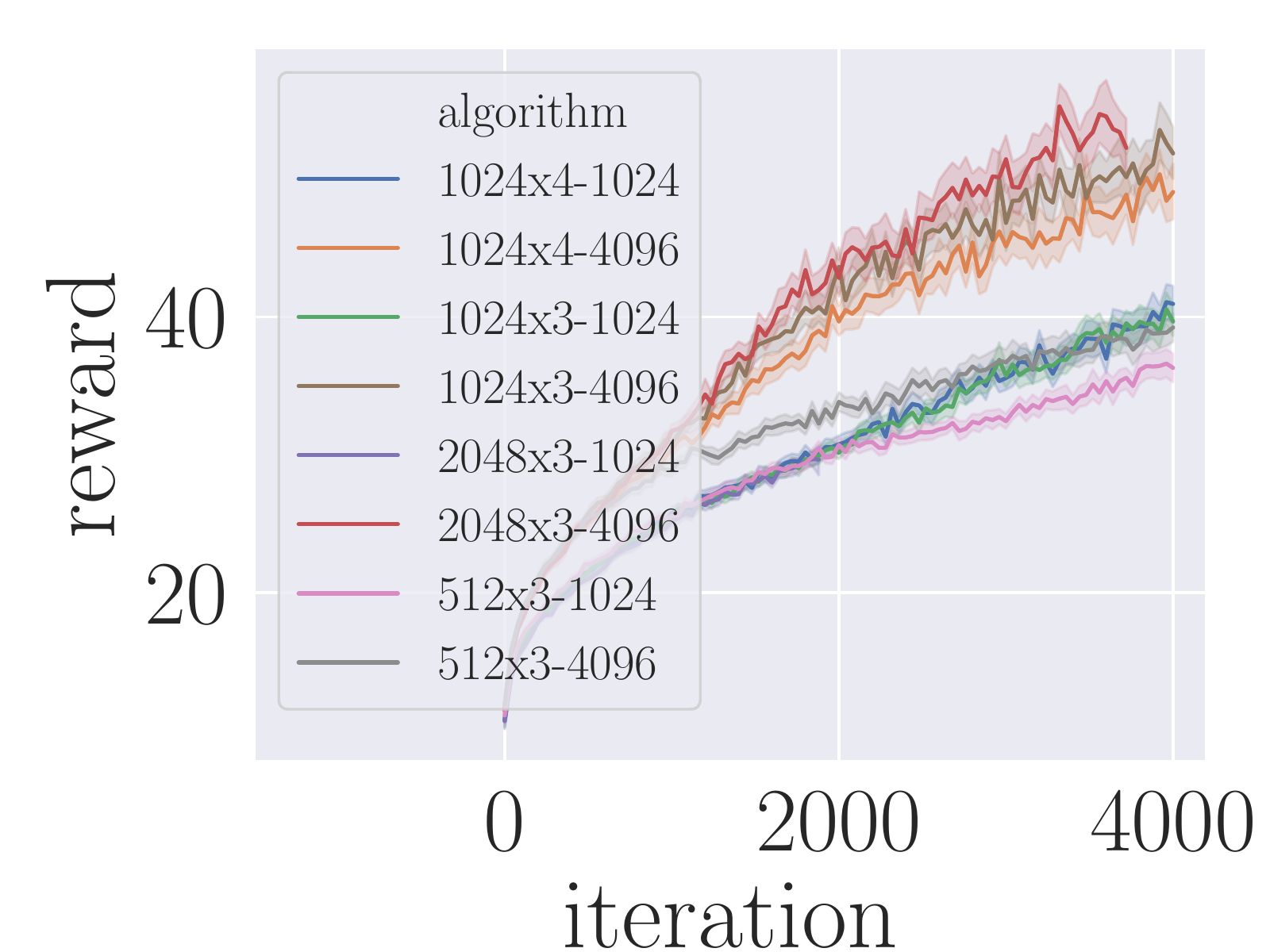}
    \end{subfigure}
    \caption{{Ablation study on model variants and network structures. In figure (c), we represent experiments in the format of "HiddenLayerSize"x"NumberOfLayers"-"NumberOfAgents". For example "1024x4-4096" represents an experiment on a network with 4 hidden layers of size 1024, trained with 4096 agents.}}
    \label{fig:variant_ablation}
\end{figure*}
\textbf{\cite{chentanez2018physics}}:
\cite{chentanez2018physics} is similar to PD-based methods,
but the observation function of the policy network contains additional long-term information.
A concatenation of future frames with $0, 4, 16, 64$ time-step offsets are fed as observations into the tracking network, which outputs PD targets to control the humanoid.
We do not include a separate recovery agent as in the original paper,
since it can be applied to any algorithm being compared here.
We later show that a separate recovery network is not necessary with the existence of a powerful low-level controller.

\textbf{DeepMimic-Onehot and DeepMimic-Variable}:
The original DeepMimic algorithm supports multi-motion training through the use of a one-hot vector to encode motion information into the observation function, which we name as \textbf{DeepMimic-Onehot}.
Since the number of motions during training can be quite large,
we also include another variant called \textbf{DeepMimic-Variable},
where we feed the ID (normalized between 0 and 1 for all motions in the train and test sets)
of the motion to the observation function.
\subsection{Training Performance}\label{section:results_learning_performance}
In figure~\ref{fig:multitask_performance},
we show the performance of our executor and the baseline methods on 6 datasets introduced in section~\ref{section:data_drive}.
Our executor consistently obtains better sample efficiency and performance across the datasets.
Results with the PD-based method and~\cite{chentanez2018physics} show that
while PD-controllers are very good at reproducing single or few motions,
they perform worse than torque-based controllers on large motion datasets.
We note that the PD-based method,
which only uses a small number of future frames in the observation functions,
actually performs better than~\cite{chentanez2018physics}.
This could potentially be a cause of ~\cite{chentanez2018physics} feeding too much information into the network,
some of which can contain information too far into the future that confuses the network and slows down the training process.
% "?} and makes the network slower to train.
% much of which is useless and confusing future information \kelly{
%this sounds kind of harsh - maybe something like "some of which can contain information too far into the future and can confuse the network and slow down the training process"?} and makes the network slower to train.
We see poor performance from the Kinematic-State baseline,
indicating that open-loop control is not enough for complex multi-motion animation tasks.
We also see that both DeepMimic-Onehot and DeepMimic-Variable obtain poor performance,
indicating their lack of model capacity.

\subsubsection{Testing Performance}
In this section, we study the generalization of algorithms to unseen motions.
In figure~\ref{fig:test_performance}, we show the performance of our method and the baseline algorithms on the test set.
The PD-based method and~\cite{chentanez2018physics} reach their performance plateau very quickly on the test set, while training performance is still increasing.
This over-fitting behaviour is most obvious in the two datasets shown in the figure.
We suspect that this is due to the use of PD-controllers.
Over-fitting in deep reinforcement learning is still an under-explored topic,
and we do not explore further into this direction.
% In~\cite{greydanus2017visualizing}, the authors discuss one interesting observation,
% where a reinforcement learning controller applied to solve Atari games actually utilizes the scoreboard and timer as features to generate the policy.
% This indicates that it can be potentially very dangerous to add external help or features to our controller. 
% In our case, the PD neural corrective controller can very likely be over-fitting and using PD control targets as important features.
% Empirically, the PD control target must be combined with meaningful corrective offsets generated by the neural controller to perform successful animation.
% The neural controller overfits to the PD control targets seen during training,
% and has poor performance when it is fed with unseen PD control targets during testing.
% In our empirical experiments, algorithms including that use PD neural corrective controller has worse testing performance, suggesting they can be more prone to overfitting.

On the other hand, the test performance of Kinematic-State, DeepMimic-Onehot and DeepMimic-Variable shows almost no improvement throughout the training process, demonstrating serious over-fitting as well.
%\kelly{I'm not sure the reference to the Atari example is clear to me. It might help to indicate that the atari game learned to rely on the scoreboard and timer, and overfit to it? (if I understand it correctly)}
%  \kelly{to do what? to be successful?}
% Therefore, during testing, the change in PD control targets which are unseen during training,
% During testing, the PD control targets are unseen during training,
%\kelly{may be helpful to emphasize the change is due to unseen motions during training}
% \kelly{learned during training?}, resulting in poor performance.

\begin{figure}[!h]
    \centering
    \begin{subfigure}{.23\textwidth}
        \centering
        \quad All
        \includegraphics[width=\linewidth]{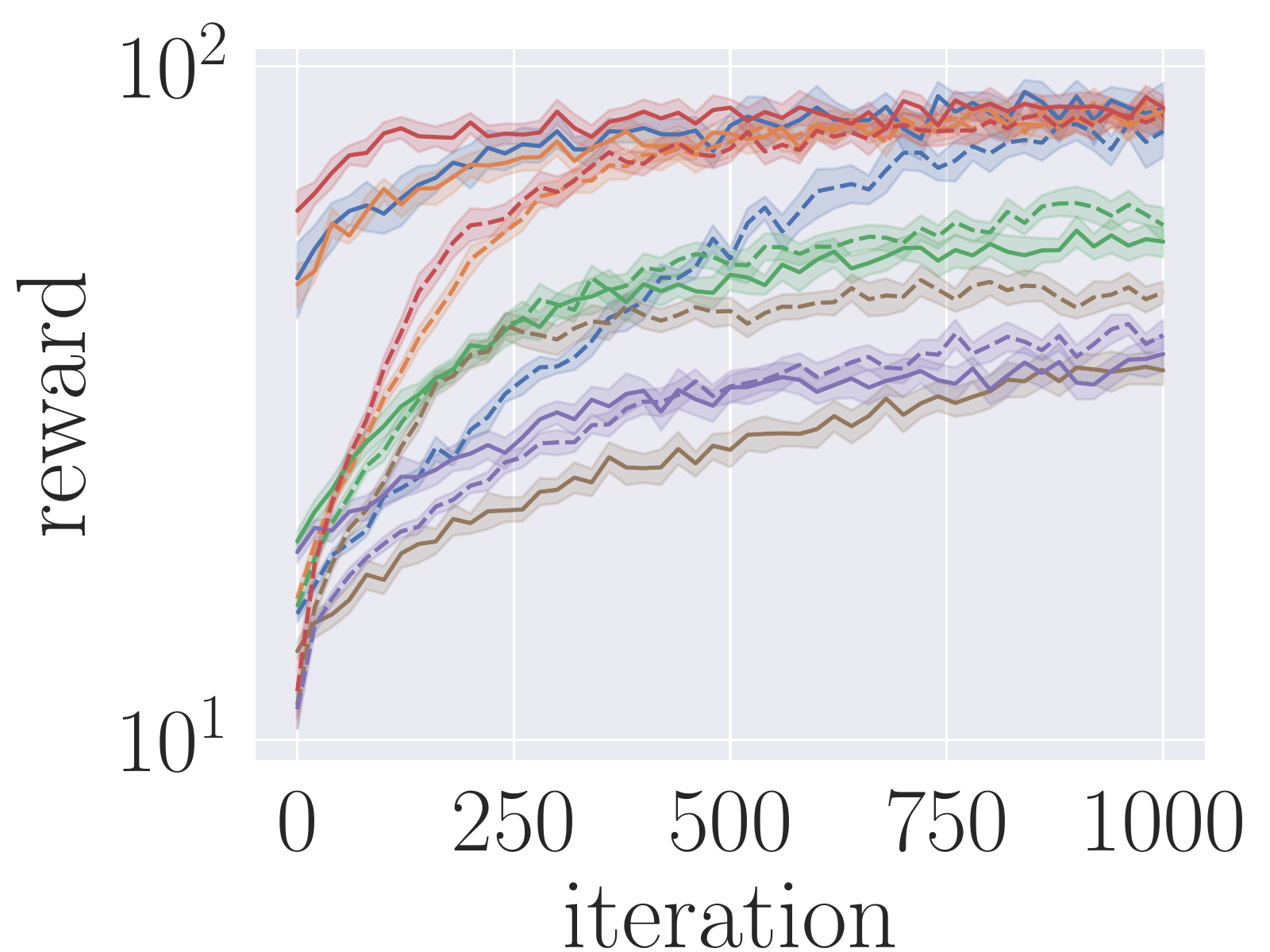}
    \end{subfigure}
    \begin{subfigure}{.23\textwidth}
        \centering
        \quad Walk
        \includegraphics[width=\linewidth]{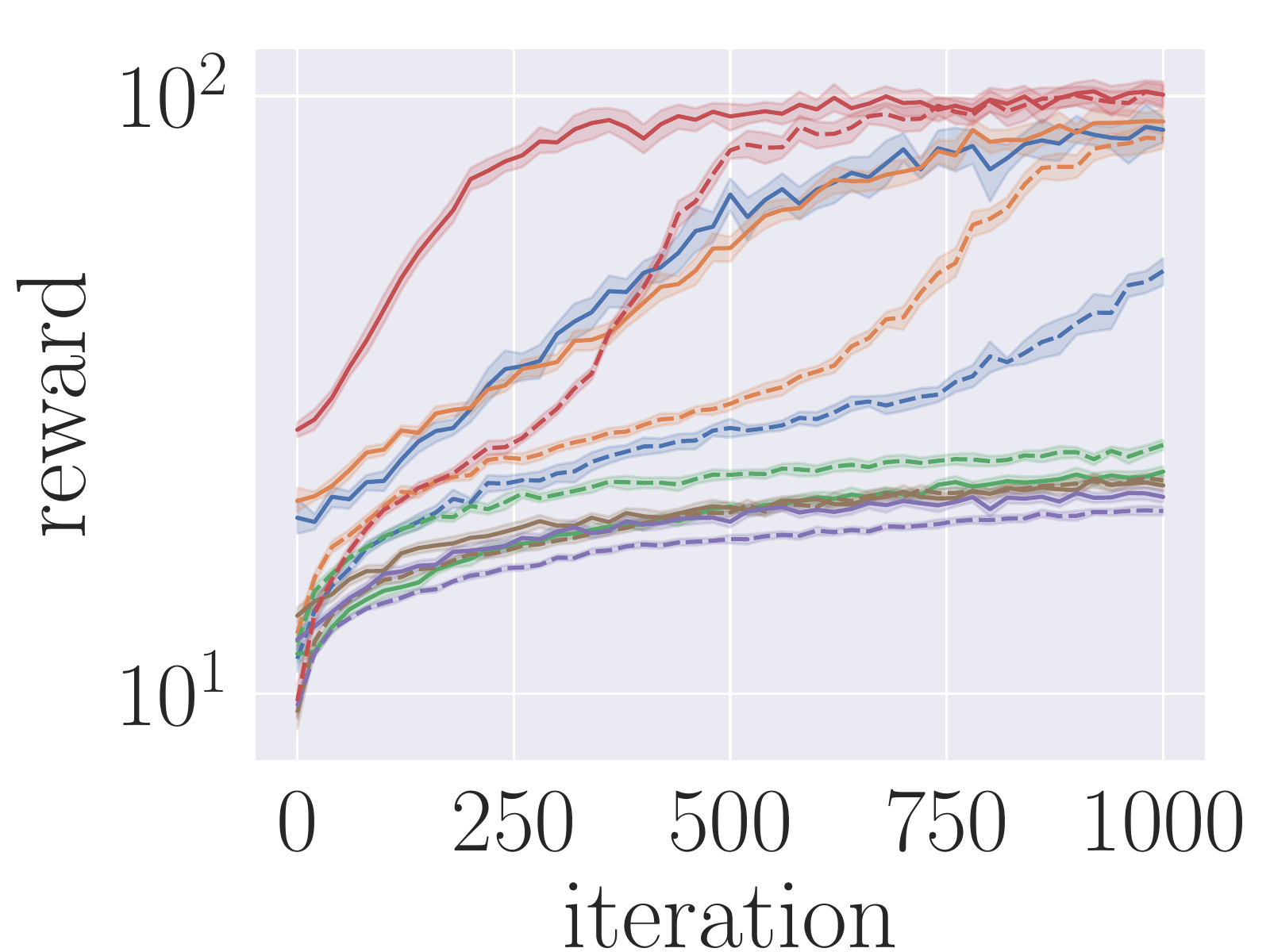}
    \end{subfigure}
    \begin{subfigure}{.48\textwidth}
        \centering
        \includegraphics[width=\linewidth]{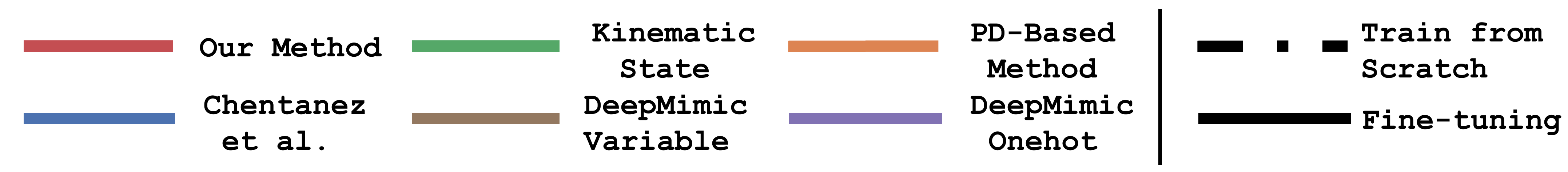}
    \end{subfigure}
    \caption{{The performance of transfer learning for different controllers,
    with and without a pretrained model.}}
    \label{fig:transfer_learning}
\end{figure}
\subsection{Transfer Learning with Fine-tuning}
Since the test set was not seen during training, for academic purposes only,
we can reuse the test-set to study how the controllers' learnt features can be transferred to a dataset.
In figure~\ref{fig:transfer_learning}, dashed lines represent models trained from scratch,
and solid lines represent models with pre-trained knowledge.
%that are fine-tuned from the training set. \kelly{fine-tuned from training or test? it's slightly confusing}
We show that our low-level controller is highly transferable,
suggesting potential use cases where we can first train a model on an enormous dataset,
then fine-tune the model on smaller datasets for better specialization.
The PD-based method and~\cite{chentanez2018physics} also have slightly worse but reasonable transferability,
while Kinematic-State, DeepMimic-Onehot and DeepMimic-Variable show no or even negative transferability.
\begin{table*}[!h]
\resizebox{0.98\textwidth}{!}{
    \begin{tabular}{l|c|c|c|c|c|c|c|c|c|c|c|c}
        \toprule
          & Speed-1.6        & Speed-1.5        & Speed-1.4      & Speed-1.3      & Speed-1.2       & Speed-1.1       & Speed-0.9       & Speed-0.8       & Speed-0.7       & Speed-0.6       & Speed-0.5       & Speed-0.4       \\
\midrule
DeepMimic & 21.5\%              & 20.2\%              & 21.5\%            & 22.5\%            & 27.4\%             & 38.1\%             & 36.3\%             & 22.2\%             & 15.0\%             & 13.7\%             & 10.8\%             & 8.1\%              \\
\midrule
Ours      & 65.8\%              & 76.9\%              & 88.7\%            & 95.9\%            & 98.5\%             & 98.0\%             & 99.2\%             & 96.6\%             & 91.7\%             & 70.0\%             & 47.4\%             & 34.5\%             \\
\bottomrule
\toprule
          & heavy & light & Proj-1/1Hz & Proj-1/5Hz & Proj-1/10Hz & Proj-1/20Hz & Proj-1/30Hz & Proj-1/40Hz & Proj-1/50Hz & Proj-1/60Hz & Proj-1/70Hz & Proj-1/80Hz \\
\midrule
DeepMimic & 65.7\%              & 33.5\%              & 13.2\%            & 27.7\%            & 37.5\%             & 51.7\%             & 60.0\%             & 63.0\%             & 62.0\%             & 65.2\%             & 63.9\%             & 68.8\%             \\
\midrule
Ours      & 77.9\%              & 98.2\%              & 31.6\%            & 69.9\%            & 86.8\%             & 95.9\%             & 97.1\%             & 98.0\%             & 98.7\%             & 98.7\%             & 98.9\%             & 99.2\% \\
\bottomrule
\end{tabular}

}
\caption{In this table we show the zero-shot robustness of our algorithm compared with DeepMimic~\cite{peng2018deepmimic}.
    The keyword \texttt{Speed} represents the experiments where we modify the speed of the reference motion with certain ratio,
    and \texttt{Proj} represents the experiments where we modify how frequently the projectiles are thrown at the agents (i. e. how many time-steps there are between two projectiles are thrown at the agent). \texttt{heavy} and \texttt{light} represent the two experiments where we use different humanoid models.
    In the table we show the relative performance compared to the original performance.
    }\label{table:robustness}
%\vspace{-0.3cm}
\end{table*}

\subsection{Ablation Study on Low-level Executor}\label{section:ablation_lowlevel_executor}
In figure~\ref{fig:variant_ablation},
we study how the choice of target states and network structures affects performance of the network.
We design two variants, the \texttt{lookahead} and \texttt{stack}.
In \texttt{lookahead}-$k$ variant, we use the target state from the $k$th future frame, instead of the next frame. %  \kelly{did I understand this correctly?} \TW --> yeah i think so
In \texttt{stack}-$k$, we include all target states in $k$ future frames into the observation function.
As we can see from figure~\ref{fig:variant_ablation} (a), (b),
the number of future frames from the high-level scheduler (i. e. $\tau$),
does not visibly affect the performance curve of the low-level executor.
We also observe that including information too far into the future can be detrimental to both training and testing performance.
Furthermore, for applications such as real-time video streams, it is not possible to generate future frames for $\tau > 1$.
Therefore, in section~\ref{section:results_interactive},
we always set $\tau = 1$, which is essentially an inverse dynamics controller.
In figure~\ref{fig:variant_ablation} (c),
we also note that in contrast with single-motion or few motion trainings such as~\cite{peng2018deepmimic,bergamin2019drecon},
where narrow (layer width 256, 512 for example) neural networks with 2 hidden-layers are used,
our algorithm requires much wider and deeper neural network structures.
Empirically, we use 3 layers of dimension 1024, trained with 4096 agents, due to limitations of computational resources.

\begin{figure}[!t]
    \centering
    \begin{subfigure}{.23\textwidth}
        \centering
        \quad All
        \includegraphics[width=\linewidth]{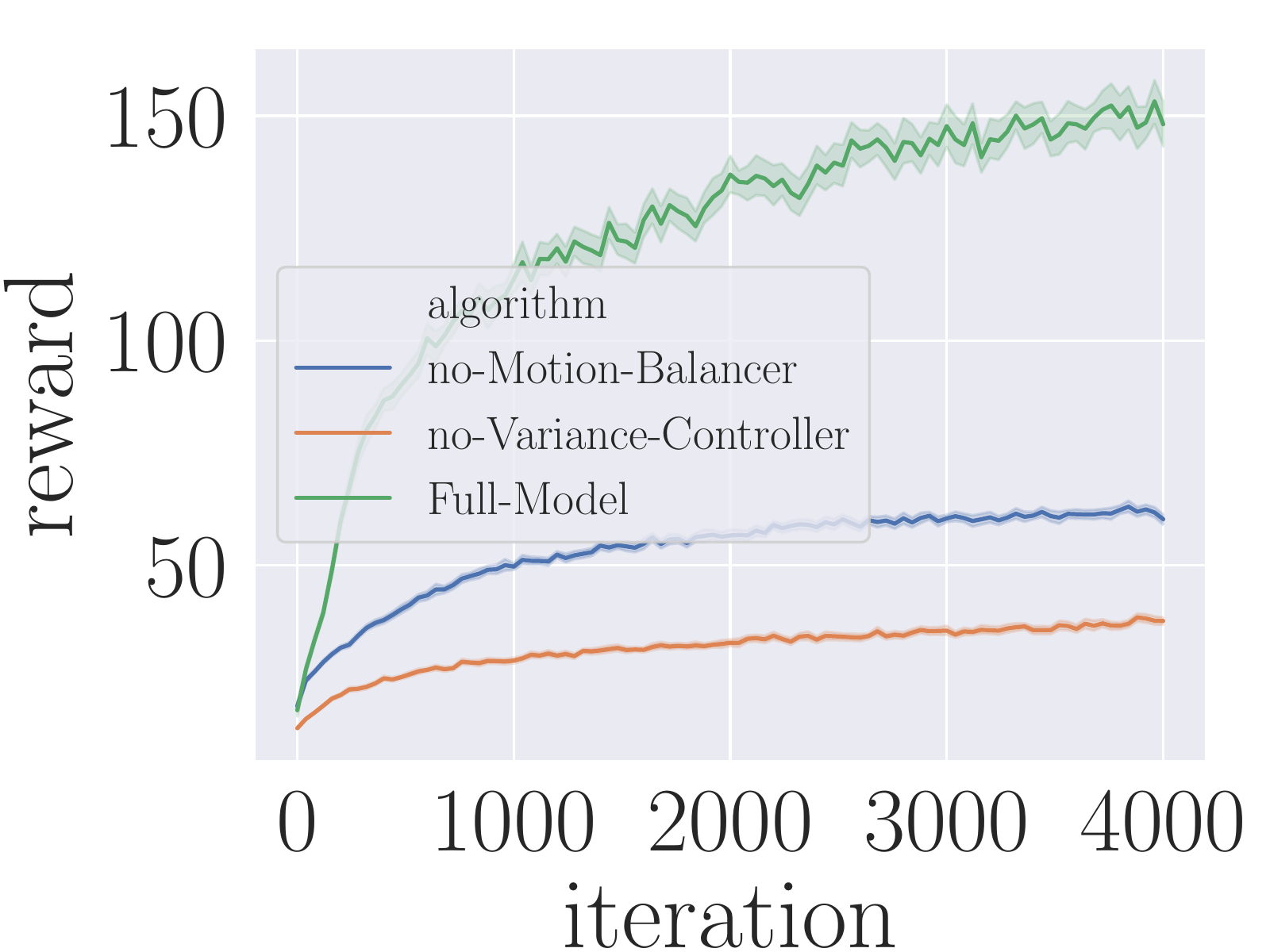}
    \end{subfigure}
    \begin{subfigure}{.23\textwidth}
        \centering
        \quad\,\, RunJumpWalk
        \includegraphics[width=\linewidth]{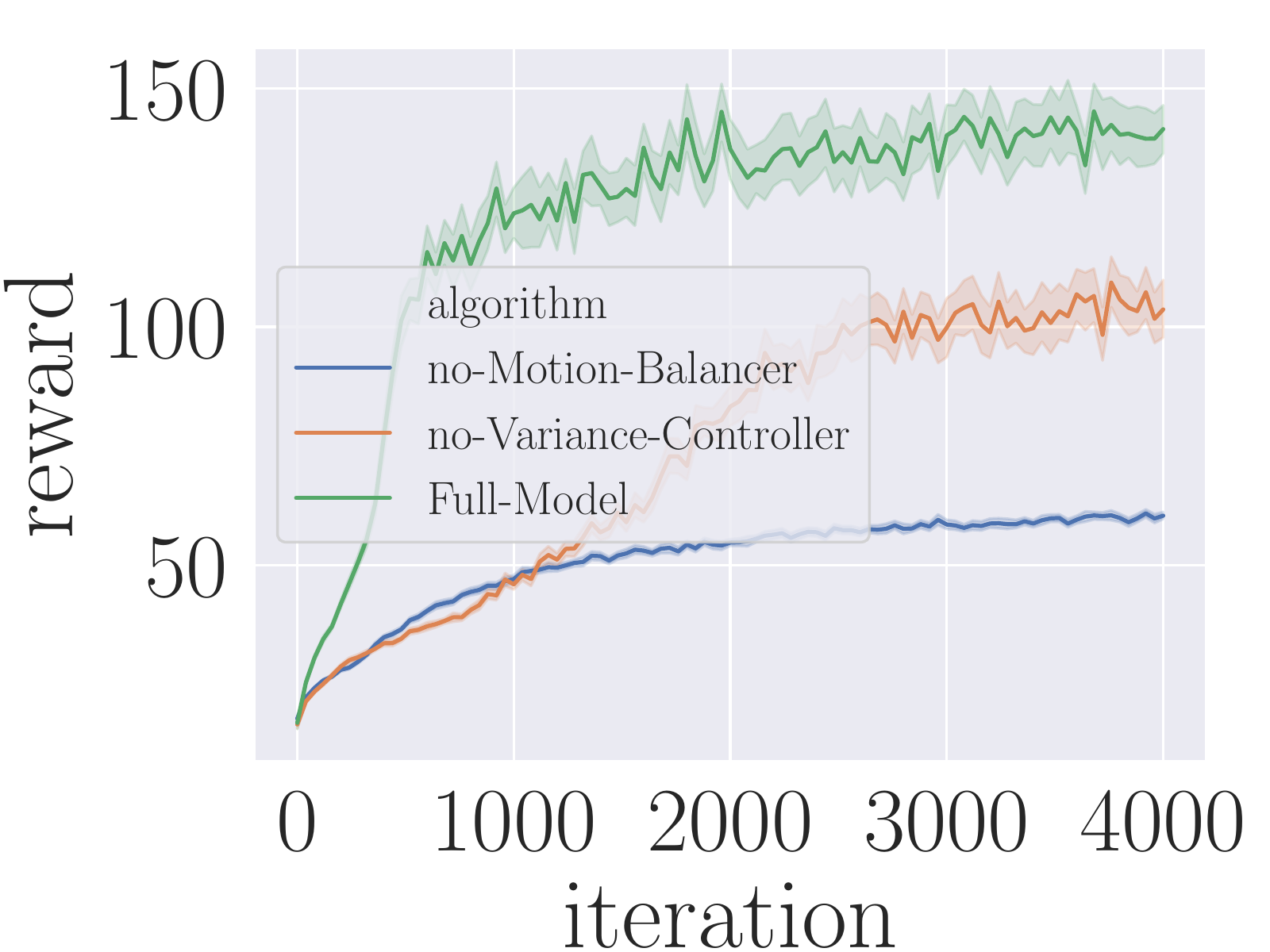}
    \end{subfigure}
    \caption{{This ablation study shows performance with and without variance controller and motion balancer.}}
    \label{fig:ablation_module}
\end{figure}

In figure~\ref{fig:ablation_module},
we show that the motion balancer and the variance controller are essential components to the success of training.
Removing any of the two modules from our algorithm will cause a big performance gap.

\subsection{Zero-Shot Robustness}\label{section:results_robustness}
%\subsection{Zero-shot Robustness}\label{section:zero-shot_robustness}
%Traditionally, projectiles and force perturbation are used
%to show the robustness of a physics-based controller.
%In this section, we include the common retargeting tasks,
%and extend the idea of robustness into \textbf{zero-shot robustness},
%as introduced in section~\ref{section:zero-shot_robustness}.

Previously in~\cite{peng2018deepmimic,bergamin2019drecon},
results are shown where the agent can resist perturbation from projectiles,
or retarget to models with different weight distributions.
However, we argue that this leaves a gap between actual applications and what is demonstrated.
More specifically,
the retargeting is done by retraining a new model,
and the robustness to projectiles, as we show later in the experiments, is still very limited.
% the results are shown by explicitly having projectiles in the training.
% \kelly{I think deepmimic mentioned they didn't have projectiles in training, not too sure about drecon}
% There are several drawbacks to training with perturbations:
% 1) In applications such as video games,
% the perturbation can complex and difficult to model.
% Training the agent with a specific perturbation type could lead to overfitting to a particular type of perturbation,
% instead of learning a robust controller towards different perturbations.
% 2) Some of the perturbations cannot be easily implemented and added during training,
% such as interference from other agents.
% 3) Different models may be required to be trained with different perturbations, and network
%  performance can be sacrificed to combat perturbation.

In this project, we introduce \textbf{zero-shot robustness},
where the agent never sees the perturbation or retargeting information during training,
and is asked to perform tasks under perturbations or using different humanoid models with varying masses.
We argue that a robust controller with the ability to combat unseen perturbations and retargeting problems
will have a much broader potential for real-life applications.
More specifically, we have the following zero-shot robustness tasks:

\textbf{Zero-shot Pertubation Robustness}:
In this case, the agents are trained without projectiles being thrown at them.
During testing, the agents are required to perform the tasks under projectiles.
We include experiments with a full range of projectile density and frequencies,
and use the reward as the metric to numerically evaluate the performance.

\textbf{Zero-shot Speed Robustness}:
Traditionally the motion data samples have a certain fixed speed.
In DeepMimic, a phase variable is used to encode the time information of the motion.
In this experiment, the agents are trained with motions at the original speed, but during testing, the agents are required to reproduce motions at different speeds.

\begin{figure}[!t]
    \centering
    \begin{subfigure}{0.15\textwidth}
        \centering
        \includegraphics[width=\linewidth]{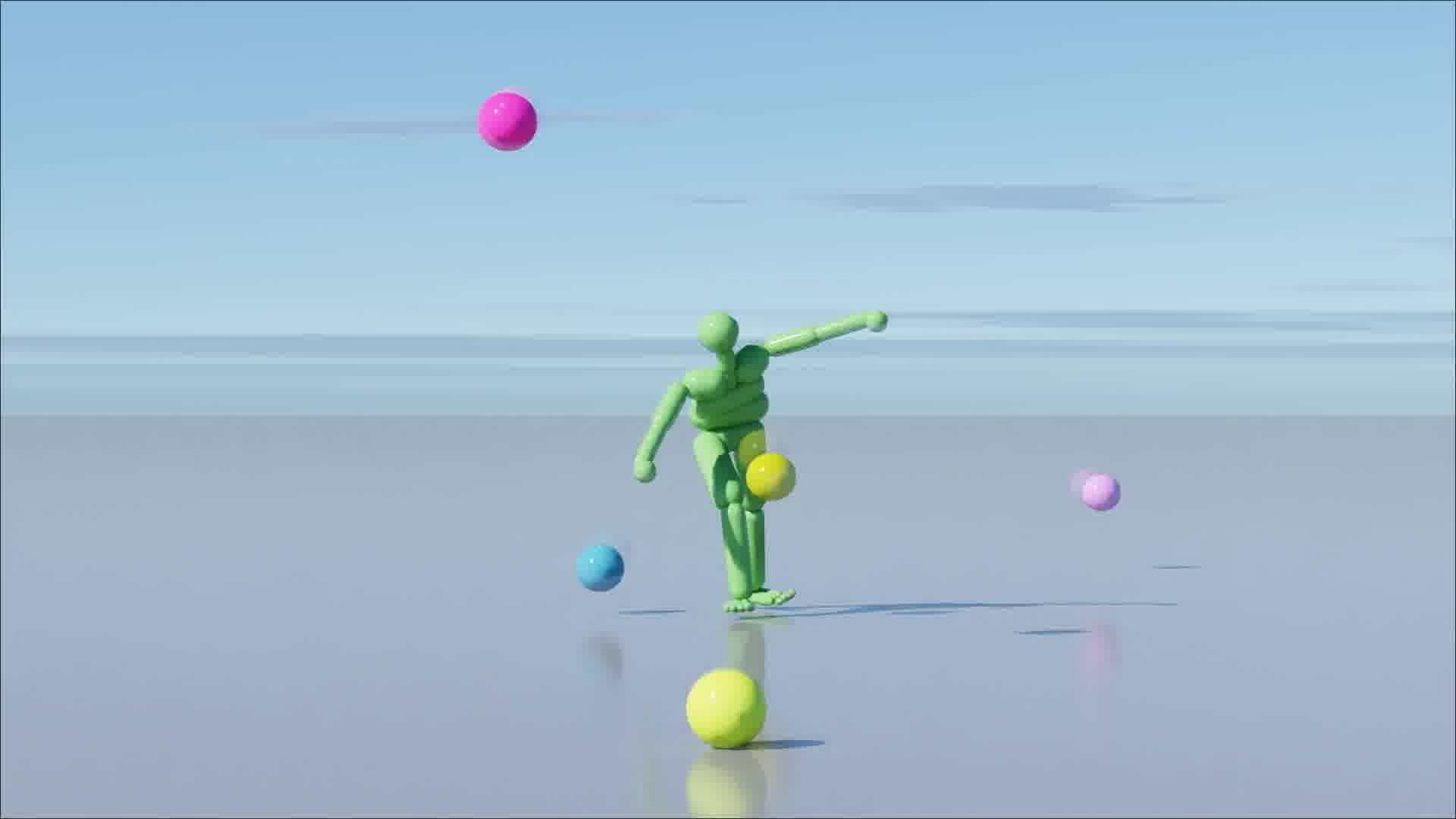}
    \end{subfigure}
    \begin{subfigure}{0.15\textwidth}
        \centering
        \includegraphics[width=\linewidth]{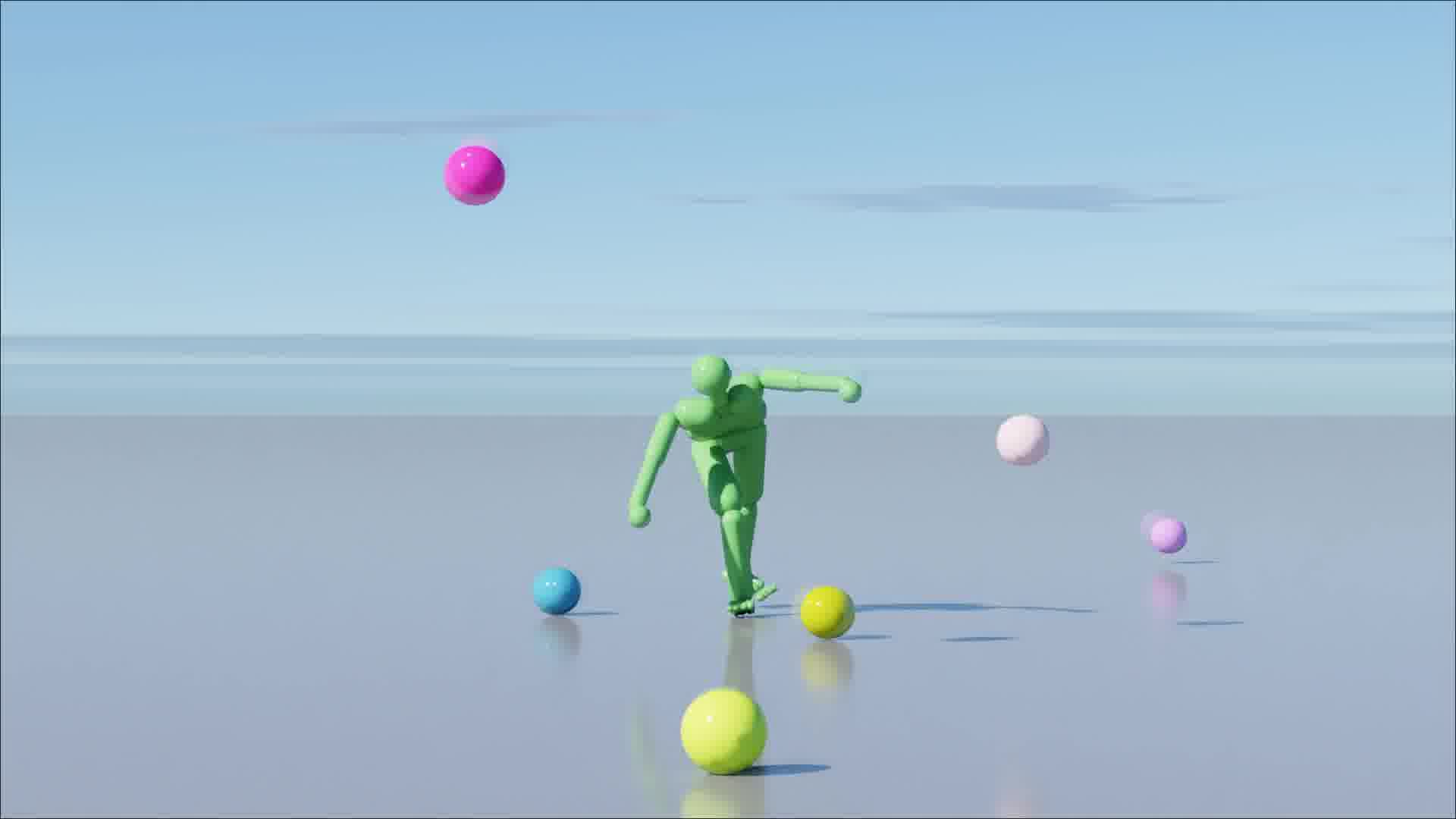}
    \end{subfigure}
    \begin{subfigure}{0.15\textwidth}
        \centering
        \includegraphics[width=\linewidth]{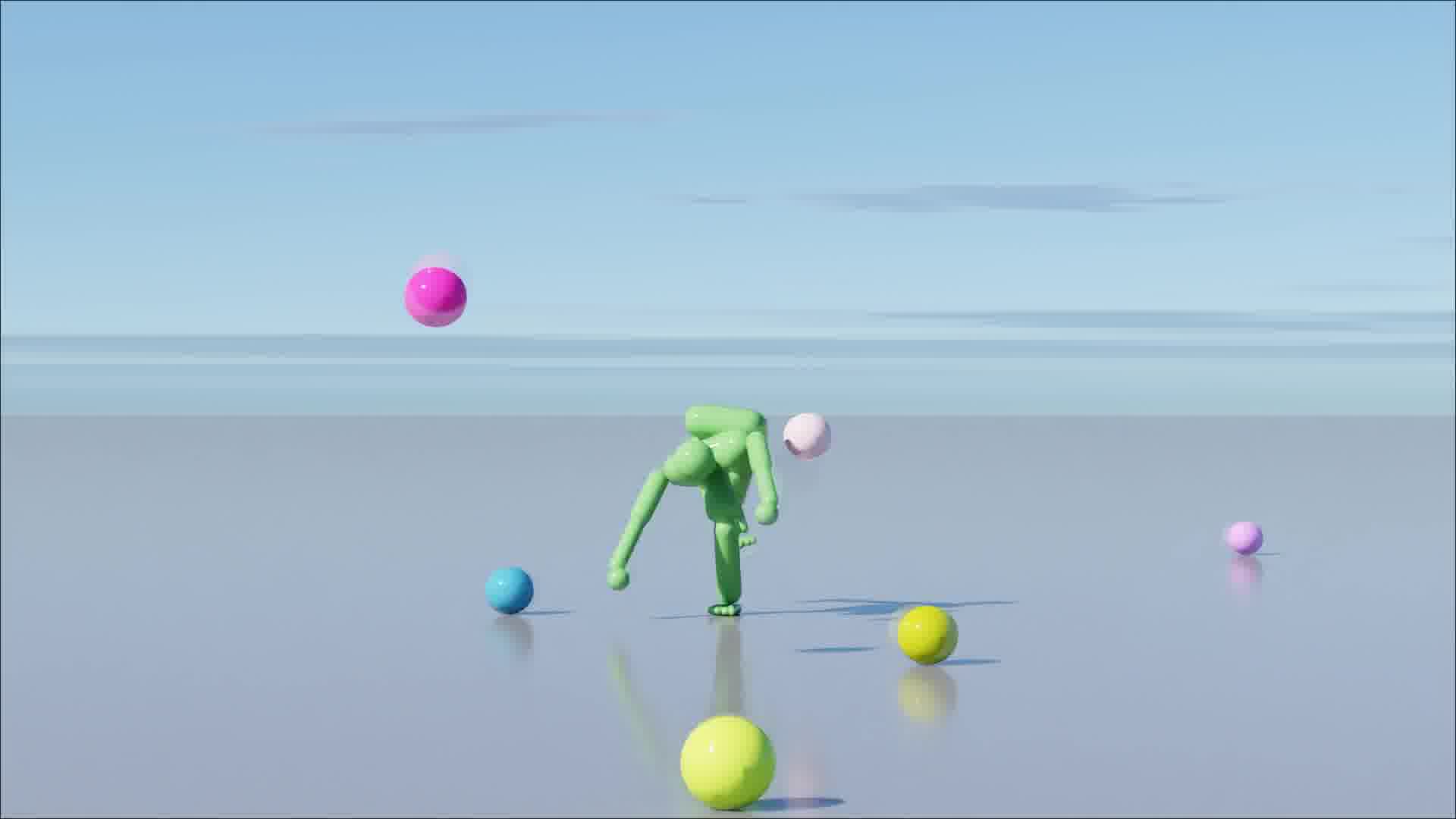}
    \end{subfigure}
    \begin{subfigure}{0.15\textwidth}
        \centering
        \includegraphics[width=\linewidth]{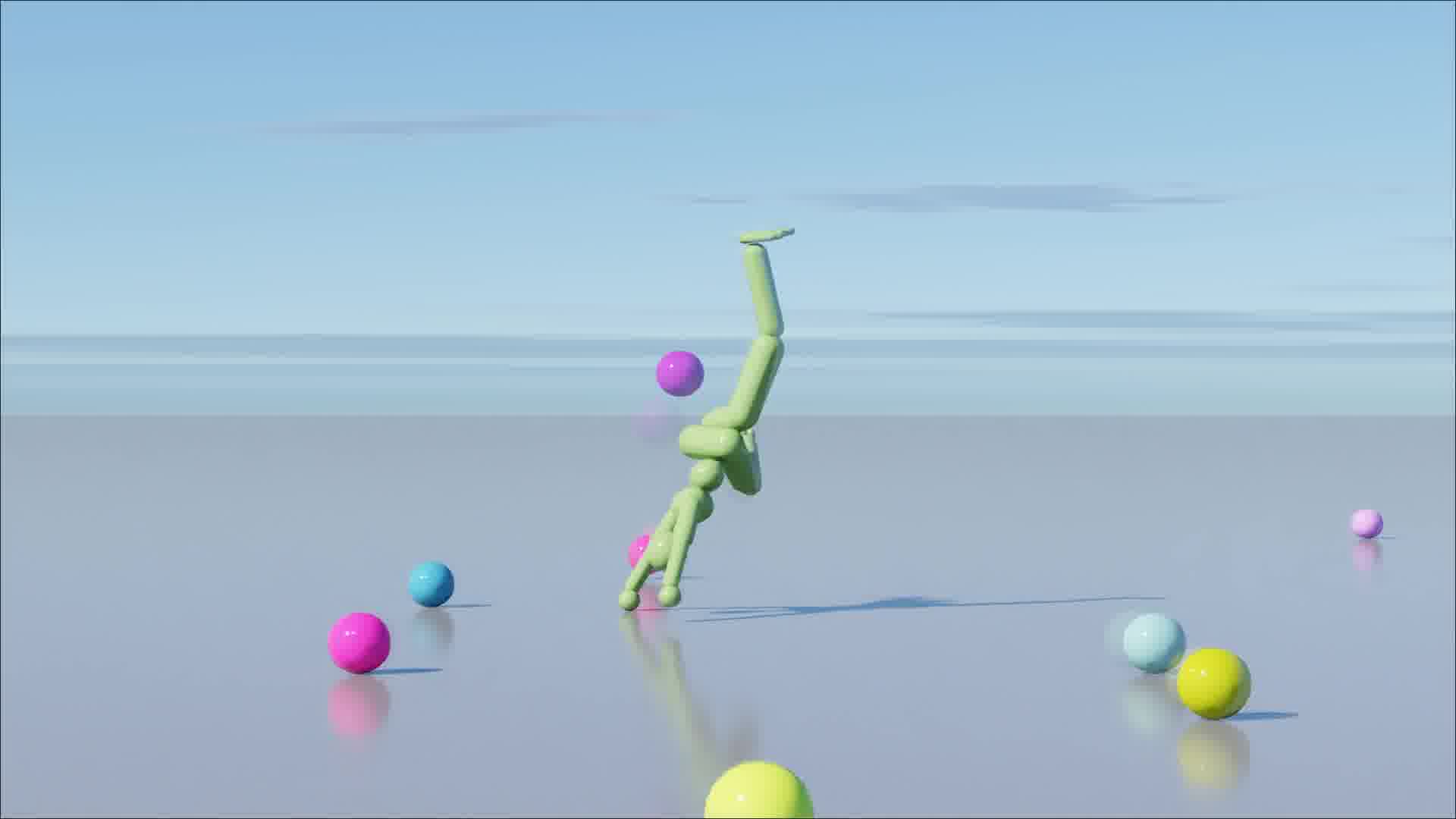}
    \end{subfigure}
    \begin{subfigure}{0.15\textwidth}
        \centering
        \includegraphics[width=\linewidth]{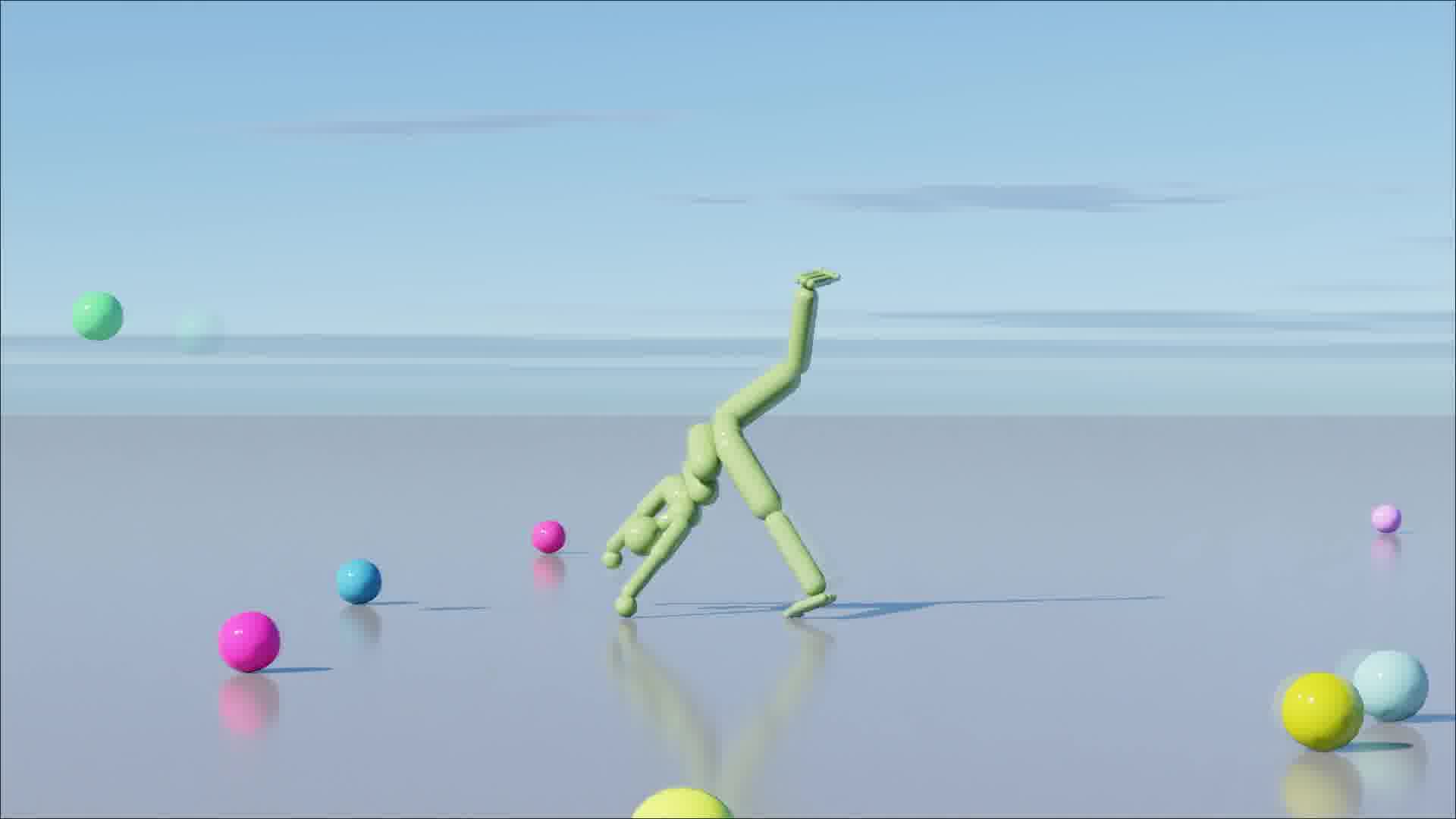}
    \end{subfigure}
    \begin{subfigure}{0.15\textwidth}
        \centering
        \includegraphics[width=\linewidth]{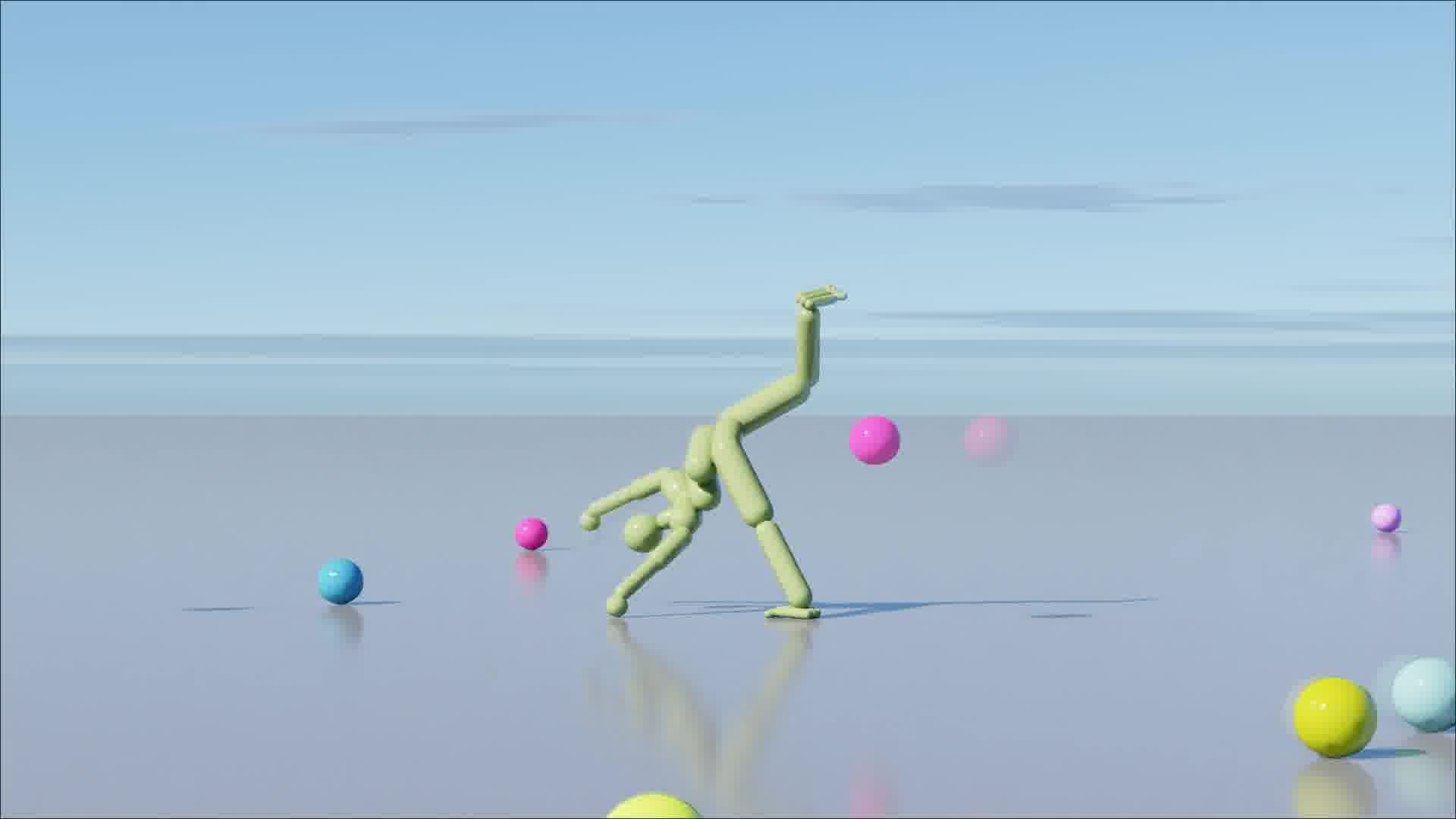}
    \end{subfigure}

    \caption{{Snapshots of our agents under projectiles. We refer the readers to the attached videos for more details.}}
    \label{fig:projectile_robustness}
\end{figure}
\textbf{Zero-shot Model Retargetting Robustness}:
In this case, instead of retraining on the new humanoid models as in DeepMimic,
we ask the agent to reproduce the motion using a different humanoid model,
which it has never seen during training.
We use humanoid models that are 25\% lighter and 25\% heavier.

To numerically analyze the performance,
we use the relative performance compared to the original performance without perturbation,
projectiles or model-mismatch.
As we can see from table~\ref{table:robustness},
our method performs significantly better than DeepMimic.
DeepMimic demonstrates very limited zero-shot robustness,
while our method can still obtain almost $95\%$ performance in the majority of experiments.
We did not include PD-based method and~\cite{chentanez2018physics} in this comparison,
as their performance on the evaluated cartwheel motion has a large gap compared to DeepMimic and our method, where they fail to reproduce the specific cartwheel motion when training with a large dataset.
We also refer to the attached videos for more details.

\begin{figure}[!t]
    \centering
    \begin{subfigure}{0.15\textwidth}
        \centering
        \includegraphics[width=\linewidth]{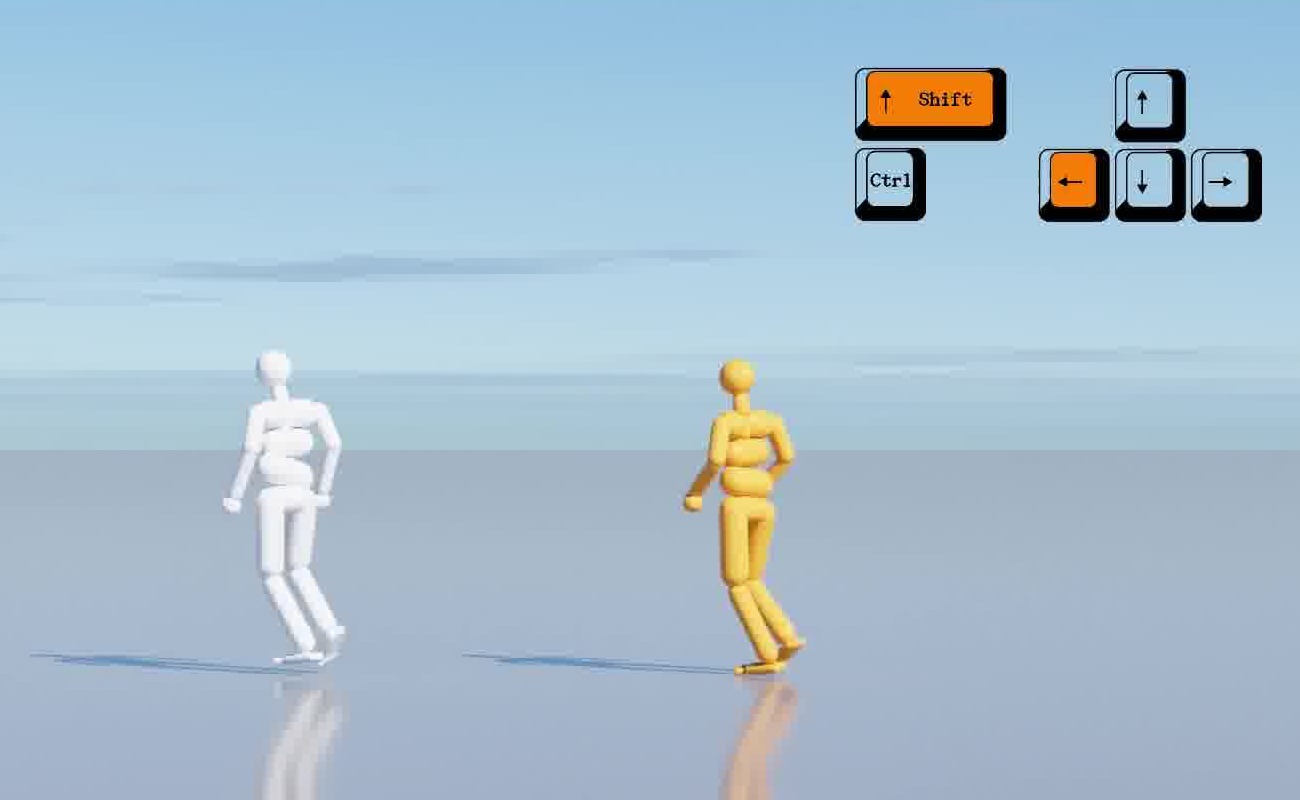}
    \end{subfigure}
    \begin{subfigure}{0.15\textwidth}
        \centering
        \includegraphics[width=\linewidth]{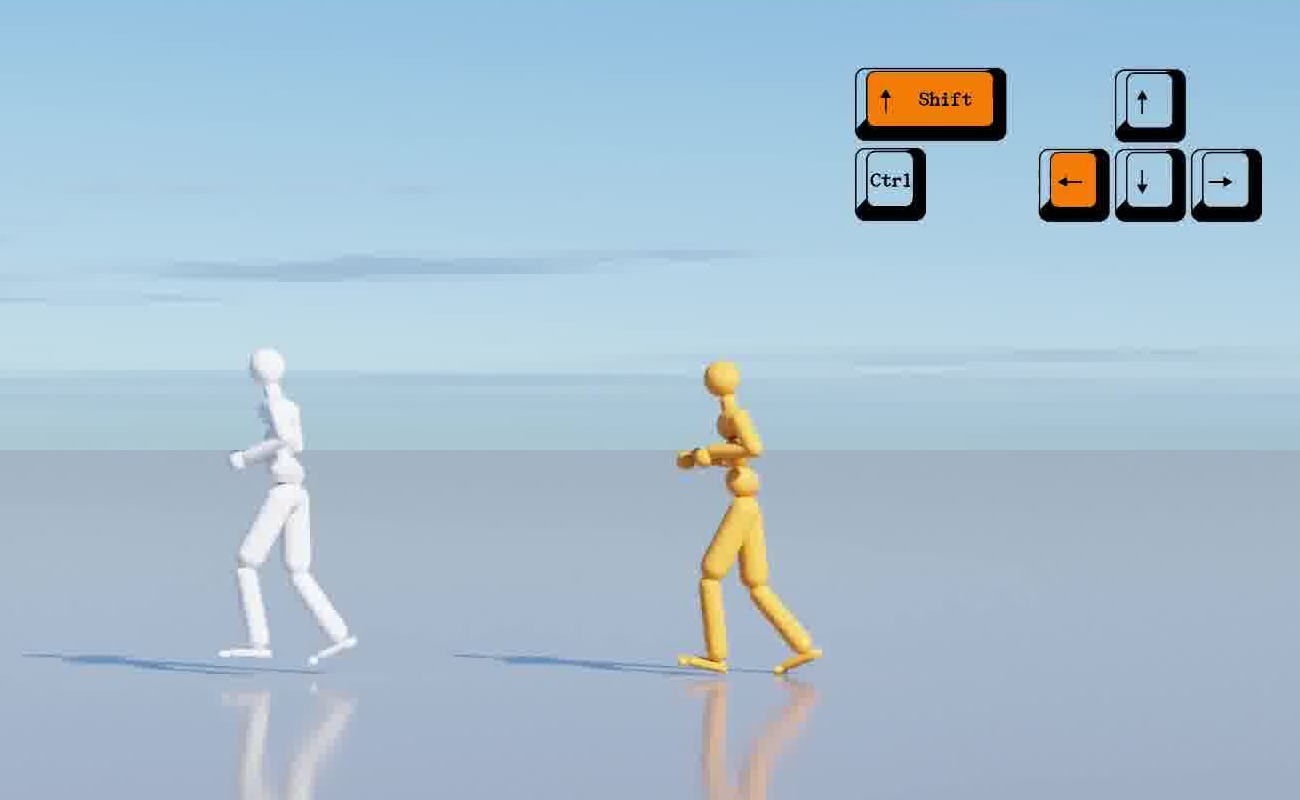}
    \end{subfigure}
    \begin{subfigure}{0.15\textwidth}
        \centering
        \includegraphics[width=\linewidth]{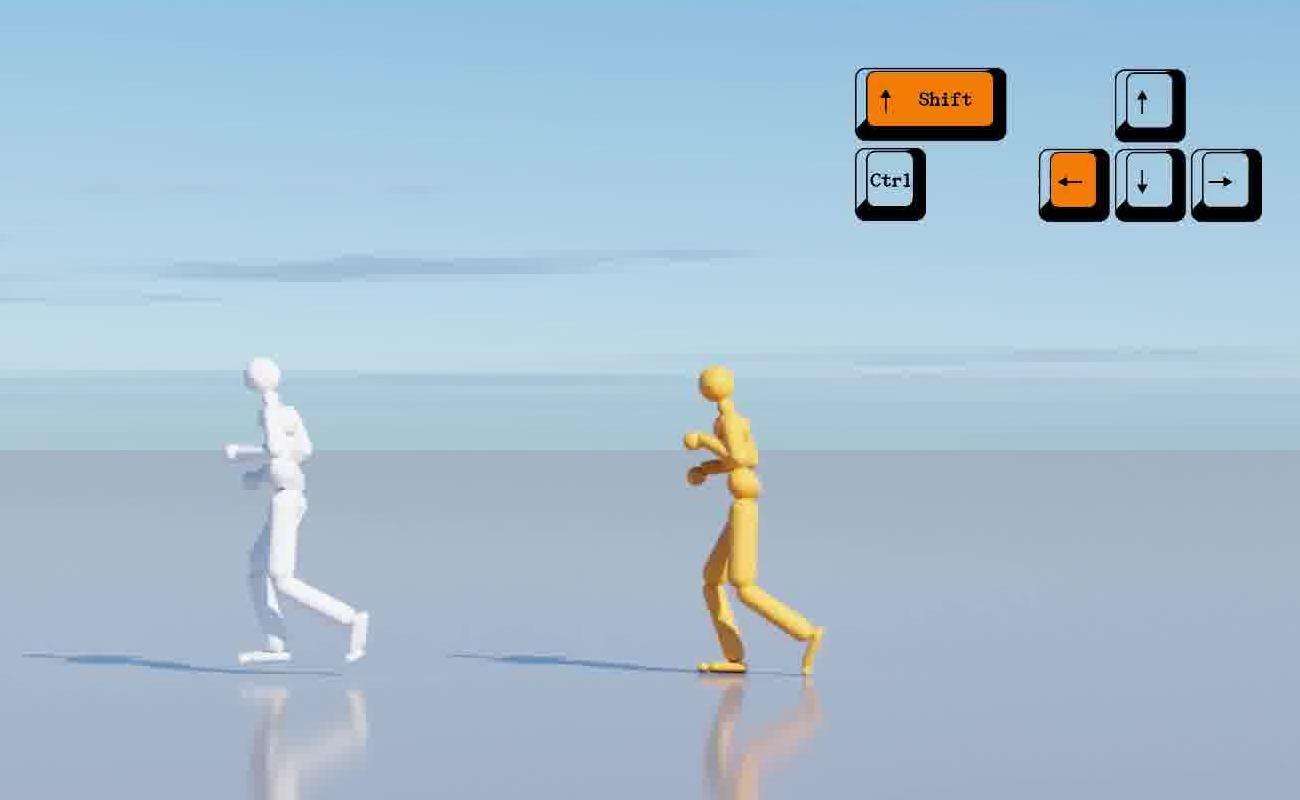}
    \end{subfigure}

    \begin{subfigure}{0.15\textwidth}
        \centering
        \includegraphics[width=\linewidth]{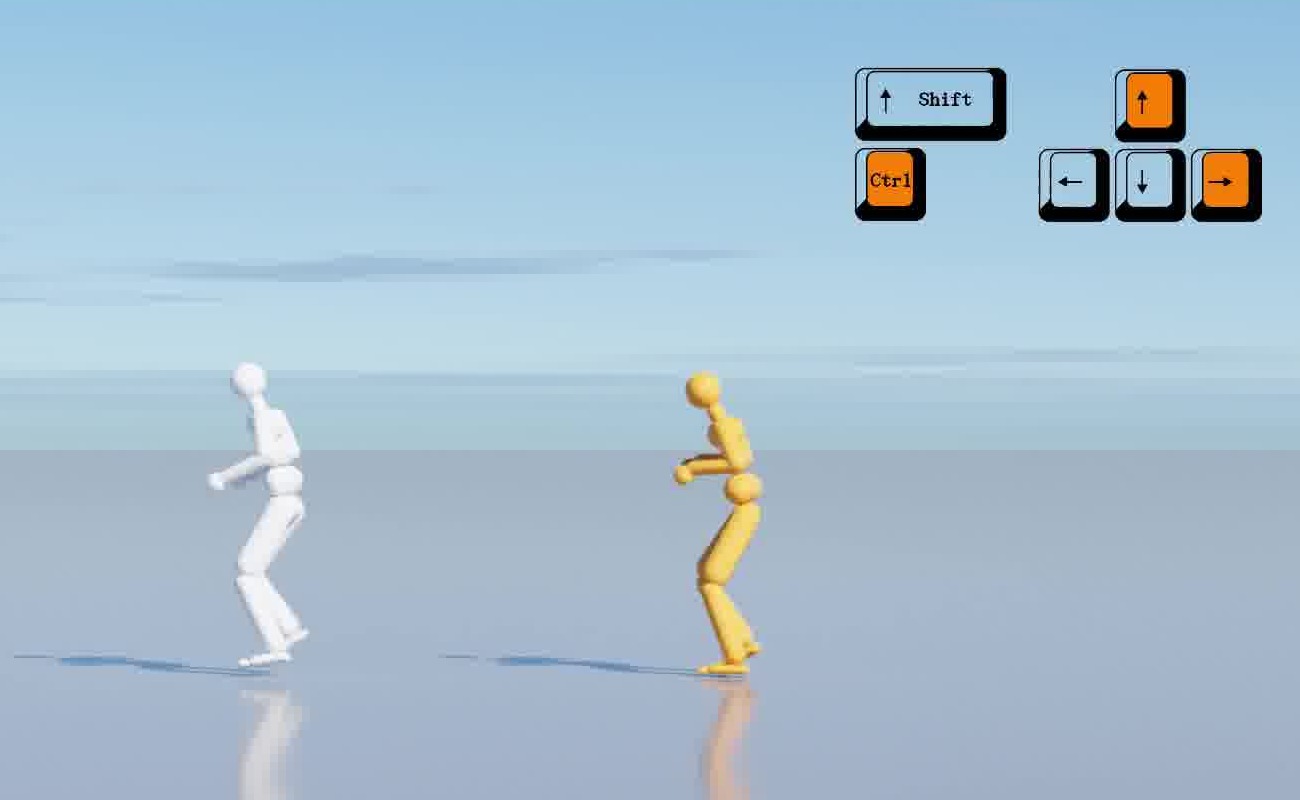}
    \end{subfigure}
    \begin{subfigure}{0.15\textwidth}
        \centering
        \includegraphics[width=\linewidth]{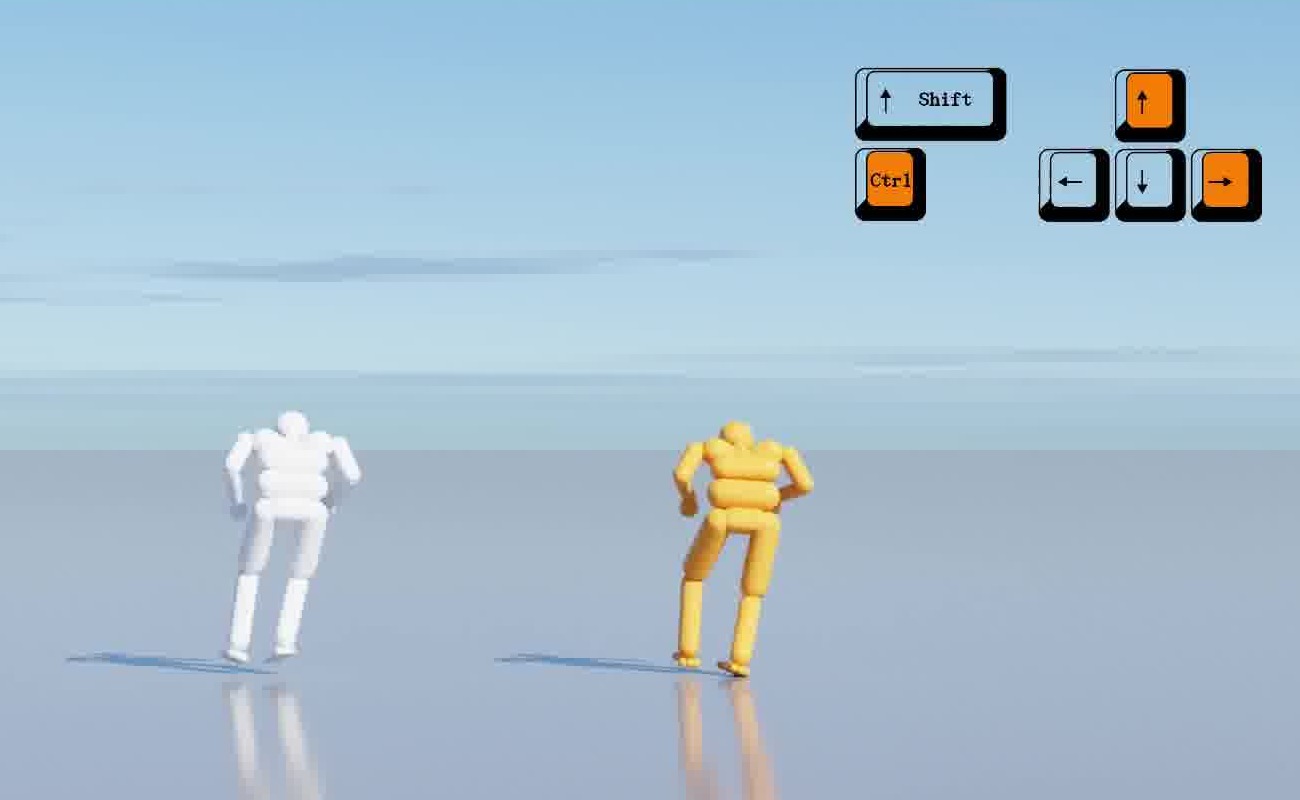}
    \end{subfigure}
    \begin{subfigure}{0.15\textwidth}
        \centering
        \includegraphics[width=\linewidth]{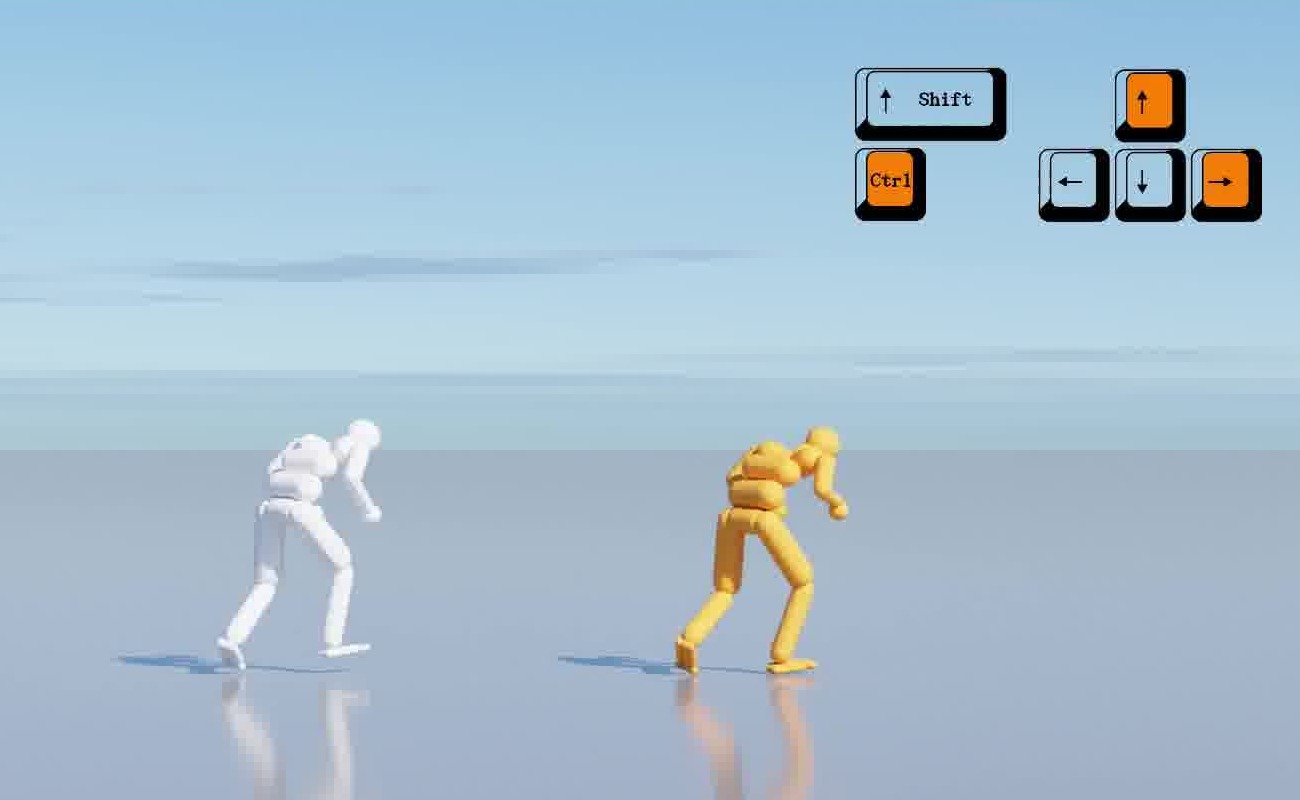}
    \end{subfigure}

    \begin{subfigure}{0.15\textwidth}
        \centering
        \includegraphics[width=\linewidth]{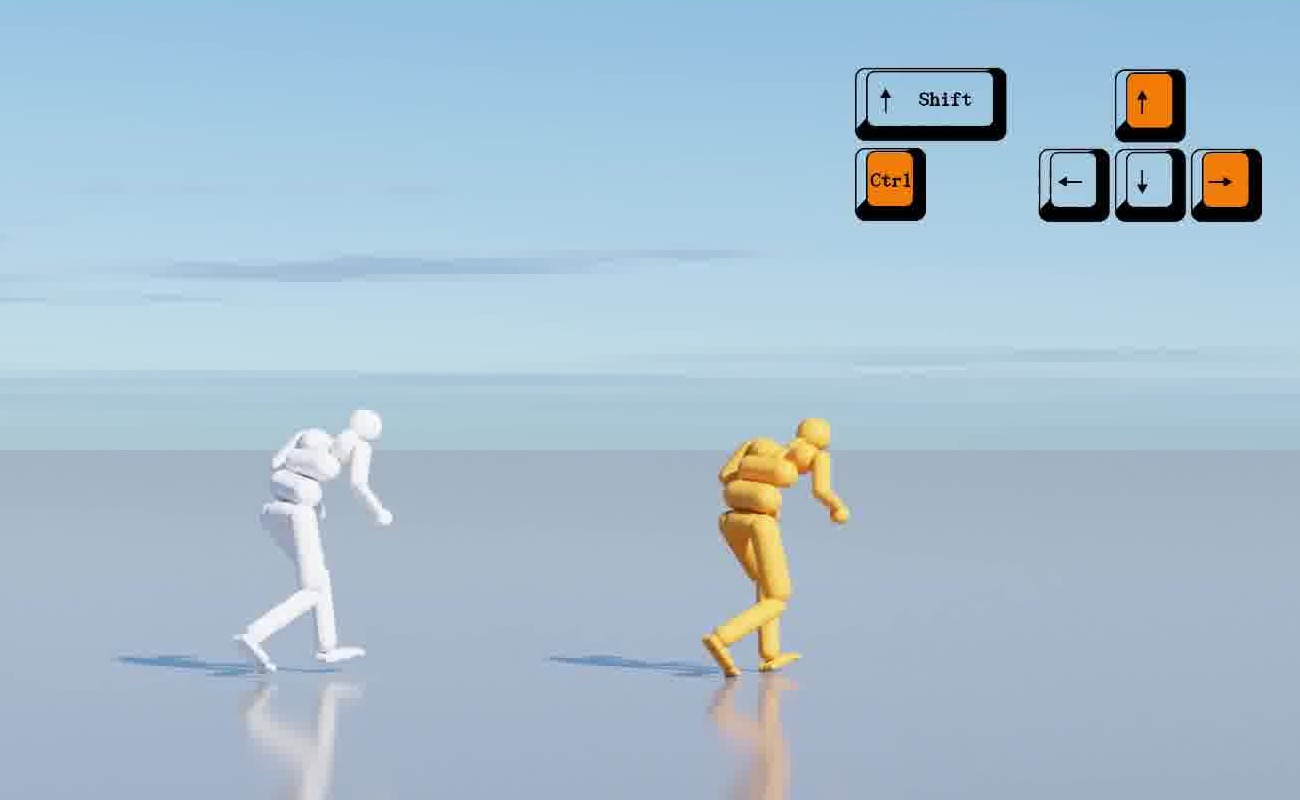}
    \end{subfigure}
    \begin{subfigure}{0.15\textwidth}
        \centering
        \includegraphics[width=\linewidth]{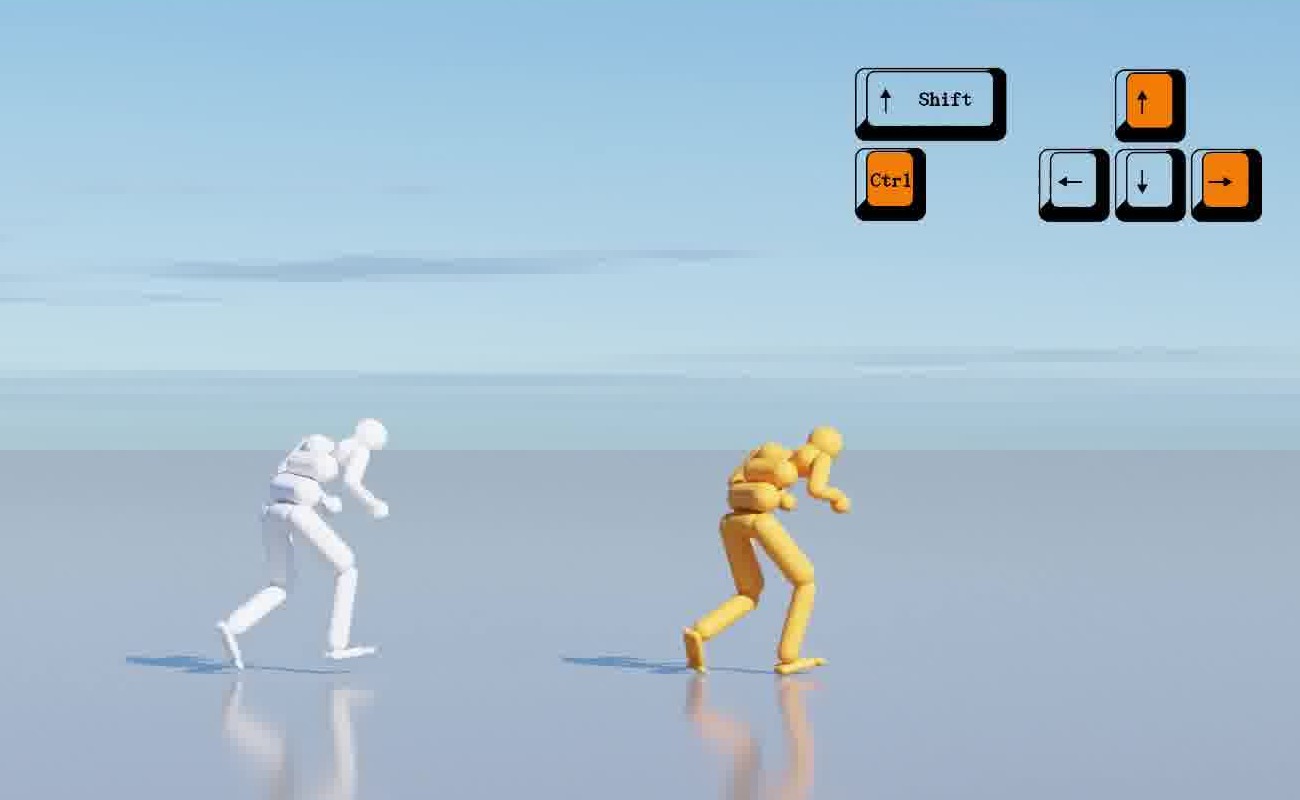}
    \end{subfigure}
    \begin{subfigure}{0.15\textwidth}
        \centering
        \includegraphics[width=\linewidth]{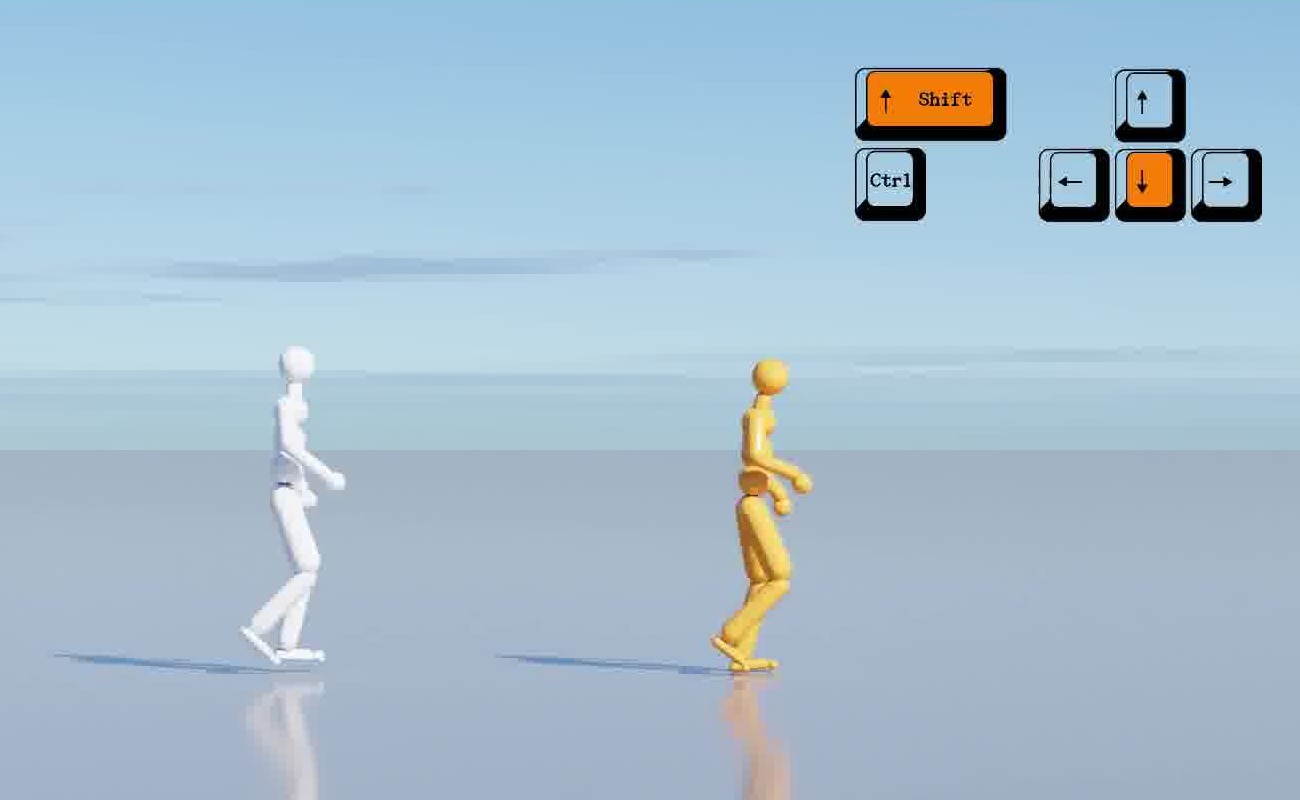}
    \end{subfigure}

    \begin{subfigure}{0.15\textwidth}
        \centering
        \includegraphics[width=\linewidth]{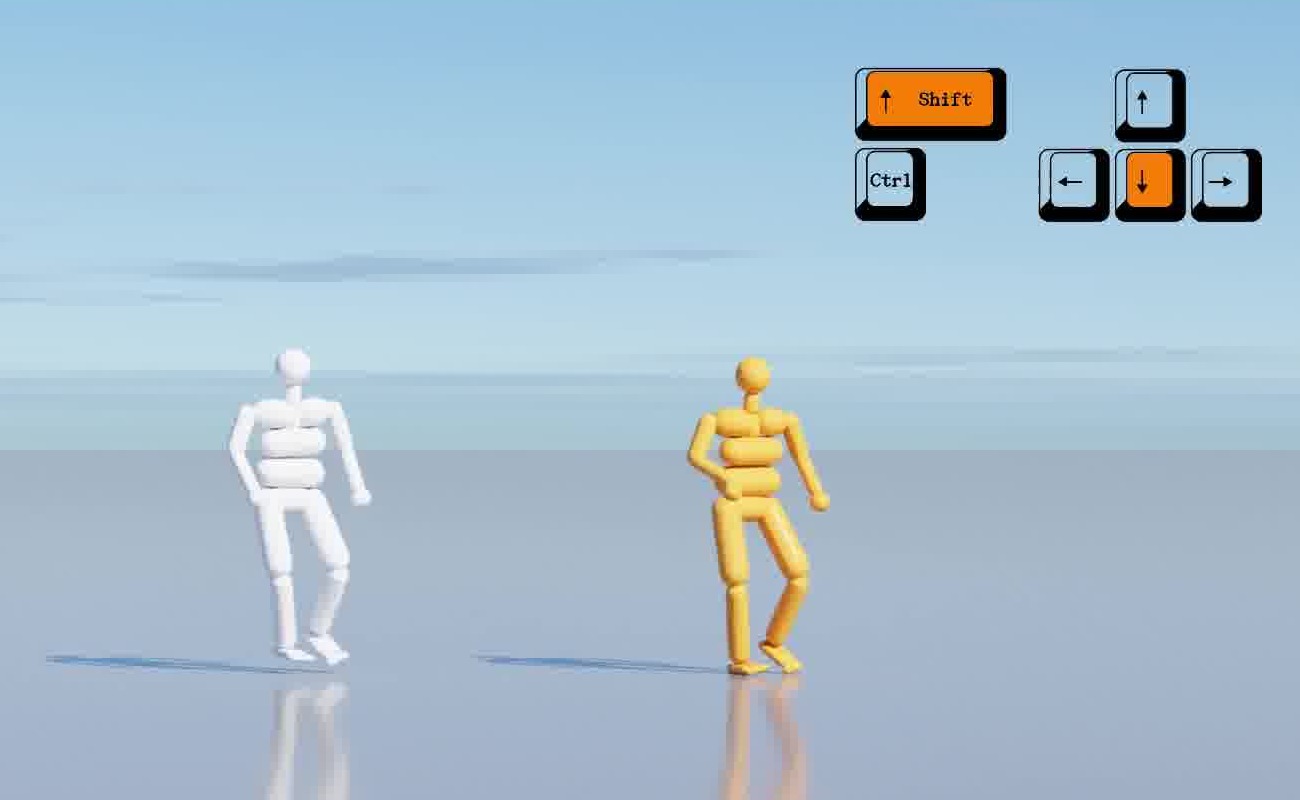}
    \end{subfigure}
    \begin{subfigure}{0.15\textwidth}
        \centering
        \includegraphics[width=\linewidth]{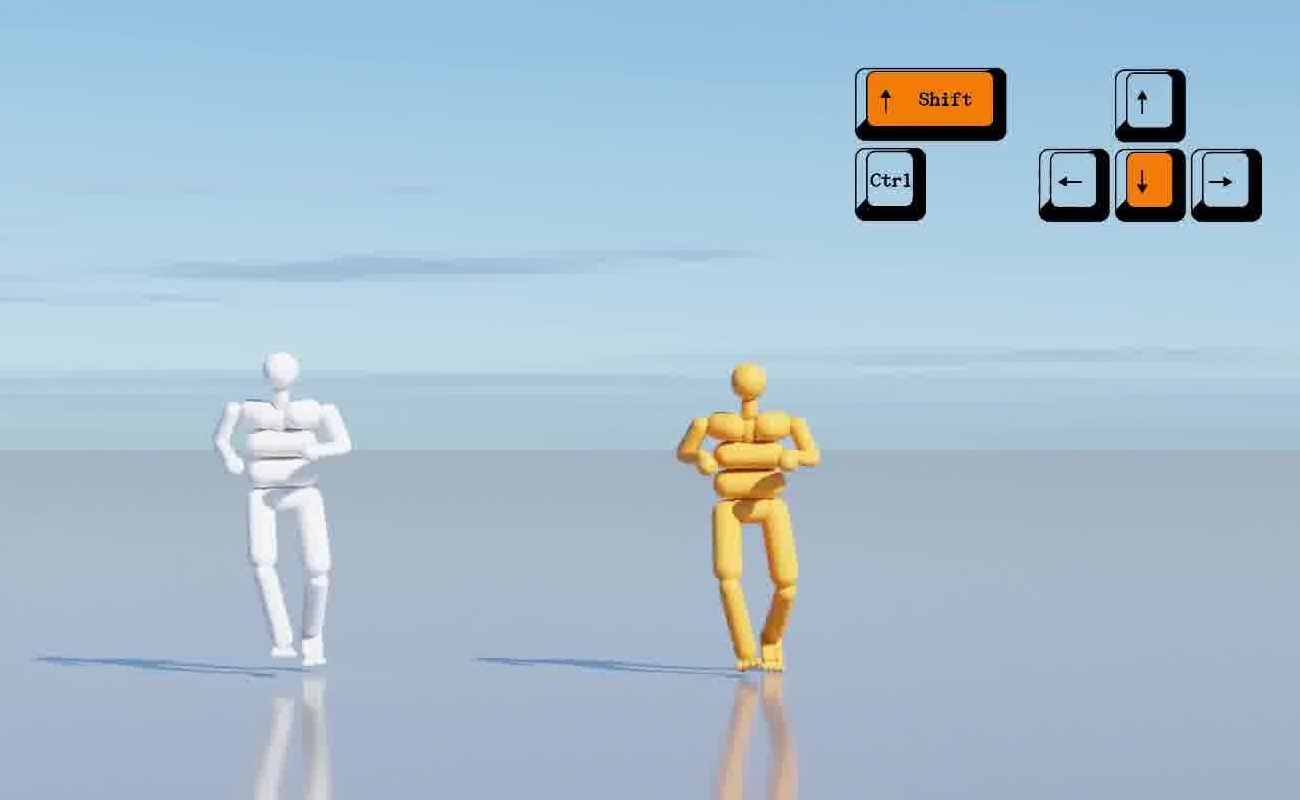}
    \end{subfigure}
    \begin{subfigure}{0.15\textwidth}
        \centering
        \includegraphics[width=\linewidth]{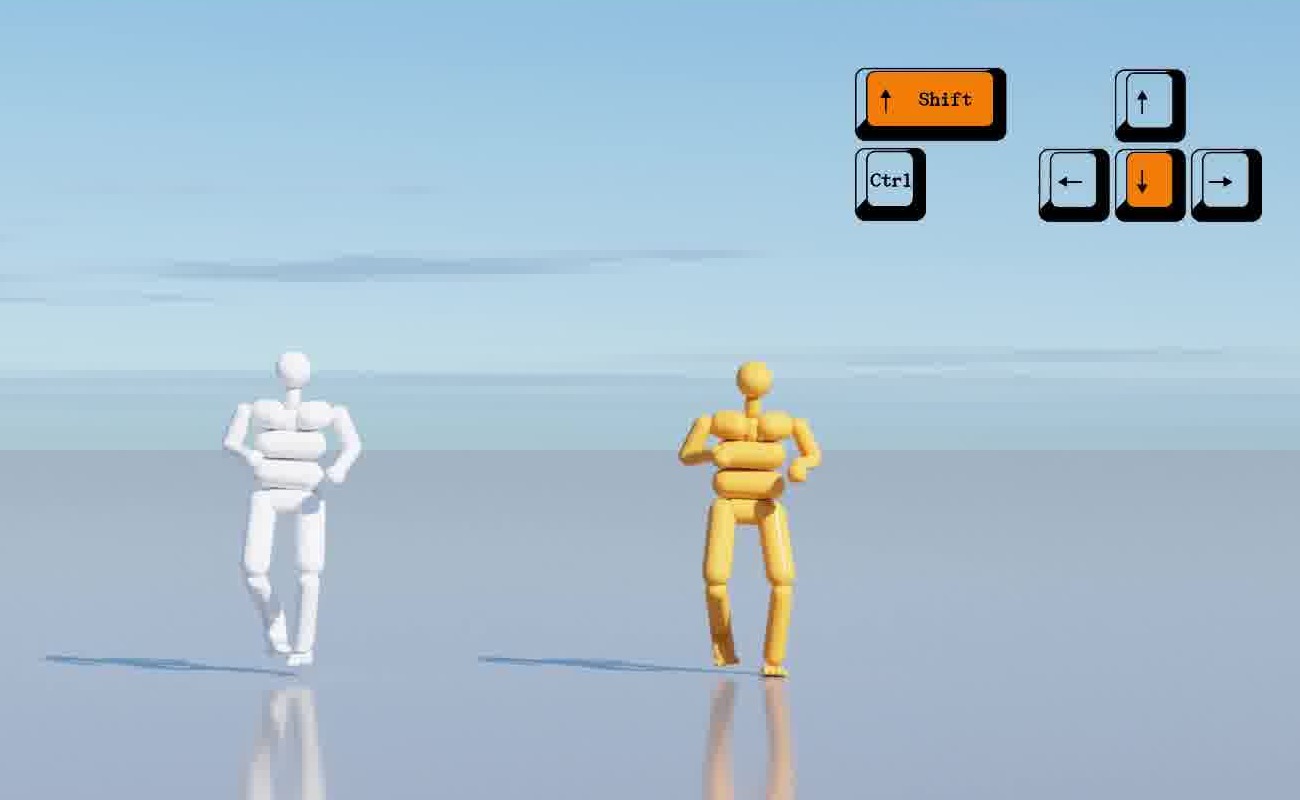}
    \end{subfigure}

    \begin{subfigure}{0.15\textwidth}
        \centering
        \includegraphics[width=\linewidth]{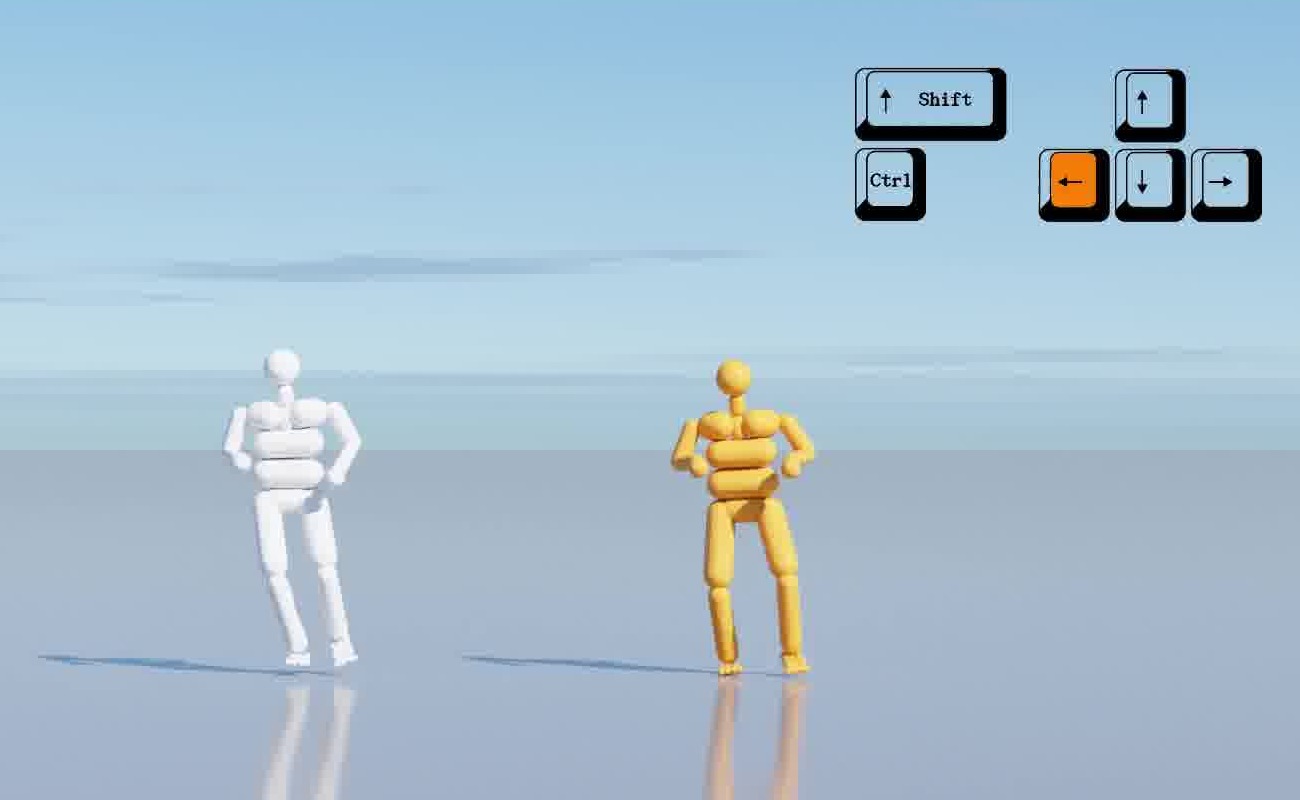}
    \end{subfigure}
    \begin{subfigure}{0.15\textwidth}
        \centering
        \includegraphics[width=\linewidth]{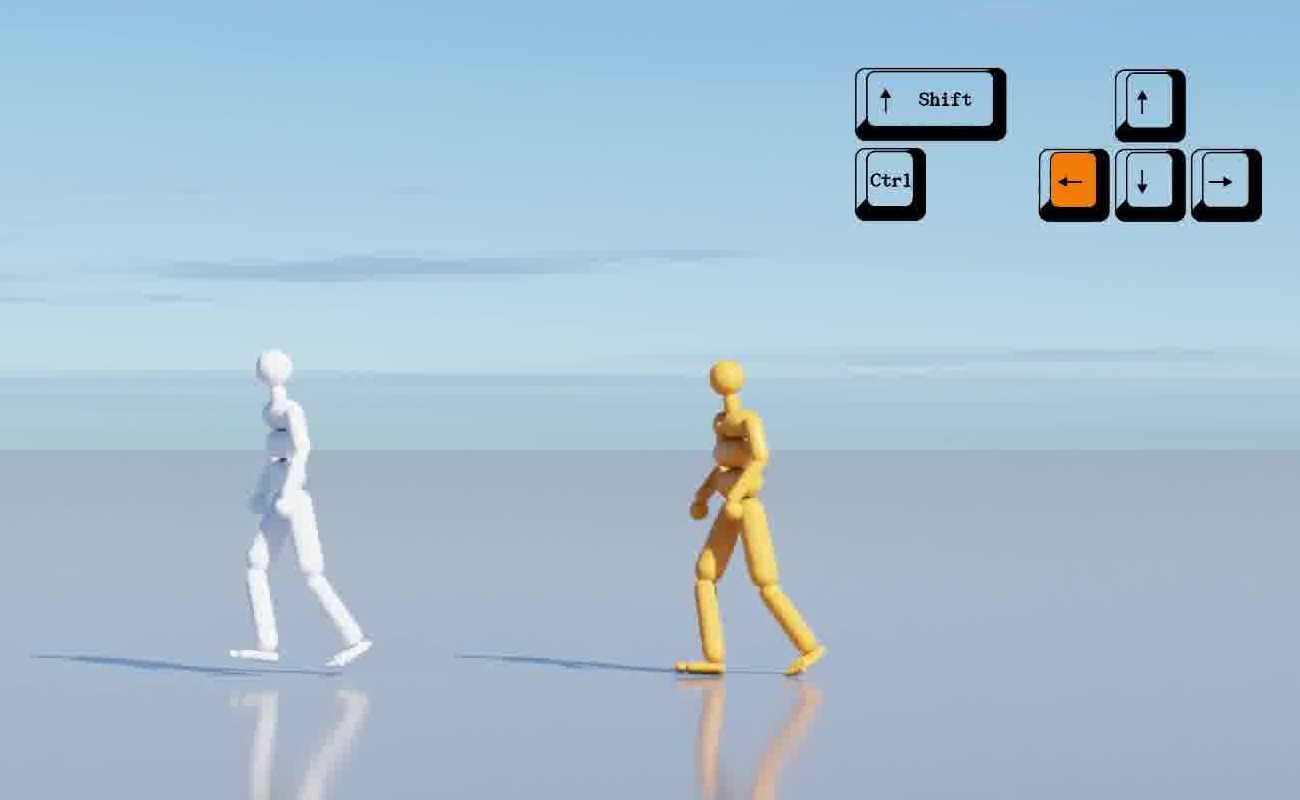}
    \end{subfigure}
    \begin{subfigure}{0.15\textwidth}
        \centering
        \includegraphics[width=\linewidth]{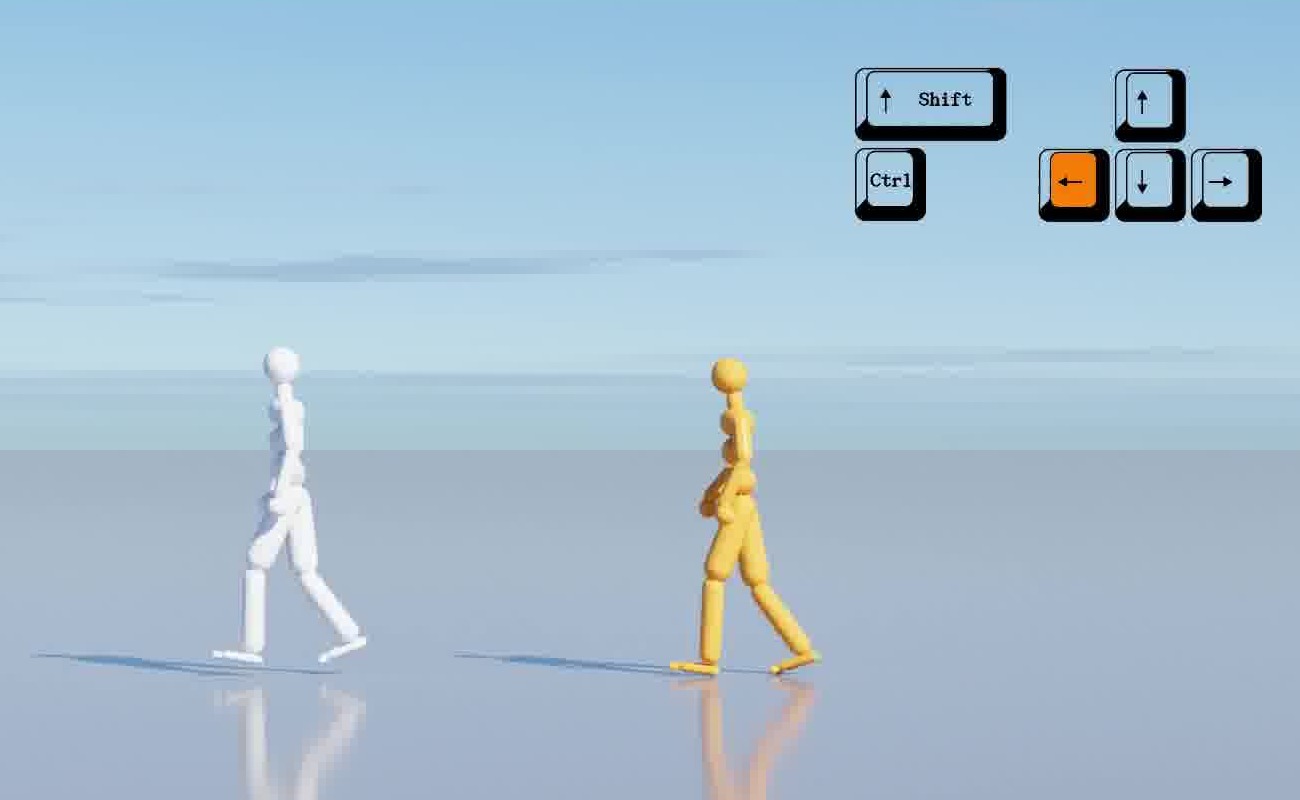}
    \end{subfigure}

    \begin{subfigure}{0.15\textwidth}
        \centering
        \includegraphics[width=\linewidth]{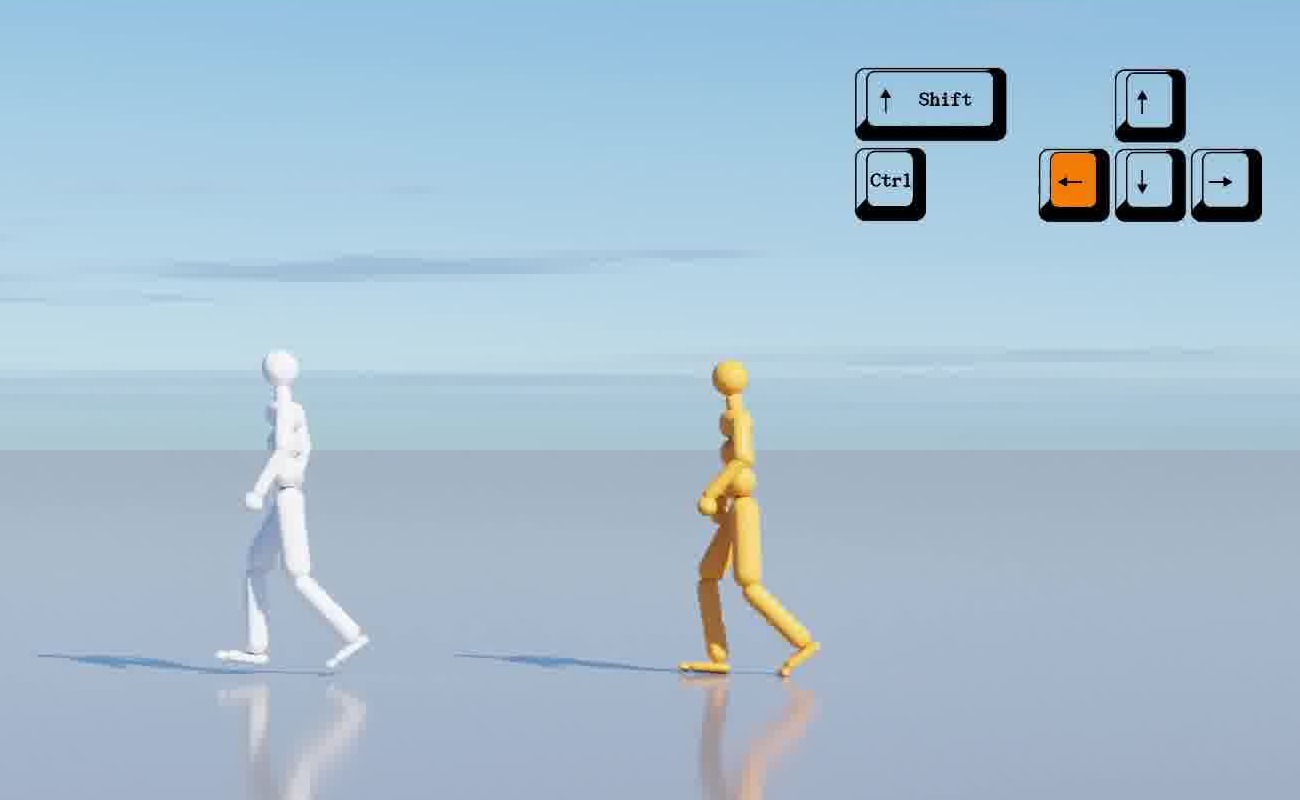}
    \end{subfigure}
    \begin{subfigure}{0.15\textwidth}
        \centering
        \includegraphics[width=\linewidth]{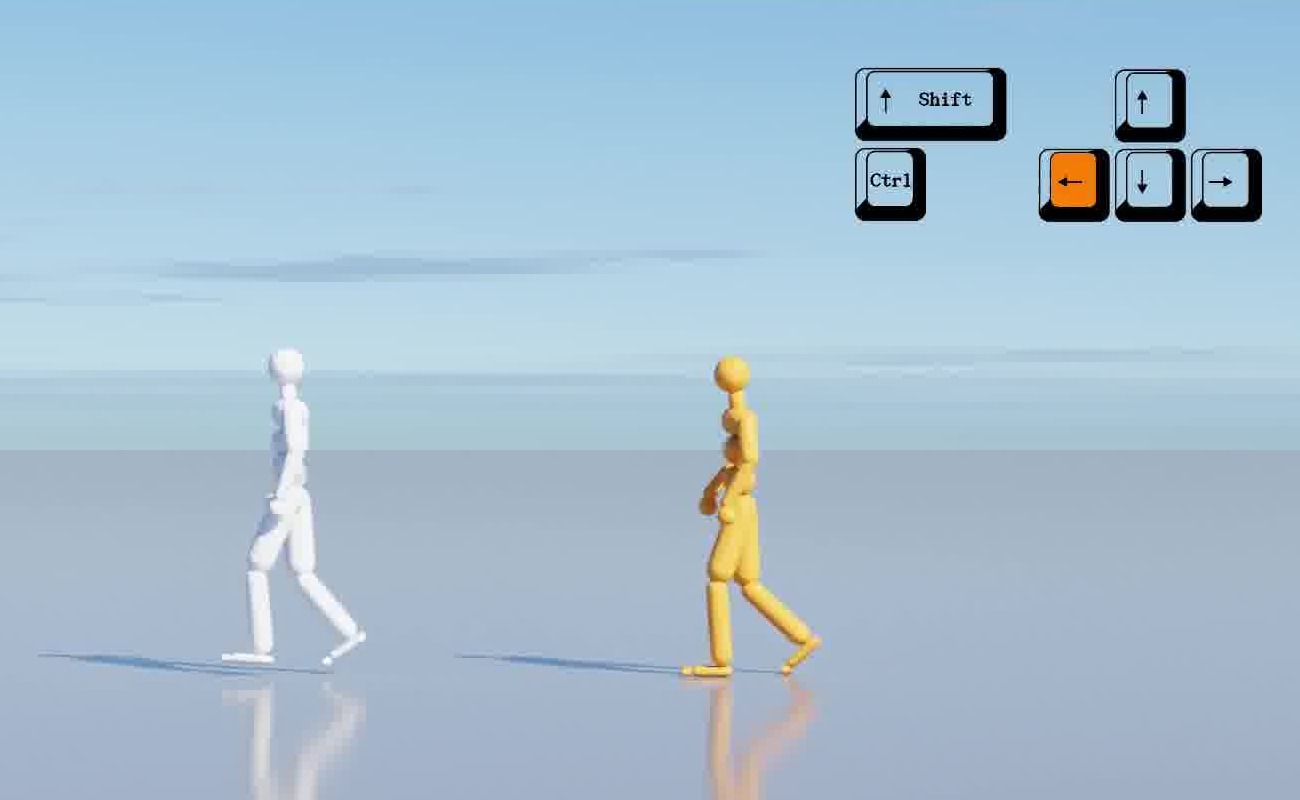}
    \end{subfigure}
    \begin{subfigure}{0.15\textwidth}
        \centering
        \includegraphics[width=\linewidth]{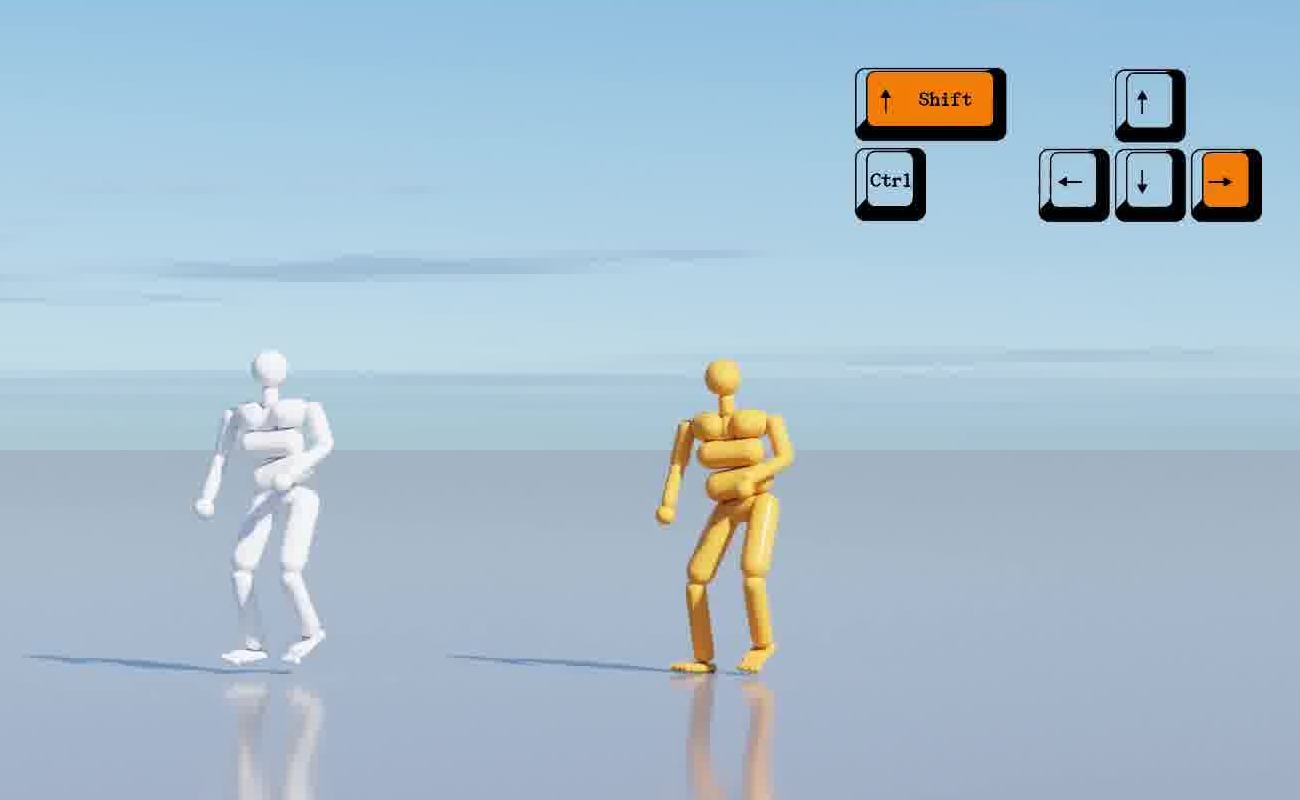}
    \end{subfigure}

    \begin{subfigure}{0.15\textwidth}
        \centering
        \includegraphics[width=\linewidth]{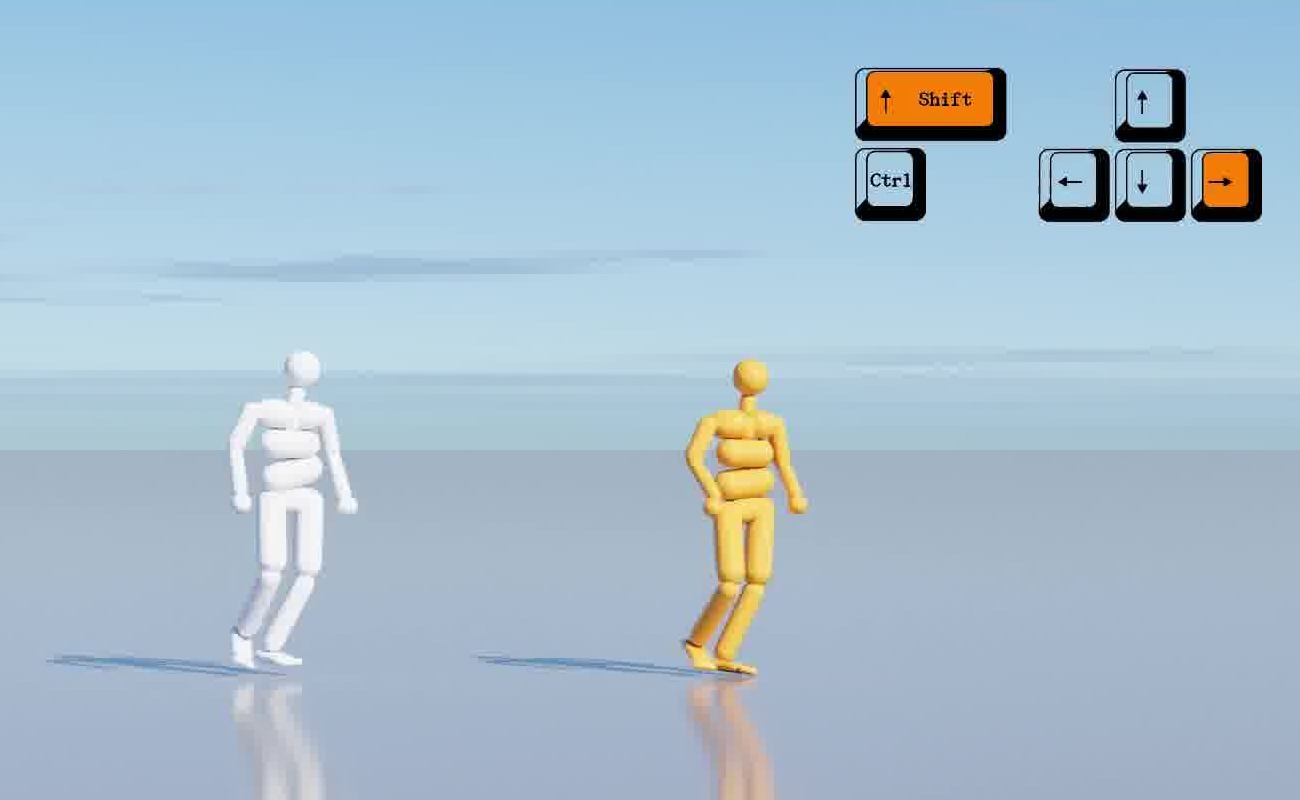}
    \end{subfigure}
    \begin{subfigure}{0.15\textwidth}
        \centering
        \includegraphics[width=\linewidth]{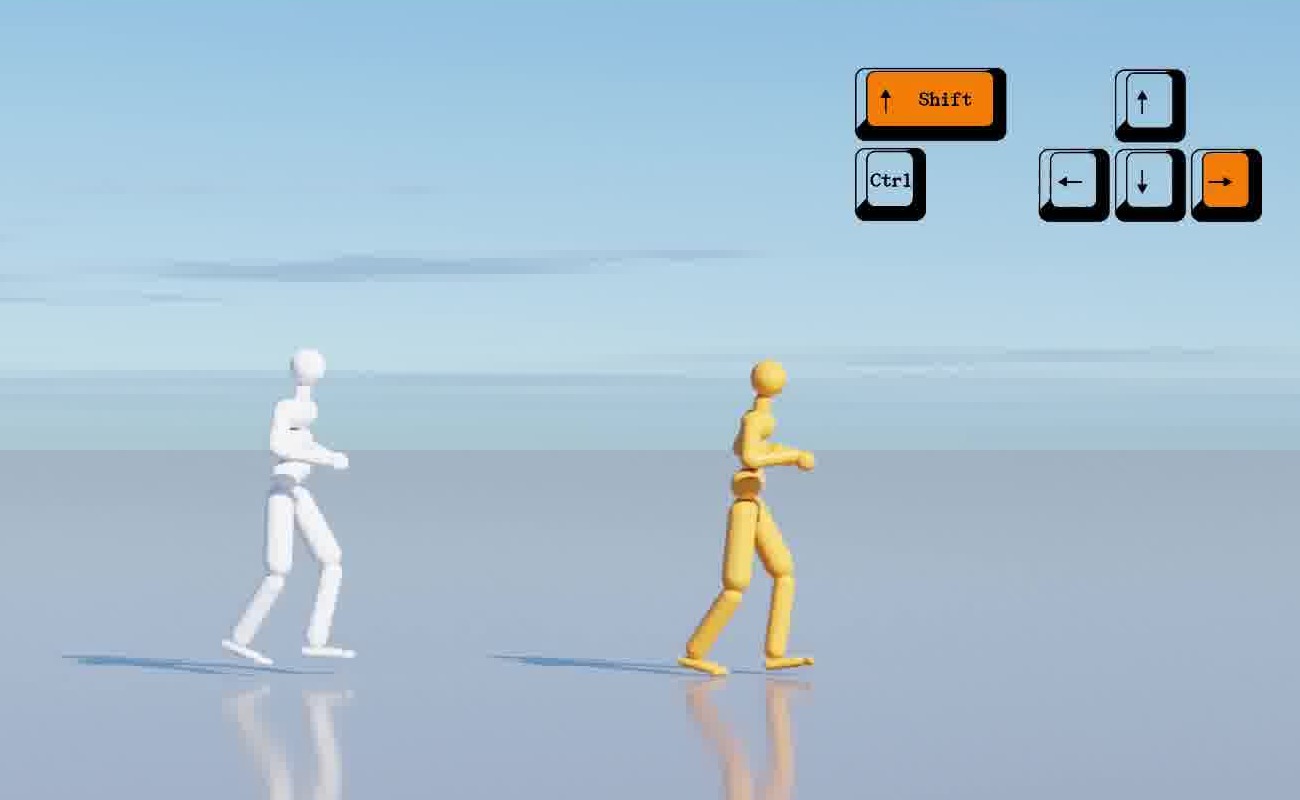}
    \end{subfigure}
    \begin{subfigure}{0.15\textwidth}
        \centering
        \includegraphics[width=\linewidth]{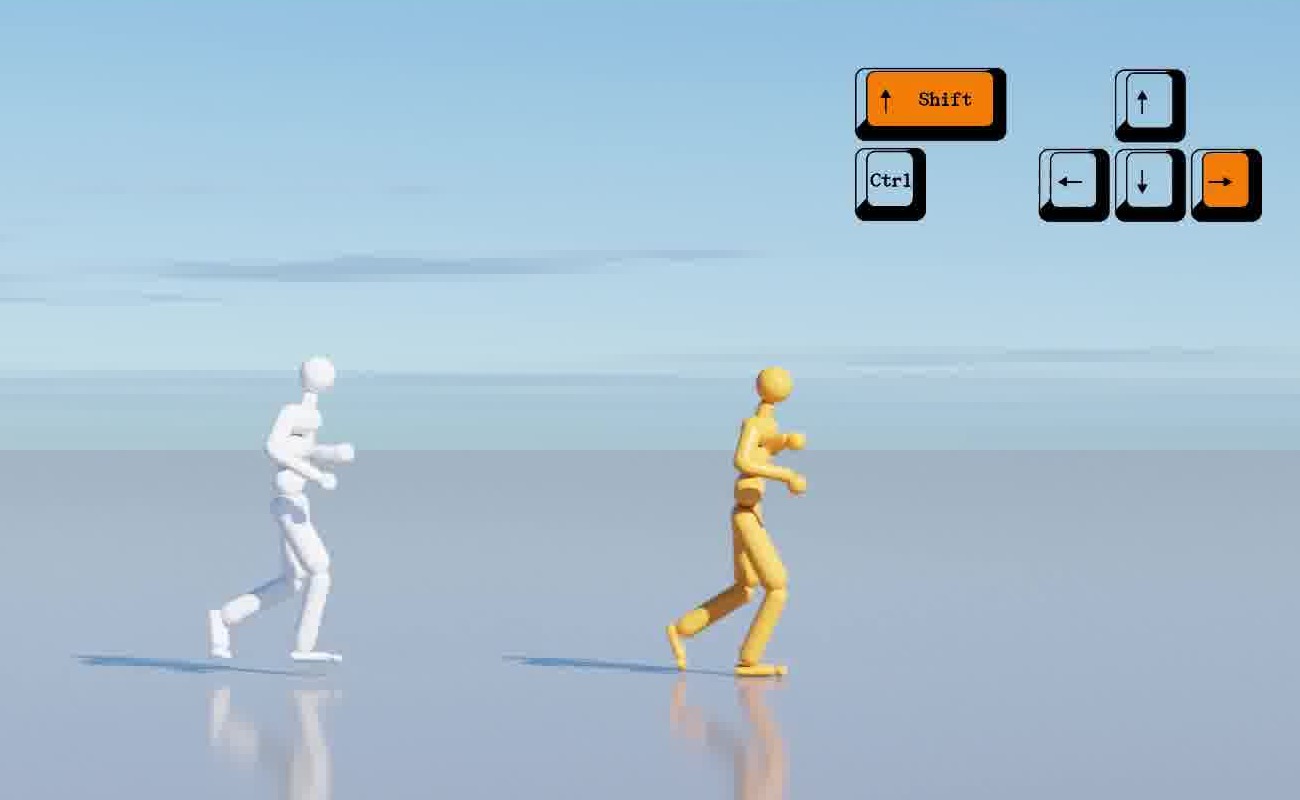}
    \end{subfigure}
    \caption{{Snapshots of the keyboard driven interactive control application. Note that our controller is real-time responsive to keyboard commands. On the left is the target agent states, and on the right, the yellow agents represent ones that are physically simulated by our algorithm.}}
    \label{fig:keyboard_driven_results}
\end{figure}
\begin{figure*}[!t]
    \centering
    \begin{subfigure}{0.15\textwidth}
        \centering
        \includegraphics[width=\linewidth]{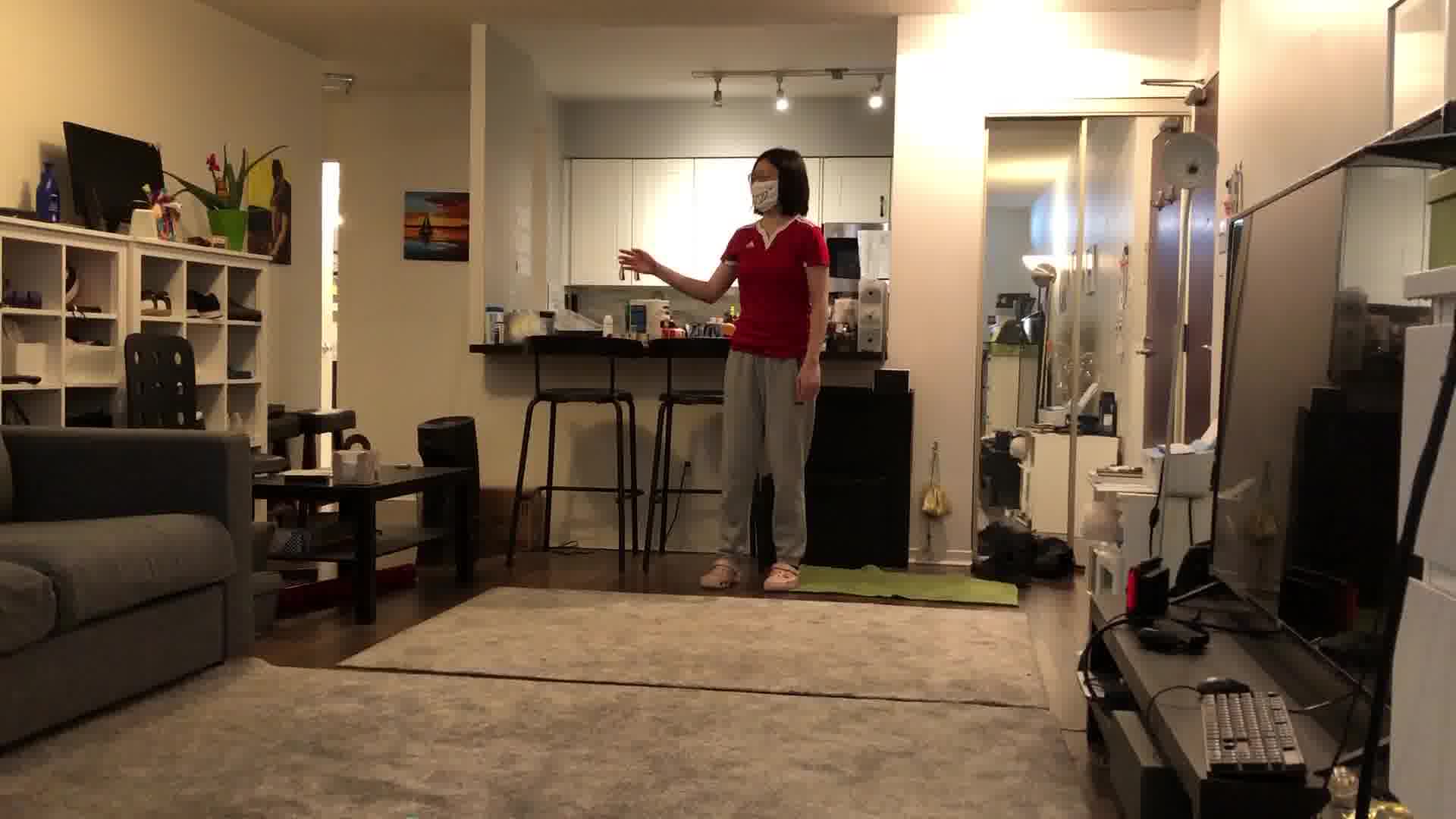}
    \end{subfigure}
    \begin{subfigure}{0.15\textwidth}
        \centering
        \includegraphics[width=\linewidth]{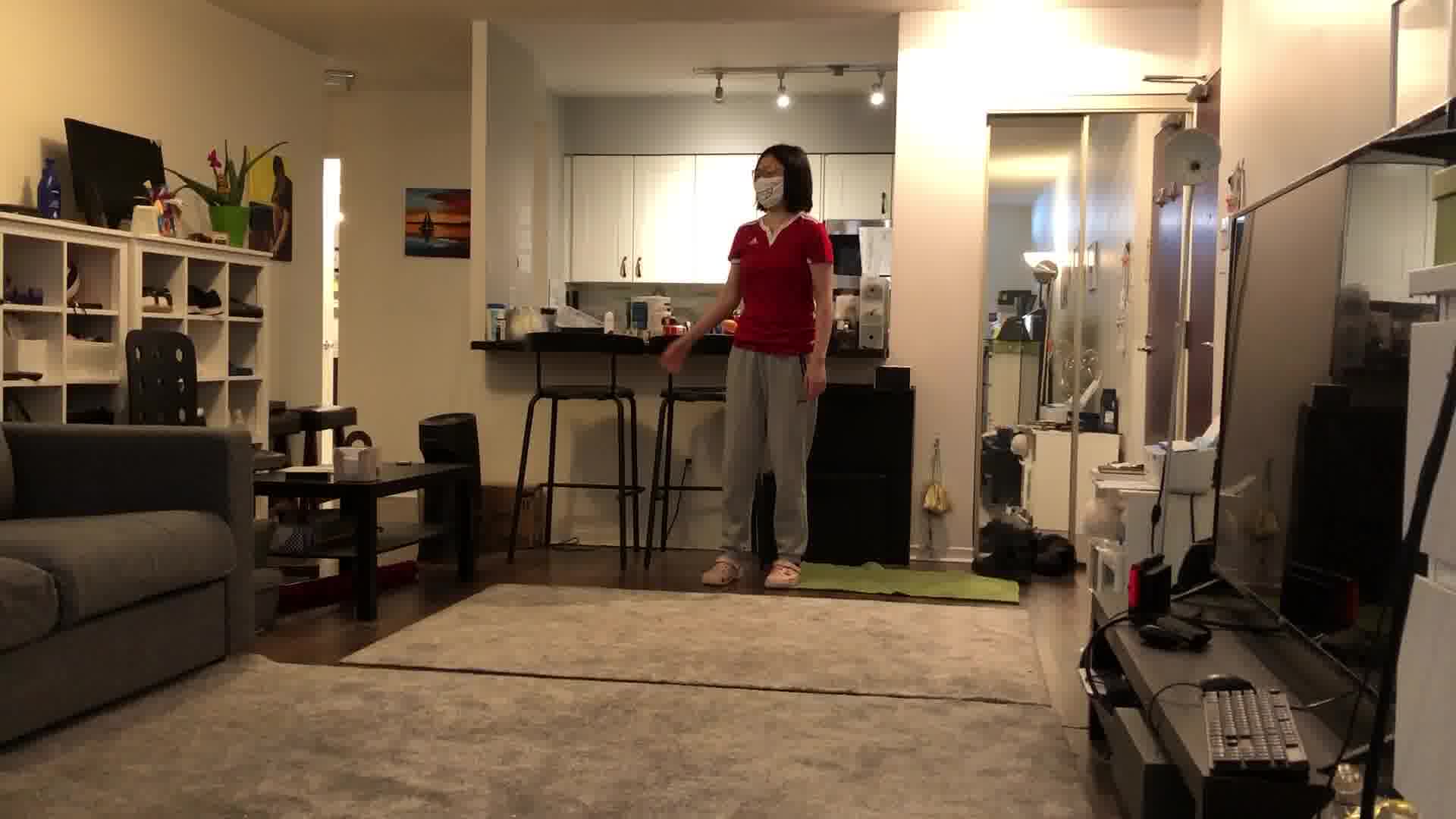}
    \end{subfigure}
    \begin{subfigure}{0.15\textwidth}
        \centering
        \includegraphics[width=\linewidth]{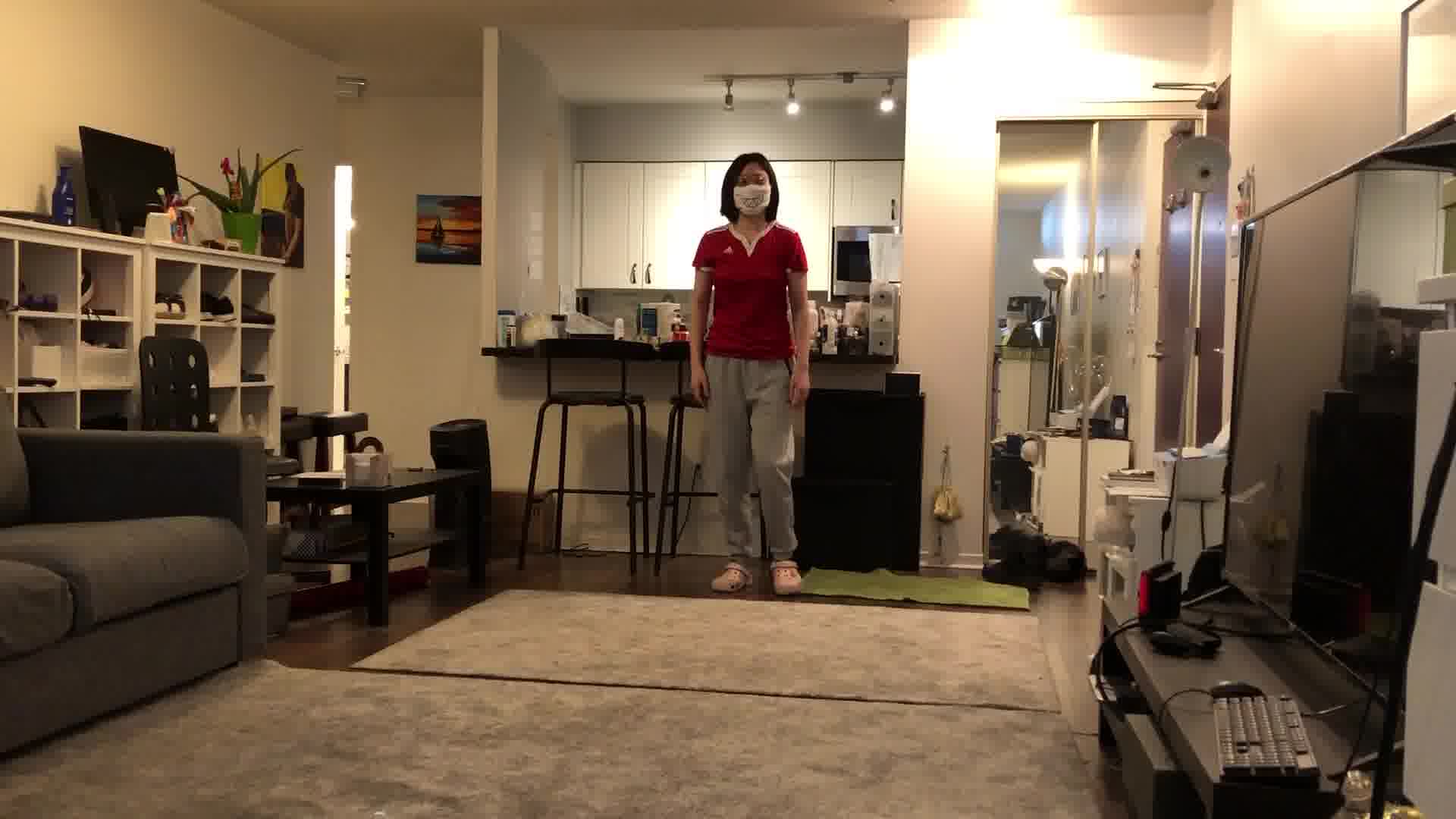}
    \end{subfigure}
    \begin{subfigure}{0.15\textwidth}
        \centering
        \includegraphics[width=\linewidth]{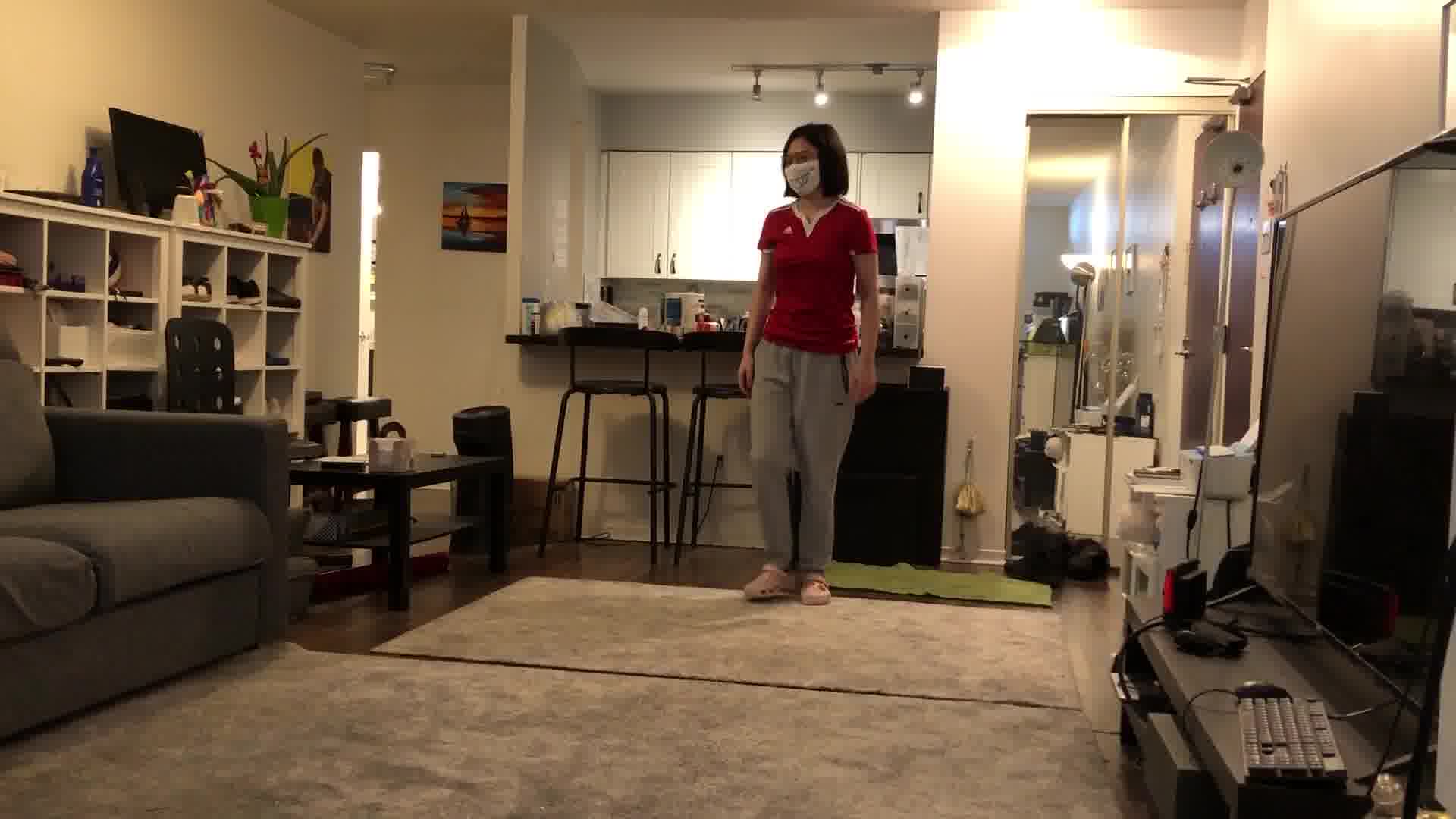}
    \end{subfigure}
    \begin{subfigure}{0.15\textwidth}
        \centering
        \includegraphics[width=\linewidth]{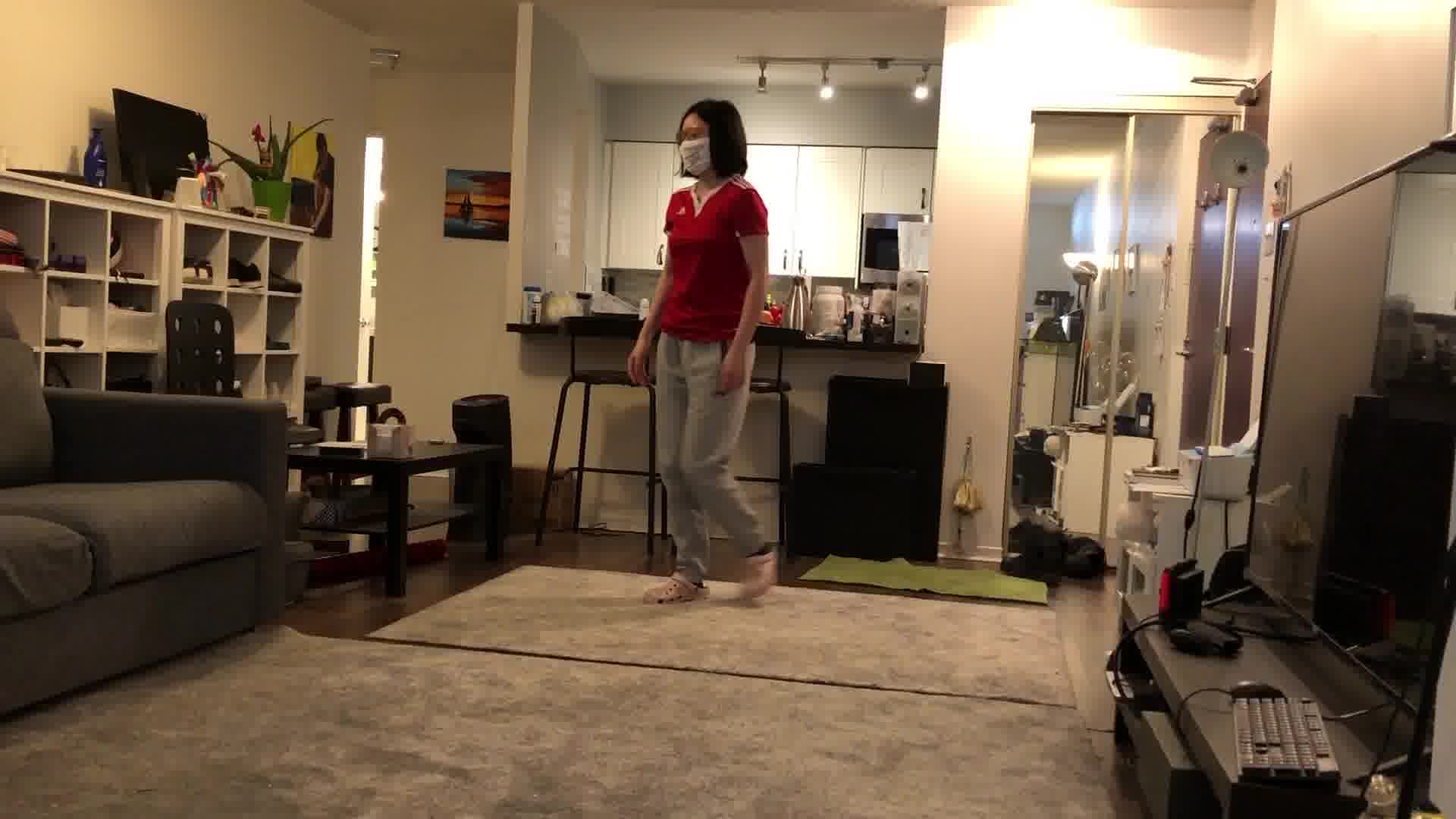}
    \end{subfigure}
    \begin{subfigure}{0.15\textwidth}
        \centering
        \includegraphics[width=\linewidth]{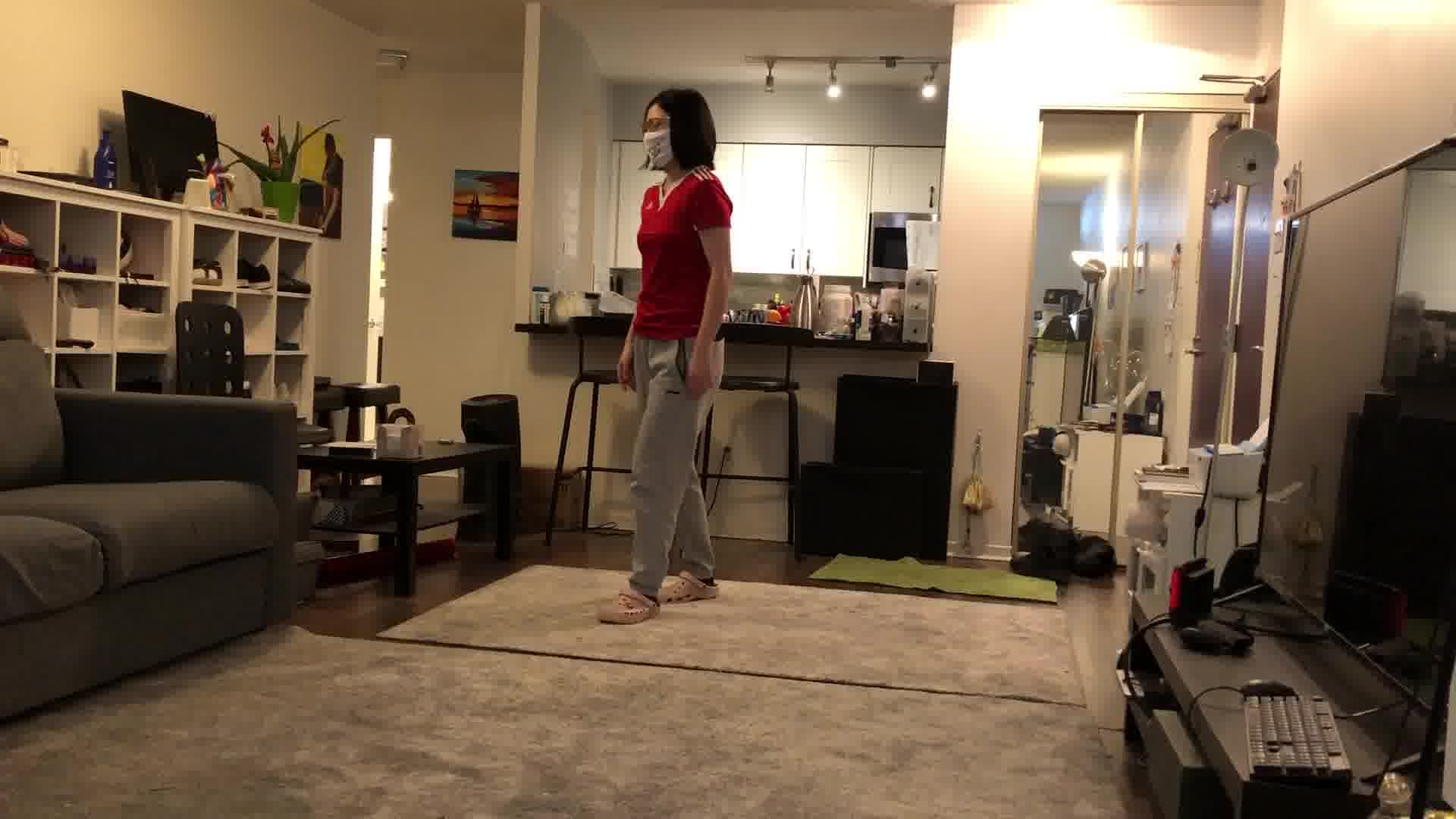}
    \end{subfigure}

    \begin{subfigure}{0.15\textwidth}
        \centering
        \includegraphics[width=\linewidth]{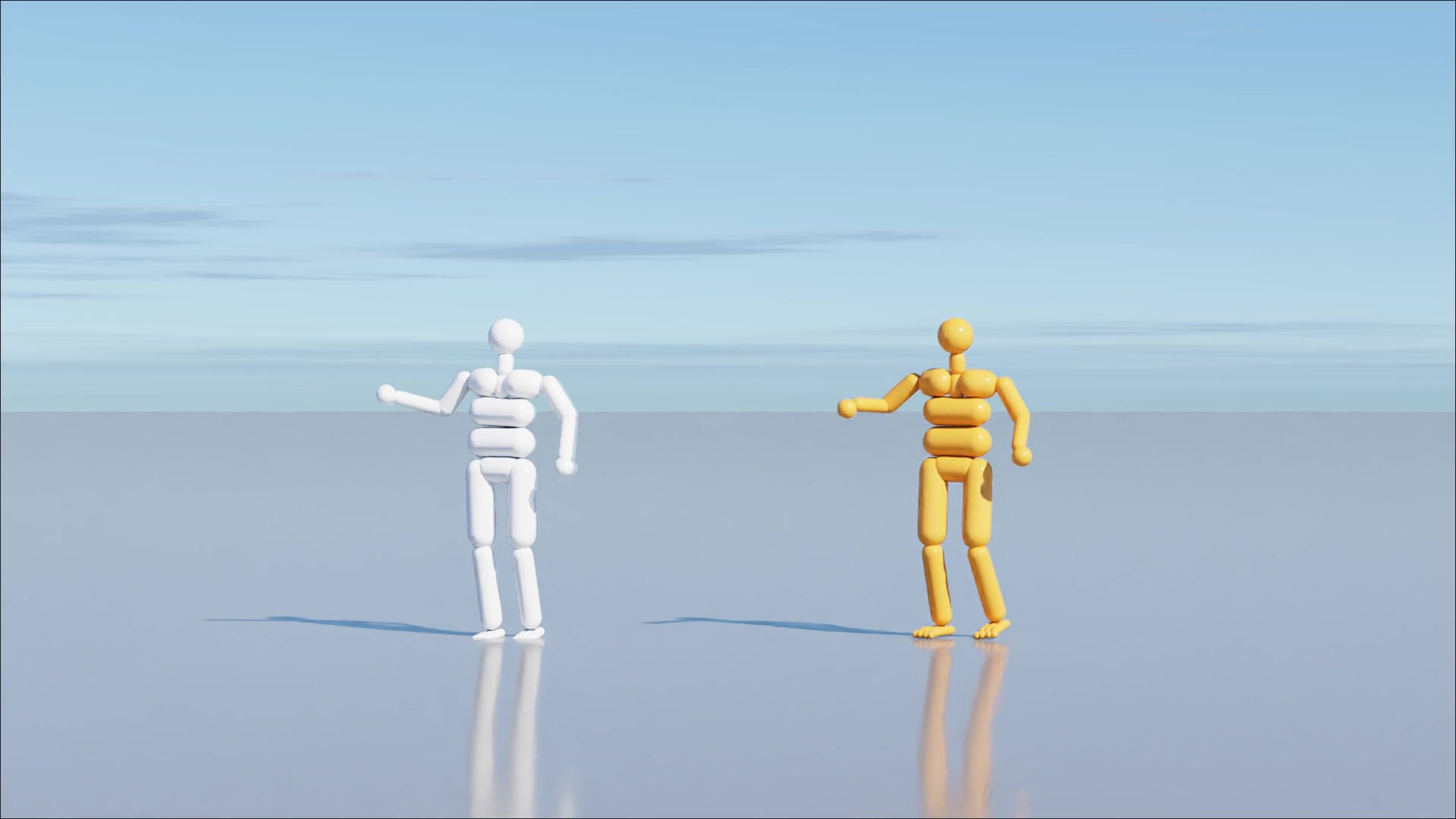}
    \end{subfigure}
    \begin{subfigure}{0.15\textwidth}
        \centering
        \includegraphics[width=\linewidth]{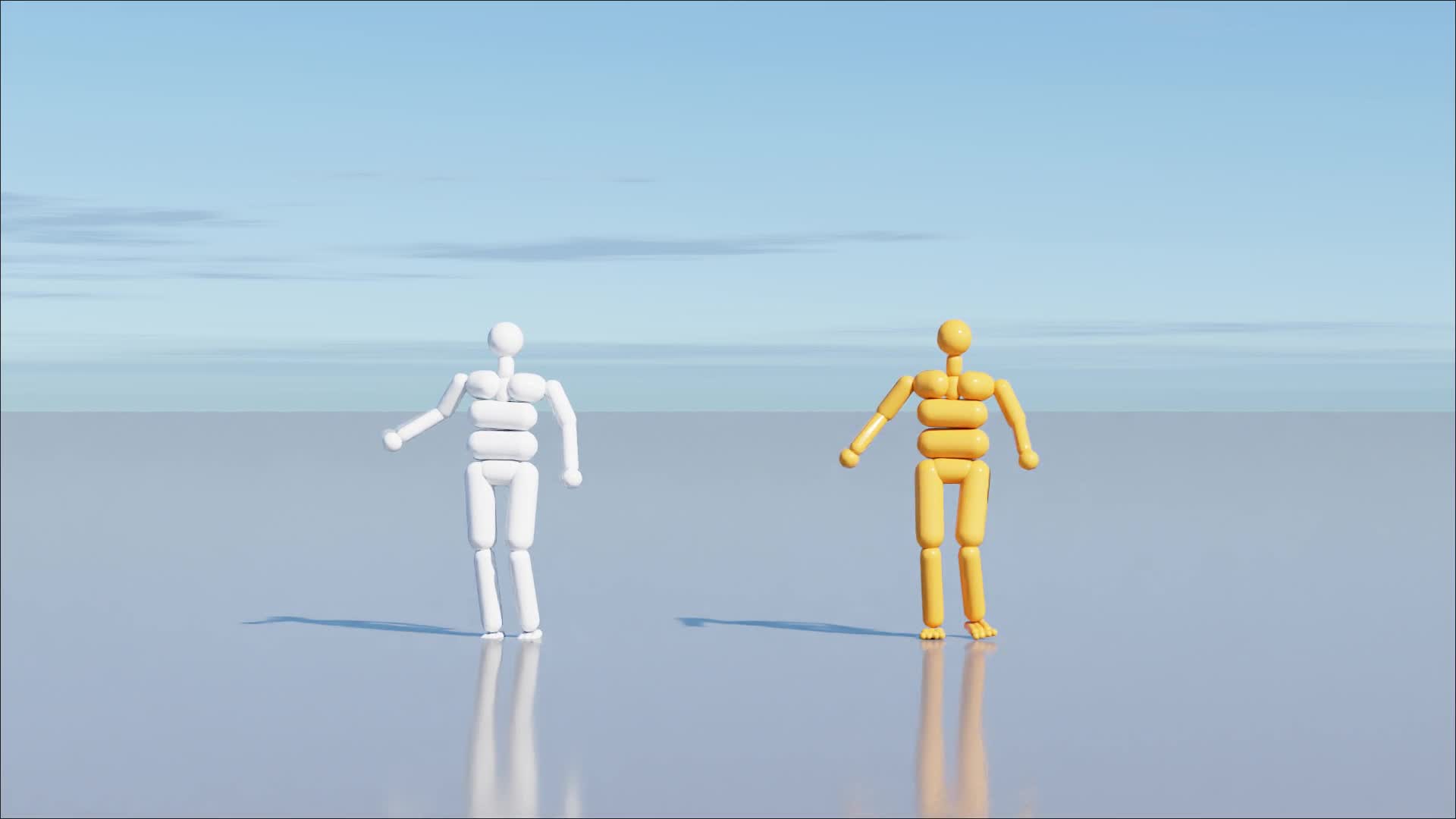}
    \end{subfigure}
    \begin{subfigure}{0.15\textwidth}
        \centering
        \includegraphics[width=\linewidth]{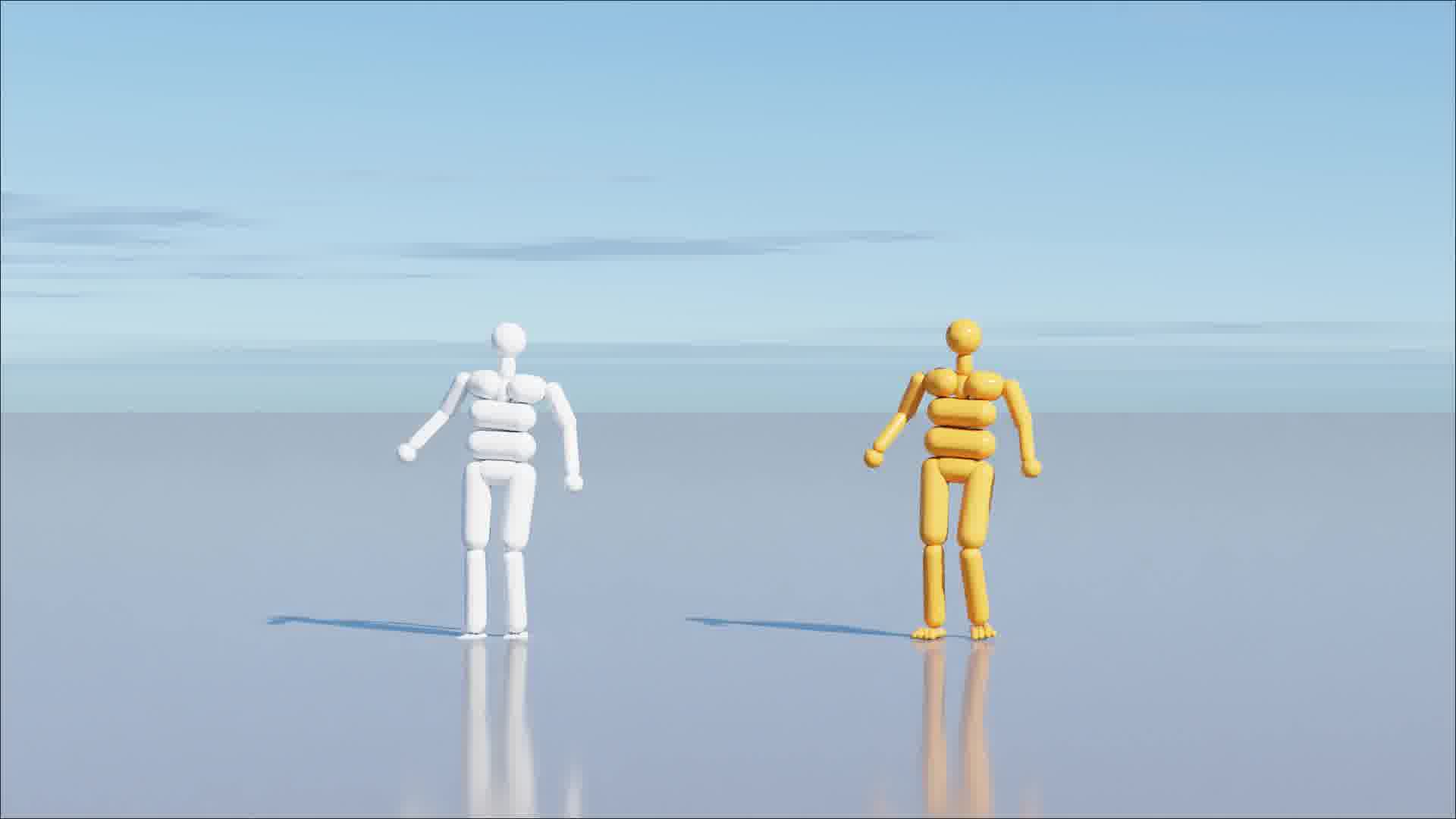}
    \end{subfigure}
    \begin{subfigure}{0.15\textwidth}
        \centering
        \includegraphics[width=\linewidth]{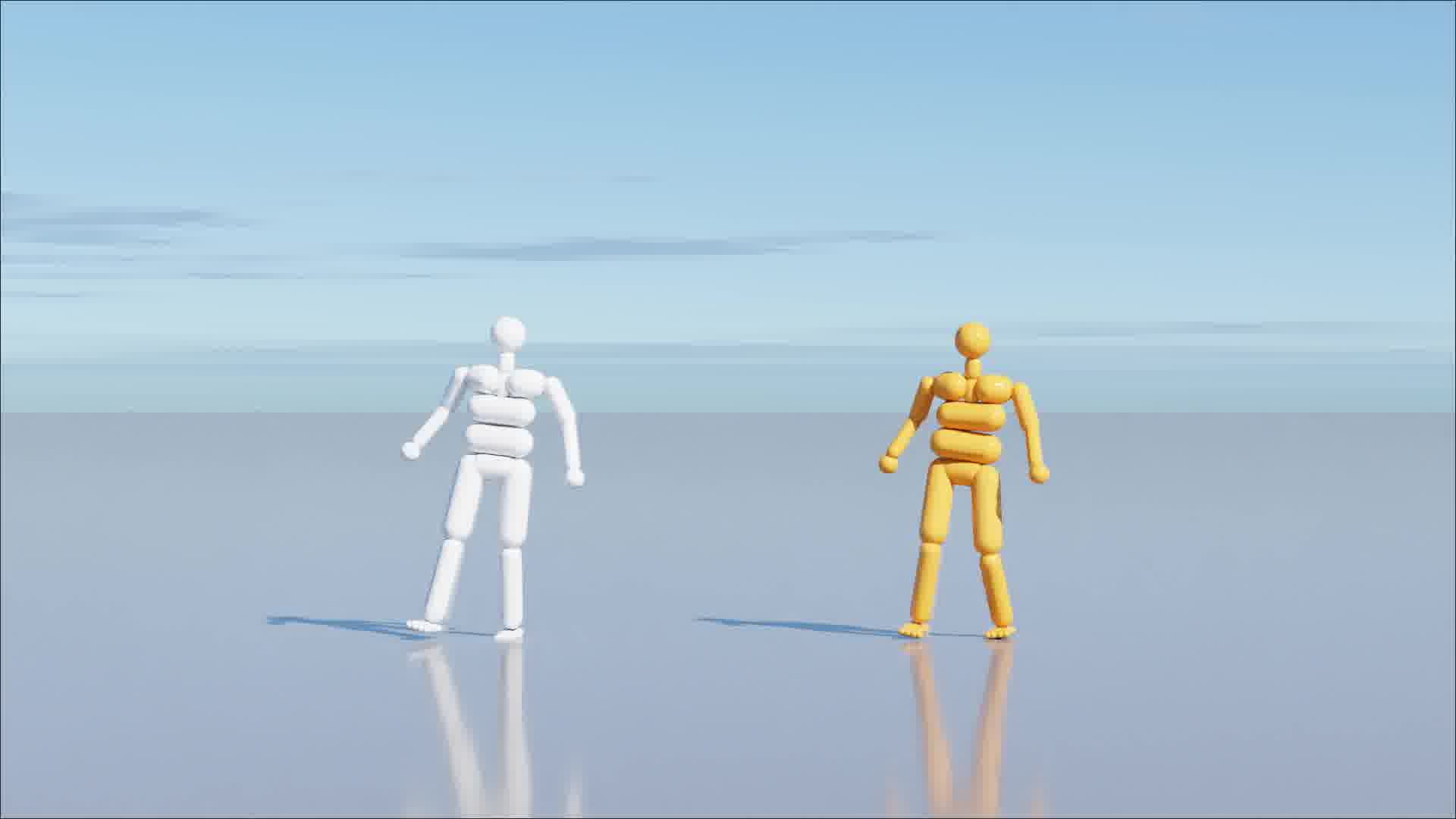}
    \end{subfigure}
    \begin{subfigure}{0.15\textwidth}
        \centering
        \includegraphics[width=\linewidth]{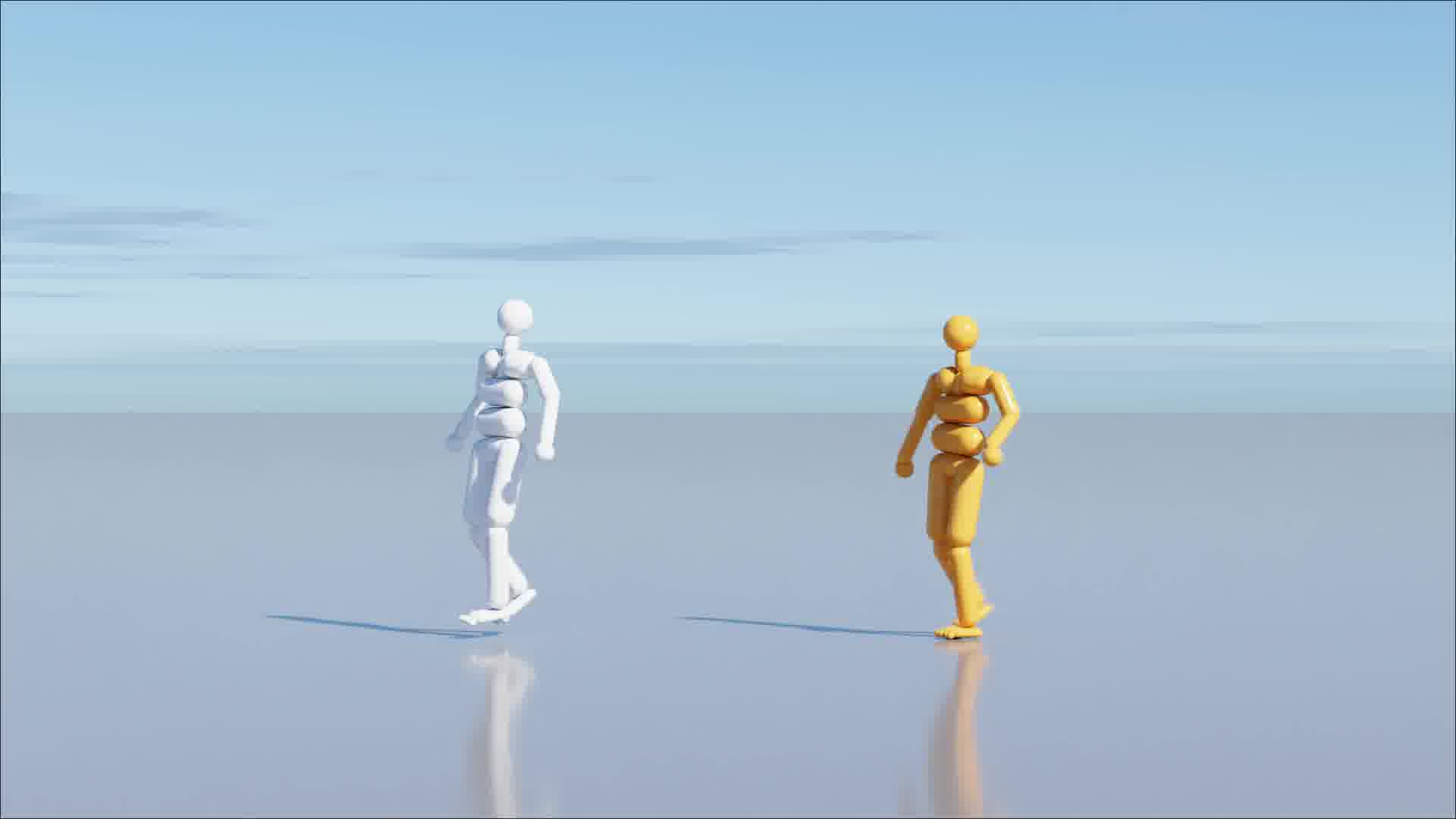}
    \end{subfigure}
    \begin{subfigure}{0.15\textwidth}
        \centering
        \includegraphics[width=\linewidth]{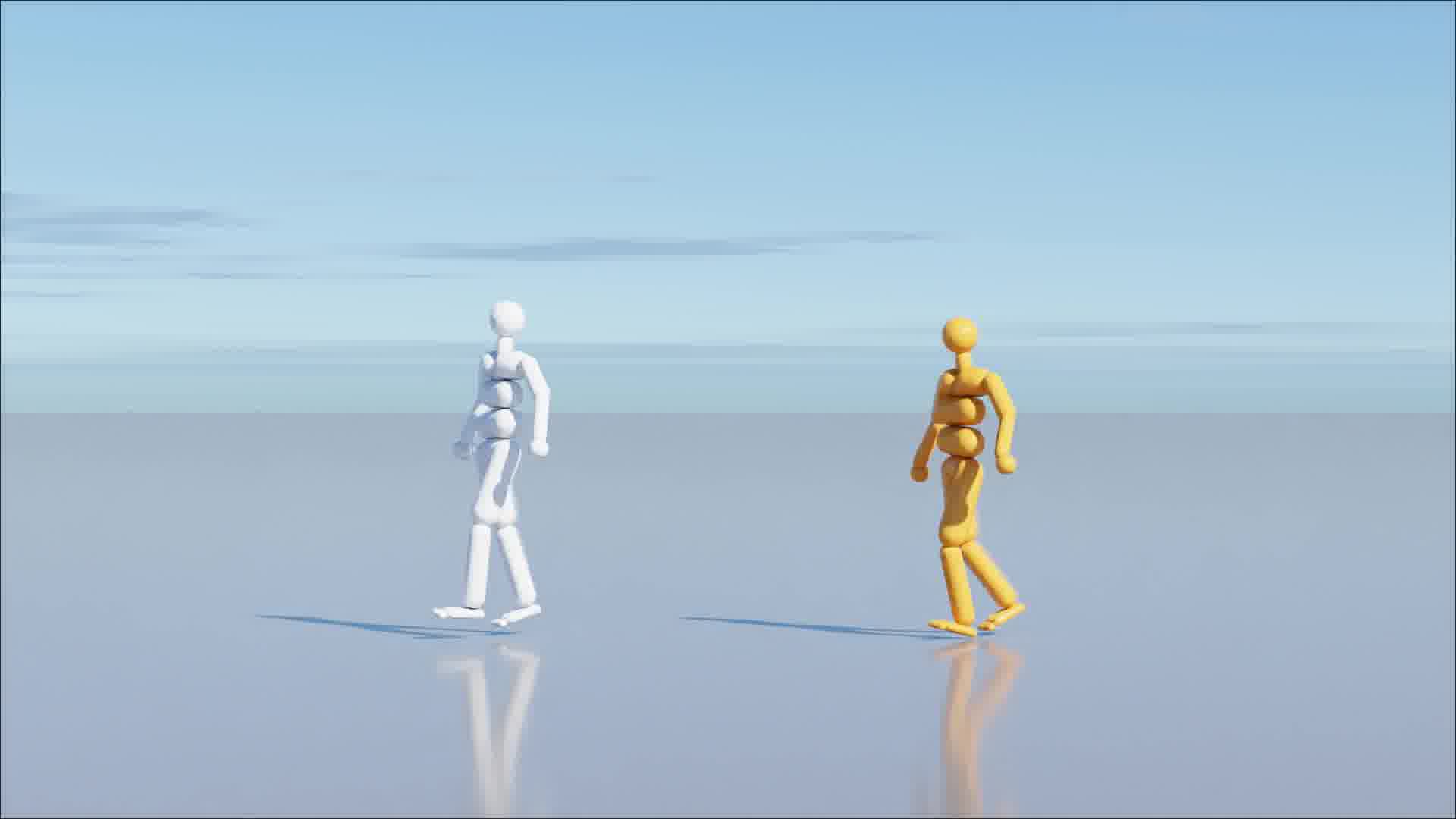}
    \end{subfigure}

    \begin{subfigure}{0.15\textwidth}
        \centering
        \includegraphics[width=\linewidth]{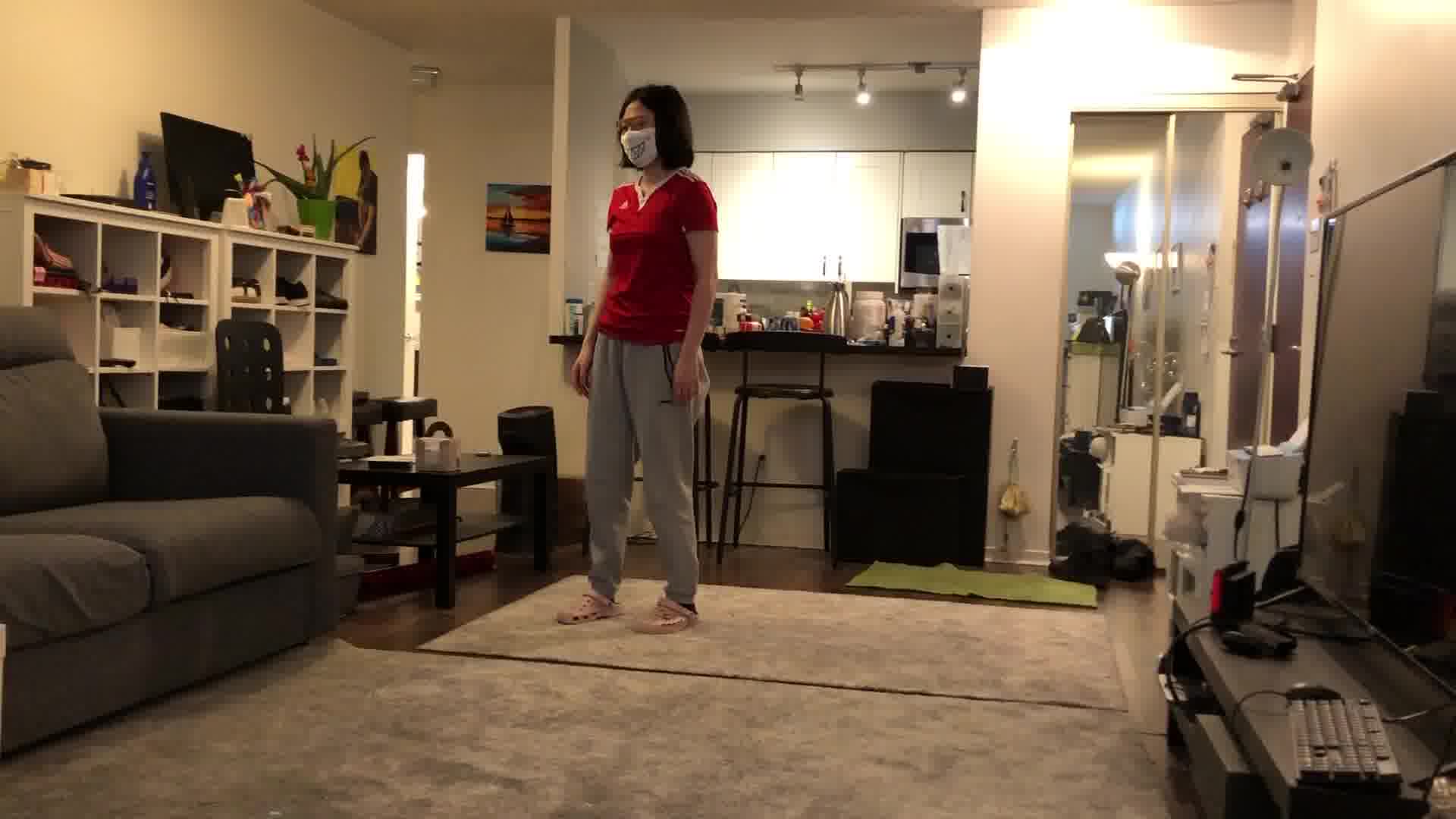}
    \end{subfigure}
    \begin{subfigure}{0.15\textwidth}
        \centering
        \includegraphics[width=\linewidth]{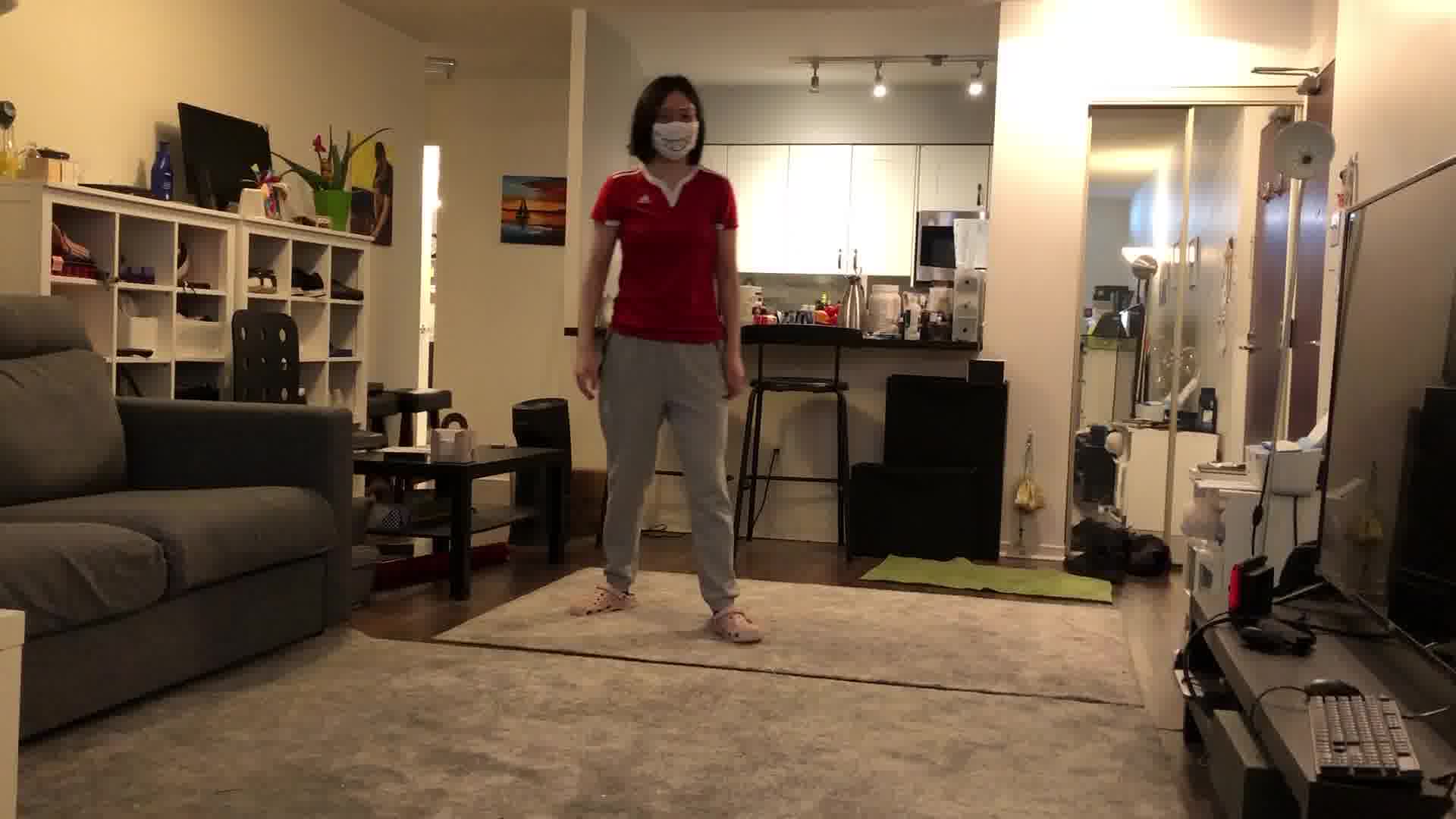}
    \end{subfigure}
    \begin{subfigure}{0.15\textwidth}
        \centering
        \includegraphics[width=\linewidth]{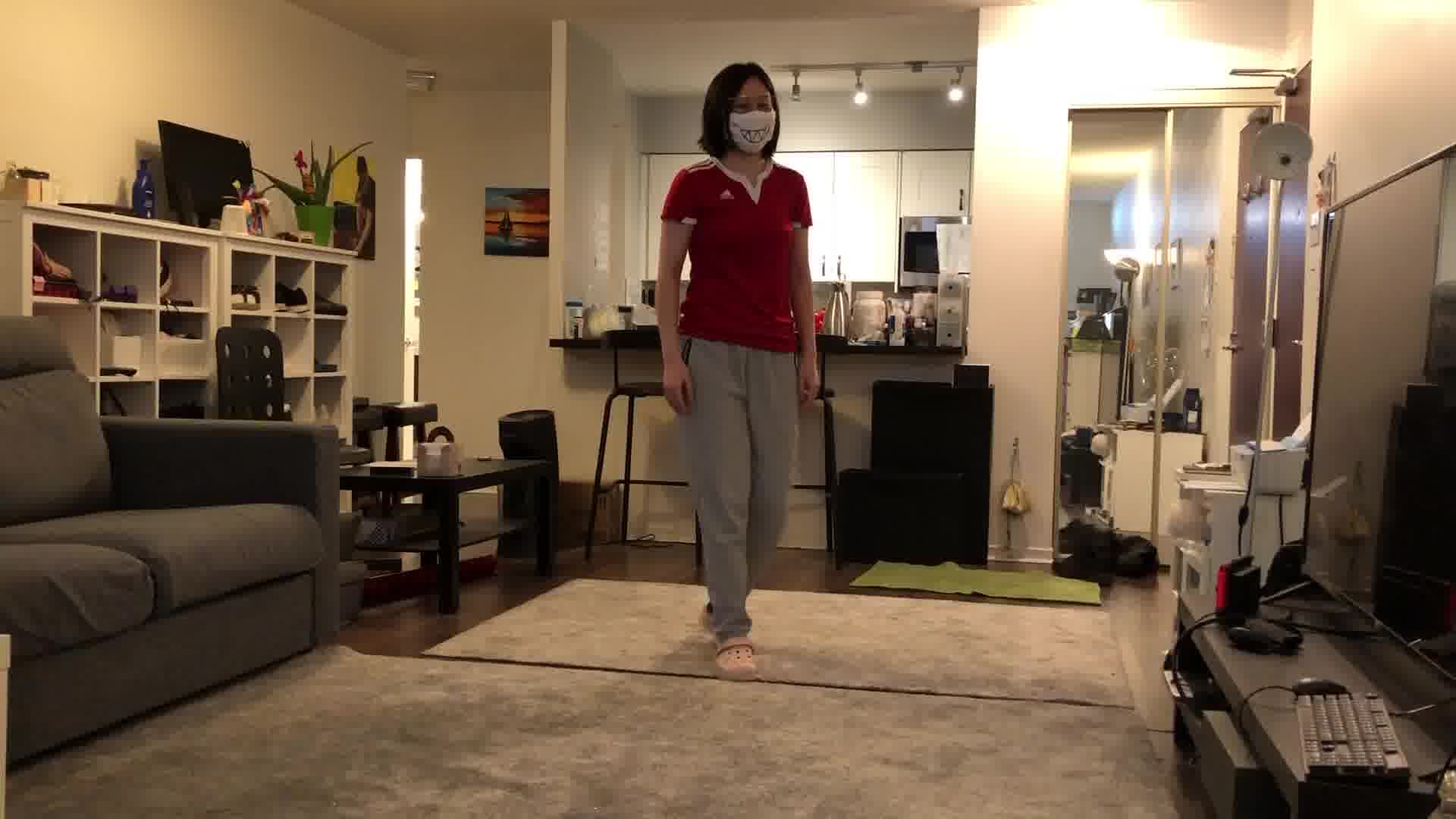}
    \end{subfigure}
    \begin{subfigure}{0.15\textwidth}
        \centering
        \includegraphics[width=\linewidth]{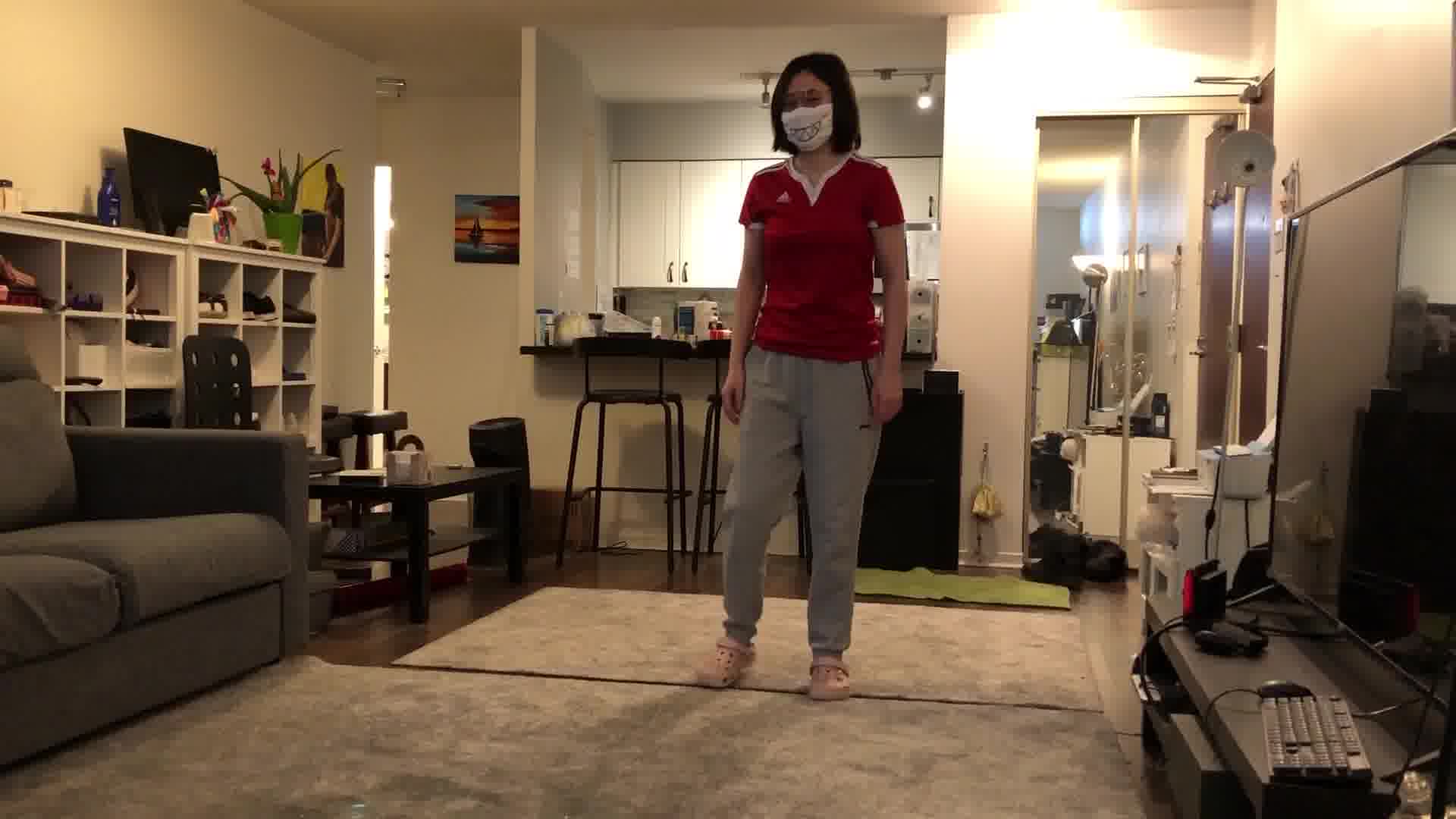}
    \end{subfigure}
    \begin{subfigure}{0.15\textwidth}
        \centering
        \includegraphics[width=\linewidth]{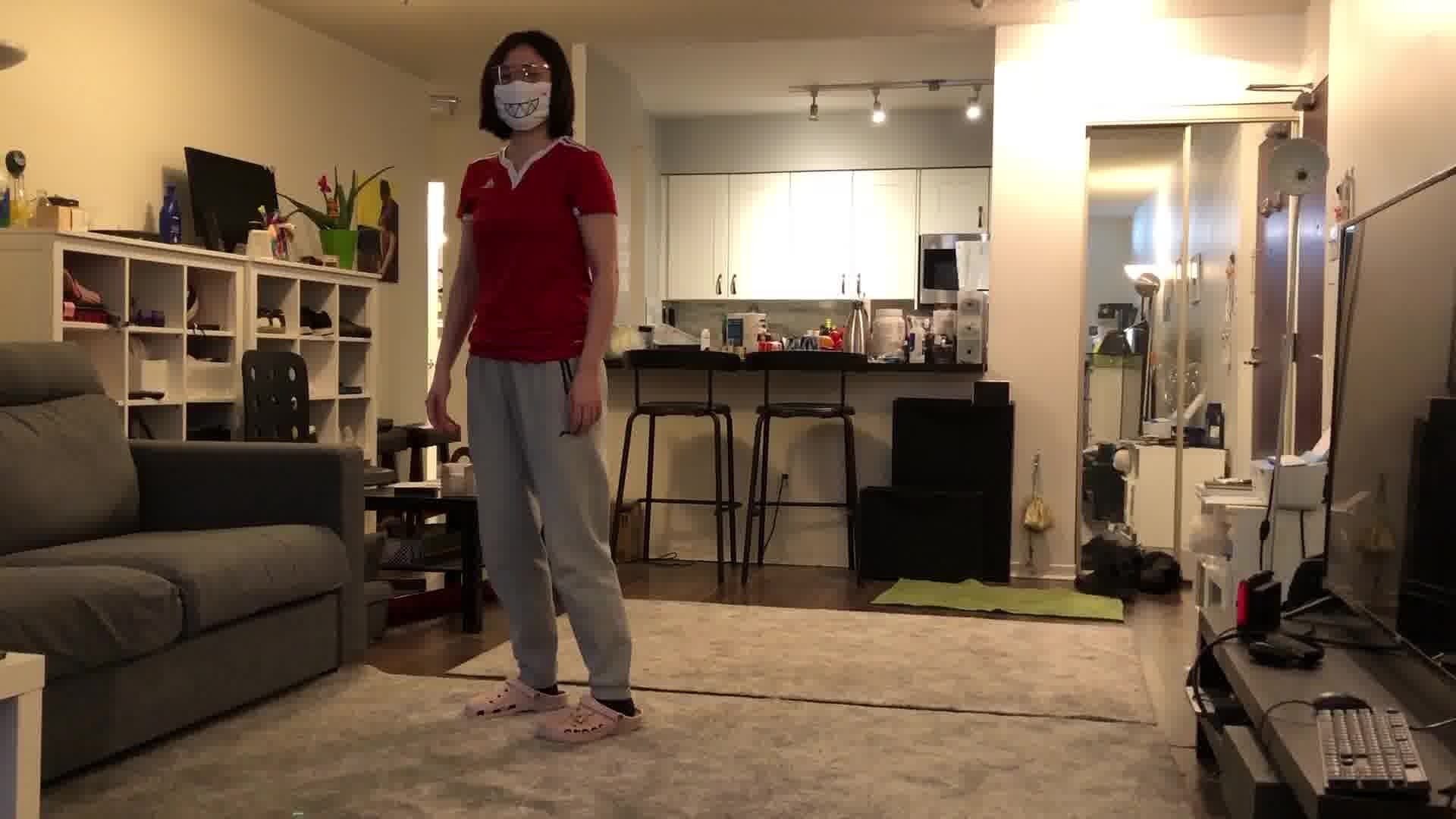}
    \end{subfigure}
    \begin{subfigure}{0.15\textwidth}
        \centering
        \includegraphics[width=\linewidth]{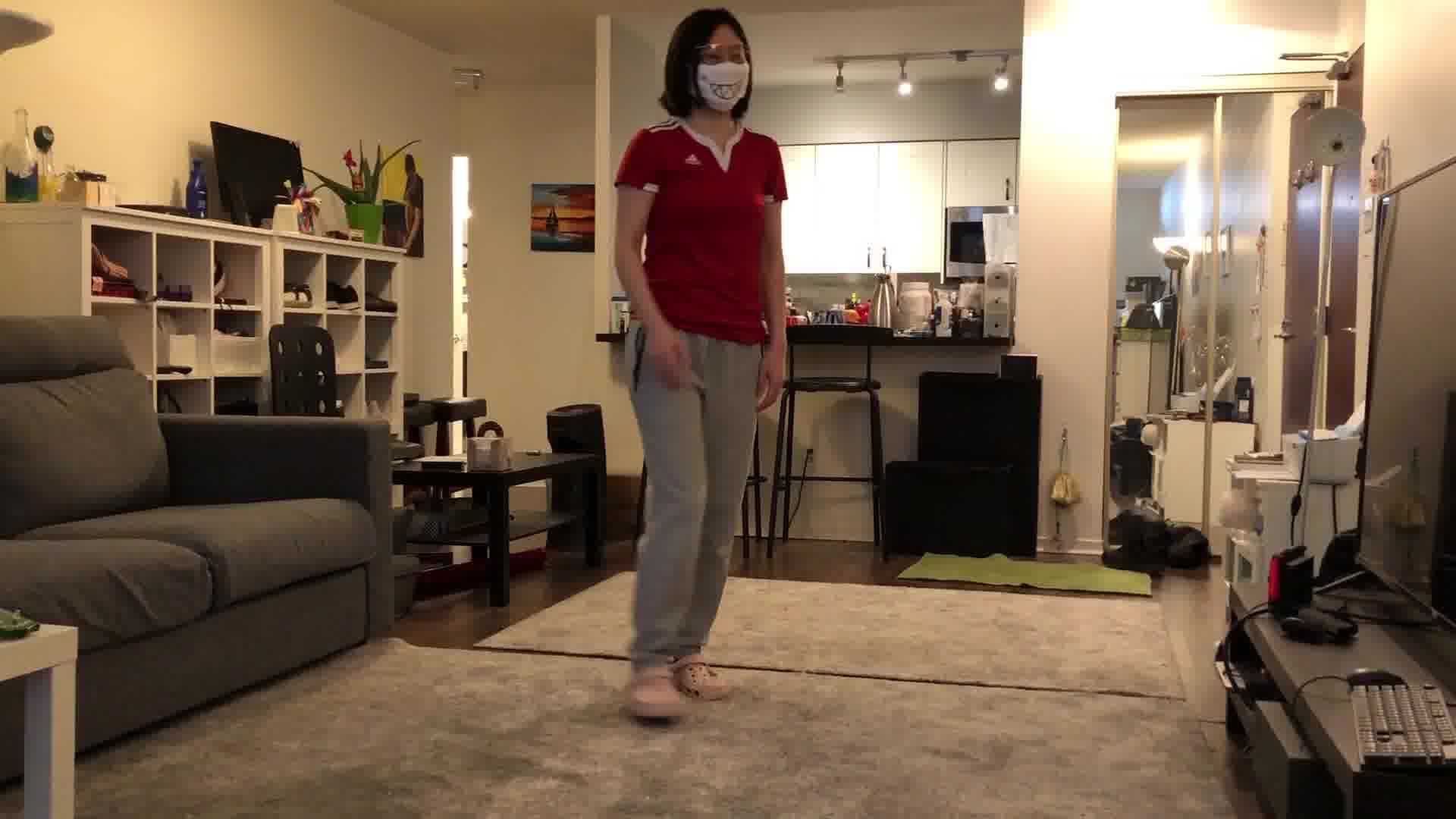}
    \end{subfigure}

    \begin{subfigure}{0.15\textwidth}
        \centering
        \includegraphics[width=\linewidth]{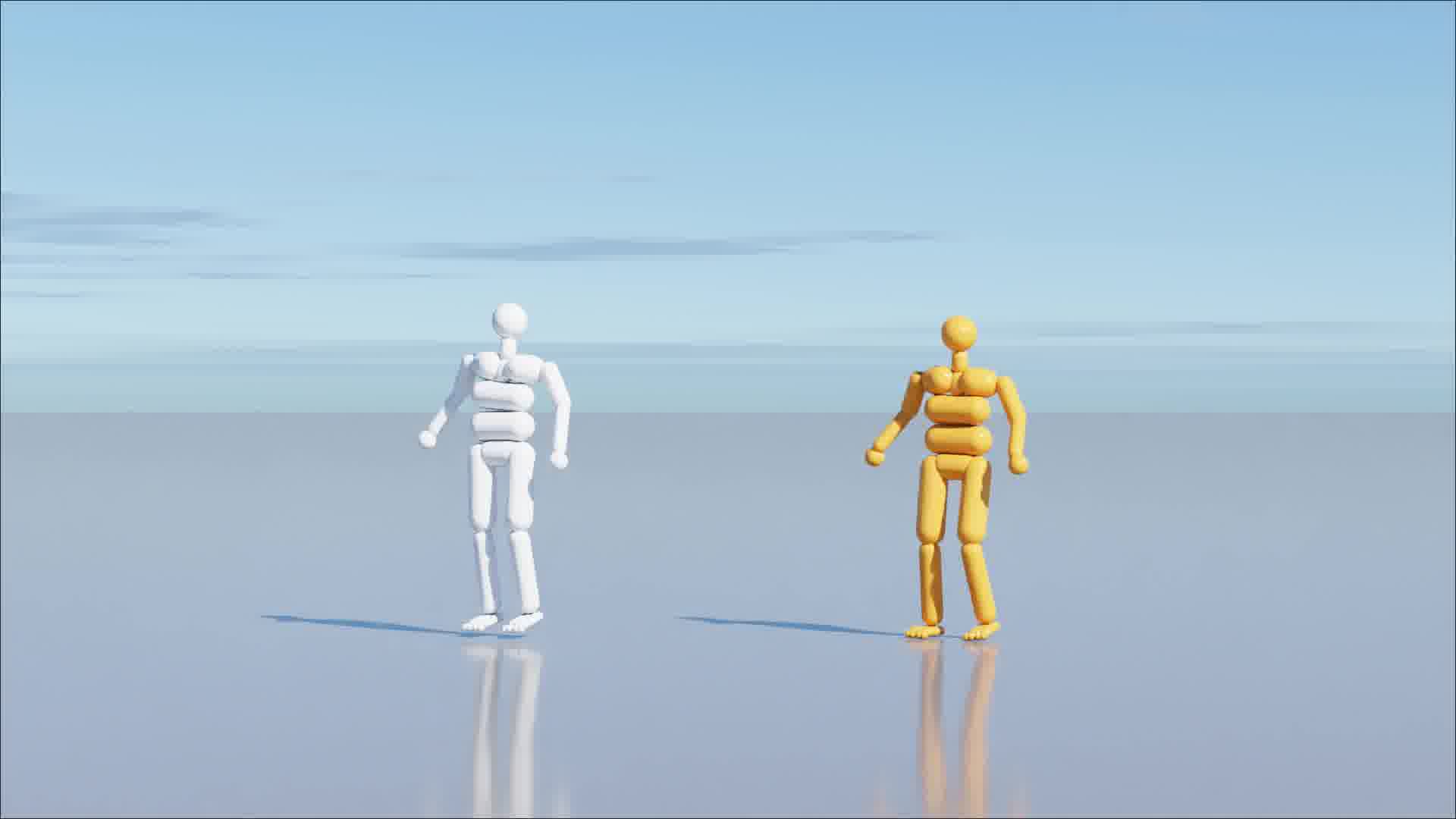}
    \end{subfigure}
    \begin{subfigure}{0.15\textwidth}
        \centering
        \includegraphics[width=\linewidth]{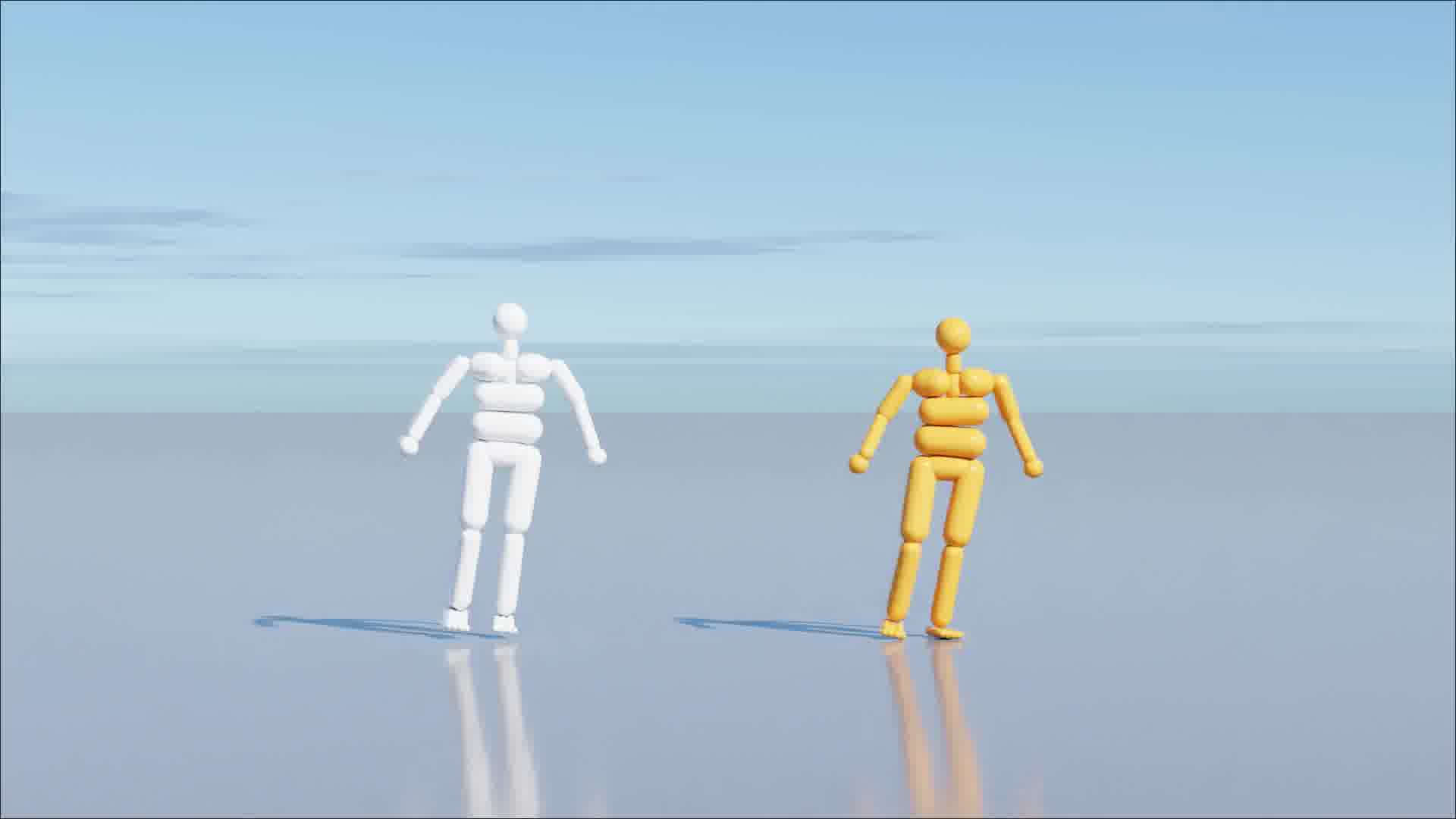}
    \end{subfigure}
    \begin{subfigure}{0.15\textwidth}
        \centering
        \includegraphics[width=\linewidth]{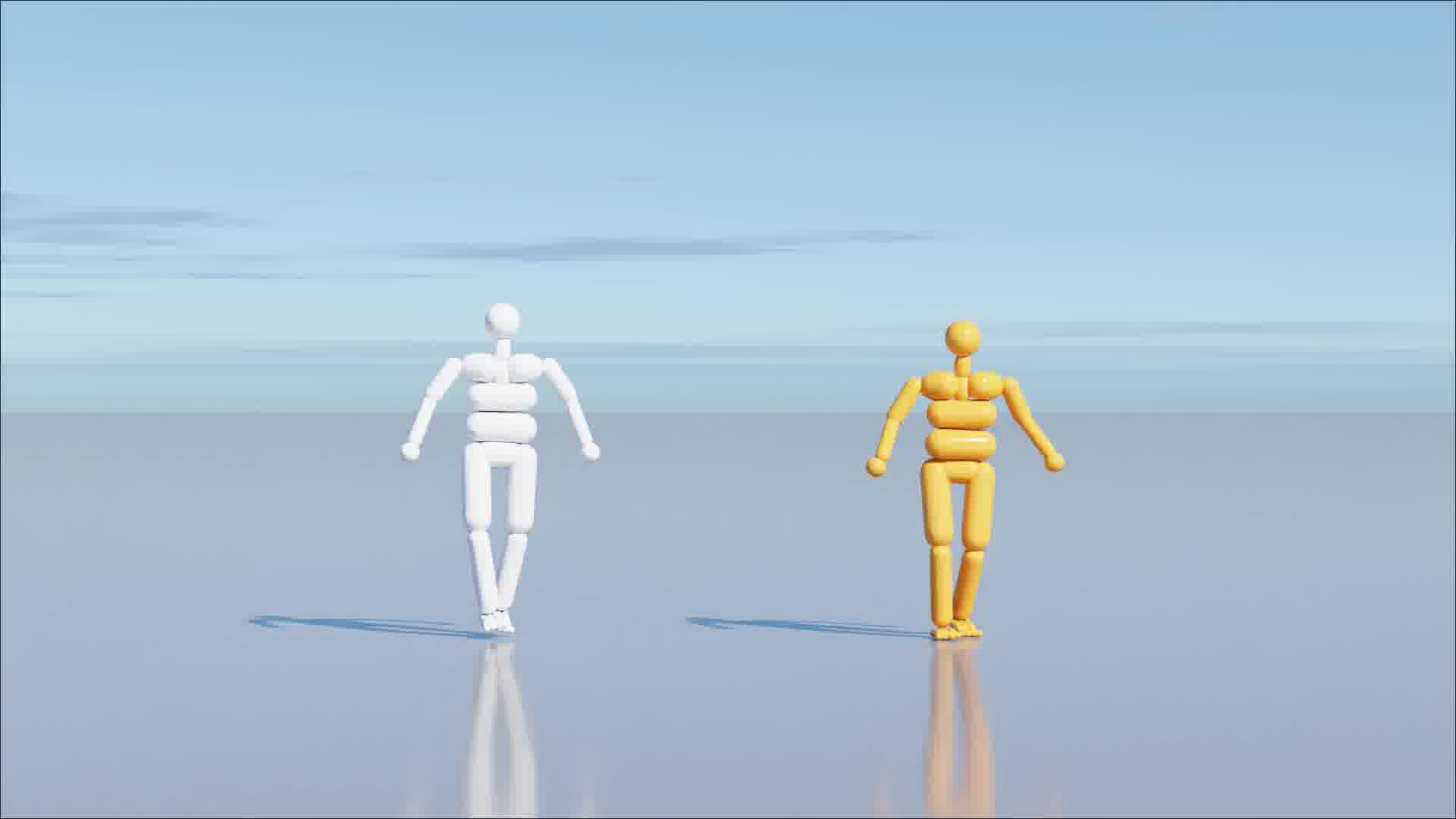}
    \end{subfigure}
    \begin{subfigure}{0.15\textwidth}
        \centering
        \includegraphics[width=\linewidth]{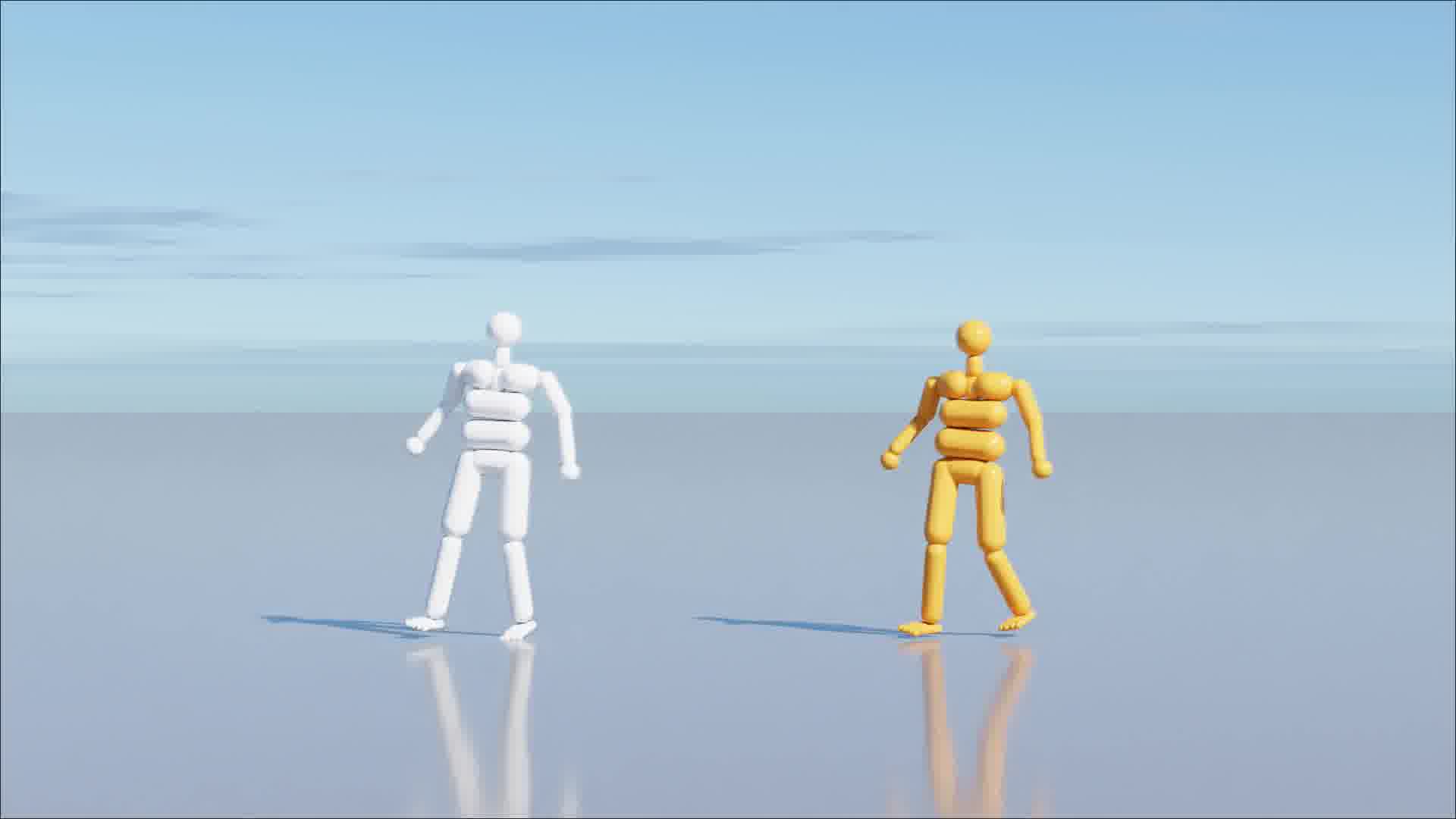}
    \end{subfigure}
    \begin{subfigure}{0.15\textwidth}
        \centering
        \includegraphics[width=\linewidth]{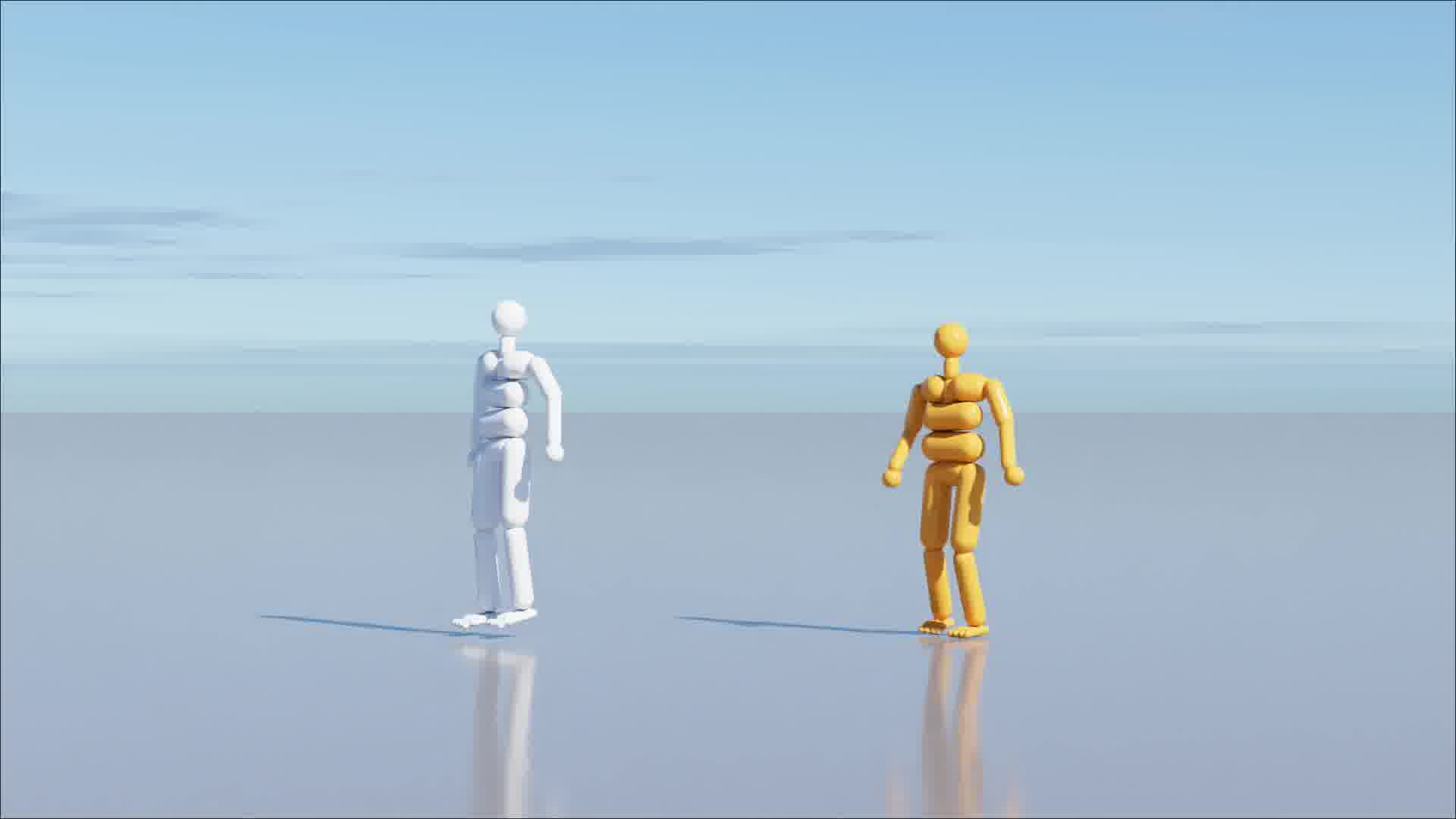}
    \end{subfigure}
    \begin{subfigure}{0.15\textwidth}
        \centering
        \includegraphics[width=\linewidth]{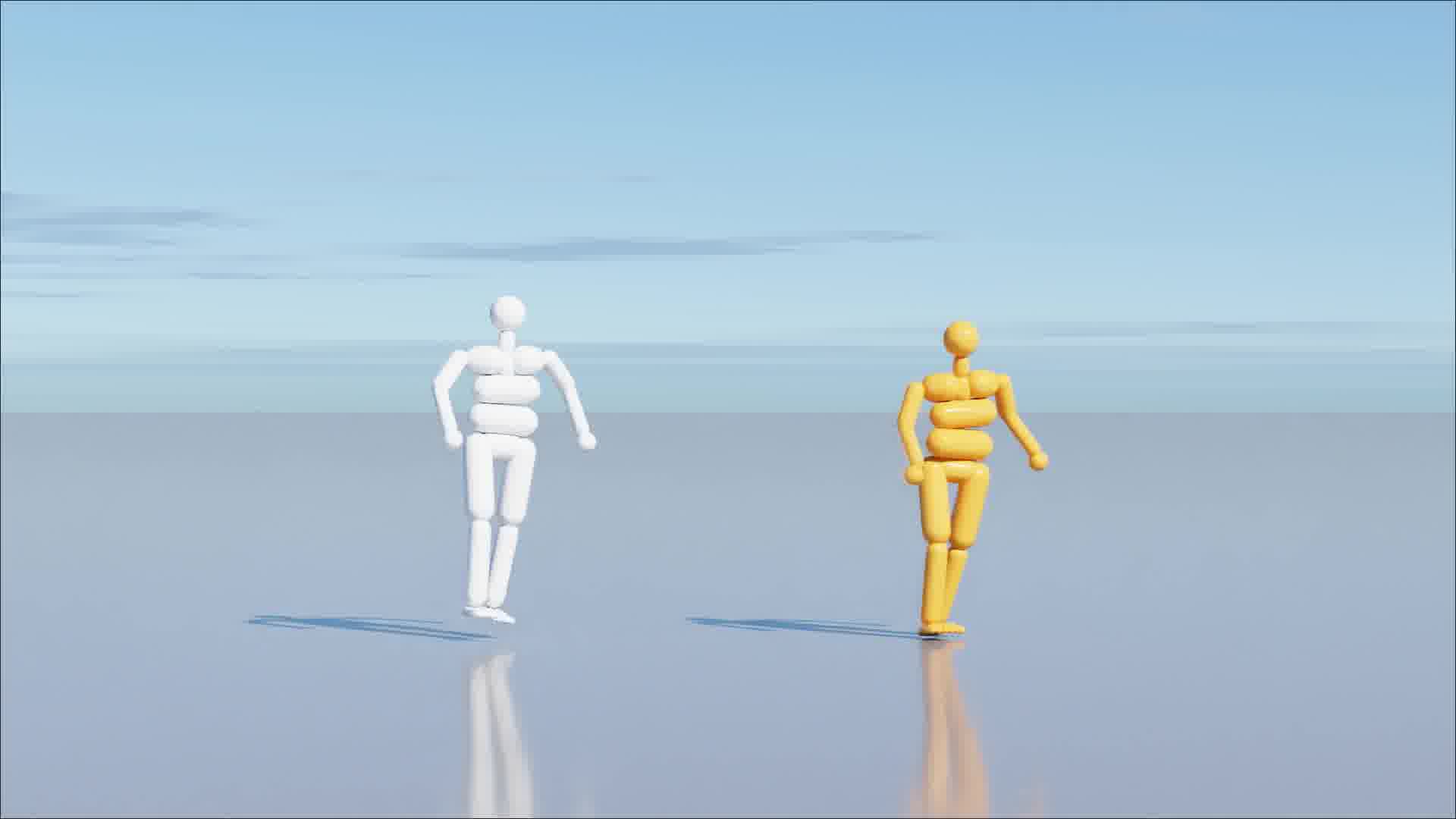}
    \end{subfigure}

    \begin{subfigure}{0.15\textwidth}
        \centering
        \includegraphics[width=\linewidth]{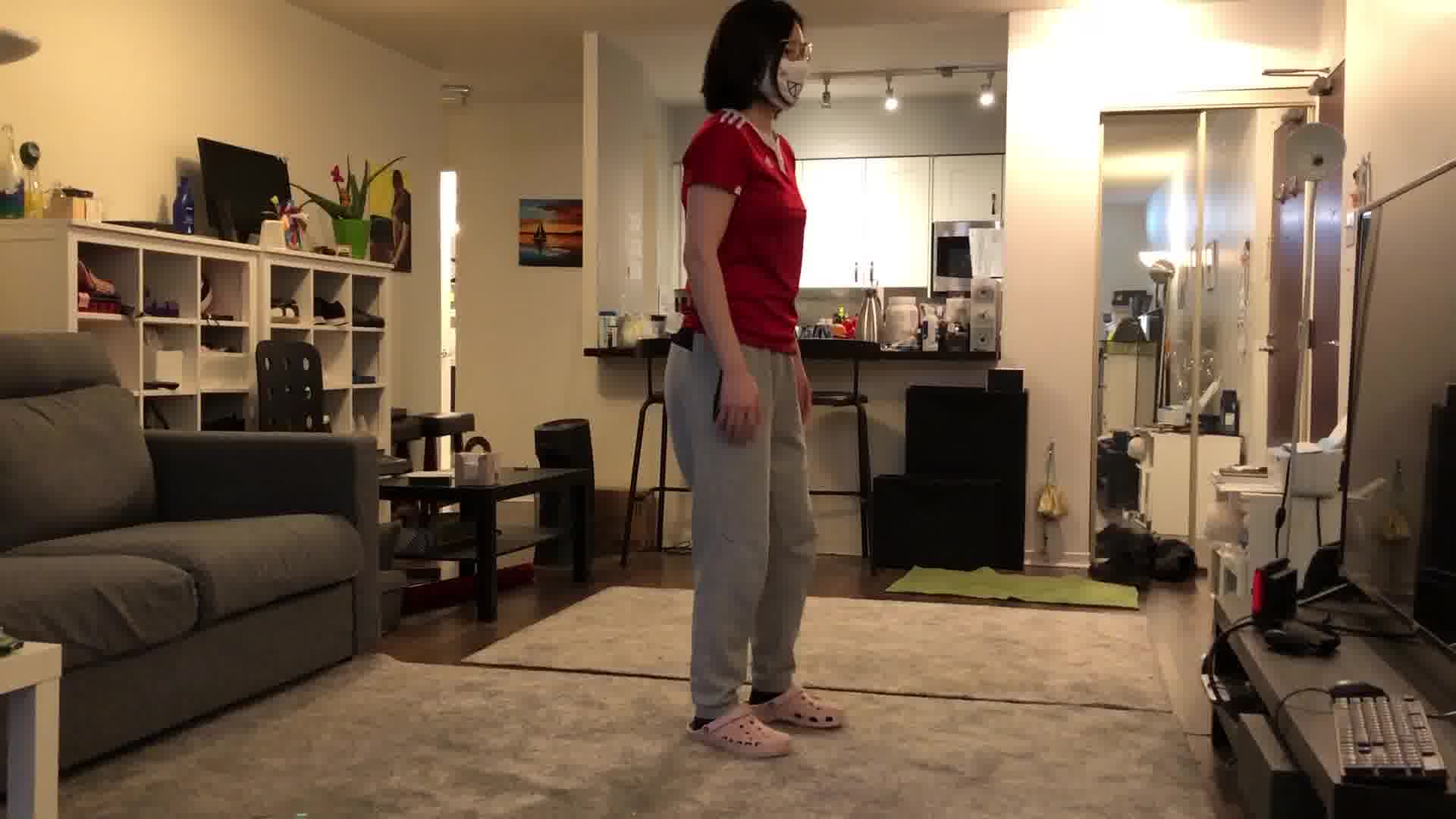}
    \end{subfigure}
    \begin{subfigure}{0.15\textwidth}
        \centering
        \includegraphics[width=\linewidth]{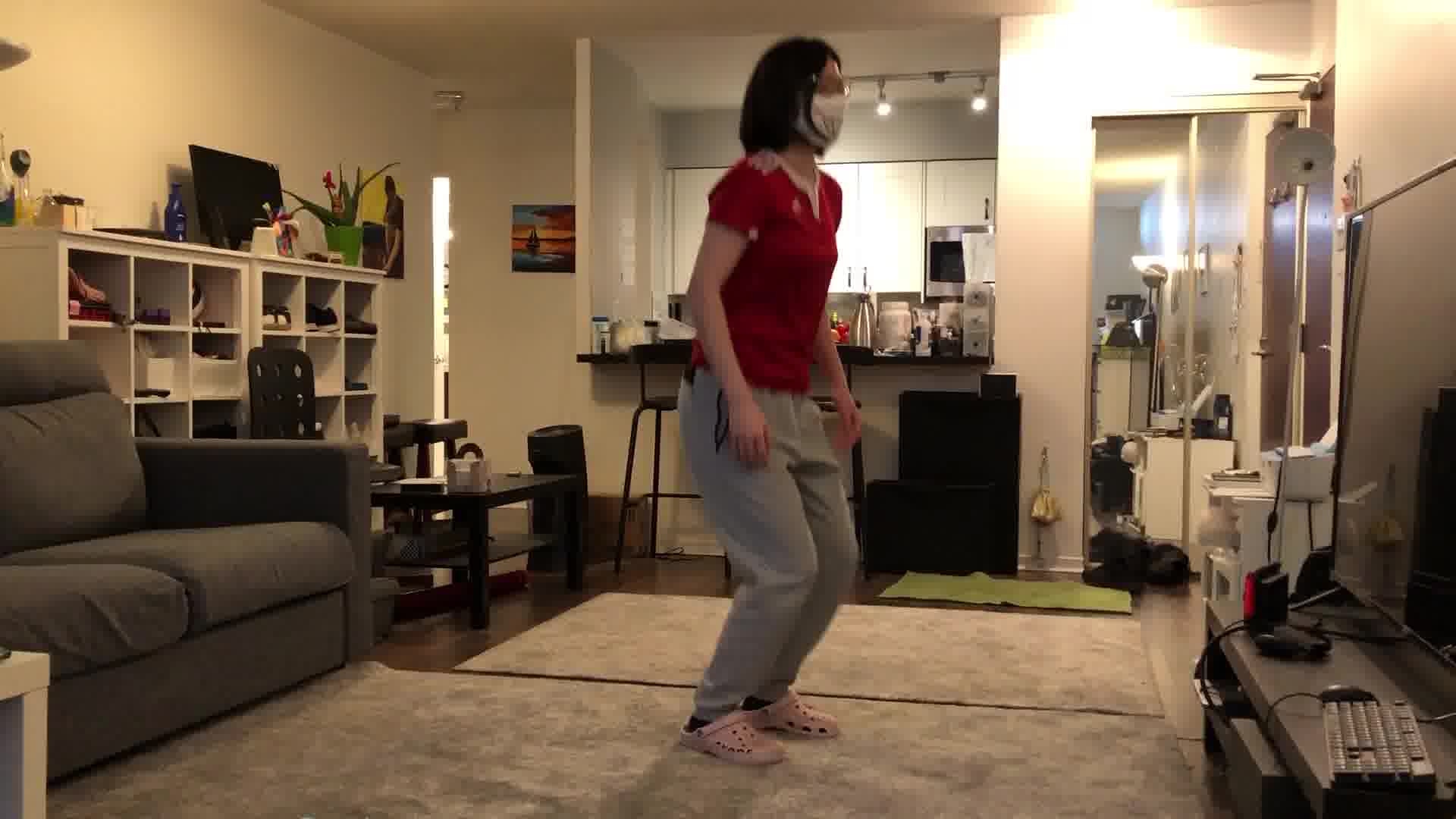}
    \end{subfigure}
    \begin{subfigure}{0.15\textwidth}
        \centering
        \includegraphics[width=\linewidth]{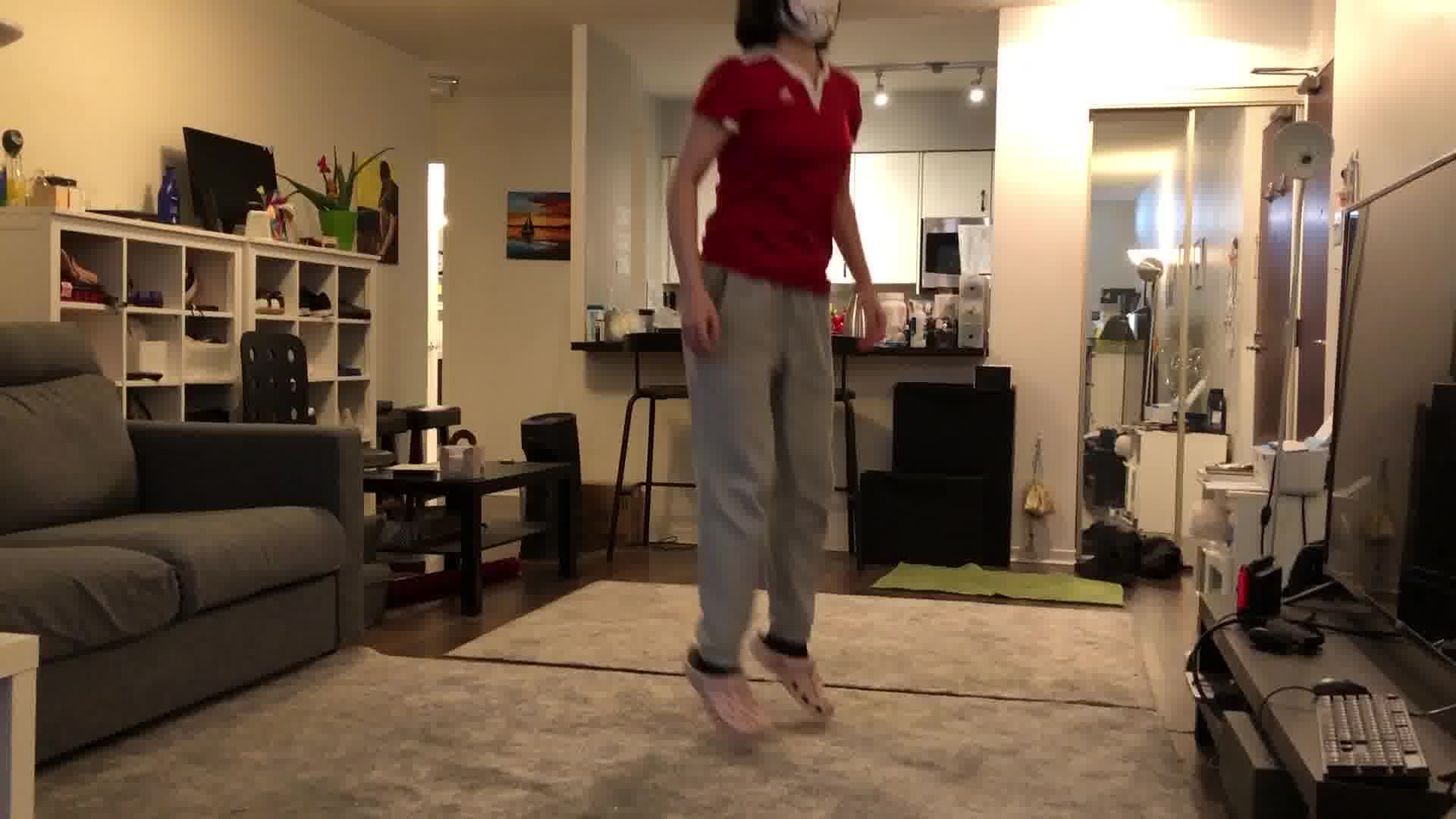}
    \end{subfigure}
    \begin{subfigure}{0.15\textwidth}
        \centering
        \includegraphics[width=\linewidth]{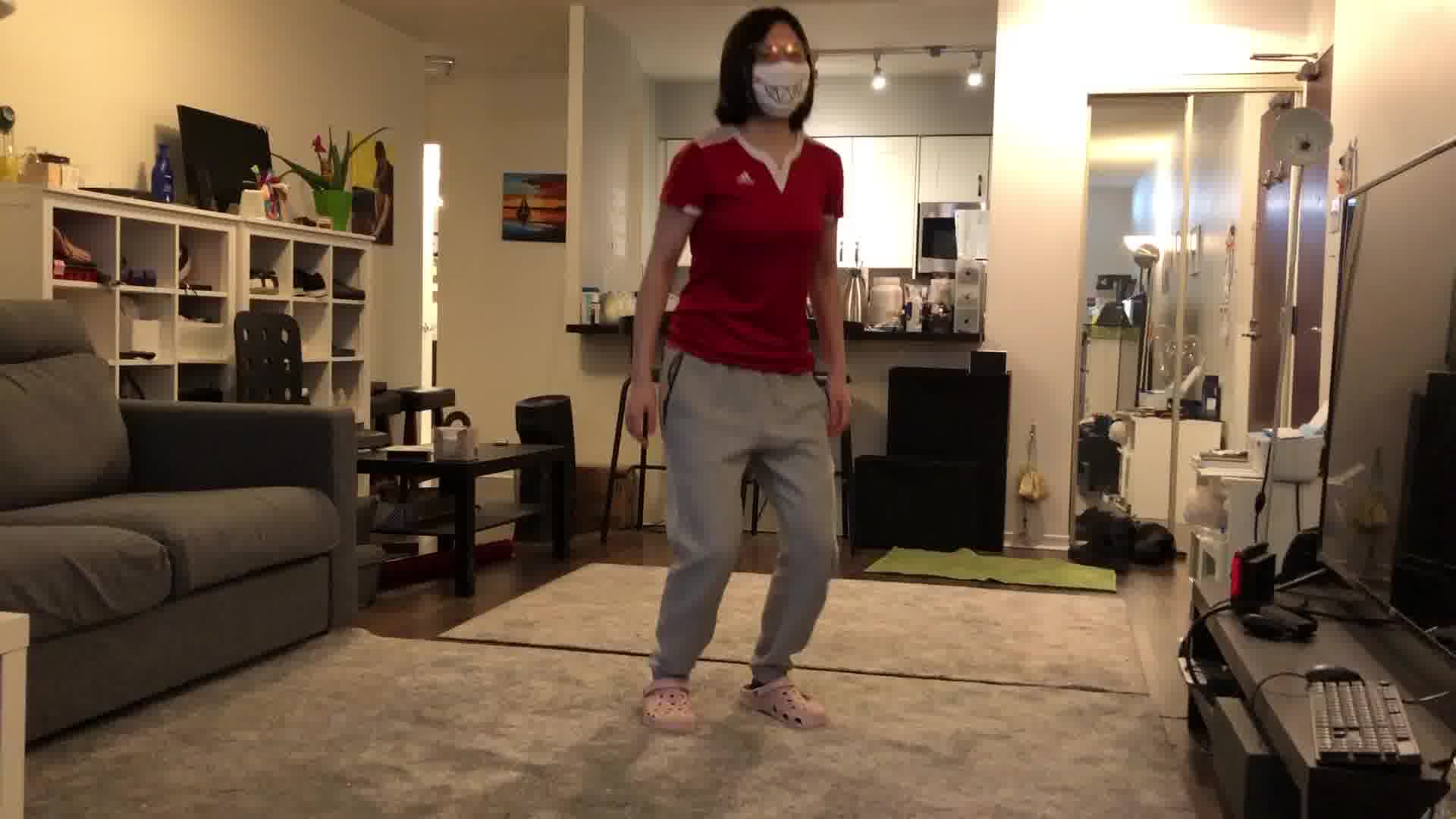}
    \end{subfigure}
    \begin{subfigure}{0.15\textwidth}
        \centering
        \includegraphics[width=\linewidth]{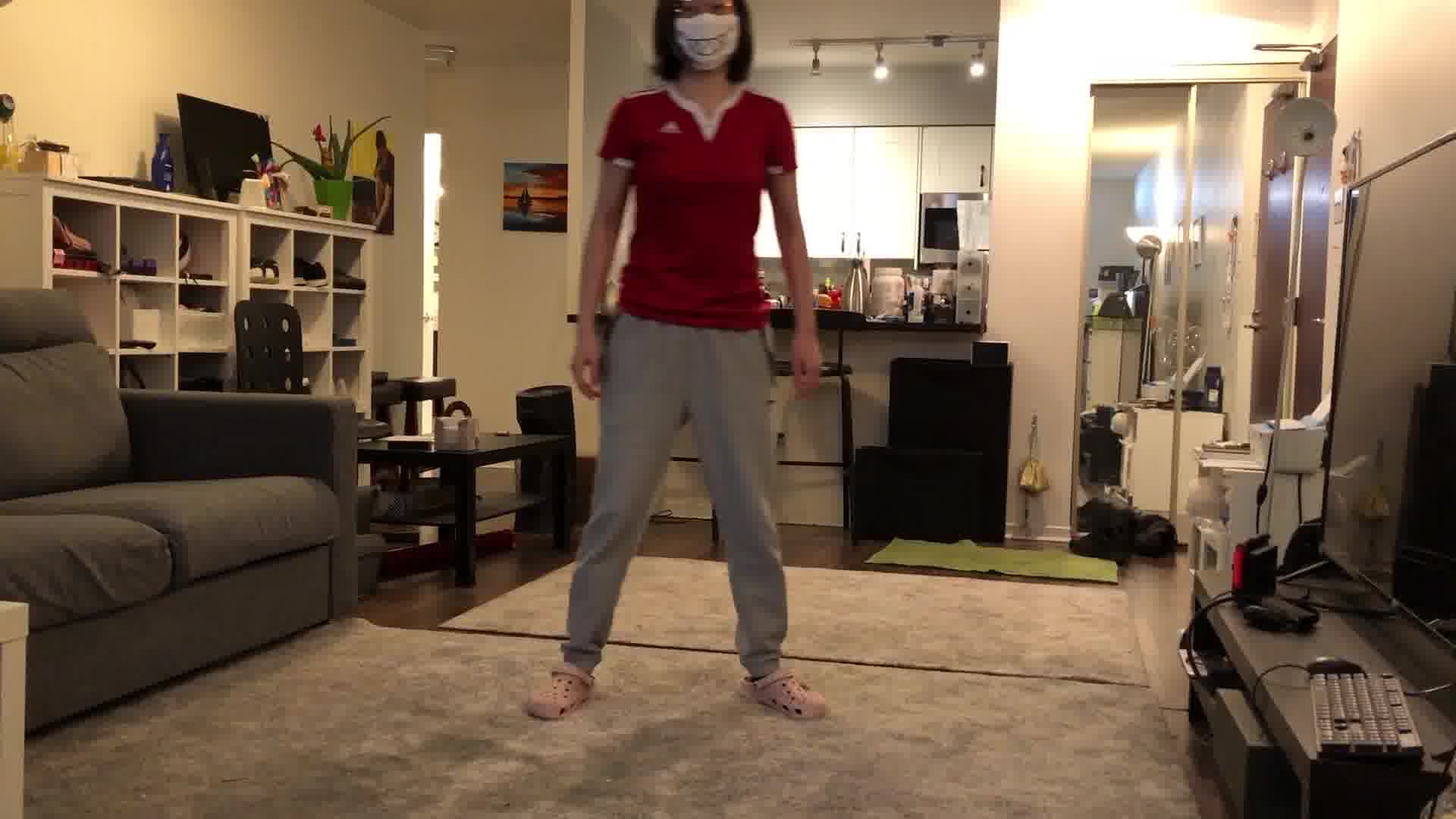}
    \end{subfigure}
    \begin{subfigure}{0.15\textwidth}
        \centering
        \includegraphics[width=\linewidth]{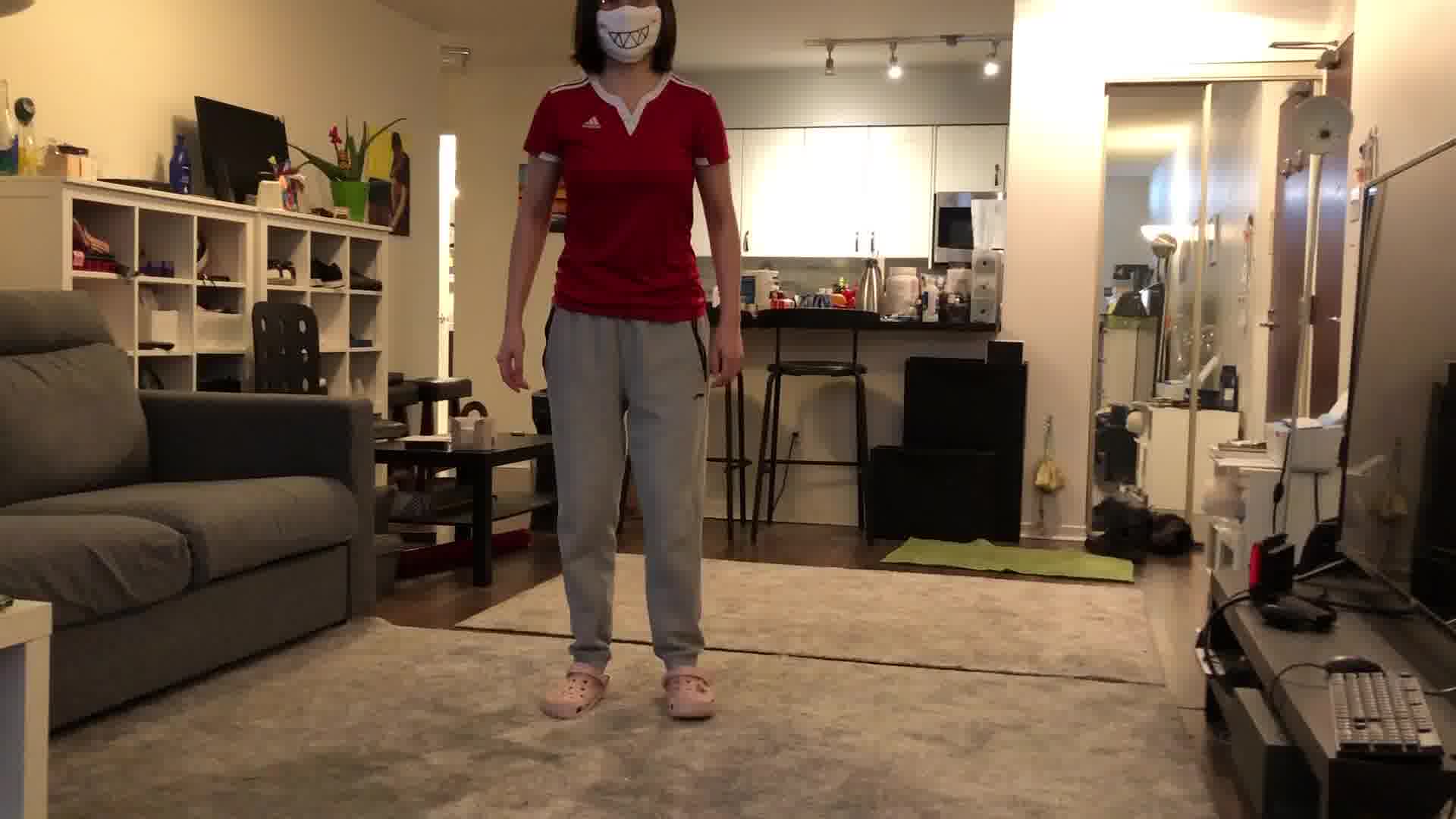}
    \end{subfigure}

    \begin{subfigure}{0.15\textwidth}
        \centering
        \includegraphics[width=\linewidth]{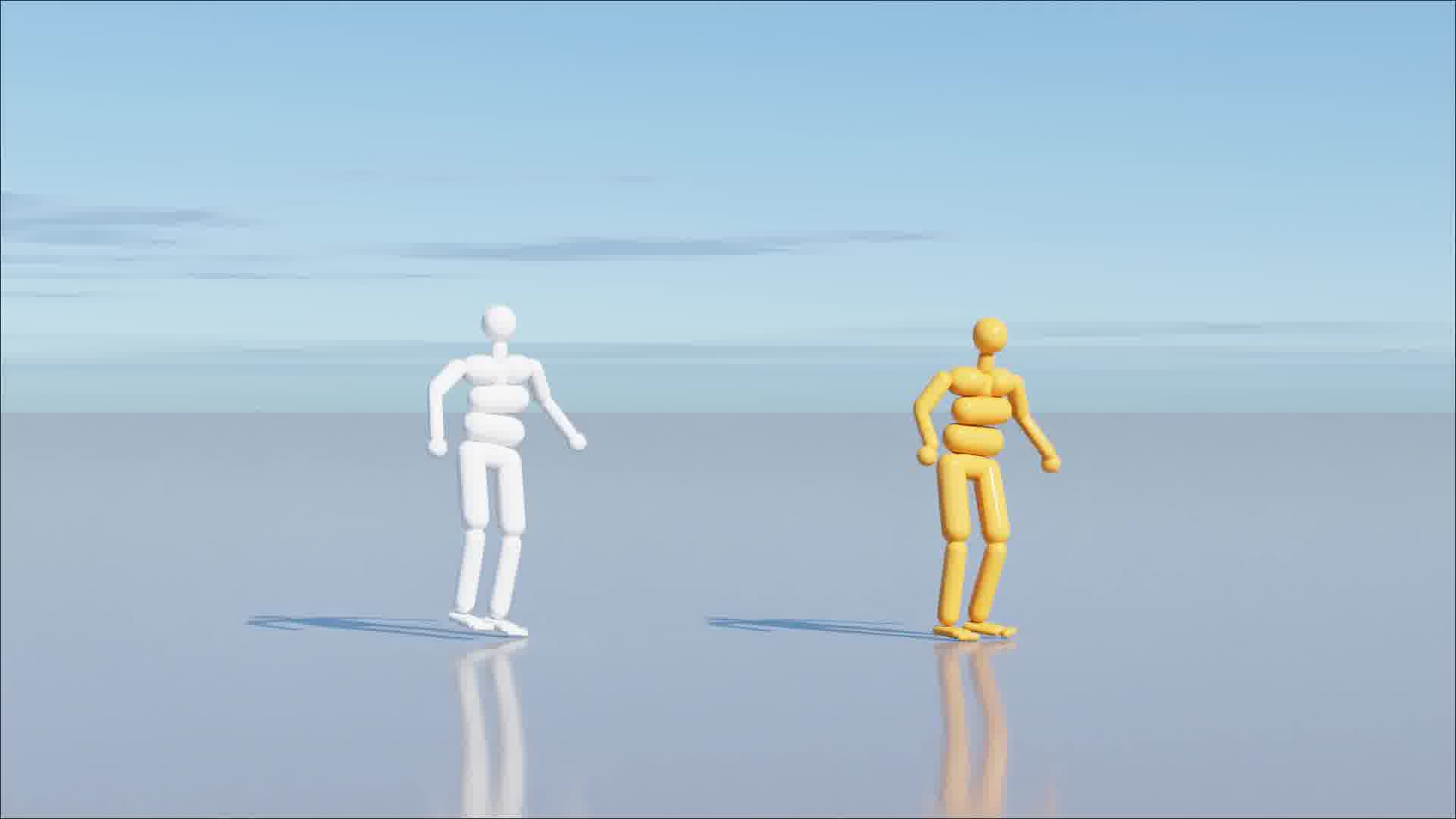}
    \end{subfigure}
    \begin{subfigure}{0.15\textwidth}
        \centering
        \includegraphics[width=\linewidth]{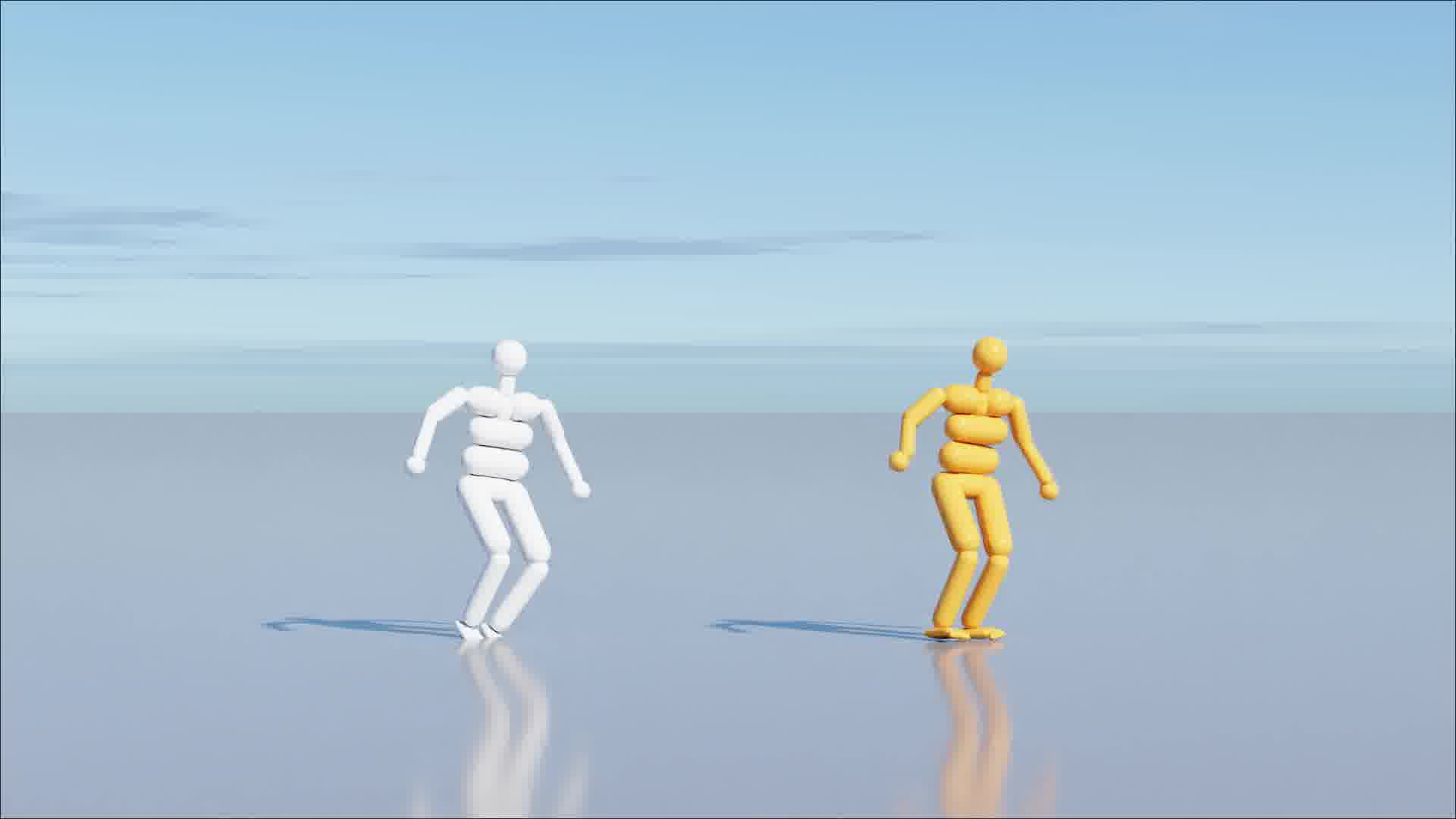}
    \end{subfigure}
    \begin{subfigure}{0.15\textwidth}
        \centering
        \includegraphics[width=\linewidth]{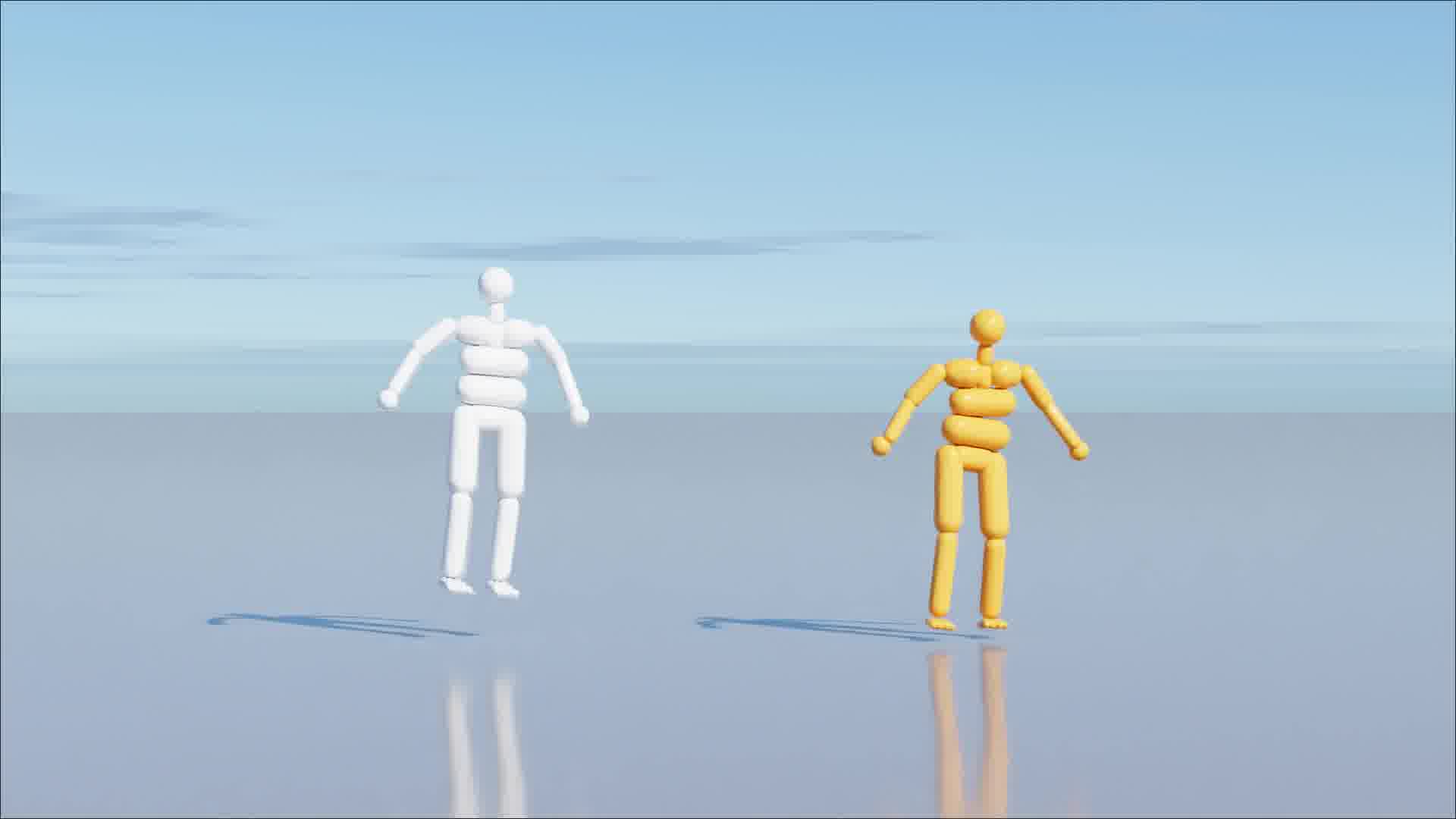}
    \end{subfigure}
    \begin{subfigure}{0.15\textwidth}
        \centering
        \includegraphics[width=\linewidth]{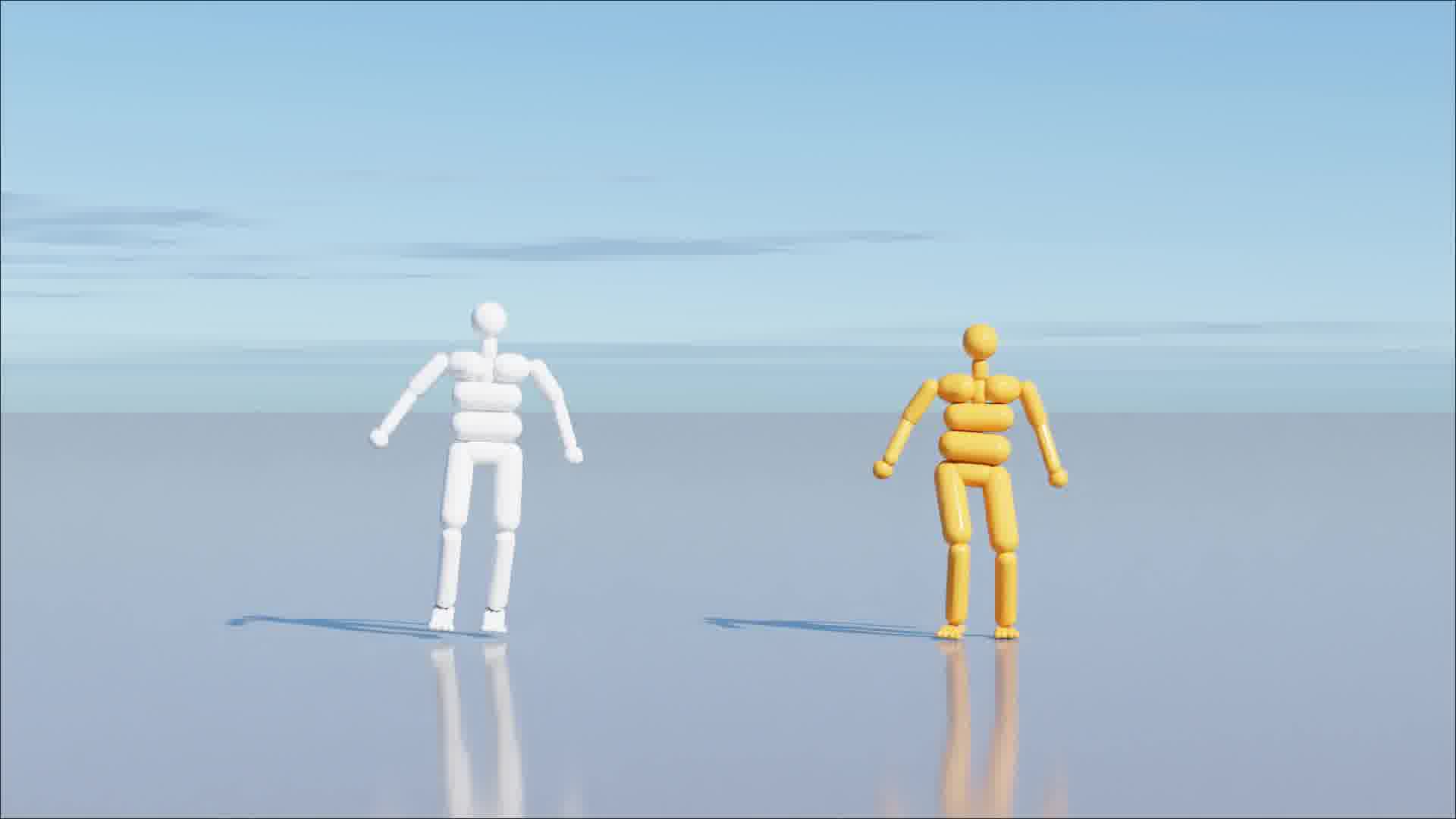}
    \end{subfigure}
    \begin{subfigure}{0.15\textwidth}
        \centering
        \includegraphics[width=\linewidth]{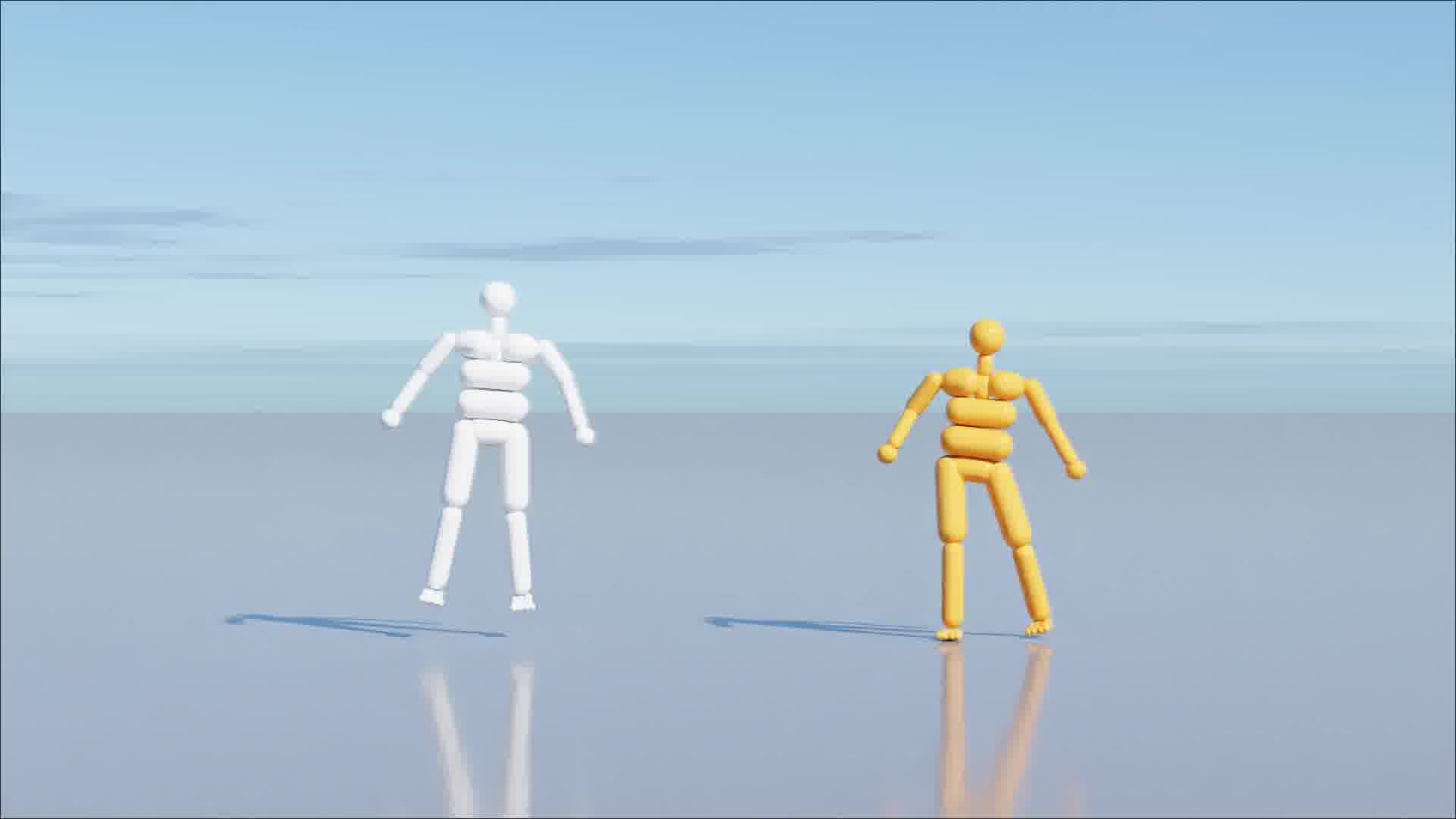}
    \end{subfigure}
    \begin{subfigure}{0.15\textwidth}
        \centering
        \includegraphics[width=\linewidth]{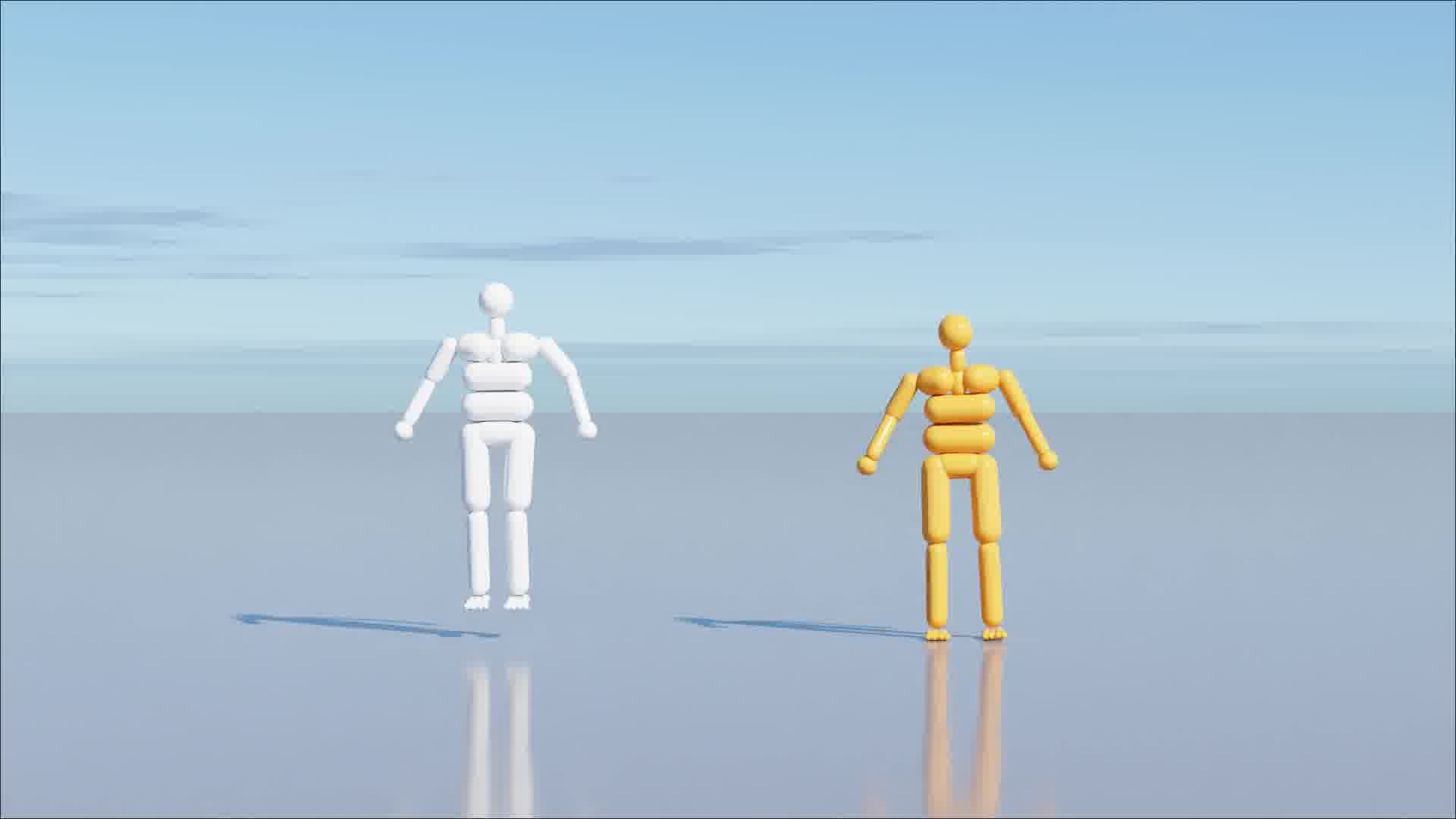}
    \end{subfigure}

    \caption{{In this figure, we show how our controller reacts real-time to the remote host captured by a camera. It successfully reproduces waving, walking, turning and jumping behaviors.}}
    \label{fig:motion_video_drive}
\end{figure*}

\begin{figure*}[!t]
    \centering
    \begin{subfigure}{0.15\textwidth}
        \centering
        \includegraphics[width=\linewidth]{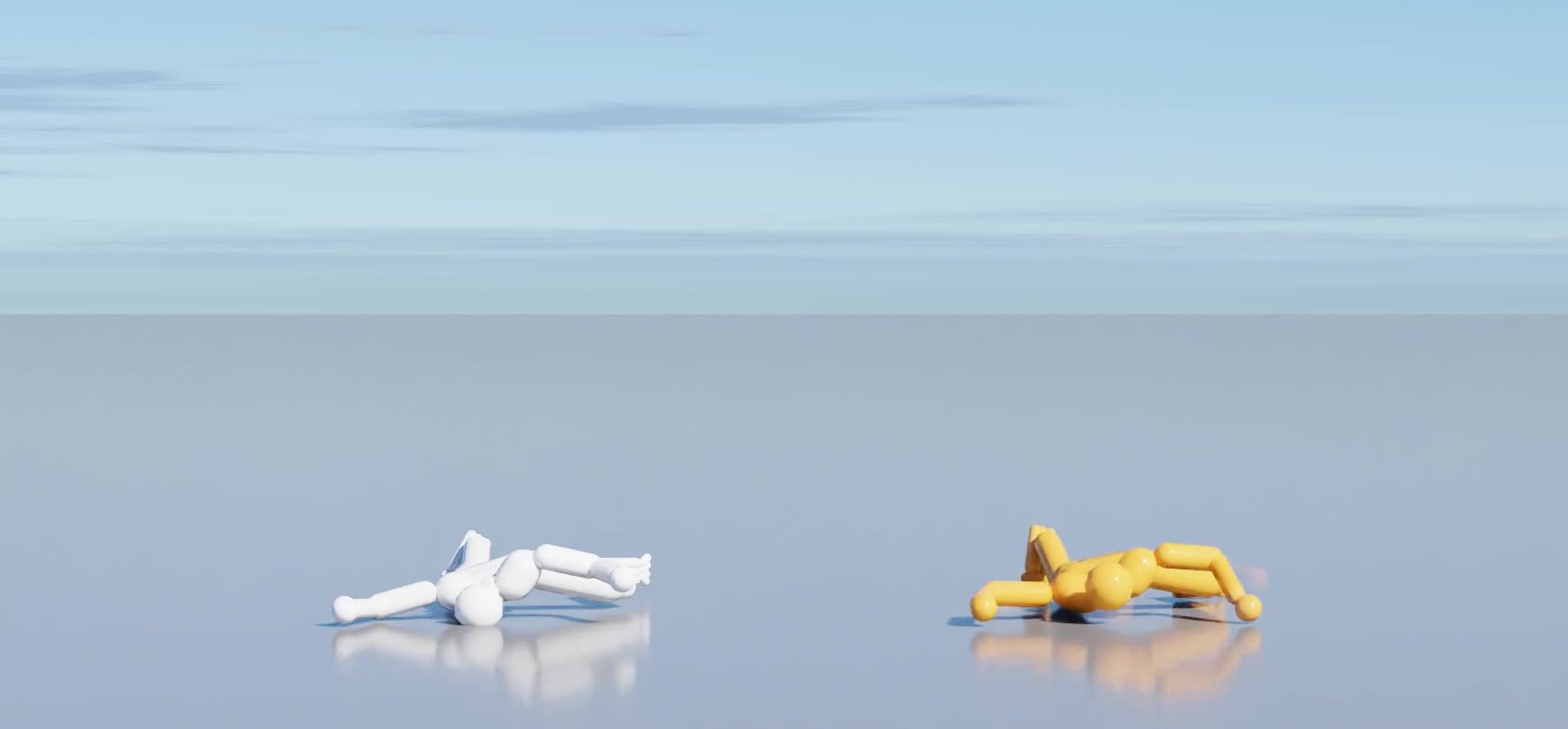}
    \end{subfigure}
    \begin{subfigure}{0.15\textwidth}
        \centering
        \includegraphics[width=\linewidth]{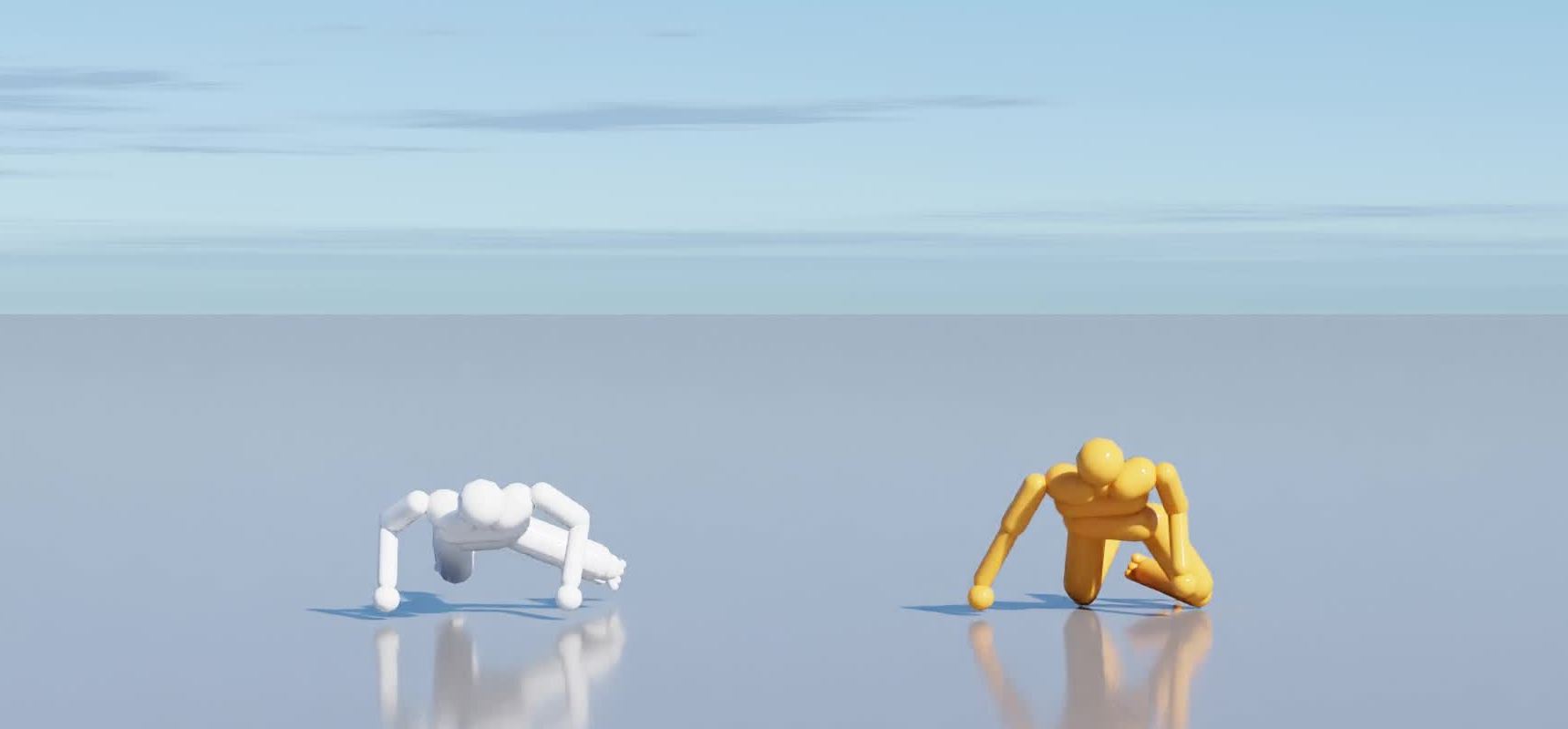}
    \end{subfigure}
    \begin{subfigure}{0.15\textwidth}
        \centering
        \includegraphics[width=\linewidth]{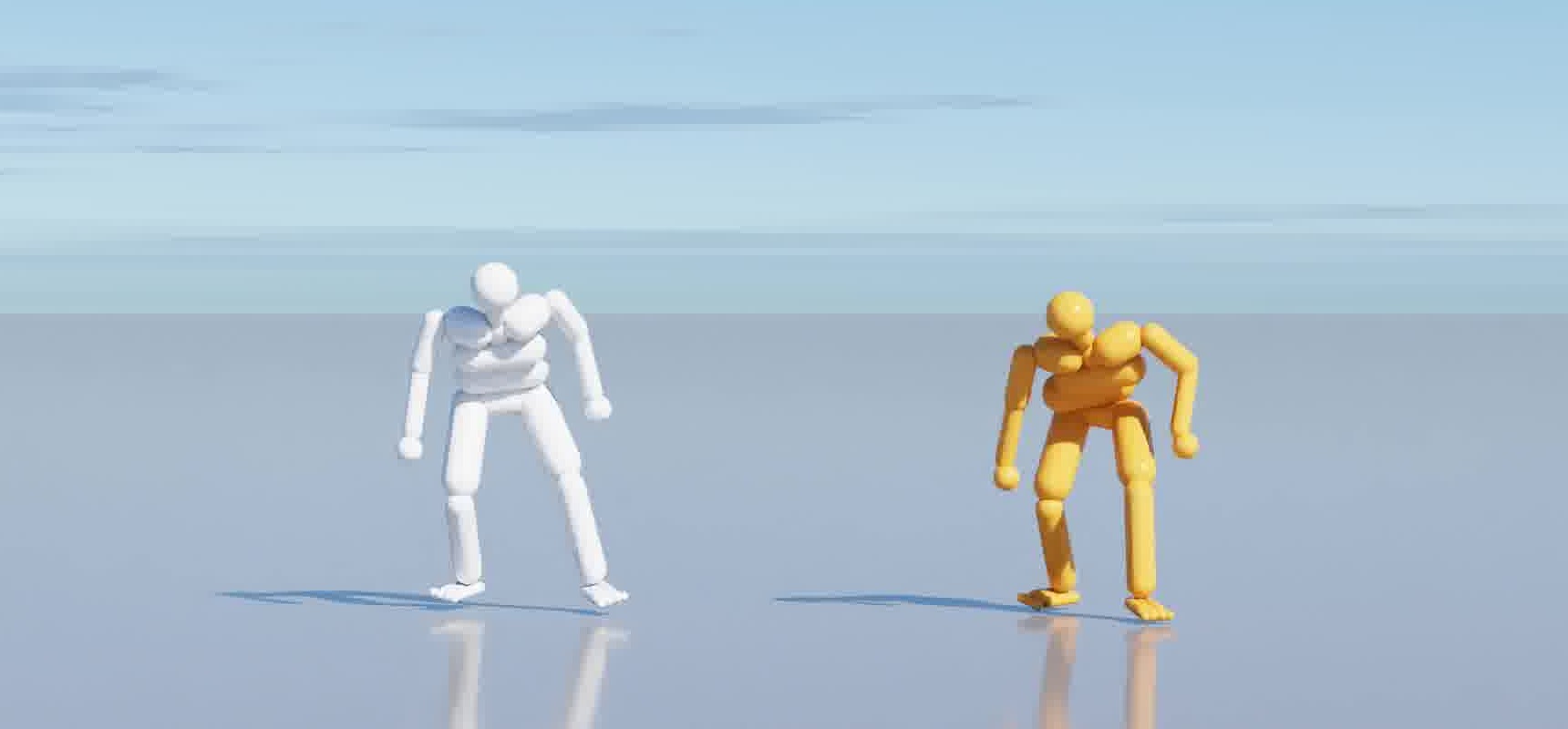}
    \end{subfigure}
    \begin{subfigure}{0.15\textwidth}
        \centering
        \includegraphics[width=\linewidth]{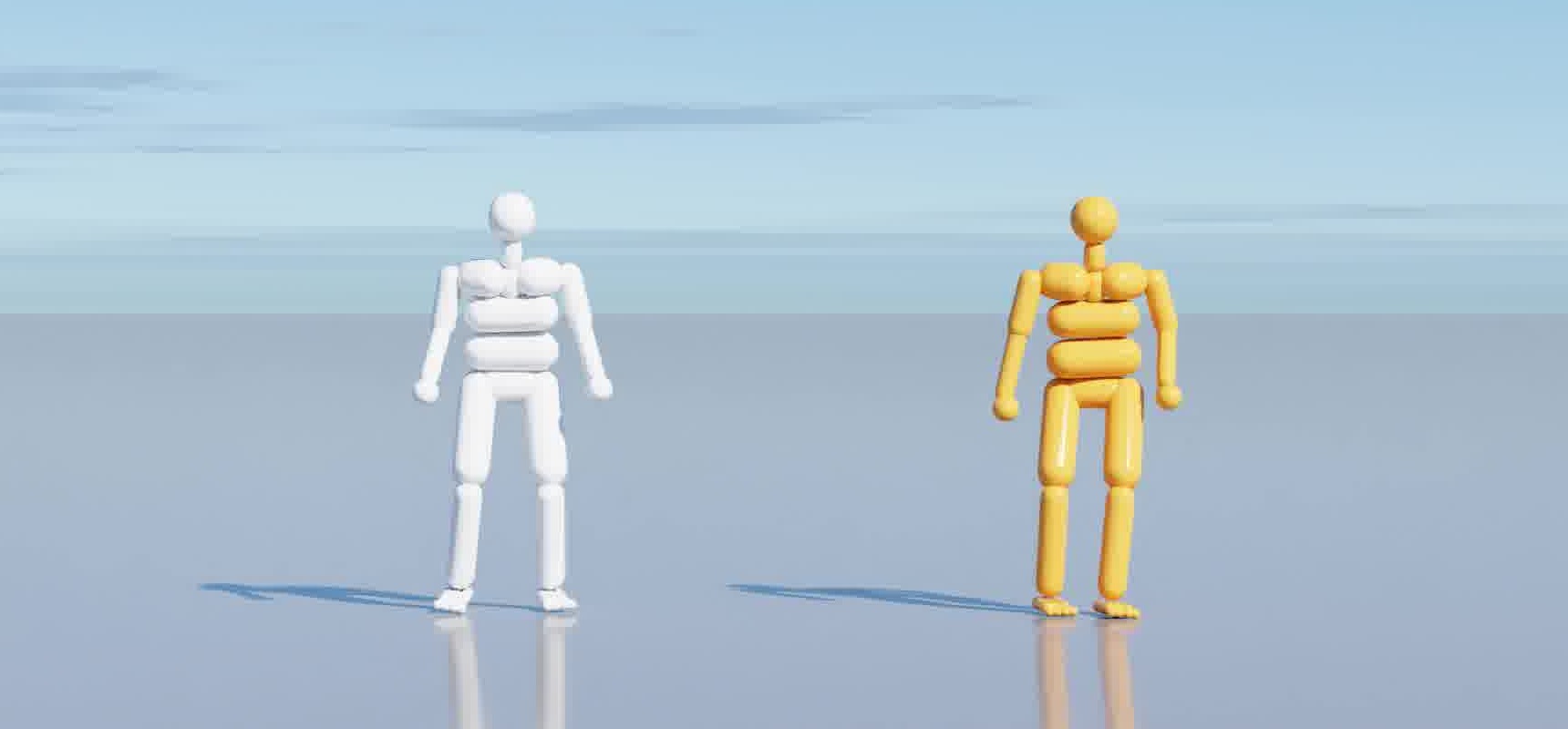}
    \end{subfigure}
    \begin{subfigure}{0.15\textwidth}
        \centering
        \includegraphics[width=\linewidth]{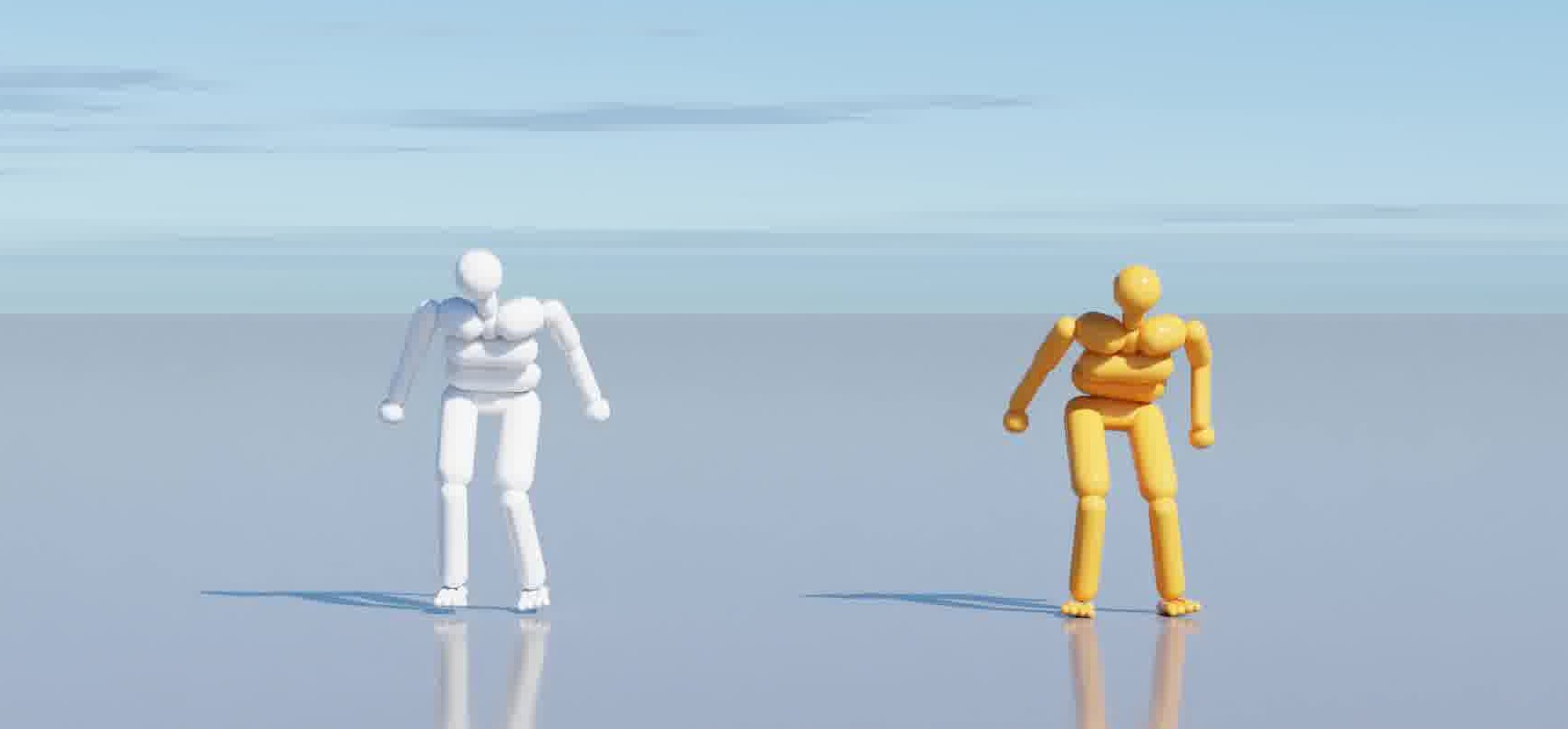}
    \end{subfigure}
    \begin{subfigure}{0.15\textwidth}
        \centering
        \includegraphics[width=\linewidth]{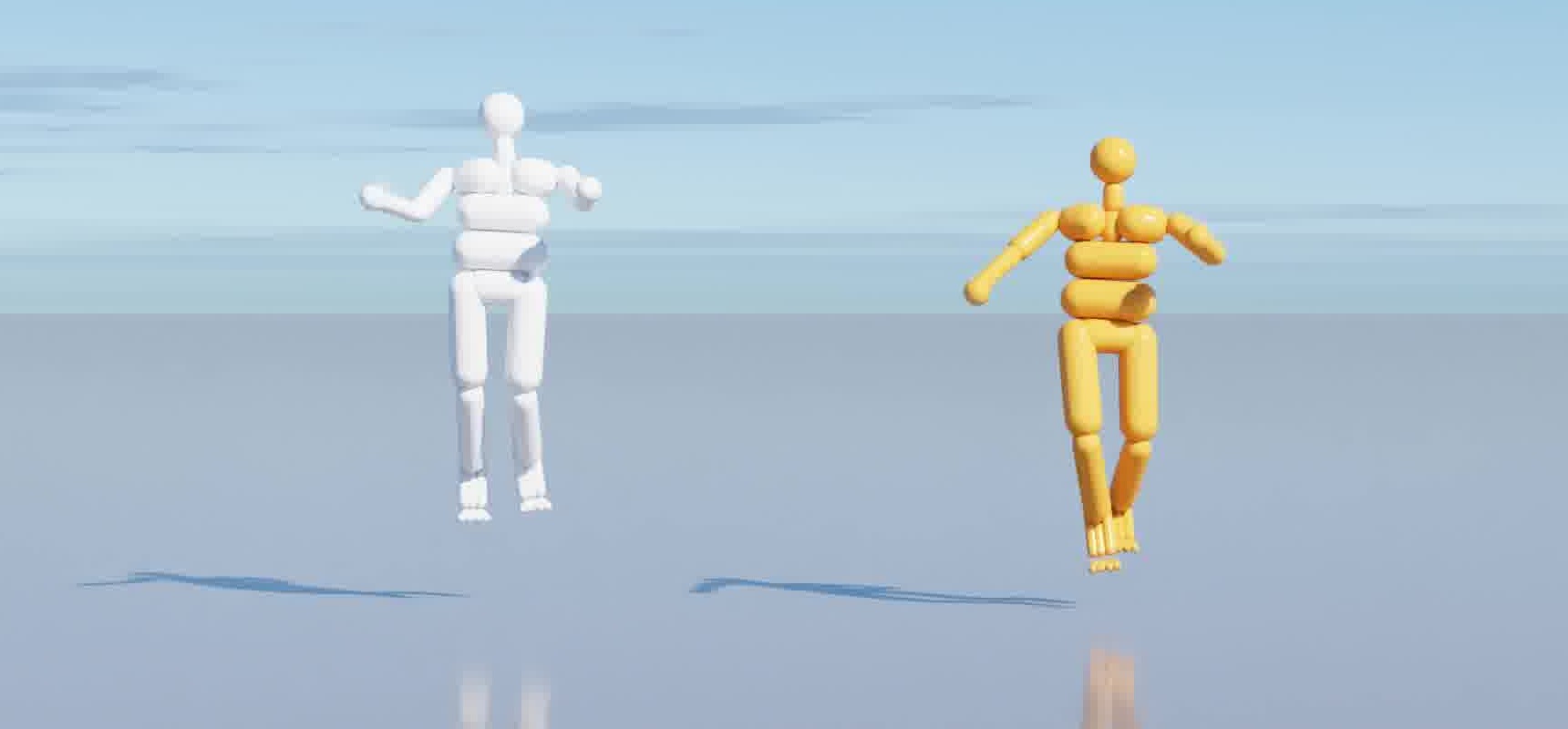}
    \end{subfigure}

    \begin{subfigure}{0.15\textwidth}
        \centering
        \includegraphics[width=\linewidth]{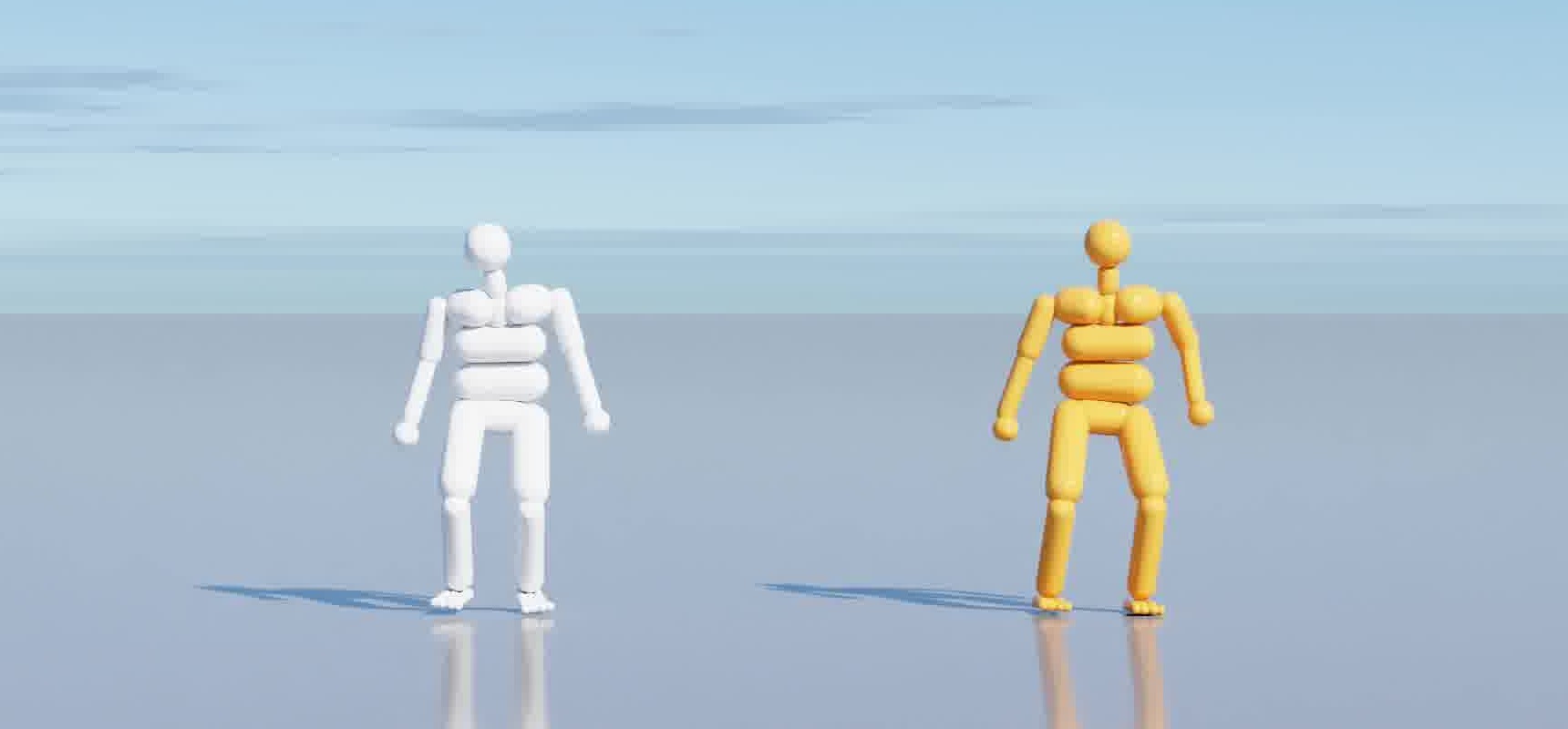}
    \end{subfigure}
    \begin{subfigure}{0.15\textwidth}
        \centering
        \includegraphics[width=\linewidth]{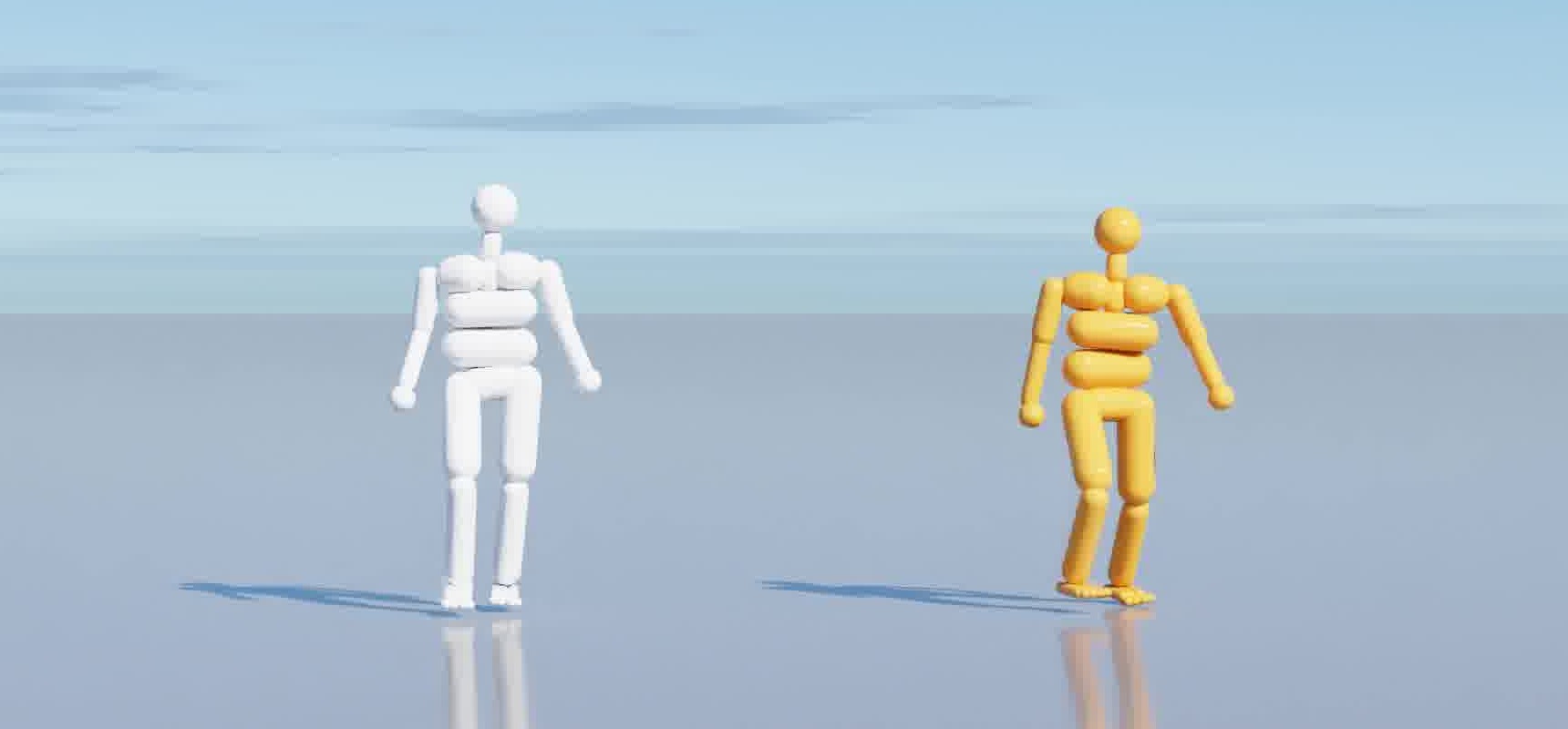}
    \end{subfigure}
    \begin{subfigure}{0.15\textwidth}
        \centering
        \includegraphics[width=\linewidth]{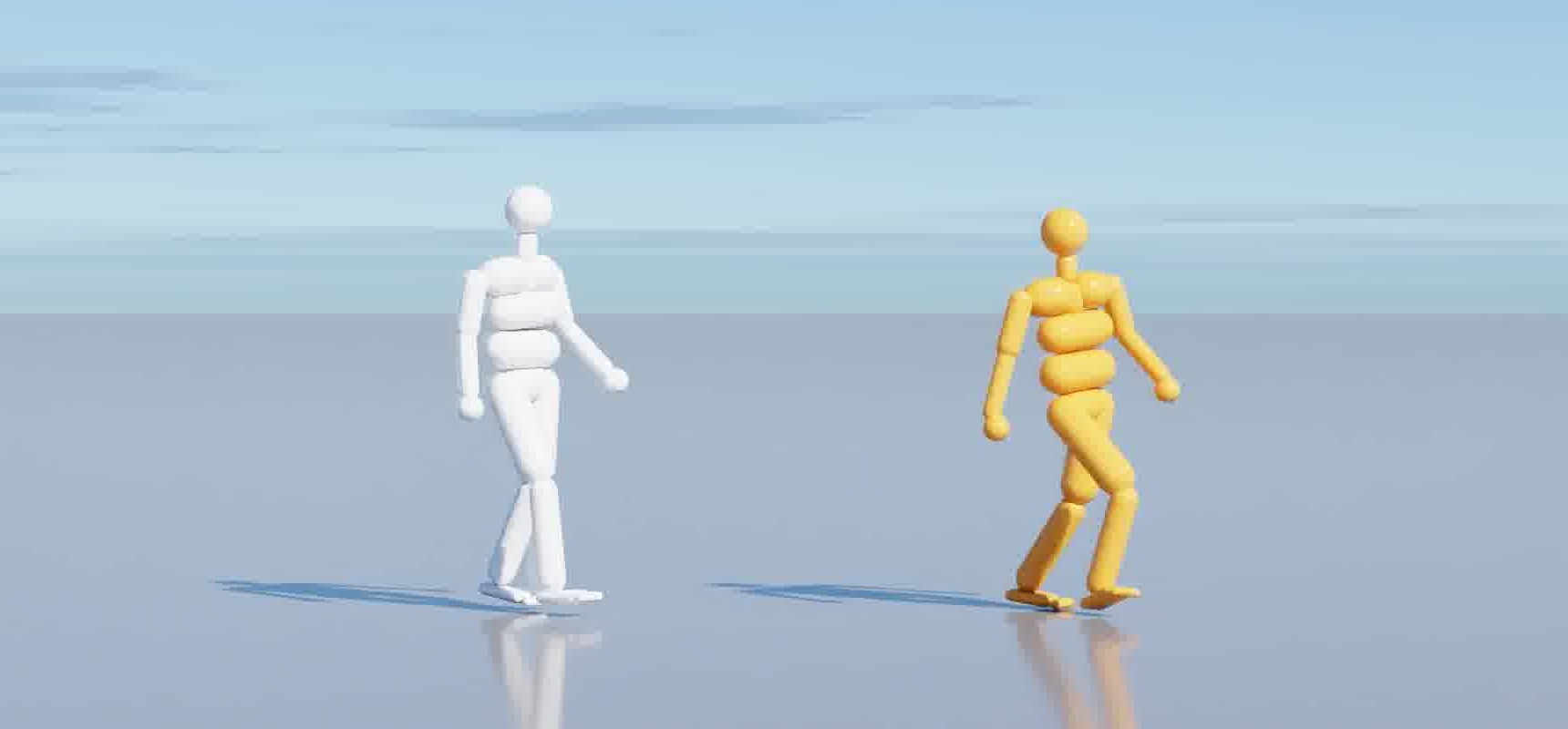}
    \end{subfigure}
    \begin{subfigure}{0.15\textwidth}
        \centering
        \includegraphics[width=\linewidth]{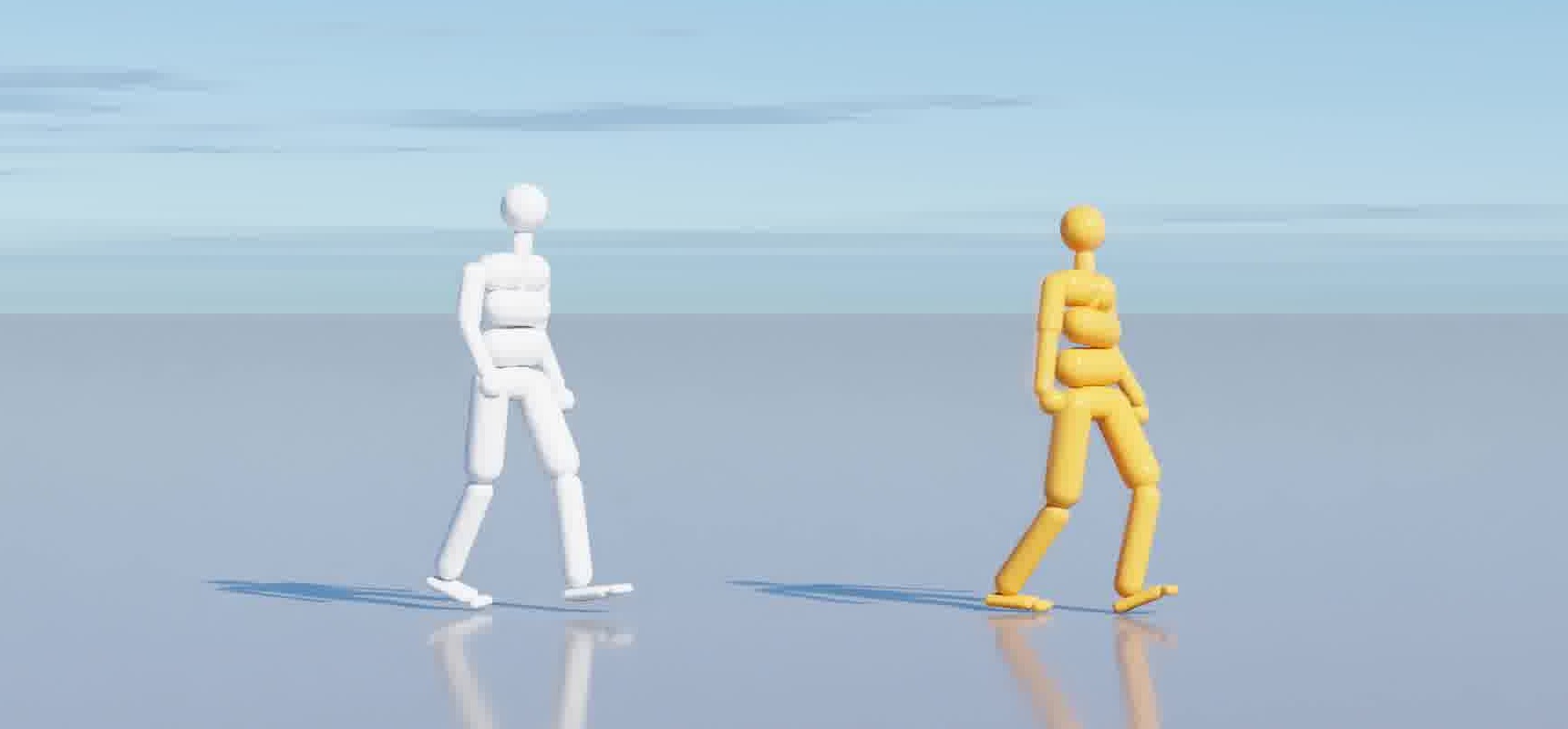}
    \end{subfigure}
    \begin{subfigure}{0.15\textwidth}
        \centering
        \includegraphics[width=\linewidth]{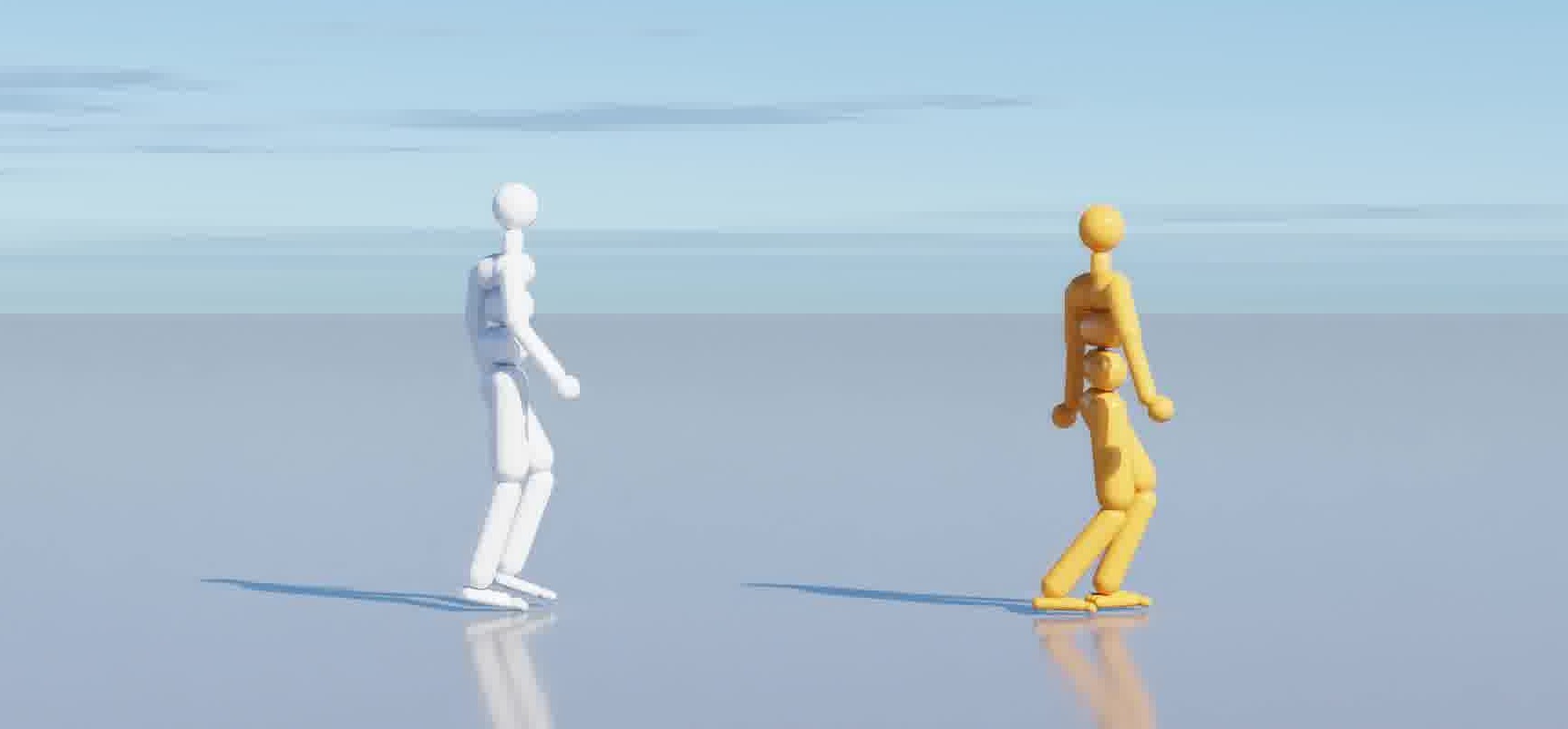}
    \end{subfigure}
    \begin{subfigure}{0.15\textwidth}
        \centering
        \includegraphics[width=\linewidth]{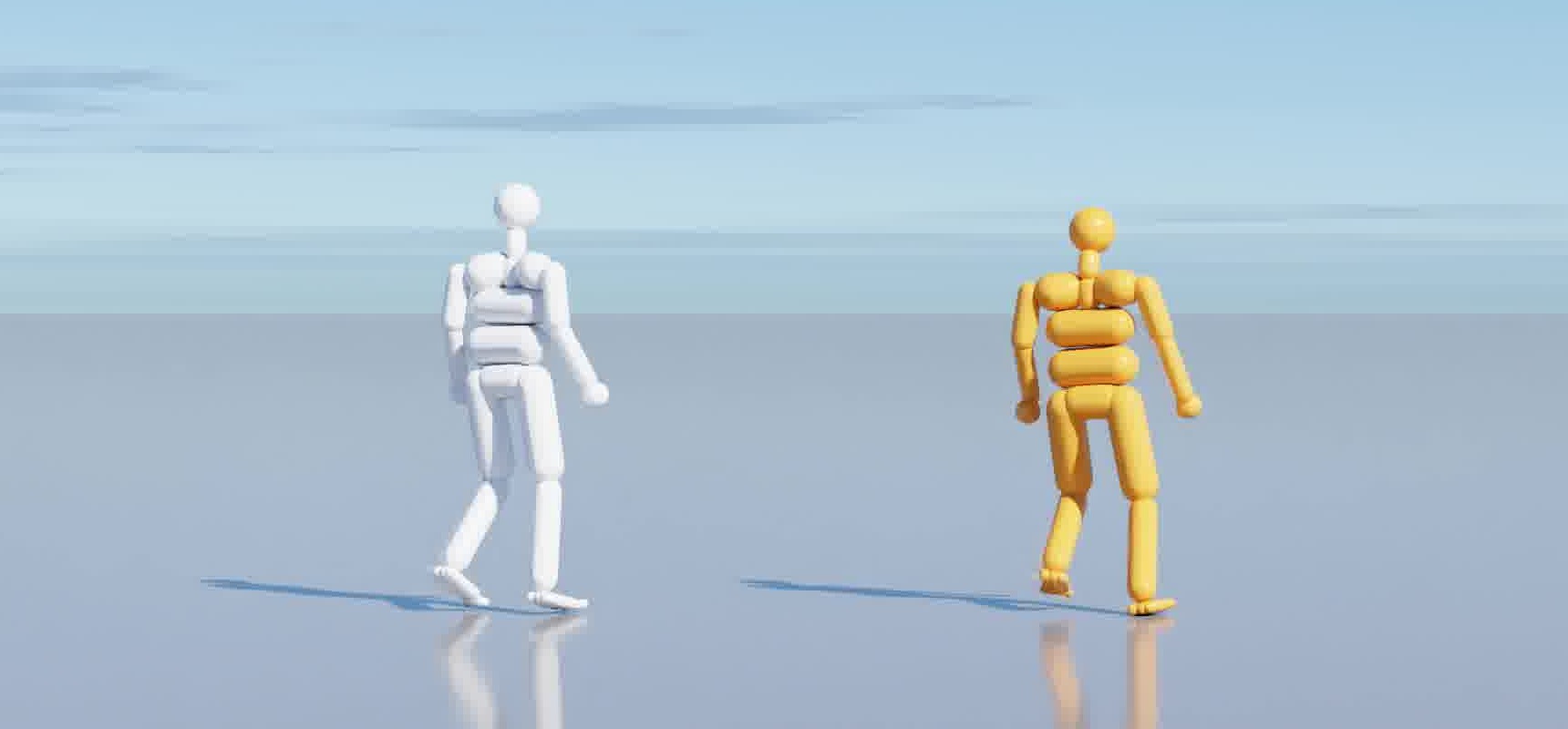}
    \end{subfigure}

    \begin{subfigure}{0.15\textwidth}
        \centering
        \includegraphics[width=\linewidth]{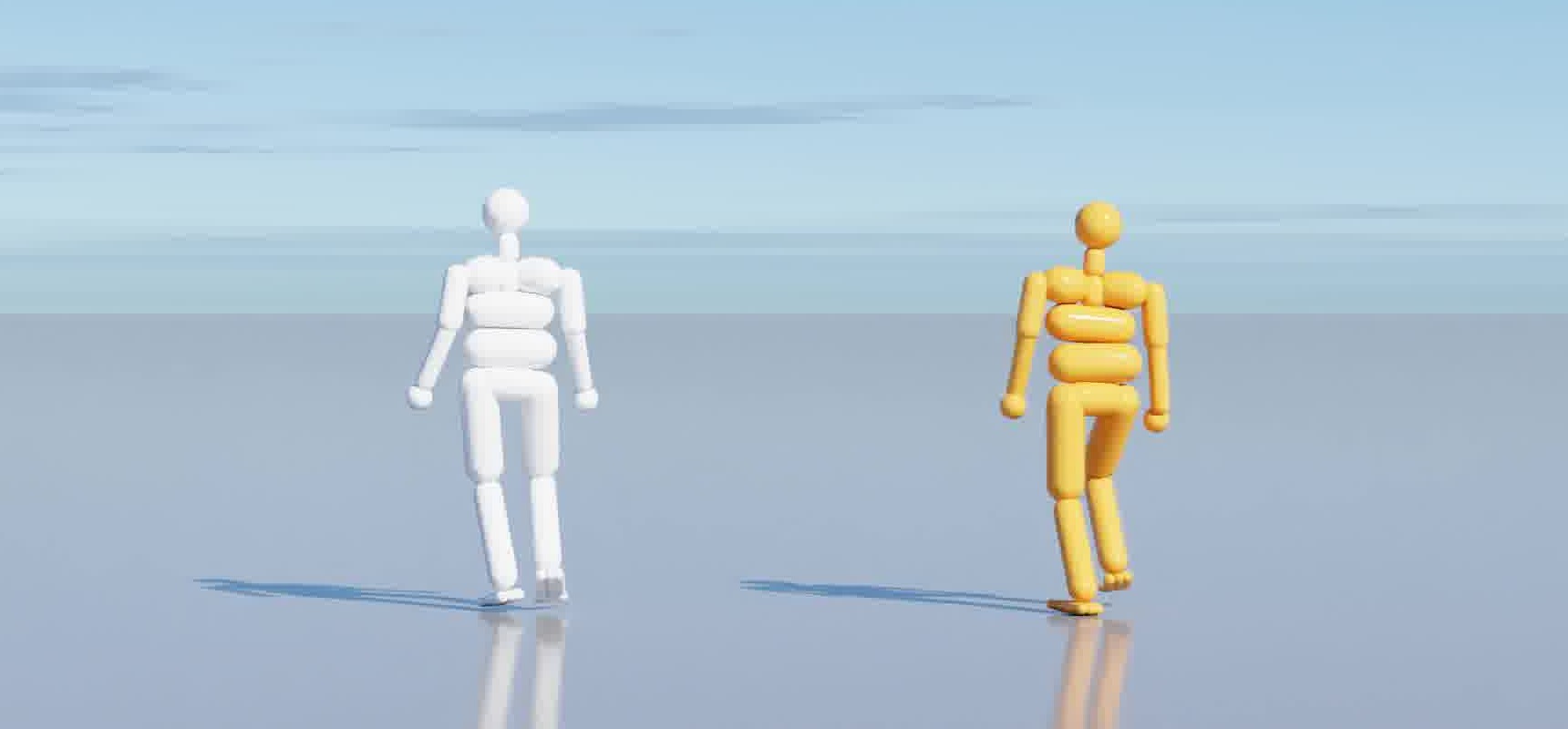}
    \end{subfigure}
    \begin{subfigure}{0.15\textwidth}
        \centering
        \includegraphics[width=\linewidth]{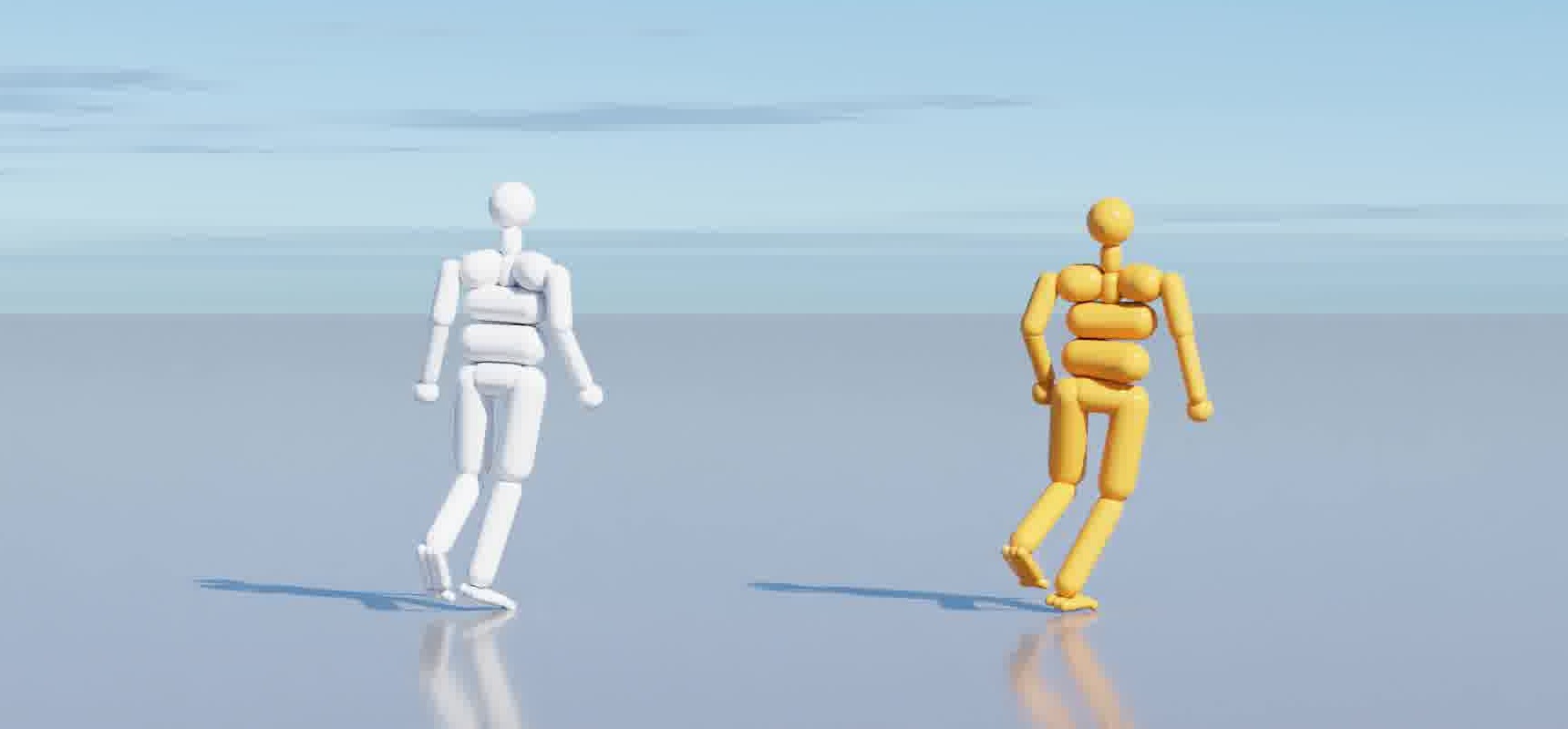}
    \end{subfigure}
    \begin{subfigure}{0.15\textwidth}
        \centering
        \includegraphics[width=\linewidth]{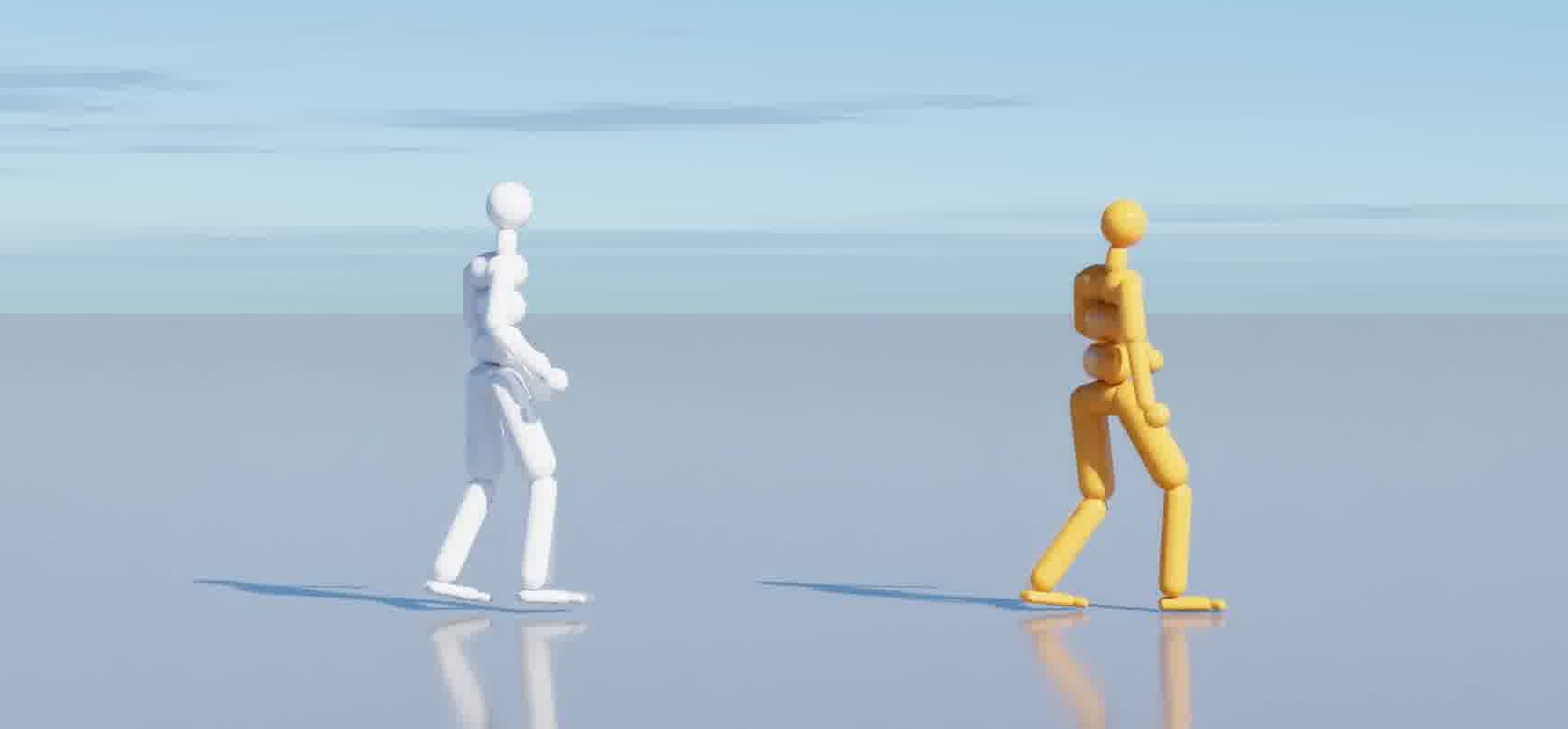}
    \end{subfigure}
    \begin{subfigure}{0.15\textwidth}
        \centering
        \includegraphics[width=\linewidth]{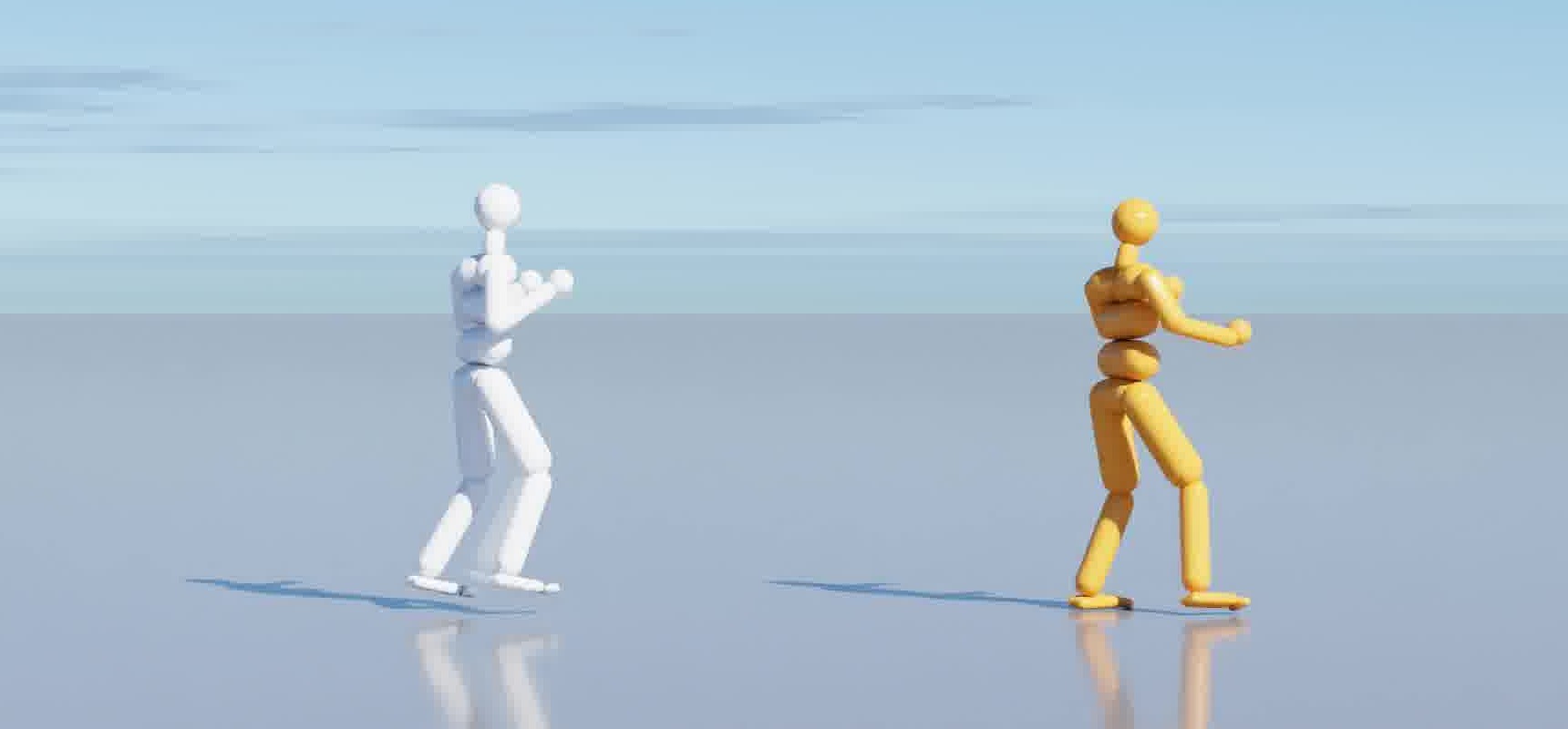}
    \end{subfigure}
    \begin{subfigure}{0.15\textwidth}
        \centering
        \includegraphics[width=\linewidth]{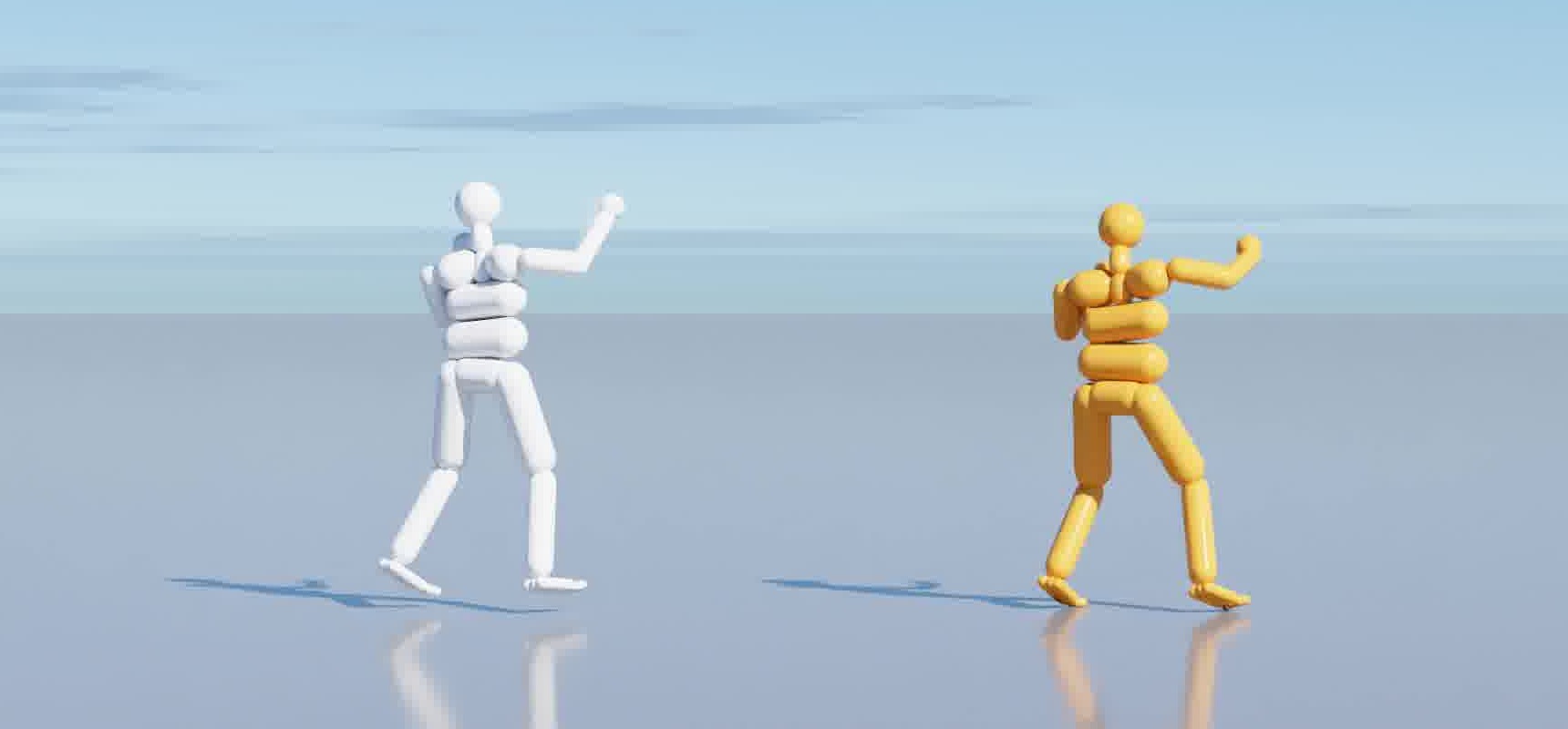}
    \end{subfigure}
    \begin{subfigure}{0.15\textwidth}
        \centering
        \includegraphics[width=\linewidth]{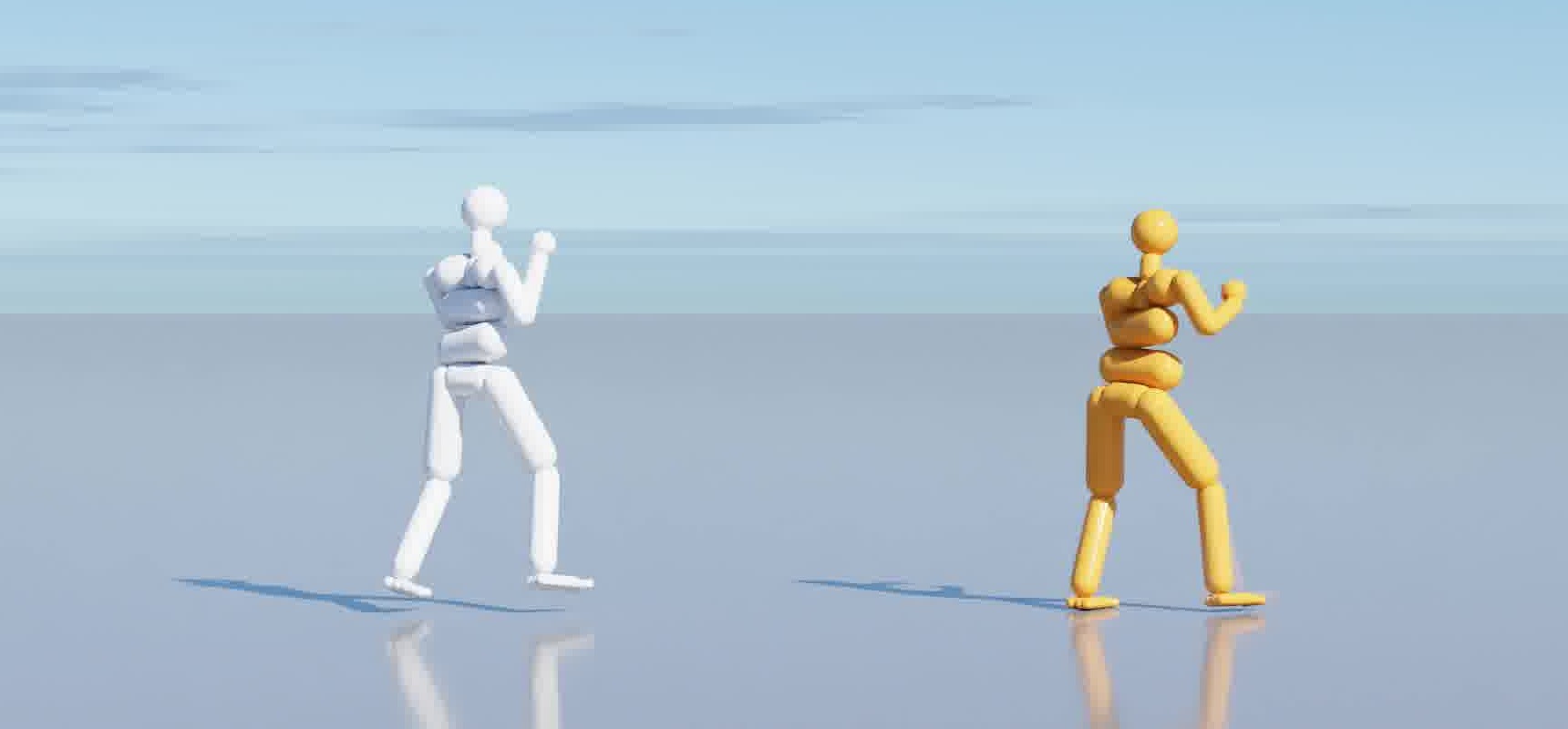}
    \end{subfigure}

    \begin{subfigure}{0.15\textwidth}
        \centering
        \includegraphics[width=\linewidth]{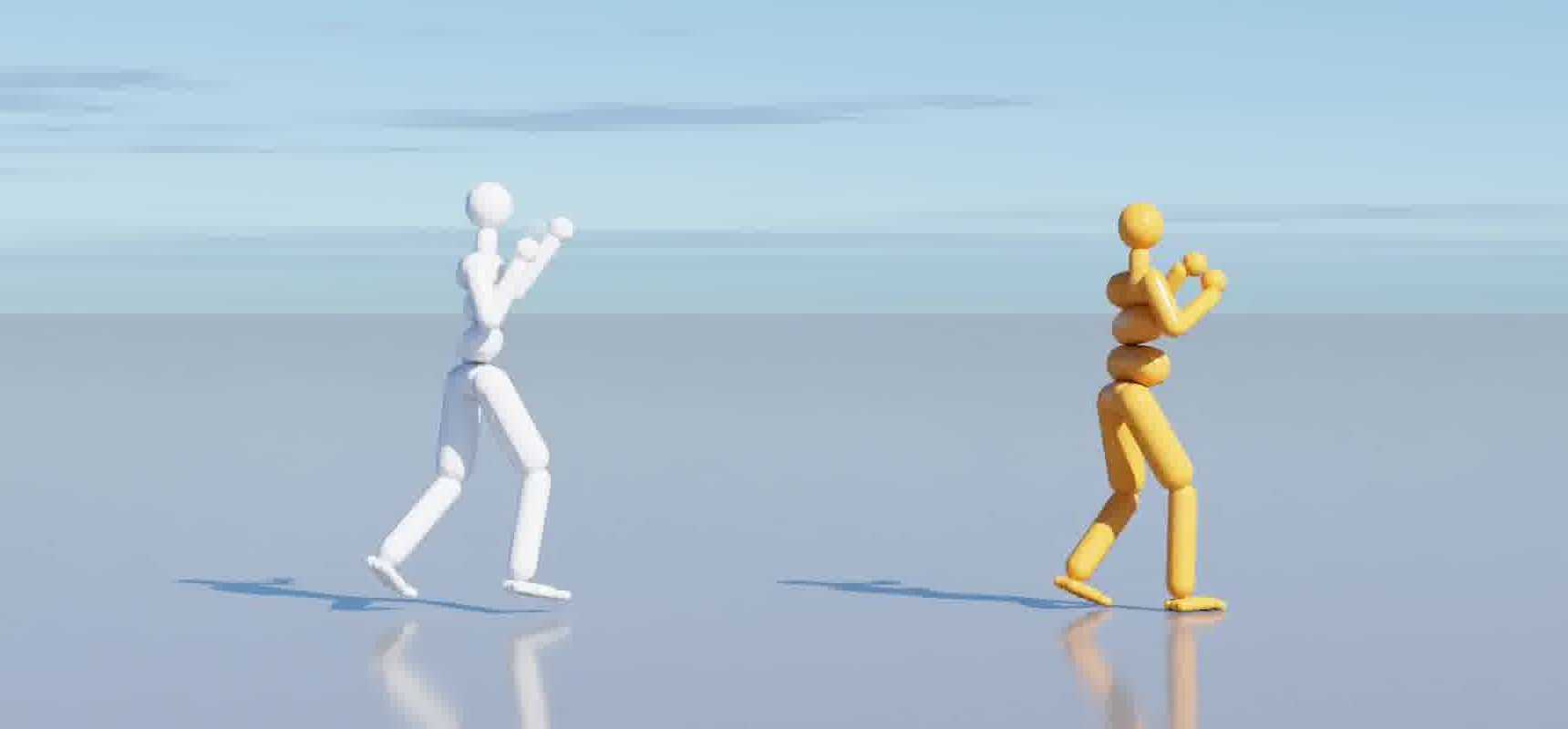}
    \end{subfigure}
    \begin{subfigure}{0.15\textwidth}
        \centering
        \includegraphics[width=\linewidth]{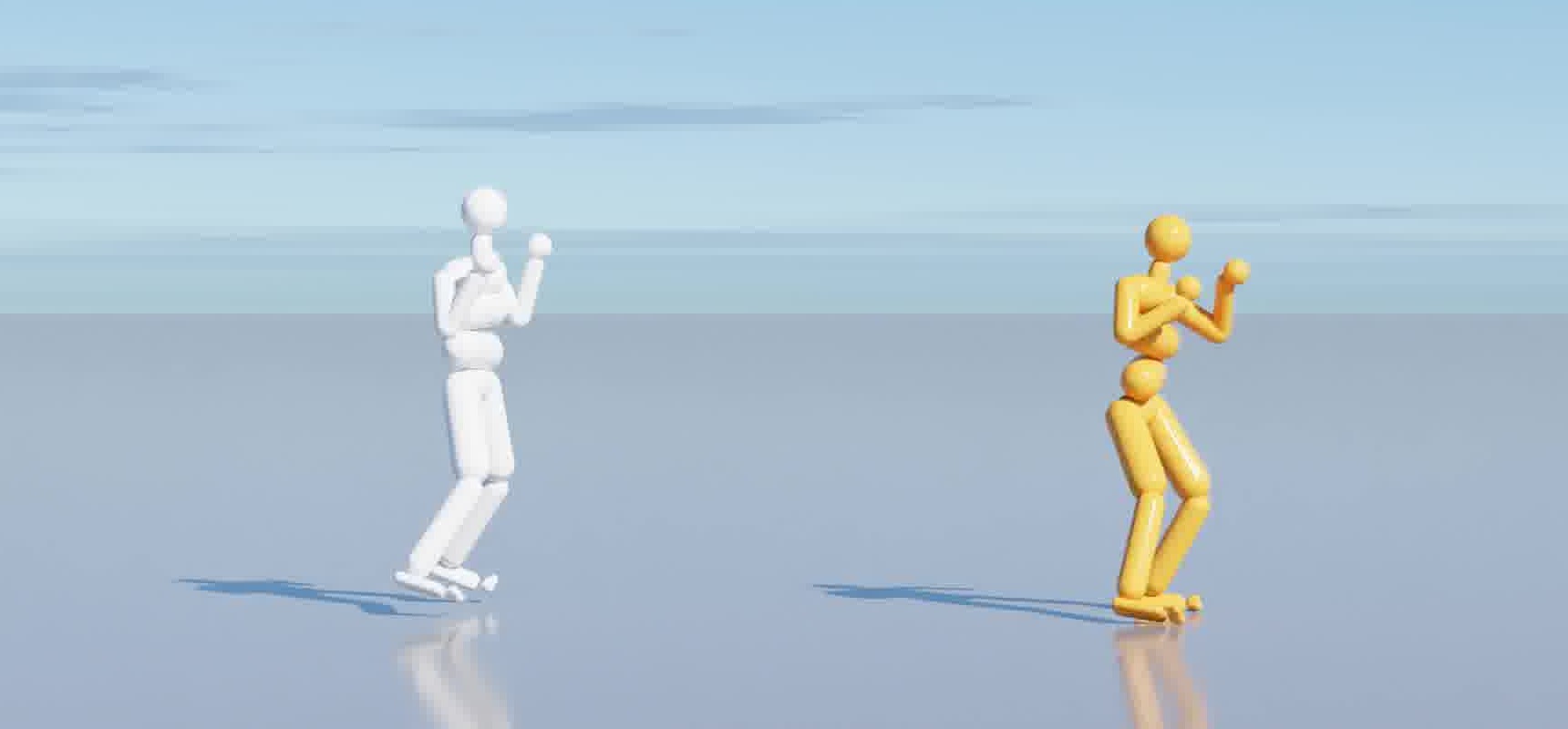}
    \end{subfigure}
    \begin{subfigure}{0.15\textwidth}
        \centering
        \includegraphics[width=\linewidth]{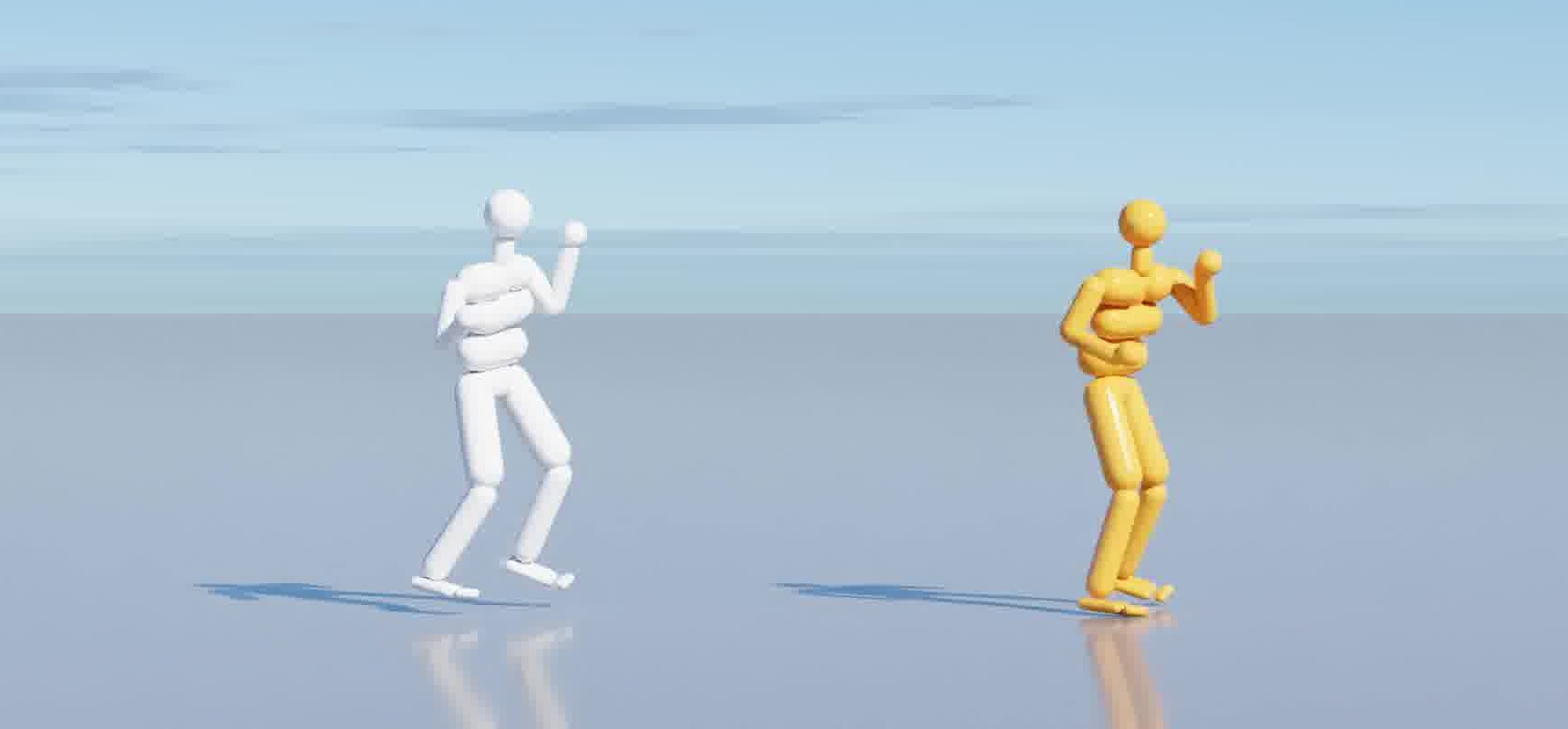}
    \end{subfigure}
    \begin{subfigure}{0.15\textwidth}
        \centering
        \includegraphics[width=\linewidth]{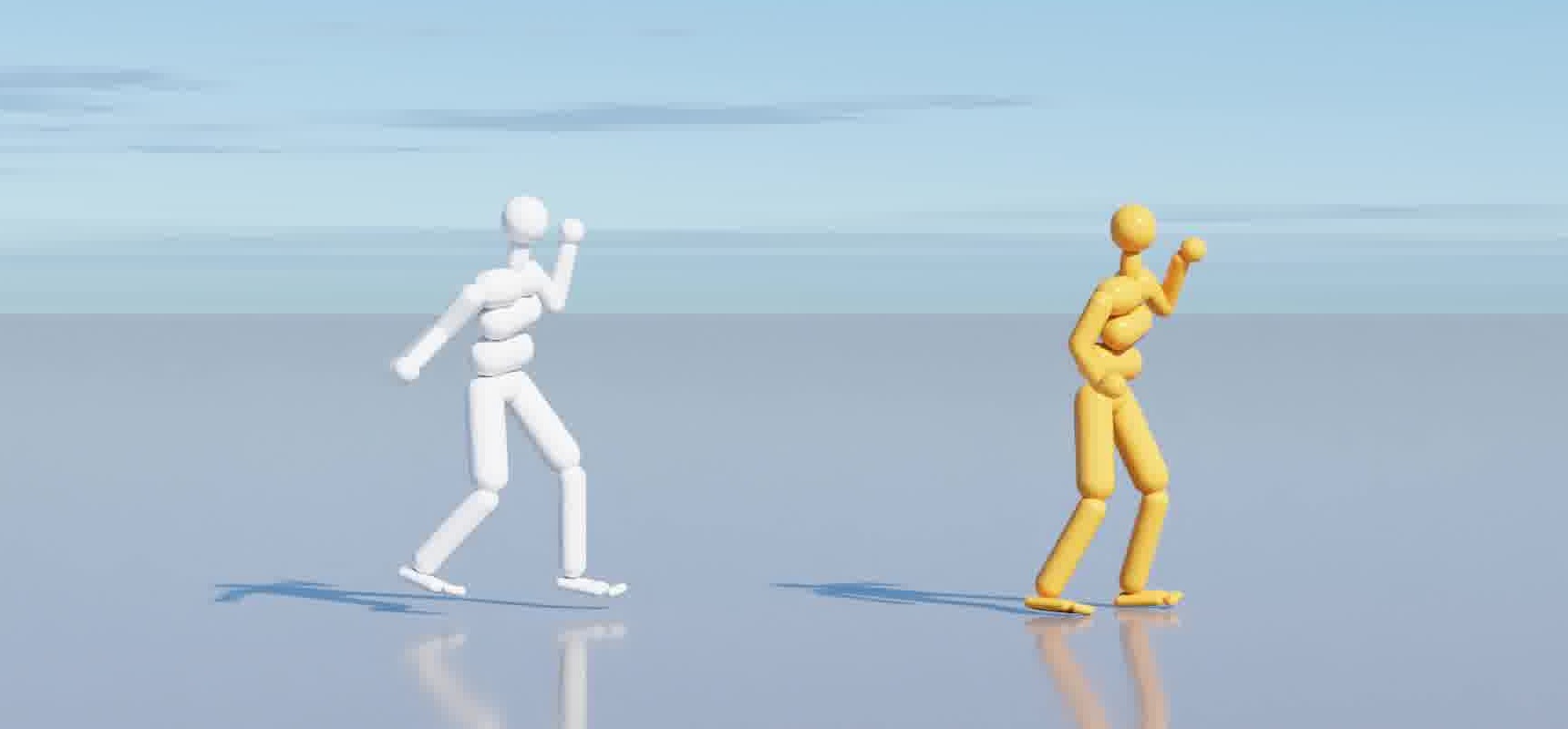}
    \end{subfigure}
    \begin{subfigure}{0.15\textwidth}
        \centering
        \includegraphics[width=\linewidth]{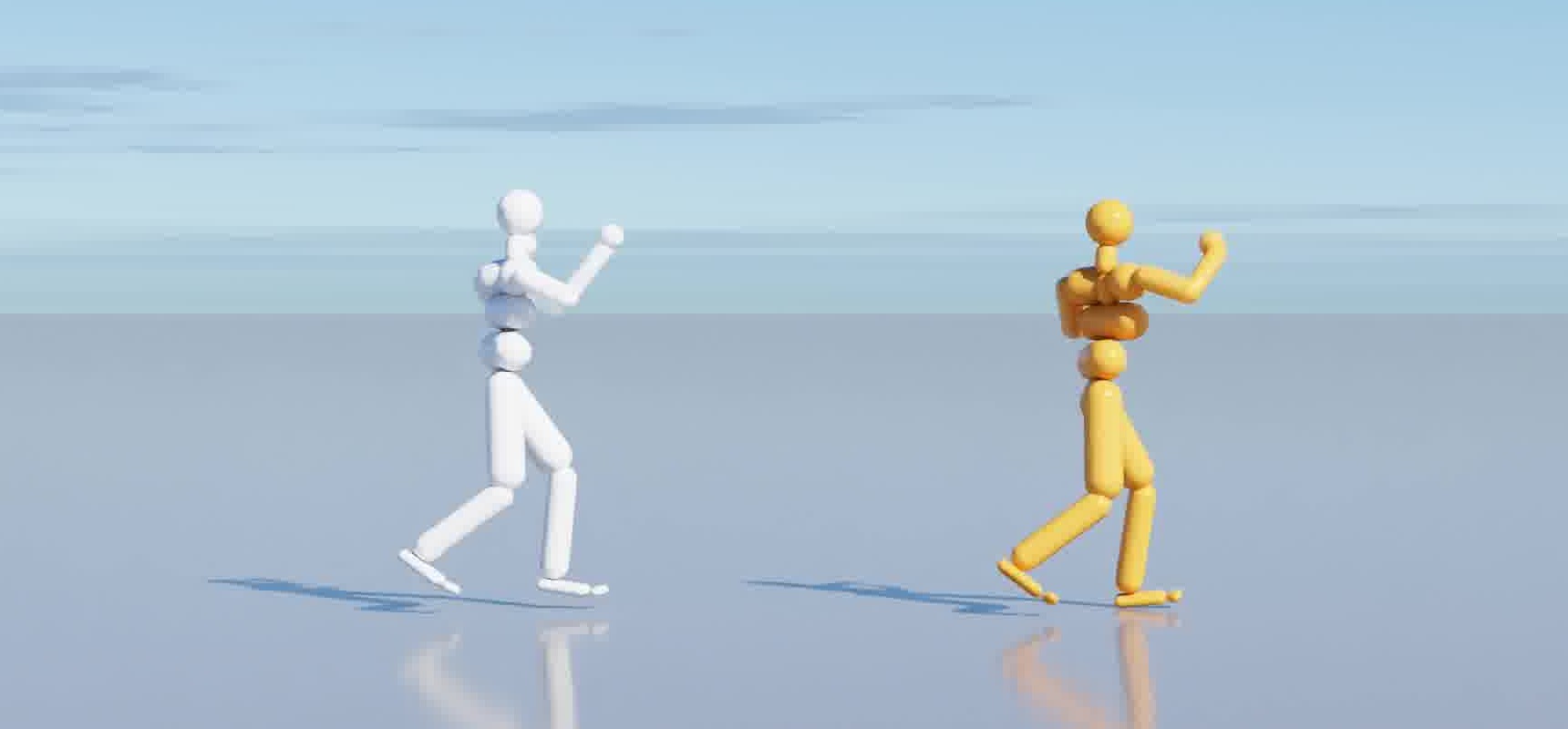}
    \end{subfigure}
    \begin{subfigure}{0.15\textwidth}
        \centering
        \includegraphics[width=\linewidth]{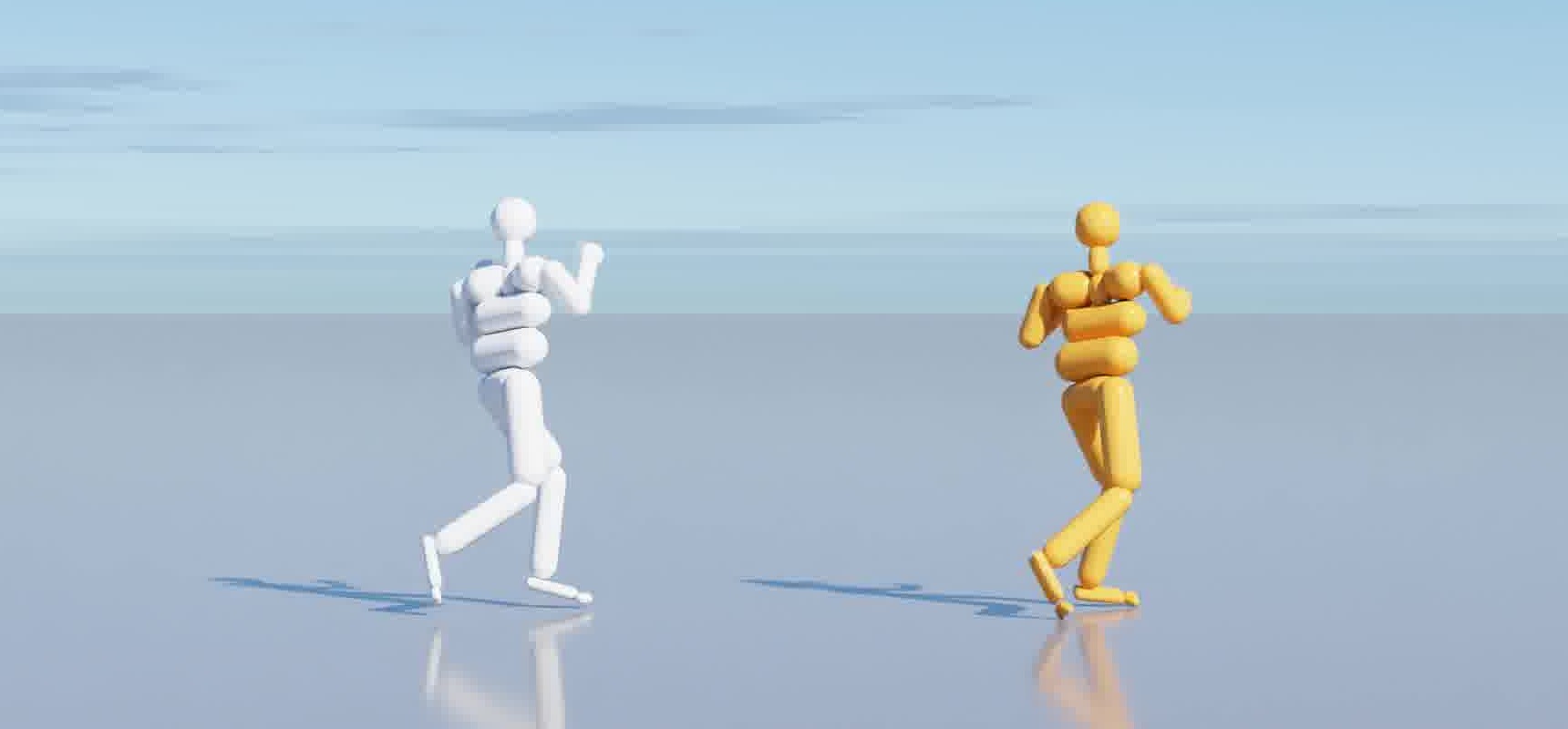}
    \end{subfigure}
    \caption{{Motion stitching scheduler performance on unseen motions. Our controller can react real-time to generate getting up, walking and boxing behaviors.}}
    \label{fig:motion_stitching_performance}
\end{figure*}

\section{Interactive Applications}\label{section:results_interactive}
In this section, we combine different high-level motion schedulers with our trained low-level executor (section~\ref{section:lowlevel}), showcasing a number of interactive applications. We emphasize that all of the presented applications are real-time interactive and do not require any additional training or fine tuning of our low-level executor, which can be used on-the-fly in all of these settings. Since snapshots cannot fully demonstrate motion, we refer readers to the demo video.

\subsection{Keyboard Driven Interactive Control}
Here we show the application using the high-level planner described in section~\ref{section:keyboard_driven}.
Figure~\ref{fig:keyboard_driven_results} shows the snapshots of our agents controlled by the keyboard command (inset), with our agent shown in yellow and PFNN reference motion shown in white on the left.
% \masha{Please clarify the sentences following. It seems to say our model causes sliding and floating?..}
Note that we do not use inverse kinematics to force the agent's feet to touch the ground in PFNN.
This increases the engineering efficiency of PFNN module,
and yet we show that the yellow agent generated by \ourmodel{} can still demonstrate realistic physics-based motions in a real-time fashion.
% at the cost of more obvious sliding and floating behavior.
We are almost able to perfectly follow the target states generated by PFNN.

Due to varying engines and humanoid models,
we cannot fully reproduce the original PFNN with more motion gaits and uneven terrain.
However, our universal framework's efficiency and simplicity indicates that the motion quality and variety generated by our algorithm is only bottlenecked by the quality of the high-level scheduler used. Our algorithm has the potential to adapt to different schedulers with varying designs and implementations,
such as the ones used in basketball and soccer video games.
% We also believe our model can be used on systems with specific applications,
% such as basketball and soccer video game motion systems. \kelly{not sure if this sentence is very informational}
\subsection{Interactive Motion Stitching}
In this section,
we discuss the results where we randomly select a motion from a motion dataset and our algorithm will respond to that real time.
In the cover image figure~\ref{fig:teaser},
the agent demonstrates the master of much more interactive complex skills compared to DeepMimic.
In the attached video,
\ourmodel{} also demonstrates strong \textbf{emergent physics-based transition},
where we show that, in contrast with some existing methods which design or interpolate realistic transitions,
our system can smoothen sharp transition,
where animation principles such as anticipation, ease-in \& ease-out,
are automatically satisfied with our low-level executor.
We also note that the transition skills are not recorded in the dataset (not learnt from motion data),
and they are mastered by {\ourmodel} by generalizing from other skills.

In figure~\ref{fig:motion_stitching_performance},
we further demonstrate the effectiveness and extreme transferability of our algorithm,
by forcing our agent to react to motions it has never seen before.
Our agent can still generate high-quality physics-based animation as shown in the figure~\ref{fig:motion_stitching_performance}.

It is worth mentioning that motion stitching can be viewed as the simplest motion graph system,
which completely ignores generating smooth transitions between motions from the scheduler.
However, the low-level executor is able to naturally generate smooth physics-based transitions on its own.
Since our algorithm can demonstrate realistic motions using the simplest motion graph system,
we believe it can also utilize better designed motion graph systems vastly available both in the research and engineering communities.
The transferability skills on unseen motions demonstrated by our model also suggest potential use for motion systems with large motion variety. 

\subsection{Interactive Video Controlled Animation}
In figure~\ref{fig:motion_video_drive},
we show how our algorithm can be used to teleport the motions captured from a remote host, 
to its physics-based avatars in the simulated environment real-time.
Note that different from~\cite{peng2018sfv},
our system is real-time and does not require a high-fidelity pose-estimator,
or hours of online re-training.
In figure~\ref{fig:motion_video_drive},
indoor behaviors such as walking, turning, waving and jumping can be animated efficiently.
Despite the mismatch in frame rate between the pose-estimator and our simulated environment,
and the visually obvious pose estimation errors,
our agent generates realistic real-time physics-based motions.

We expect the algorithm can be further improved,
and combined with the vast available online datasets generated from video websites such as YouTube.

\section{Conclusion and Discussion}
In this paper, we proposed a universal neural controller for a variety of real-time interactive control applications.
We closely study physics-based motion control on a large-scale dataset 
where novel techniques including constrained multi-objective reward optimization,
motion balancing, and variance control are essential for the success of our framework.
The controller we propose obtains much better robustness and generalization
compared with existing research,
where training and testing are generally performed on the same motion distribution.
Once trained, our method does not require further online retraining and can be applied on-the-fly
to various applications, such as real-time interactive control from keyboard,
videos and motion stitching.

We also identify limitations of our framework and potential research topics for the future.
One such topic is exploring how we can further improve the capacity of the neural controller,
so that it can master even more skills
without worrying about asymptotic performance drop for each motion.
% when compared to training only on that motion.
Most of the applications we demonstrate are research driven.
The high-level schedulers used still have areas for improvement compared with those used in high-quality video games today.
It remains to be explored how well our framework can be combined with an AAA video-game motion system.
We also did not explore the ability of our method on varying terrain types and physical interactions with objects in the scene. All of these topics present an exciting avenue for future work. 

\iffalse
Nevertheless, we hope that this project will inspire future research and push the limit of physics-based animation even further.
From the algorithm's perspective,
the recent advancements in network capacity, model-based planning and neural inductive bias,
such as graph neural networks,
can be regarded as promising guidelines for improvement of neural animation controllers.
From the applications' perspective,
the universal framework proposed in this paper and the metrics used throughout this paper such as testing performance and zero-shot robustness,
can be applied to different application scenarios.
We are interested in exploring the limits of our controller from both perspectives.
\fi

\bibliography{ref}\bibliographystyle{ACM-Reference-Format}
\end{document}